\RequirePackage{fix-cm} 
\documentclass[twoside,12pt]{report}

\usepackage[american]{babel}

\usepackage{amsmath,amsthm,amssymb} 

\usepackage{slashed,bm,siunitx,mathtools} 
\usepackage{cite, graphicx,subcaption,longtable,float} 
\usepackage[export]{adjustbox}
\graphicspath{ {images/} }

\usepackage[tableposition=below,font=small]{caption} 
\usepackage{acronym} 

\usepackage{indentfirst} 

\usepackage{setspace} 
\usepackage{pdfpages} 

\usepackage{chngpage} 
\usepackage{makecell}

\usepackage[colorlinks = true,
            linkcolor = blue,
            urlcolor  = blue,
            citecolor = blue,
            anchorcolor = blue]{hyperref} 
\usepackage[top=3cm, bottom=2.2cm, left=3.31cm, right=2.65cm]{geometry} 
\usepackage[T1]{fontenc}
\usepackage[utf8]{inputenc}
\usepackage{newtxtext,newtxmath}

\usepackage{titlesec} 
\titleformat{\section}
  {\normalfont\Large\bfseries}{\thesection}{1em}{}

\usepackage{fancyhdr}

\titleformat{\chapter}[display]
{\normalfont}
{\fontsize{75}{10}\selectfont \filleft ~\thechapter}
{0.7cm}
{\fontsize{29}{25}\selectfont \it \bfseries \filleft}
\titlespacing*{\chapter}{0cm}{2.2cm}{1.5cm}

\usepackage[version=4]{mhchem} 

\usepackage{xcolor} 
\definecolor{MBcomment}{rgb}{0.35,0.7,0}

\newcommand{\ket}[1]{\left| \, #1 \, \right\rangle}

\usepackage{ifthen} 

\newcommand{\tr}[2][]{%
  \ifthenelse{\equal{#1}{}}{%
\mathrm{tr} \hspace*{-0.06em}\left[ #2 \right]
}{%
\mathrm{tr} \hspace*{-0.06em}[\!\begin{multlined}[t]
#1 {} \\
{} #2] 
\end{multlined}
}%
}

\newcommand{\vectorB}[1]{\mathbf{#1}}

\DeclareMathOperator{\Ree}{Re}
\DeclareMathOperator{\Imm}{Im}

\newcommand{\refEq}[1]{~(\ref{#1})}

\definecolor{stoppedhere}{rgb}{1,0,0}
\newcommand{\stoppedhere}{\textcolor{stoppedhere}{\noindent
\kern-0.63cm \%\%\%\%\%\%\%\%\%\%\%\%\%\%\% \newline
\noindent \%\%\%\%\%\%\%\%\%\%\%\%\%\%\%\% \newline
\noindent \%\%\%\%\%\%\%\%\%\%\%\%\%\%\%\%\% \newline
\noindent \%\%\%\%\%\%\%\%\%\%\%\%\%\%\%\%\%\% \newline
\noindent \%\%\%\%\%\%\%\%\%\%\%\%\%\%\%\%\% \newline
\noindent \%\%\%\%\%\%\%\%\%\%\%\%\%\%\%\% \newline
\noindent \%\%\%\%\%\%\%\%\%\%\%\%\%\%\% \newline
}%
}

\usepackage{dsfont} 

\usepackage{array,booktabs} 

\usepackage{enumitem} 

\usepackage{bbold}

\newenvironment{blockquote}[1][0cm]{%
  \par%
  \medskip
  \addtolength{\leftskip}{#1}%
  \noindent\ignorespaces}{%
  \par\medskip}

\begin{document}

\numberwithin{equation}{chapter}

\includepdf[pages=-,noautoscale = true,scale=1.02]{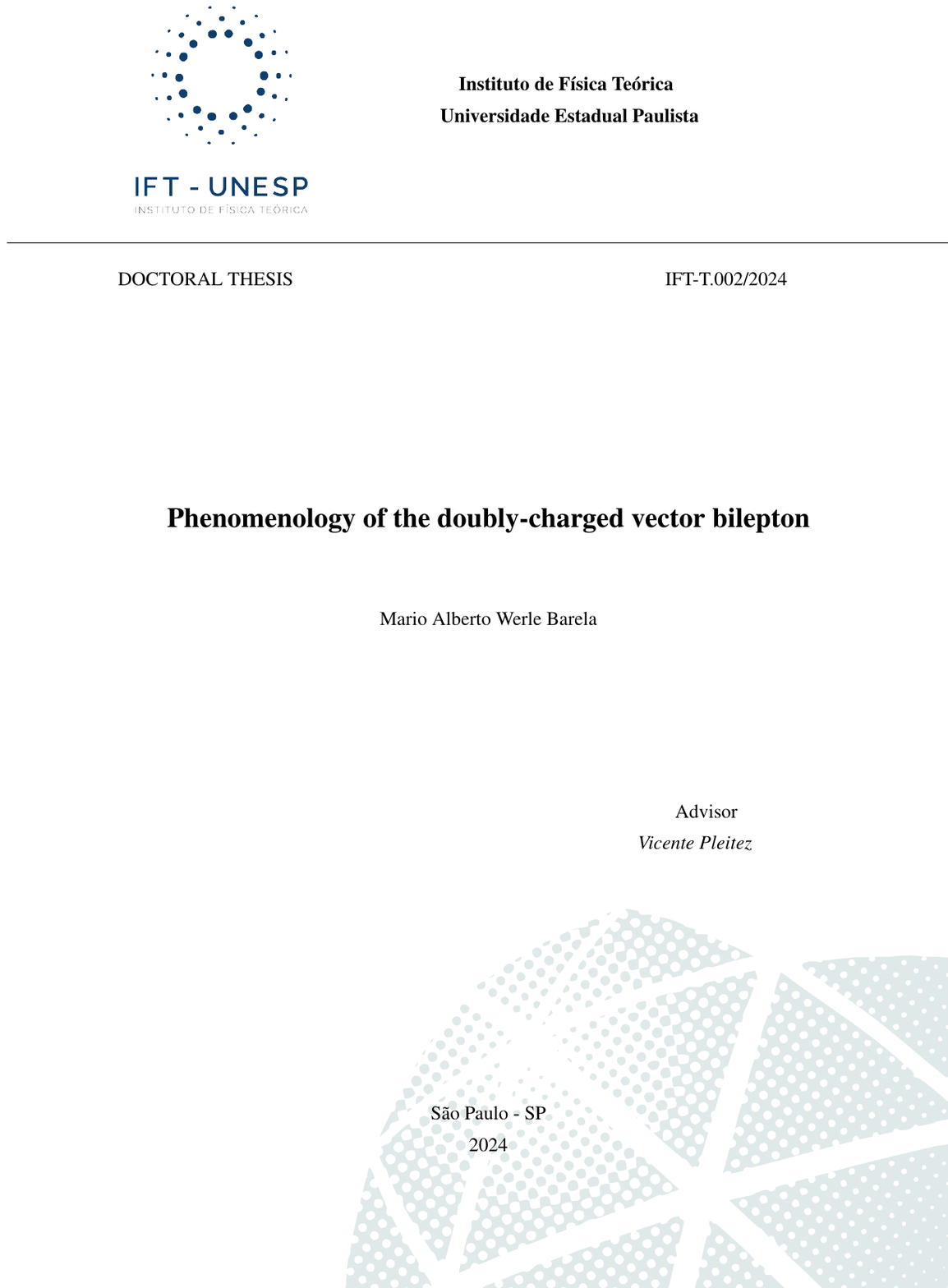}

\includepdf[pages=-]{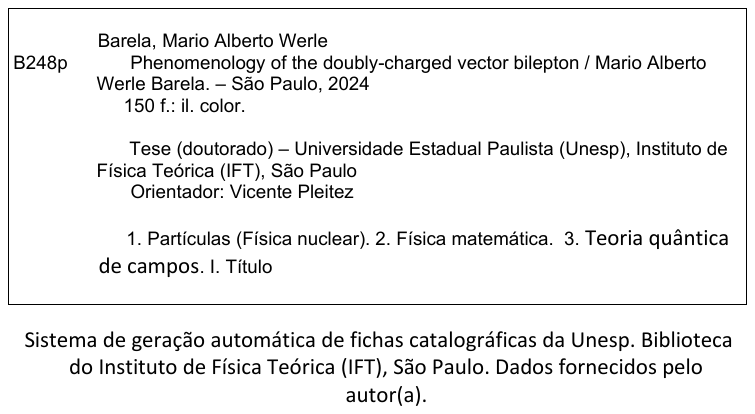}

\includepdf[pages=-]{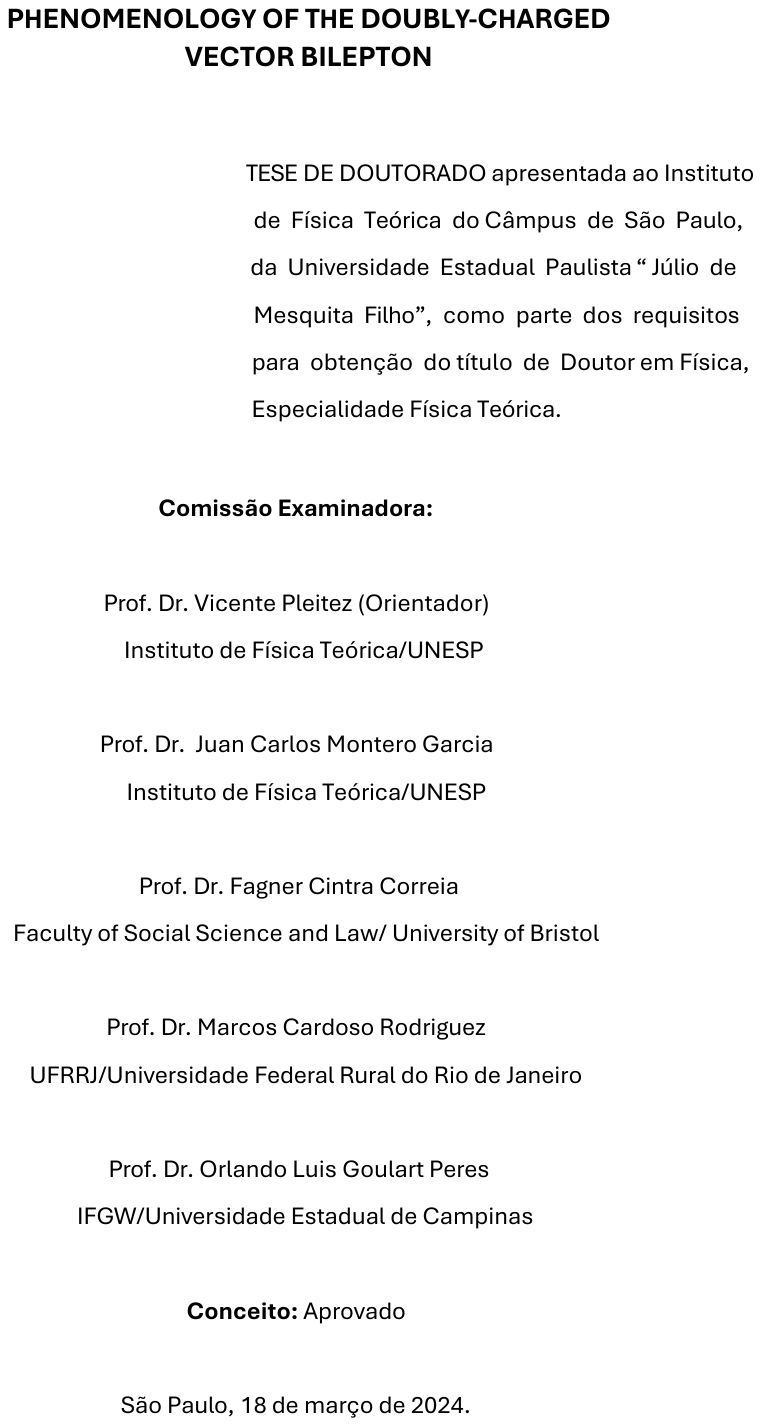}

\pagebreak

\pagenumbering{roman}

\newpage
\newpage

{
\makeatletter
\@twosidefalse
\makeatother

\vspace*{19cm}

\begin{blockquote}[9cm]
{\fontsize{14pt}{6.8mm}\selectfont

{\it A Mario, Joana, Alberto e Ogeni,}

{ \it a quem devo quase tudo, }

{\it obrigado. } 

}
\end{blockquote}

\pagebreak

\newpage
\newgeometry{top=2cm, bottom=2.2cm, left=3.31cm, right=2.65cm}
\makeatletter
\@twosidefalse
\makeatother

\begin{center}
{\LARGE \bf Resumo}
\end{center}
\vskip 0.7cm
O Modelo Padrão (MP) foi a primeira teoria completa (excluída a gravidade), fundamental da natureza a ser estabelecida. A conquista dessa posição se deve às previsões revolucionárias, eventualmente confirmadas, surgidas do modelo, além de sua extraordinária capacidade de acomodar dados de precisão e de inédita complexidade. Por outro lado, o MP é incapaz de explicar algumas observações universais -- além de apresentar características incômodas. Em decorrência disso, nova física é necessária, o que, no reino de altas energias, com poucas exceções, implica em novas partículas. Nesta tese exploraremos uma tal espécie exótica: o bóson vetorial (ou, simplesmente, vetor) bileptônico duplamente carregado. Em especial, focamos em seus efeitos como mediador de Violação de Sabor Leptônico Carregado, o que, por sua vez, pode representar uma `{\it smoking gun}' para sua descoberta ou para a limitação de seu espaço de parâmetros. Esse tipo de processo ainda não foi ceticamente explorado, e a mistura leptônica em sua interação com o $U^{\pm \pm}$ não foi alvo de consideração focada. 

Nosso primeiro passo é investigar que limites um processo trimuônico, no LHC, pode impor no espaço de parâmetros dessa partícula. Os resultados nos motivaram a buscar por dados de origens alternativas que pudessem, também, gerar limites úteis. Os sete canais de decaimento leptônico de três corpos se mostraram ótimos candidatos, e efetuamos uma análise detalhada e independente de modelo do que eles podem implicar sobre o $U^{\pm \pm}$ e outras duas espécies exóticas. Atenção especial é dada às interferências relevantes entre as novas contribuições. 

Por fim, seguimos para construir uma análise profunda da evolução dos acoplamentos do Modelo 3-3-1, principal teoria a conter um vetor bileptônico, inspecionando de que modo ela é ameaçada por polos de Landau no seu acoplamento abeliano. Encontramos explicitamente as contribuições de 1-loop das partículas exóticas aos runnings, obtendo que o regime perturbativo do modelo é, de fato, consideravelmente maior do que usualmente suposto.

\vspace*{3.5mm}

Durate a produção desta tese, o autor contribui para os seguintes trabalhos originais: 

\begin{itemize}
\item M.W. Barela and V. Pleitez, {\it Trimuon production at the LHC}, \href{https://doi.org/10.1103/PhysRevD.101.015024}{Phys. Rev. D 101(2020) 015024}.

\item M.W. Barela and J. Montaño Domínguez, {\it Constraints on exotic particle masses
from flavor violating charged lepton decays and the role of interference}, \href{https://doi.org/10.1103/PhysRevD.106.055013}{Phys.Rev. D 106 (2022) 055013}.

\item M.W. Barela, {\it On the 3-3-1 Landau pole}, \href{https://doi.org/10.1016/j.nuclphysb.2024.116475}{Nuclear Physics, Section B, 116475 (2024)}.

\item M.W. Barela and R. Capdevilla, {\it Di-Higgs Signatures in Neutral Naturalness}, \href{https://doi.org/10.1007/JHEP02(2024)050}{J. High Energ. Phys. 2024, 50 (2024)}.
\end{itemize} 

\noindent Este texto é baseado, principalmente, nos três primeiros.

\vskip 0.7cm

\noindent
{\bf Palavras-chaves}: Física de Partículas; Extensões do Modelo Padrão; Vetor Bileptônico Duplamente Carregado; Modelo 3-3-1; Violação de Sabor Leptônico Carregado.

\noindent
{\bf Áreas do conhecimento}: Física de partículas; Física de altas energias; Fenomenologia do LHC.
\vskip 0.5cm

\pagebreak

\begin{center}
{\LARGE \bf Abstract}
\end{center}
\vskip 0.7cm
The Standard Model stands as the first established, complete (gravity aside), fundamental theory of nature. This position was merited by the confirmation of many revolutionary predictions the model implied, besides its ability to fit high complexity precision data to an extreme degree. In the other hand, the Standard Model is unable to account for some universal observations -- besides possessing a number of undesirable characteristics.  Thus, there is a pressing need for new physics, which, in the high energy realm, usually imply new particles. In this thesis, we delve into one such particle: the doubly-charged vector bilepton. This particle is a singular feature of Beyond the Standard Model theories and its phenomenology is still incipient. We focus, in particular, on its power as a mediator of Charged Lepton Flavour Mediation, a type of process which, in turn, may represent a smoking gun with respect to its discovery or constraining of its parameters. Such processes have not been skeptically explored, and the lepton mixing in their interaction with the $U^{\pm \pm}$ not been properly focused on. 

The first step we take is investigating what bounds an LHC trimuon process may impose on the parameter space of this particle. The results motivate us to look for alternative experimental sources of limits. The seven 3-body lepton decay channels delivered a great prospect, and we perform a detailed, model independent analysis of what they can imply for the $U^{\pm \pm}$ and other two exotic species. Special attention is paid to occurring interferences between new contributions. 

Finally, we proceed to construct a thorough analysis of the evolution of couplings in the Minimal 3-3-1 Model, the main theory incorporating a vector bilepton, examining how it is threatened by Landau poles in its abelian coupling. We explicitly obtain the 1-loop contributions of exotic particles to the runnings, finding that the perturbative regime of the model is in fact considerably larger than conventionally assumed.

\vspace*{3.5mm}

During the production of this thesis, the author have contributed to the following original works: 

\begin{itemize}
\item M.W. Barela and V. Pleitez, {\it Trimuon production at the LHC}, \href{https://doi.org/10.1103/PhysRevD.101.015024}{Phys. Rev. D 101(2020) 015024}.

\item M.W. Barela and J. Montaño Domínguez, {\it Constraints on exotic particle masses
from flavor violating charged lepton decays and the role of interference}, \href{https://doi.org/10.1103/PhysRevD.106.055013}{Phys.Rev. D 106 (2022) 055013}.

\item M.W. Barela, {\it On the 3-3-1 Landau pole}, \href{https://doi.org/10.1016/j.nuclphysb.2024.116475}{Nuclear Physics, Section B, 116475 (2024)}.

\item M.W. Barela and R. Capdevilla, {\it Di-Higgs Signatures in Neutral Naturalness}, \href{https://doi.org/10.1007/JHEP02(2024)050}{J. High Energ. Phys. 2024, 50 (2024)}.
\end{itemize} 

\noindent This text is mainly based on the first three.

\vskip 1.0cm

\noindent
{\bf Key-words}: Particle Physics; Beyond the Standard Model; Doubly-charged Vector Bilepton; 3-3-1 Model; Charged Lepton Flavour Violation.

\noindent
{\bf Branches of knowledge}: Particle physics; High energy physics; LHC phenomenology.

\restoregeometry

\makeatletter
\@twosidefalse
\makeatother
\pagebreak
\pagebreak


\begin{center}
{\LARGE \bf Agradecimentos}
\end{center}
\vskip 2.0cm

\begin{blockquote}

{
\fontsize{12.4pt}{16.6pt}\selectfont

\noindent À minha família, por tudo. {\it Amo vocês}. 

\vspace*{0.5cm}

\noindent Ao meu orientador, Vicente Pleitez, pelo conhecimento, gentileza e compreensão.

\vspace*{0.5cm}

\noindent Ao meu colaborador, Rodolfo Capdevilla, um co-orientador espontâneo, por tudo que me ensinou.

\vspace*{0.5cm}

\noindent Aos meus amigos -- {\it Gui, Junior, Polar, Vitor, Gadeia, Lucas} e {\it João Paulo} --, que, a maioria desde a infância, integram o conjunto (de baixíssima cardinalidade) das coisas que importam.

\vspace*{0.5cm}

\noindent À Emanuele, pelo amor, pelo presente e pelo futuro.

\vspace*{0.5cm}

\noindent E, principalmente, a Deus, pelas graças não merecidas e, em especial, por não ter desistido deste filho.

\vspace*{0.5cm}

\noindent E ao CNPq, pelo suporte financeiro.

}

\end{blockquote}

\onehalfspace

\pagebreak

\begin{center}
{\LARGE \bf Notations and conventions}
\end{center}
\vskip 1cm

\begin{itemize}

\item Throughout this thesis, repeated Lorentz indices are always summed, as usual:

\begin{equation}
a_\mu b^\mu \coloneqq \sum_{\mu=1}^4 a_\mu b^\mu. \nonumber
\end{equation}
Non-Lorentz indices are sometimes implicitly summed, which should be clear from the context. The choice between matrix and component notation is also made at convenience and in an obvious manner. 

\item We use a mostly minus signature for the metric:

\begin{equation}
\eta_{\mu \nu} = \mathrm{diag}\left(+1,-1,-1,-1 \right). \nonumber
\end{equation}

\item The gauge covariant derivative is written with a minus sign:

\begin{equation}
D_\mu = \partial_\mu - i g A_\mu^a T_a. \nonumber
\end{equation}
This is consistent with the following definitions for the field strength tensor,

\begin{equation}
F_{\mu \nu} \equiv \frac{i}{g}\left[D_\mu, D_\nu \right] = \partial_\mu A_\nu^a - \partial_\nu A_\mu^a + gf^{abc}A_\mu^b A_\nu^c, \nonumber
\end{equation}
and the transformation rules,

\begin{equation}
\begin{split}
\Psi &\to U \Psi, \\
A_\mu &\to UA_\mu U^{-1} - \frac{i}{g}(\partial_\mu U) U^{-1}. \nonumber
\end{split}
\end{equation}%
for $\Psi$ in the fundamental and $A_\mu$ in the adjoint representation.

\item As for the Standard Model, we normalize the weak hypercharge without the traditional 1/2 factor, {\it i.e.}, the right-handed charged lepton singlets have $Y=-1$. Moreover, the vacuum expectation value of the Higgs doublet, sometimes denoted $v_W$ and other times $v_H$, is given by 

\begin{equation}
v_H = \frac{|\mu|}{\sqrt{\lambda}} \approx \SI{246}{GeV}, \nonumber
\end{equation}
without the factor of $1/\sqrt{2}$ that is occasionally implemented.

\item Finally (and of little importance), we are idiosyncratic with regard to relation signs, using the ones below in the corresponding situations:

\begin{itemize}
\item $\coloneqq$ \hspace*{0.7mm} is used when a defined property is being explained;

\item $\equiv$ \hspace*{0.7mm} is used to define a symbol;

\item $\approx$ \hspace*{0.7mm} is used to denote a numerical approximation;

\item $\simeq$ \hspace*{0.7mm} symbolizes an analytical approximation;

\item $\sim$ \hspace*{0.7mm} denotes a loose similarity, be it depicting a functional dependence or a numerical proximity or order of magnitude;

\item $\propto$ \hspace*{0.7mm} signifies a proportionality.
\end{itemize}

\end{itemize}

\pagebreak

{
\hypersetup{hidelinks}

\tableofcontents

\pagebreak
\listoffigures
}

{
\makeatletter
\@twosidetrue
\makeatother

\pagebreak

\pagenumbering{arabic}

{
\fontsize{12.2pt}{15.1pt}\selectfont

\newpage

\pagestyle{fancy}
\fancyfoot{}
\fancyhead[R]{\thepage}
\fancyhead[L]{{\it \leftmark}}
\fancypagestyle{plain}{ %
  \fancyhf{} 
  \renewcommand{\headrulewidth}{0pt} 
  \renewcommand{\footrulewidth}{0pt}
}


\pagebreak

\chapter{Introduction} \label{Chapter1}

It could be argued that the most impressive success of physics throughout history was the discovery of the Higgs boson, in 2012, independently confirmed by two teams, at the Large Hadron Collider (LHC). The reason this feat is so astounding is that it confirms a prediction made forty years earlier, based purely on the -- relatively new at the time -- mechanism of gauge invariant renormalizability. This observation was the last experimental piece expected in order to stabilize the Standard Model of Elementary Particles and Fundamental Interactions (SM) as the accepted established general theory of particle physics, although understood to be flawed as a complete fundamental description of reality. 

However, the arising low TeV scale has proven to be cryptic, and the period that followed 2012 is marked, in collider and general experimental particle physics, by a desert in which neither paradigm changing nor much guiding evidence of new physics have been observed. Some of the most interesting facts stemming from data collected is this period are the ratification of the SM predictions for some Yukawa couplings~\cite{CMS:2018uag,ATLAS:2019nkf}; and the recently obtained $W$-boson mass~\cite{CDF:2022hxs}, in tension with the SM expectation. With insufficient data-driven insight to guide new constructions, the theoretical struggle to find solutions to the various SM issues is made difficult.

This thesis is another attempt at investigating and restricting the Beyond the Standard Model (BSM) theory space while this research context is not altered by new discoveries, trying to improve our understanding and intuition regarding the alternative models, hopefully selecting their best aspects and constraining their parameter space. The main object of our efforts is the doubly-charged vector bilepton, perhaps the rarest exotic spin-1 particle to exist within an ultraviolet complete theory with a non-simple Gauge Group. 

The Minimal 3-3-1 Model (m331) is the only theory of this type that features such a bilepton $U^{\pm \pm}$, and predicts that it couples to charged leptons through a unitary mixing matrix $V_U$. To our knowledge, the effects of this matrix had not received focused consideration yet. The first tale we shall tell is motivated by this, and corresponds to a study of the trimuon $pp \to e3\mu$ LHC process. This reaction is free of irreducible background and allows for an exploration of lepton mixing through $U^{\pm\pm}$ mediation. Our analysis is model independent to some extent but evokes the m331 when a benchmark is necessary, and our results comprise exclusion contours on a bi-dimensional parameter space formed by the bilepton mass and one of the free angles that parametrize $V_U$. Our findings demonstrate that the trimuon process provides bounds in agreement with the existing literature, but only in a small sector of the observed parametric region, in the rest of which bileptons of very small masses remain possible by this process. 

Once Charged Lepton Flavour Violation (CLFV) proved to be a possible smoking gun for the vector bilepton, we were prompted to look for alternative such channels to which the $U^{\pm \pm}$ can, in theory, contribute. In this spirit, our second tale corresponds to a model independent study of the purely leptonic CLFV 3-body decays. Among the several benefits of this set of channels as a phenomenological guide is the fact that their branching ratios are simple well known quantities and have sensitivity expected to increase. More importantly, the exploration of such processes avoid the worries with statistical treatment, hadronic and collider physics. We investigate the parameter space of every exotic degree of freedom that can, in principle, contribute to these decay channels: a doubly-charged vector bilepton, a doubly-charged scalar boson, and an exotic neutral scalar. Besides seeking to uncover the strongest conservative bounds on the contributing particles that this data is able to originate, we contemplate the following question:

\vspace*{2.8mm}

\begin{adjustwidth}{1.1cm}{1.1cm}
\textit{Exclusion contours on BSM parameters are usually derived from a minimal model that isolates some exotic species. Would the resulting allowed regions be relevantly different if a second new particle, subdominant within the process at hand, was included to interfere with the primary contribution?}
\end{adjustwidth} 
 
\vspace*{3mm}
 
The last research avenue we shall thread is model specific. As already defined, the main protagonist of our work, the $U^{\pm \pm}$, is a rare feature of BSM theories contained most notably in the m331. We thus proceed to deliver a focused investigation into this theory through a Renormalization Group analysis of its perturbativity regime. The standing result is that the model, without additional mechanisms, breaks down at scales of a few TeV because of a Landau pole in its abelian gauge coupling. We carefully check this claim, verifying its premises and implications. The position of the Landau pole in $g_X$ is rederived through an explicit, brute force approach, and is shown to appear at higher scales than originally thought.

This thesis is organized as follows:  In Chapter~\ref{Chapter2} we start with a historic account of the construction of the Standard Model and general particle physics, and follow to a necessary review of the SM; Chapter~\ref{Chapter3} is entirely devoted to the in-depth description of the most prominent flaws of the SM, which justify the necessity of BSM physics; In Chapter~\ref{Chapter4}, a complete exposition of the m331 is undertaken; In Chapter~\ref{chap:LHCTrimuons}, the study of the LHC trimuon process as a source of bounds for the $U^{\pm \pm}$ is described; In Chapter~\ref{Chapter6}, we describe the theory behind the 3-body lepton decay phenomenology that this thesis develops, explaining the definitions, choices, calculational and computational methods; In Chapter~\ref{Chapter7} the results of the study are presented; Chapter~\ref{Chapter8} corresponds to our last front of investigation, in which the structure of the Renormalization Group of the gauge couplings of the m331 is investigated; And our conclusions are presented in Chapter~\ref{Chapter9}.

\cleardoublepage


\pagebreak

\chapter{The Standard Model} \label{Chapter2}

The current picture of our understanding of the universe may be, at the most fundamental level, be subsumed to the paradigms of quantum field theory (QFT) and general relativity. With the unification of these frameworks within a single theory seeming, still, a distant dream, we are able to apply QFT in order to generate predictions regarding phenomena with characteristic energies much smaller than the Planck scale $m_p \sim \SI{10^{19}}{GeV} $. For this, however, an additional model is required -- which is now understood to be usually composed by: A symmetry structure, including the properties of the vacuum; A representation of that symmetry, which is commonly a sum of various irreducible representations and corresponds to the particle content; And an additional, {\it a priori} arbitrary (except for its symmetry), potential functional. This chapter is devoted to reviewing the characteristics of the Standard Model of Particle Physics, the ruling theory of Elementary Particles and Fundamental Interactions, starting with an attempt of recollection of the enlightening process of its foundation. We make an effort to catalogue the original references, hoping that this Section could provide a useful consultation resource. This Chapter then follows to a light-speed review of the model itself.

\section{History and development}

By 1925\footnote{Take the dates claimed within this section with a grain of salt, as distinct sources differ by one or even two years regarding some events.}, the reality of the highly non-intuitive nature of physics in the quantum scales was unavoidable, and the understanding of the nature of quanta itself had already matured through several phenomenological observations. In order to produce a retrospective notion, we recall some notable ones, such as the double-slit experiment of Thomas Young (1801); The advancements in the spectrometry of hydrogen by Balmer (1885) and Rydberg (1888); The observation of the Zeeman effect (1896); The explanation of the Larmor Precession (1897); The progress on black-body radiation, through the suggestion by Max Planck that the emitted energy was quantized (1900); The photon hypothesis, correctly proposed by Albert Einstein in an effort to describe the photoelectric emission spectrum, together with his theory of special relativity, both in 1905; The Stark effect (1913); The Stern-Gerlach experiment (1922), which resulted in the proposition of the quantization of microscopic angular momentum; The extension of the wave-particle duality to particles, by Louis de Broglie (1923); And the Pauli exclusion principle (1924).\footnote{These are just a few unequivocally crucial developments within the history of quantum physics -- there are, of course, dozens of additional key historic events which could be cited.} 

Simultaneously, there was a somewhat disjoint `basic knowledge of the structure of matter'. This corresponded to the understanding of a minimal set of ingredients that would evolve to a theoretical substrate and eventually become the standard particle theory. The most basic component of this model was the electron, whose existence was recognized since 1897, when J. J. Thompson determined the composition of cathode rays. The development of this matter continued and, in 1911, Ernest Rutherford, Ernest Marsden and Hans Geiger established that the atom, knew to be neutral, contained a positively charged core, the nucleus, through their famous experiment which scattered alpha particles (Helium nuclei) through a thin gold plate. In 1919, the study of the interaction of the same type of particles with the nitrogen in air led Rutherford to find that the occurring reaction responsible for the scintillation was \ce{^{14}_{7}N + ^{4}_{2}$\alpha$ -> ^{17}_{8}O + ^{1}_{1}p}, thus discovering the positively charged subatomic particles within nuclei, the protons.

It was in this scientific context that, in 1925, Werner Heisenberg, Max Born and Pascual Jordan wrote the first specific account of matrix quantum mechanics, after which the development of the field quickly accelerated. In 1926 Erwin Schrodinger postulated his wave equation, which would become the central quantitative mathematical postulate of non-relativistic quantum mechanics. In 1927, Werner Heisenberg formulated the uncertainty principle and, together with Niels Bohr, proposed the Copenhagen interpretation of wavefunctions, the standard perspective of quantum physics. Finally, in 1927, motivated by an attempt to purge the negative probability solutions to the Klein-Gordon equation, Paul A. M. Dirac proposed a first order differential equation which was simultaneously compatible with the established quantum theory and special relativity. The revolutionary quality of Dirac equation cannot be overstated: not only it formally accounted for the previously \textit{ad hoc} introduced spin, as it led to the prediction of antimatter, one of the greatest achievements in the history of physics. 
 
By that point, the ideas of quantum mechanics (QM) had already began to be applied to systems with an infinite number of degrees of freedom, as the quantization of fields (or second quantization) seemed to naturally solve issues such as causality or multiparticle states. This was a crucial development and perhaps the effective start of the model building efforts that would culminate in the SM. After the neutrino was hypothesized by Pauli, in 1930, in order to explain the continuous spectrum of beta decays, in 1933 Enrico Fermi invented his famous weak force theory to explain the same phenomenon. In 1935, Hideki Yukawa analysed the force binding nuclei together and, from typical atomic nuclei radius, predicted it to be carried by intermediate `\textit{heavy quanta}' mesons (in analogy to the `\textit{light quanta}' which carried electromagnetic forces), with a mass of $\sim \hspace*{-0.28em}\SI{100}{MeV}$. In the 1930s, the menu of subatomic particles empirically increased, with the observation of the neutron, by James Chadwick, in 1932, and the observation of the positron (confirming Dirac theory), in 1932, by Carl Anderson, and of the muon, in 1936, by Carl Anderson and Seth Neddermeyer. 

The birth of quantum field theory followed to be marked, throughout most of the process, by a grave theoretical impediment: the omnipresent infinities which arose in evaluations of physical quantities. Although even today there is some reserve against QFT among formalists, the advancements of the late forties were providential to eventually assuage the apprehension of most theoreticians with the construction. Effective methods to deal with the divergences and save perturbative expansions appeared between 1946 and 1948, when the scientific community was finally vibrant again after World War II. An approach of perturbative renormalization based purely on the familiar operator formalism was introduced independently by Sin-Itiro Tomonaga~\cite{Tomonaga:1946zz} and Julian Schwinger~\cite{Schwinger:1948iu,Schwinger:1948yk,Schwinger:1948yj,Schwinger:1949ra}. In the same year, Richard Feynman stated his path integral formulation of QM~\cite{Feynman:1948ur,Feynman:1949zx}, a groundbreaking alternative complete approach to operational quantum physics, which allowed profound theoretical matters to be analysed directly, and represented another method of renormalization. One year later, in 1949, Freeman Dyson proved that the two distinct approaches, one of Schwinger and Tomonaga and the other of Feynman, are, in fact, equivalent~\cite{Dyson:1949bp,Dyson:1949ha}.

Renormalization cemented quantum electrodynamics (QED) as a good theory, the first block of the SM. Its tremendous success could be visualized through its achievements on the understanding of the anomalous magnetic moment of the electron and the Lamb shift in the hydrogen spectrum, which motivated the search for a QFT of other phenomena, specially beta decay and nuclear forces. In the year before and in the fourteen years following the finalization of QED, extra experimental findings favoured the theoretical efforts: in 1947, had occurred the discovery of the pion (thought to be the meson predicted by Yukawa) by Lattes, Powell and Occhilini, and of the strange kaon meson, by Rochester and Butler. Then, in 1956, Cowan and Reines found the electron neutrino, while the muon one was observed in 1962 by Lederman, Schwartz and Steinberger. In the side of theory, this period was mainly marked by two events: in 1953, Kazuhiko Nishijima and Tadao Nakano~\cite{Nakano:1953zz} first quoted an empirical relation between the quantum numbers of barion number, strangeness (earlier defined quantity, seemly conserved during collisions but not during decays) and isospin, independently found in 1956 by Murray Gell-Mann~\cite{Gell-Mann:1956iqa}; And the postulation, in 1957, of neutrino flavour oscillation by Bruno Pontecorvo~\cite{Pontecorvo:1957cp}. 

One of the most important steps in the popularization of the modern model building strategies came in 1962, when Gell-mann~\cite{Gell-Mann:1961omu} and Yuval Ne'eman~\cite{Neeman:1961jhl} independently classified known hadrons according to the Eightfold Way\footnote{A reference to the buddhist doctrine of the Noble Eightfold Path.}. The idea amounted to the grouping of all known mesons and baryons according to representations of $SU(3)$, with strangeness and electric charge as the quantum numbers. In total, nature seemed to contain a scalar meson octet, a scalar meson singlet, a fermionic baryon octet, and a spin-$\frac{3}{2}$ baryon decuplet. The Eightfold Way led to the presentation, in 1964, of the quark model, made independently by Gell-Mann~\cite{Gell-Mann:1964ewy} and George Zweig~\cite{Zweig:1964ruk}. The quark model postulated that the known hadrons were not elementary, but composed by smaller particles, the quarks. The eightfold way was thus justified, with the group representations arising therein as composite representations induced by a fundamental triplet, with the \textit{up}, \textit{down} and \textit{strange} quarks as components.

Although the quark model helped make sense of the known hadron list (whose non-`\textit{elementarity}' was not without evidence~\cite{Bjorken:1968dy,Bloom:1969kc,Breidenbach:1969kd,Feynman:1969ej,Friedman:1972sy}), the theory was somewhat discredited by the fact that these quarks had never been observed in isolation. The idea of symmetry as a fundamental dynamical principle, however, already occupied a large room in the particle physicist imaginarium. As QED could have been regarded as a $U(1)$ gauge theory, in the 1950s Chen Ning Yang and Robert Mills generically proposed a non-abelian gauge group~\cite{Yang:1954ek}, and the strategy to describe fundamental interactions through symmetric minimal couplings became standard. Models of this sort could be proposed to replace, for instance, the Fermi theory of beta decay, which was known to be non-renormalizable by the formalism of Tomonaga-Schwinger-Feynman. From the beginning, however, there existed a crucial obstacle to this sort of theory: masses. 

The only massless vector boson known at the time was the photon, and it had been supposed that any other one would surely have been already observed. The problem is then stated, because a mass term for a gauge boson put in by hand would destroy the symmetry and, as was eventually realized~\cite{Komar:1960af,Kamefuchi:1961sb,Salam:1962aq,Veltman:1968ki}, is not renormalizable. The symmetry properties of the vacuum then naturally became focus of attention, as it was believed that a theory with non-symmetric vacua (a spontaneously broken symmetry) would present itself in nature as an approximate symmetry -- as was known to be the case of the eightfold way, for instance, as the masses of quarks within a same multiplet were not exactly identical. It was, then, disappointing, when Jeffrey Goldstone stated a theorem~\cite{Goldstone:1961eq}, in 1961, which claimed that for every dimension of the symmetry which is not respected by the vacuum, arise a massless scalar boson, none of which was seen in the universe.

Following the usual scientific route, this setback would ultimately lead to another one of the greatest achievements in history. In 1964, Peter Higgs~\cite{Higgs:1964ia,Higgs:1964pj}, and  François Englert collaborating with Robert Brout~\cite{Englert:1964et}, independently found a mechanism to obtain the benefits of spontaneous symmetry breaking without its burdens. They showed how, if the broken symmetry is regarded to be local, the unwanted goldstone bosons may be `swallowed' by the transversal gauge vector bosons to grant them a mass, turning into their longitudinal helicity component. Not only did this get rid of the unseen scalars, but also achieved a massive force carrier, as was wanted in order to model the nuclear forces by a Yang-Mills theory.

Seeking after underlying symmetry principles for the observed forces was popular at the mid sixties for yet another, disjoint (or opposite, even), reason: the hypothesis which suggested the pion as carrier meson of the nuclear forces had recently been tuned to a successful construction. The theory with a chiral isotopic symmetry $SU(2)_L \times SU(2)_R$, explicitly broken by quark mass terms and spontaneously broken at low energies by non-perturbative quark condensation, produced the pion as a pseudo-Nambu-Goldstone boson. This represented a groundbreaking accomplishment and resolved a few previously unavoidable issues mentioned in this summary. In particular, this advancement diluted the interest in the Higgs mechanism, as it diminished the desire to get rid of the Goldstone bosons.

Different constructions of the same idea were tested, still in attempts to describe the strong force. The most popular tried to fit known mesons as the gauge bosons of the vector and axial vector symmetries, with the pion appearing as a leftover goldstone boson. Eventually, however, Abdus Salam and John Clive Ward employed this program in the context of the electroweak interactions~\cite{Salam:1964ry}. In 1964, they proposed a $SU(2)\times U(1)$ group (which was already mentioned by Sheldon Glashow in 1961~\cite{Glashow:1961tr}) together with a manual breaking of symmetry, predicting three massive vector bosons plus the photon. Independently, in 1967 Steven Weinberg~\cite{Weinberg:1967tq} proposed the same group, now spontaneously broken. The first test the theory faced was the issue of renormalizability, known since the twenties to be a necessity of an exact fundamental model. This challenge was overcome in a series of works by t'Hooft~\cite{tHooft:1971qjg}, Veltmann and t'Hooft~\cite{tHooft:1972tcz}, Lee and Zinn-Justin~\cite{Lee:1972fj,Lee:1972ocr,Lee:1972yfa,Lee:1973fn} and, finally, Becchi, Rouet, Stora and Tyutin~\cite{cmp/1103899003}, with the BRST formalism. After this theoretical matter was resolved, the theory could be confronted with newly acquired data. In 1973, the neutral electroweak currents were firstly observed at the Gargamelle bubble chamber~\cite{GargamelleNeutrino:1973jyy}, at CERN, and, ten years later, after the availability of the Super Proton Synchroton, the $W$ and $Z$ bosons were in the UA1 experiment~\cite{UA1:1983crd,UA1:1983mne,UA1:1983ylf}. The standard theory of electroweak interactions was thus established, and there was a single ingredient missing confirmation, to which we shall briefly come back.

A detour from model structure must be made in order for its particle components to be assessed. In 1964, a fourth quark had already been proposed by James Bjorken and Glashow~\cite{Bjorken:1964gz}, for the purely aesthetic reason of parallelism with the leptonic sector, within which four components existed. Then, in 1970, when the one-family electroweak theory was already well known, Glashow, John Iliopoulos and Luciano Maiani showed how the existence of a fourth quark, called charm, would cancel the neutral currents responsible for the decay of the strange kaon meson $\overline{K}^0$ into muons, which, without this mechanism, should be much more common than observed~\cite{Glashow:1970gm}. Four years after the charm quark existence had been confidently claimed by the three collaborators, the charmed $J/\psi$-meson was simultaneously observed by Burton Richter's group at SLAC~\cite{SLAC-SP-017:1974ind} and Samuel Ting's at BNL~\cite{E598:1974sol}. 

With the electroweak theory consolidated on two families of quarks and leptons, it remained to describe the force which bounded nuclei, stable cores of positive and neutral particles, together. Following the trend of the 1960s and the example of the electroweak theory, a crucial step was taken when -- through the renormalization group methods of Gell-Mann, Low, Callan, Symanzic, Coleman and Jackiw~\cite{PhysRevD.2.1541,Symanzik:1970rt,CALLAN197042} -- Frank Wilczek and David Gross~\cite{Gross:1973id}, and David Politzer~\cite{Politzer:1973fx}, independently, found out in 1973 that non-abelian gauge interactions become arbitrarily weak at high energies. If the strong force was indeed an interaction dictated by a non-abelian gauge theory, asymptotic freedom not only allowed perturbation theory to be sensible at high energies, as it predicted that the strength of the corresponding force does not fall with distance, motivating an explanation for the mysterious refusal of the quarks to appear isolated. Furthermore, the necessity of a new quantum number was already known by, for example, the requirement of the Pauli exclusion principle applied to equal constituents baryons, and was expected to be three by fittings of some observed meson decay rates. Because of this, it was natural when, still in 1973, Harald Fritzsch, Heinrich Leutwyler and Gell-Mann proposed that the new quantum number, \textit{color}, corresponded to a conserved $SU(3)$ gauge group~\cite{Fritzsch:1973pi}. The reason as to why the eight extra massless spin-1 bosons had never been observed was soon understood to be another facet of confinement: since they interact strongly with particles and, in particular, themselves, the gluons are trapped inside color singlets~\cite{PhysRevLett.31.494,Gross:1973ju,Gross:1974cs,Fritzsch:1973pi}.

To finalize the proposition of the particle content of the SM, some blocks of nature had yet to be recognized. The first individual of the third generation to appear was the tau lepton, discovered between 1974 and 1977, by Martin Lewis Perl's group at SLAC~\cite{Perl:1975bf,Perl:1976rz}. In 1977, Leon Lederman's team found the bottom quark, already predicted four years earlier as a mechanism to allow the experimentally observed $\mathit{CP}$-violation~\cite{Christenson:1964fg} by Kobayashi and Maskawa~\cite{Kobayashi:1973fv}. The top quark, much heavier and needed by the electroweak symmetry structure, was discovered only eighteen years later by the CDF collaboration~\cite{CDF:1995wbb}. Finally, the tau-neutrino was firstly experimentally inferred in 2000, by the DONUT experiment at Fermilab~\cite{DONUT:2000fbd}.

The SM was thus defined as a $SU(3)_c \times SU(2)_L \times U(1)_Y$ gauge theory, with three generations of leptons, each composed of a negatively charged, colorless, massive Dirac fermion and a massless Weyl neutrino; And three generations of 3-colored quarks, each containing a $Q=2/3$ and a $Q=-1/3$ Dirac fermion. The model also features a fundamental scalar doublet of a complex positively charged and a complex neutral field, which, through the carefully designed scalar and Yukawa potential, triggers spontaneous symmetry breaking and grants a mass to the $W$ and $Z$ boson and all the fermions, keeping the gluons, photons and neutrinos massless. While the electroweak sector is well understood, the $SU(3)_c$ dynamics was known from the beginning to be highly non-perturbative at low energies, and is to this day a vibrant research field, even within the SM context.

The glorious last piece of the puzzle would come to its place in 2012, when, after exhaustive search at the LEP and Fermilab, the spin-0 neutral mass eigenstate, leftover physical degree of freedom of the SM scalar doublet, was observed as an excess of events around \SI{125}{GeV}, by the Atlas and CMS groups, at the LHC. This was the crowning achievement of the SM and of the theoretical advancements made during its pursuit, and the last experimental observation to revolutionize particle physics.

\section{Review}\label{sec:SMReview}

With the familiarity gained through the historical approach above, we are exceptionally prepared to briefly review the SM~\cite{Quigg:2013ufa,Donoghue:1992dd,Schwartz:2014sze}. Now, the SM (or, at least, its basic facts) is a topic extensively understood, which is why we shall try to subsume it quickly, in an effort not to bore the reader with unnecessary discussions. 

The SM is a model built around the gauge group 

\begin{equation}\label{eq:ChargeSM}
SU(3)_c \times SU(2)_L \times U(1)_Y.
\end{equation}
%
The representation content is, as dictated by the usual programme, built with \textit{a priori} knowledge of the electric charge of the degrees of freedom that form the basis of the representations. The electric charge operator is then given in terms of $Y$ and of the third component of $SU(2)_L$ isospin (the diagonal generators) as\footnote{Notice that many authors chose the alternative normalization $Y \to 2Y$, in which case $Q = I_3 + \frac{1}{2}Y$.}

\begin{equation}
Q = I_3 + Y.
\end{equation}

Each generation within the leptonic sector is formed by left-handed doublets of a charged lepton grouped with the corresponding neutrino, plus the right-handed charged singlet: 

\begin{equation}
L_\ell \equiv \begin{pmatrix} \nu_\ell \\ \ell \end{pmatrix}_{L} \thicksim (\mathbf{2},-1/2), \;\;\;\; \ell_R \thicksim (\mathbf{1},-1), \;\;\;\; \ell=e,\mu,\tau.
\end{equation}

The color sector presents three families of left-handed doublets containing an up- and a down-type quark, plus a right-handed singlet for every flavour. Every particle is a color triplet, as seen below
 
\begin{equation}
\begin{gathered}
Q_i \equiv \begin{pmatrix} u_i \\ d_i \end{pmatrix}_{L} \thicksim \left(\mathbf{3},\mathbf{2},\frac{1}{6}\right) \\[3mm]
u_{iR} \thicksim \left(\mathbf{3},\mathbf{1},\frac{2}{3}\right), \;\;\;\;\; d_{iR} \thicksim \left(\mathbf{3},\mathbf{1},-\frac{1}{3}\right)  \\[3mm]
i = 1,2,3; \;\;\; u_i  =  u,c,t; \;\;\; d_i=d,s,b.
\end{gathered}
\end{equation}

The scalar sector proposes a single doublet

\begin{equation}
\phi \equiv \begin{pmatrix}
\phi^+ \\
\phi^0
\end{pmatrix} \thicksim \left(\mathbf{2},+\frac{1}{2} \right),
\end{equation}
whose main purpose of existence is triggering spontaneous symmetry breaking (SSB). For that end, we also require the classic `mexican hat' scalar potential

\begin{equation} \label{eq:Vscalar}
V_{\phi}= -\mu^2(\phi^\dagger \phi)+|\lambda|(\phi^\dagger \phi)^2,
\end{equation} 
with $\mu>0$. In order to enable the use, without modification, of the canonical formalism appropriate to derive physical transition rates from `unphysical' perturbative amplitudes, the creation and annihilation operators must be normalized according to the LSZ reduction formula requirement

\begin{equation}
\langle 0 | \phi | 0 \rangle=0.
\end{equation}
This implies that Eq.~(\ref{eq:Vscalar}) must be expanded around the lowest energy state of the potential. It is immediate to see that $V_{\phi}$ carries a one-dimensional manifold of degenerate vacua parametrized by

\begin{equation}
\phi_0 \in \left\{       \begin{pmatrix}
\phi^+ \\
\phi^0
\end{pmatrix}, \;\; |\phi^+|^2+|\phi^0|^2=\frac{\mu^2}{2|\lambda|}\right\}.
\end{equation} 
By charge conservation and Lorentz invariance, we favor a point in the pure neutral direction and rewrite $\phi$ as

\begin{equation}
\phi = \frac{1}{\sqrt{2}}\begin{pmatrix}
\phi_1 + i \phi_2\\
v_H + h + i \phi_3
\end{pmatrix},
\end{equation}
where $v_H \equiv |\mu|/\sqrt{\lambda}$ is chosen real and a possible source of $\mathit{CP}$ violation is ignored. It is easy to verify that this vacuum is only left invariant by transformations generated by the electric charge, generating the following pattern of SSB

\begin{equation}
SU(3)_c \times SU(2)_L \times U(1)_Y \to SU(3)_c \times U(1)_\mathrm{EM}.
\end{equation}

Another trivial exercise shows that the diagonalization of the gauge sector can be achieved by

\begin{table}[t!]
\centering
\caption{SM gauge groups and their relevant symbols.}\label{Tab:SMsymbols}
\begin{tabular}{cccc}
\toprule
			 & Gauge boson & Coupling constant & Generator
 \\ \midrule
   $SU(3)_c$ &  $g^a_\mu$    &  $g_s$        &  $\frac{1}{2}\lambda^a$ \\
   $SU(2)_L$ &  $W^a_\mu$    &  $g$          &  $\frac{1}{2}\sigma^a$ \\
   $U(1)_Y$  &  $b_\mu$    &  $g^\prime$   &  $Y$ \\
   $U(1)_\mathrm{EM}$ &  $A_\mu$    &  $e$   &  $Q$ \\ \toprule
\end{tabular}
\end{table}

\begin{equation}
\begin{gathered}
W_\mu^{\pm}=\frac{W_\mu^1 \mp i W_\mu^2}{\sqrt{2}}, \qquad Z_\mu=\frac{-g'b_\mu+g W_\mu^3}{\sqrt{g^2+g'^2}} \\
A_\mu=\frac{gb_\mu+g' W_\mu^3}{\sqrt{g^2+g'^2}}.
\end{gathered}
\end{equation}
where the symbols are defined in Table~\ref{Tab:SMsymbols}. The corresponding masses are given by

\begin{equation} \label{eq:rhoMP}
\begin{split}
M_W &= \frac{g v_H}{2} \\
M_Z &= \frac{\sqrt{g^2+g'^2}}{g}M_W \equiv \frac{M_W}{c_W} \\
M_A &= 0,
\end{split}
\end{equation}
where we have introduced the notation $\tan \theta_W \equiv t_W = g^\prime/g$, which entails the complementary trigonometric definitions of $s_W \equiv \sin \theta_W$ and $c_W \equiv \cos \theta_W$. With this, the strength of the photon-fermion interaction, immediately identifiable with the positive fundamental electric charge, may be written as

\begin{equation}
e=\frac{gg'}{\sqrt{g^2+g'^2}} = g s_W = g^\prime c_W.
\end{equation}

An additional Lagrangian density is necessary in order for the matter fields to acquire a mass through the electroweak SSB. This is the Yukawa potential, and is, by definition, a general renormalizable interaction bilinear on fermion and linear on scalar fields. Naively, the following expression could appear sufficient

\begin{equation}\label{eq:symYukawa}
\begin{split}
\mathcal{L}_{\mathrm{Y}}=-\sum_{\ell} \zeta^\ell_\ell [\bar{\ell}_R(\phi^\dagger L_\ell) +(\bar{L}_\ell\phi)\ell_R] & {} -\sum_{i} \zeta^{u}_i [(\bar{Q}_i\tilde{\phi}) u_{iR}+\bar{u}_{iR}(\tilde{\phi}^\dagger Q_i)] \\
& {} -\sum_{d} \zeta^{d}_i [(\bar{Q}_i\tilde{\phi}) d_{iR}+\bar{d}_{iR}(\tilde{\phi}^\dagger Q_i)],
\end{split}
\end{equation}
where $\tilde{\phi} = i\sigma_2 \phi^*$ and the $\zeta$ are couplings to be fitted. If these Yukawa couplings are put to zero, however, the SM, with the usual diagonal form and standard normalization of the kinetic terms, due to its particle content and gauge symmetry, can be seen to possess an accidental global symmetry generated by independent unitary rotations, in flavour space, of each of its fermion multiplets:

\begin{equation}
\begin{split}
Q_i &\to U^Q_{ij} \, Q_j \\
u_{iR} &\to U^u_{ij} \, u_{jR} \\
d_{iR} &\to U^d_{ij} \, d_{jR} \\
L_i &\to U^L_{ij} \, L_j \\
\ell_{iR} &\to U^\ell_{ij} \, \ell_{jR}. \\
\end{split}
\end{equation}
A sector of this symmetry is explicitly broken by the Yukawa interactions, and it is not hard to show by brute force manipulation of the rules that the resulting physical effect of such a $U(3)^5$ transformation on the Yukawa sector can be equivalently factorized as  

\begin{equation}
SU(3)^5 \times U(1)_Y \times U(1)_B \times U(1)_L \times U(1)_{\mathrm{PQ}} \times U(1)_{\ell_R},
\end{equation}
where each $SU(3)$ factor corresponds to a fermion multiplet; $U(1)_Y$ coincides with the gauged weak hypercharge group; $U(1)_{B(L)}$ is generated by the baryon(lepton) number; $U(1)_\mathrm{PQ}$ is the Peccei-Quinn symmetry, which assigns an identical quantum number to $\ell_R$ and $d_{iR}$ and the opposite one to $\phi$; and $U(1)_{\ell_R}$ rotates only the right-handed lepton singlet. 

The key observation is that the original Lagrangian is insensitive to an $U(5)$ action, and it turns out (as was already known in the sixties from strangeness violating decays and postulated by weak universality) that the correct description of the observable effects of the Yukawa sector does necessitate the violation of flavour symmetry. To summarize, the `symmetry eigenstates' do not correspond to the propagating degrees of freedom, and the Yukawa Lagrangian must be updated to 

\begin{equation}
\begin{split}
\mathcal{L}_{\mathrm{Y}}=-\sum_{\ell} Y^\ell_\ell [\bar{\ell}_R(\phi^\dagger L_\ell) +(\bar{L}_\ell\phi)\ell_R] & {} -\sum_{i} Y^{u}_{ij} [(\bar{Q}_i\tilde{\phi}) u_{jR}+\bar{u}_{jR}(\tilde{\phi}^\dagger Q_i)] \\
& {} -\sum_{d} Y^{d}_{ij} [(\bar{Q}_i\tilde{\phi}) d_{jR}+\bar{d}_{jR}(\tilde{\phi}^\dagger Q_i)],
\end{split}
\end{equation}
where all quark degrees of freedom are now understood to be symmetry eigenstates. It becomes trivial to visualize, in the unitary gauge, that all effects of the diagonalization of the quark mass matrices, given by

\begin{equation}
\mathcal{M}_{ij}^{u(d)} = Y^{u(d)}_{ij}\frac{v_H}{\sqrt{2}},
\end{equation}
can be included by the following redefinitions in Eq.~(\ref{eq:symYukawa})

\begin{equation}
\begin{split}
u_{iL(R)} &\to u_{iL(R)} \\
d_{iR} &\to d_{iR} \\
d_{iL} &\to (V_{\mathrm{CKM}})_{ij} \, d_{jL},
\end{split}
\end{equation}
where $V_{\mathrm{CKM}}$ is the Cabibo-Cobayashi-Maskawa matrix, written in terms of the biunitary transformation that diagonalizes the quark mass matrices as $V_{\mathrm{CKM}} = V_L^u V_L^{d \dagger}$.

Finally, the complete Lagrangian of the theory is then composed as 

\begin{equation}
\mathcal{L}_\mathrm{SM} = \mathcal{L}_\mathrm{kin} + \mathcal{L}_\phi + \mathcal{L}_\mathrm{Y} + \mathcal{L}_\mathrm{ghosts} + \mathcal{L}_\mathrm{gauge-fixing},
\end{equation}
where $\mathcal{L}_\mathrm{ghosts}$ comprises the ghost Lagrangian, needed, together with $\mathcal{L}_\mathrm{gauge-fixing}$, to quantize the spontaneously broken non-abelian gauge symmetry of the theory, and where the kinetic portion is naturally 

\begin{equation}
\mathcal{L}_\mathrm{kin} = \sum_f \bar{f} \slashed{D}f - \sum_X \frac{1}{4}X_{\mu\nu}^{a}X^{a\mu\nu} + (D_\mu \phi)^\dagger D^\mu \phi,
\end{equation}
with $f$ running through the matter representation content and $X$ through the set of gauge multiplets, where $X_{\mu\nu}^a = \partial_\mu X^a_\nu - \partial_\nu X^a_\mu + g f_{abc} X_\mu^b X_\nu^c$ denotes the covariant field strength tensor of the spin-1 $X^a_\mu$ field transforming in the adjoint representation of some group with structure constant $f$ (that vanishes for the abelian $U(1)$ gauge field).

The SM has thus been described as an exactly symmetric gauge theory which undergoes spontaneous symmetry  breaking through a potential whose operators have mass dimensions of two and four. It is known, then, to be {\it in principle} renormalizable. There is a caveat, however: a classical symmetry of the Lagrangian is not necessarily maintained after quantization. Such {\it anomalies} are rooted, in the functional formalism, in the non-invariance of the complete path integral measure of the effective action; And, equivalently, can be seen in the canonical field-perturbative point of view from the non-existence of a simultaneously gauge and Lorentz invariant regulator for some loop graphs with chiral vertices \cite{Bilal:2008qx,Frampton:2008zz}. There is nothing inadmissible about an anomalous global symmetry -- in fact, both baryon and lepton number are anomalous within the SM (this fact will be touched upon next chapter). A gauge symmetry, however, cannot be anomalous, as its validity is detrimental for the renormalizability of the theory. Because of its chiral nature, the SM with, for instance, a single fermion doublet is indeed anomalous. The anomaly of the 3-point function can be show to be proportional to the completely symmetric gauge theoretic quantity

\begin{equation}
\mathfrak{a}_{abc} \equiv \sum_f \tr{\gamma_5 \, t_a \{t_b,t_c\}},
\end{equation}
where here $\gamma_5$ represents the eigenvalue of $\gamma_5$ on a given fermionic degree of freedom $f$ -- {\it i.e.}, left-handed fields contribute a minus sign and right-handed ones a plus sign. $t$ is any of the 12 generators of the SM. Considering its fermion content, let us analyse the cancellation of anomalies for each combination of generators, which is equivalent to examining the anomaly of the triangle diagram with every possible arrangement of incoming gauge bosons.

\begin{enumerate}[label={\bfseries\arabic*.}]
\item $\left[SU(3)_c\right]^3$:

The anomaly of the triangle with three $SU(3)_c$ currents automatically vanishes because QCD is a vector theory, hence every right-handed contribution is exactly cancelled by a left-handed one.

\vspace*{0.85cm}

\item $\left[SU(3)_c\right]^2 \left[U(1)_Y\right]$:

Here the anomaly factor reads

\begin{equation}
\begin{split}
\mathfrak{a}_{Ybc} &= \sum_f \frac{1}{4}\tr{\gamma_5 \, Y \{\lambda_b,\lambda_c\}} = \sum_f \frac{1}{4}\gamma_5 Y\tr{\{\lambda_b,\lambda_c\}} \\
&=\sum_q \frac{1}{2}\gamma_5 Y \delta_{bc} \\
&= \frac{1}{2}\delta_{bc} \left[  (-1)\times 2 \times \left(\frac{1}{6}\right)+ 1\times 1 \times\frac{2}{3} +1 \times 1 \times \left(-\frac{1}{3}\right) \right] \\
&=0,
\end{split}
\end{equation}
where we have used the ciclicity of the trace and the normalization of the Dynkin index $\tr{t_a t_b} = \frac{1}{2}\delta_{ab}$, with $t_a = \frac{\lambda_a}{2}$. In the penultimate line, each term corresponds to a given contributing multiplet, which are the colored ones alone (notice that we have rewritten the sum to be over $q$, symbolizing that it should be restricted to quark fields). For didactic reasons, we have explicitly shown how the final numeric result is obtained, with the factors in the being ordered as 

\begin{equation}
(\mathrm{handedness})\times (\mathrm{multiplicity}) \times (Y);
\end{equation}

\vspace*{0.85cm}
\item $\left[SU(3)_c\right]^2 \left[SU(2)_L \right]$:

Since different simple factors of the SM group commute (of course, as the total group is defined through a direct product), this trace factorizes and we have 

\begin{equation}
\begin{split}
\mathfrak{a}_{abc} &= \sum_f \frac{1}{8}\tr{\gamma_5 \, \sigma_a \{\lambda_b,\lambda_c\}} \\
& = \frac{1}{8} \sum_f \gamma_5 \, \tr{\sigma_a} \tr{\{\lambda_b,\lambda_c\}} = 0,
\end{split}
\end{equation}
which vanishes since $\tr{\sigma_a}$ or, more generally, the $SU(N)$ algebra $\mathfrak{su(n)}$ is the set of hermitian {\it traceless} $n$-dimensional matrices, together with its Lie bracket.

\vspace*{0.85cm}
\item $\begin{cases}&\left[SU(3)_c\right] \left[SU(2)_L \right]^2\\
 &\left[SU(3)_c\right] \left[SU(2)_L \right]\left[U(1)_Y\right] \\
 &\left[SU(3)_c\right] \left[U(1)_Y\right]^2 \\
 &\left[SU(2)_L\right] \left[U(1)_Y\right]^2:
\end{cases}$

\vspace*{0.85cm}

These configurations vanish by the same argument, {\it i.e.}, all of them contain one of the non-abelian groups appearing on exactly one current.

\vspace*{0.85cm}
\item $\left[SU(2)_L\right]^3$:

\begin{equation}
\begin{split}
\mathfrak{a}_{abc} &= \sum_f \frac{1}{8}\tr{\gamma_5 \, \sigma_a \{\sigma_b,\sigma_c\}} = \sum_f \frac{1}{4}\gamma_5 \tr{\sigma_a\delta_{bc}} \\
&=0,
\end{split}
\end{equation}
where we have used $\{\sigma_a,\sigma_b\} = 2\delta_{ab}$.

\vspace*{0.85cm}
\item $\left[SU(2)_L\right]^2\left[U(1)_Y\right]$:

\begin{equation}
\begin{split}
\mathfrak{a}_{Ybc} &= \sum_f \frac{1}{4}\tr{\gamma_5 \, Y \{\sigma_b,\sigma_c\}} = \sum_f \frac{1}{4}\gamma_5 Y\tr{\{\sigma_b,\sigma_c\}} \\
&=\sum_f \frac{1}{2}\gamma_5 Y \delta_{bc}  \\
&= \frac{1}{2}\delta_{bc} \left[  1 \times (-1)\times 2 \times \left(-\frac{1}{2}\right)+ 3 \times(-1)\times 2\times \frac{1}{6}  \right] \\
&=0,
\end{split}
\end{equation}
where, now, only the left-handed doublets, which form a non-trivial representation of $SU(2)_L$, contribute. In this and in the next case, the factors have been ordered as

\begin{equation}
(\mathrm{color})\times  (\mathrm{handedness})\times (\mathrm{multiplicity}) \times  (Y).
\end{equation}

\vspace*{0.85cm}
\item $\left[U(1)_Y\right]^3$:

The final combination gives

\begin{equation}
\begin{split}
\mathfrak{a} &= \sum_f \gamma_5 \, Y^3  \\
&= 1\times 1\times 1 \times (-1)^3 + 1\times (-1) \times 2 \times \left(-\frac{1}{2} \right)  \\
&\quad +3\times \left[(-1)\times 2\times \left(\frac{1}{6} \right)  + 1\times 1\times \left(\frac{2}{3} \right)+ 1\times 1\times \left(-\frac{1}{3} \right)\right] \\
&=0,
\end{split}
\end{equation}

\end{enumerate} 

The 3-point functions are linearly divergent, hence the 4-point ones diverge logarithmically and a shift in the integration variable of the box diagrams may be carried without repercussions. This means that the integrals may be regularized and cannot contribute to the anomaly. Even if that was not the case, the anomaly factor of higher point functions is proportional to that of the triangle. What any of these two facts imply is that the analysis of the 3-point function is enough to guarantee that a theory is anomaly free. With this in mind, the complete assessment of the SM anomaly factors just performed shows that it is, indeed, without gauge anomalies. In particular, it may be verified that every combination of the gauge currents considered above would also vanish if a single generation was taken into account. Therefore, the SM is free of anomalies generation by generation, and is a well-defined, renormalizable\footnote{At least in the weak coupling limit of QCD.} theory. 

The grandiosity of the complexity and resulting success of the SM is not only improbable, but hard to satisfactorily narrate. It is a single, closed form theory that is more complicate (and has more intriguing aspects) than any other fundamental, supposedly exact and general, theory of nature. Its accomplishments start by the unimaginable series of predictions recalled last section, but do not end there. For instance, the theory correctly describes the plethora of recent LHC data with little deviation~\cite{ParticleDataGroup:2022pth}. Most significantly, the SM has been confidently validated by the so-called electroweak precision tests. 
One way to explore electroweak precision observables (EWPOs) in order to stress test the SM takes advantage of the vector ($g_V$) and axial-vector ($g_A$)\footnote{These parameters are usually called $v_f$ and $a_f$, respectively -- we chose the alternative naming which is more appropriate for the case when an exotic $Z^\prime$ exists, as will be the case when these quantities reappear in Section~\ref{sec:3-3-1Solution} and in Appendix~\ref{app:couplings}.} couplings of the $Zf\bar{f}$ interactions at the $Z$-pole. These enter the parametrization of the weak neutral currents, for massless fermions, as

\begin{equation}
\mathcal{L}_{Zf\bar{f}} = i\bar{\psi}_f \gamma^\mu (g^f_V - g_A^f \gamma_5) \psi_f Z_\mu.
\end{equation} 
The analysis take as typical input parameters $\alpha, \alpha_s, G_F, M_Z, M_H$ and $m_t$, which are extracted from data assuming the SM as accurate. Several precision observables are then obtained to a high accuracy, and are subsequently cast in the form of an intricate SM prediction in terms of the input and of the vector/axial-vector parameters. Some instances of such precision observables are the $Z$ decay width; $e^+e^- \to f \bar{f}$ cross-sections; The left-right assimetries of the same processes and etc. The observed $g_V,g_A$ are then compared against the SM calculation, and the accumulated results attest that the SM predictions are coherent to a high degree of precision. For great reviews on this matter, see~\cite{Erler:2019hds,Freitas:2020kcn} and references therein.

\cleardoublepage


\pagebreak

\chapter{Flaws of the Standard Model} \label{Chapter3}

Although an outstanding attempt at a fundamental theory of elementary particles (which can never be sufficiently emphasized), the Standard Model cannot be the ultimate theoretical expression of nature. The job of describing the building blocks of matter and their interactions is a \textit{difficult} one, and at this point we know of several reasons why the SM must be amended. These range from theoretical discomforts (such as the arbitrariness in the number of families) to the unacceptable inability to describe important phenomena. Notwithstanding, the problems of the SM come with a benefit: they guide us towards BSM physics, which leads to the next hypotheses, candidates to update or replace the SM. Accordingly, to understand deeply each of the issues of the theory is an imperative matter in the search for BSM physics, and that is the objective of this chapter: to lay down a rudimentary review, as self-contained as possible, of the major flaws of the SM.

\section{Neutrino masses}

The most direct and unassailable argument for the insufficiency of the SM, even within a pure particle theoretical point of view, is the absence of neutrino masses. Neutrinos are weakly interacting particles whose existence, when hypothesized last century, could only be inferred indirectly through kinematic observations, all of which indicated that its mass should be nearly zero. This, together with their highly penetrating quality, caused them to be assumed massless for several decades. 

At the same time, it was known, even from simple quantum mechanical arguments alone, that the phenomenon of neutrino oscillating between its flavours in the vacuum could only occur if at least one species had a mass. This seems obvious from the understanding that mass eigenstates are the ones that propagate and flavour the ones which are produced through interaction, so that propagation may lead to oscillation only if the two differ. Nevertheless, it is useful to put forward a simple mathematical argument which is as follows. By definition, the states which propagate are the ones that diagonalize the unitary evolution operator $U(t) = e^{-iHt}$, called mass eigenstates. These degrees of freedom do not need to correspond to symmetry eigenstates, defined as the fields whose gauge interactions are diagonal in flavour space. If probabilities are to be conserved through a change of basis, these two sets of eigenvectors must be related unitarily, that is

\begin{equation}
\ket{\nu_j, t} = \sum_\ell U_{j \ell} \ket{\nu_\ell, t},
\end{equation}
where $\ell = e, \mu, \tau$ labels flavour eigenstates and $j=1,2,3$ mass ones, and 

\begin{equation}
UU^\dagger = U^\dagger U = \mathds{1}.
\end{equation}
In the context of the simple extensions of the SM, this matrix is called Pontecorvo-Maki-Nakagawa-Sakata, and may be found to be $U_\mathrm{PMNS} \equiv (V_L^{\ell})^\dagger V_L^\nu$. With this and the law for time evolution, one may find, at once

\begin{equation}
\ket{\nu_\ell, t} = \sum_{\ell^\prime, j} U^*_{\ell j}e^{-iE_jt}U_{\ell^\prime j} \ket{\nu_{\ell^\prime},0},
\end{equation}
with which the transition amplitude for an $\ell$-flavour neutrino, at time $t=0$, to oscillate to an $\ell^\prime$ one at time $t$ is easily calculated, and, using $P\left(\nu_\ell(0) \to \nu_{\ell^\prime}(0)\right) = \delta_{\ell \ell^\prime}$, leads to a general probability

\begin{multline}
P\left(\nu_\ell(0)\right.  \left. \to \nu_{\ell^\prime}(t) \right) = \delta_{\ell \ell^\prime} + {}\\
{} - \sum_{i>j} \left(4\,\Ree\left\{ \mathcal{Q}_{\ell \ell^\prime, ij}\right\}\sin^2{\phi_{ij}} - 2\, \Imm\left\{ \mathcal{Q}_{\ell \ell^\prime, ij}\right\}\sin{2\phi_{ij}}\right),
\end{multline}
where we have defined the quartic $\mathcal{Q}_{\ell \ell^\prime,ij} \equiv U_{\ell i}U^*_{\ell^\prime i}U^*_{\ell j}U_{\ell^\prime j}$, and the phases are given by 

\begin{equation}
\phi_{ij} \equiv \frac{E_i-E_j}{2}.
\end{equation}
Since $\phi_{ij} =0 \Rightarrow P\left(\nu_\ell(0) \to \nu_{\ell^\prime}(t) \right)=0$, this finishes a simple argument to justify the claim at the beginning of this section. 

Hard evidence for neutrino oscillation was first published in 1998~\cite{Super-Kamiokande:1998kpq}, after the Super-Kamiokande experiment observed a large difference between data and expectation for the ratio of muon to electron atmospheric neutrinos. The significant muon neutrino deficit is explained by its oscillation during the course of its travel, and is better fitted by the dominant $\nu_\mu \to \nu_\tau$ hypothesis, or even by the muon neutrinos oscillating preferably to an exotic sterile flavour $\nu_\mu \to \nu_\text{s}$, but not by the simplest mode $\nu_\mu \to \nu_e$. Figure~\ref{fig:neutrinos} shows how oscillation is necessary for the collected data to be satisfactorily explained. 

\begin{figure}[t!]
\centering
\includegraphics[width=0.9\linewidth]{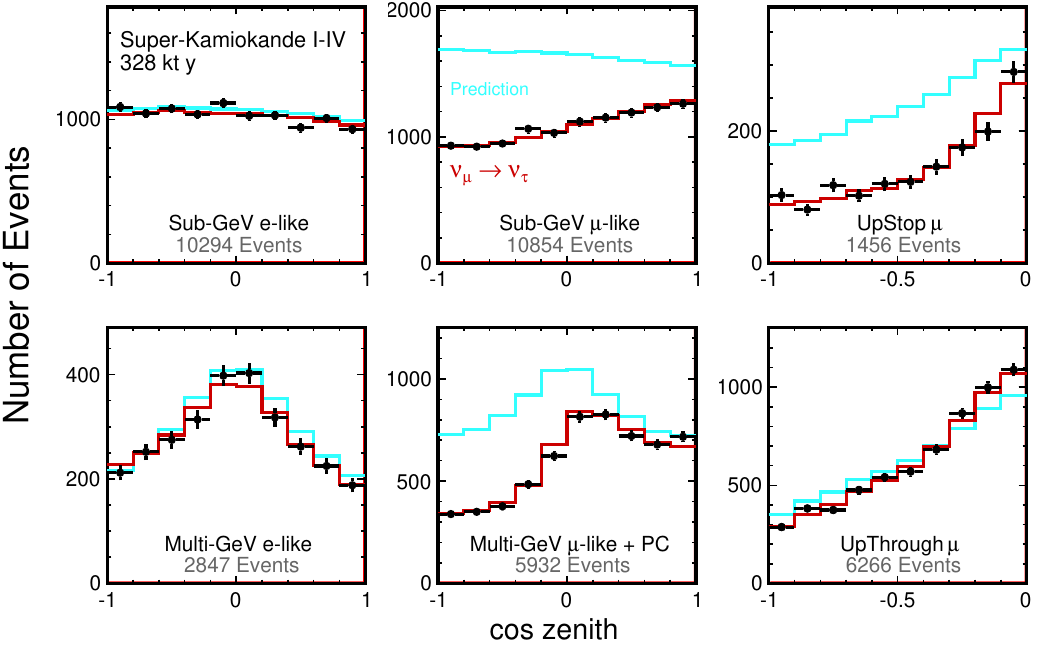}
\caption[Atmospheric neutrino events as a distribution of the zenith angle]{Number of events of atmospheric neutrinos as a distribution of the zenith angle. The blue histogram corresponds to the predictions of the non-oscillation hypothesis. The plots show how there is a clear muon neutrino scarcity and that it cannot be explained by their oscillation into electron ones. Figure taken from PDG~\cite{ParticleDataGroup:2022pth} and provided by the Super-Kamiokande Collaboration.}
\label{fig:neutrinos}
\end{figure}

The paradigm of general neutrino oscillation, unveiled through testing of the atmospheric neutrino flux, has only been reaffirmed by additional experiment designs, which investigated analogous phenomena in accelerator~\cite{K2K:2006yov}, reactor~\cite{Hsu:2006gt} and solar~\cite{SAGE:2009eeu} flux, establishing the theoretical expectations of the normal and inverted hierarchies for neutrino masses. The former postulates $m_3 \gg m_2 \gtrsim m_1$, and the latter hypothesizes $m_2 \gtrsim m_1 \gg m_3$, with $\nu_1,\nu_2,\nu_3$ being more $\nu_e,\nu_\mu,\nu_\tau$-like, respectively.

It is not difficult to manually solve this issue in the form of a minimal extension of the SM. In fact, the first possibility does not imply an increased particle content, as a neutrino mass may be obtained by the inclusion of the following dimension-5 operator in the theory

\begin{equation}
\mathcal{L}_5 = \frac{c_5}{\Lambda} \left(\overline{\tilde{L}}H\right)\left(\tilde{H}^{\dagger} L\right).
\end{equation}
However, since this interaction is non-renormalizable, it represents no solution at all and does not improve the SM as a fundamental theory. 

A better option proposes the inclusion of a right-handed neutrino singlet (consider a single generation for simplicity) $\nu_R \thicksim (\bm{1},\bm{1},0)$ to the model. This extension allows for different possibilities, each with its weaknesses:

\begin{enumerate}
\item Pure Dirac mass: 

The singlet allows for a Dirac mass to be generated by the Yukawa term

\begin{equation}
\mathcal{L}_\text{D} = y_{\nu,\text{D}} \bar{L}\tilde{\phi} \nu_R.
\end{equation}
The problem is that neutrino squared mass differences are estimated to be $\Delta m_{21}^2 \sim \SI{7.4 \times 10^{-5}}{eV^2}$ and $\Delta m_{32}^2 \sim \SI{2.5 \times 10^{-3}}{eV^2}$~\cite{DeSalas:2018rby}, and the absolute bound for the masses is usually cited, from cosmological analysis, as of the order of $\sum m_\nu< \SI{1}{eV}$. Since $m_e^2 = \SI{2.6 \times 10^{5}}{eV^2}$, this would imply a hierarchy for the dimensionless Yukawa parameters within the lepton sector already larger than the current largest hierarchy in the entire model. 

\item Pure Majorana mass:

A pure Majorana mass of the form $m_{\nu,R}\overline{\nu^c_{\hphantom{c}R}} \, \nu_R$ for the exotic degree of freedom would not solve the issue, as the right-handed field is a pure singlet and thus sterile and, therefore, could not grant a mass to the known neutrinos.

\item Seesaw Mechanism:

In the Seesaw Mechanism, with (type-II) or without (type-I) the addition of an exotic scalar triplet, the Majorana and Dirac mass terms appear together, and conspire to produce a light and a heavy mass eigenstate. This is because natural values of $y_{\nu,\text{D}}$ and $m_{\nu,R}$ produce a sufficient splitting between the two eigenvalues. Although Seesaw models are by far the most viable, the issue with type-I Seesaw is that a non-explained parametric hierarchy remains; And with the type-II mechanism is that, with the introduction of a scalar triplet, unobserved physical goldstones degrees of freedom enter the theory (which could be avoided if lepton number violation is included in the potential).
\end{enumerate}

For a nice introductory review of the theoretical and experimental aspects of neutrino masses see \cite{deGouvea:2016qpx} and \cite{Weinheimer:2013hya}, respectively.

\section{Hierarchy Problem}

Another quantum field theoretical trouble of the SM is the Hierarchy Problem (HP). Although less important than the neutrino masses, it has been traditionally used as a guide in the pursuit of new physics while new strong experimental input does not appear. The Higgs mass is radioactively corrected at 1-loop by the diagrams of Figure~\ref{fig:diagsHP}. Parametrizing the divergence in these contributions through a cutoff regulator, each of them contribute an amplitude of the form

\begin{equation}
i \mathcal{M} \sim \Lambda^2.
\end{equation}
Because of its coupling strength, the value of the fermion loop diagram with $f$ being the top quark is around two orders of magnitude larger than the others, which is why it is usually considered alone.

The usual line of thought then follows to interpret this result as physically meaningful by setting the cutoff to a scale of next physics, such as the Grand Unification $\left(\SI{10^{15}}{GeV}\right)$ or Planck $\left(\SI{10^{19}}{GeV}\right)$ scale. It is then concluded that, to be able to fit the physical mass of the Higgs, the SM implies a fine tuning of 

\begin{equation}
\frac{\delta m_H^2}{m_H^2} \sim 10^{33}.
\end{equation}
Such a 10 quadrillion fine tuning in the masses violates the principle of naturalness, which dictates that every dimensionless ratio between parameters of a same sector must be close to order one.

\begin{figure}[t!]
\centering
\includegraphics[width=0.35\linewidth]{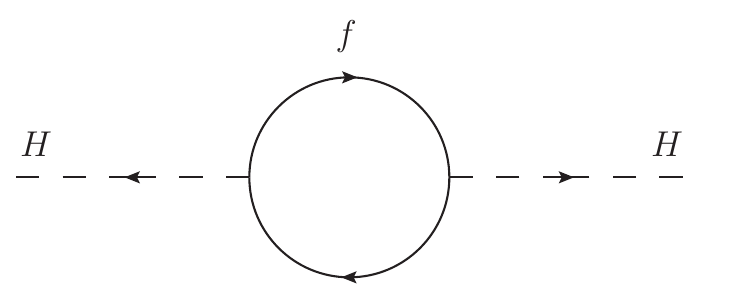}
\includegraphics[width=0.285\linewidth]{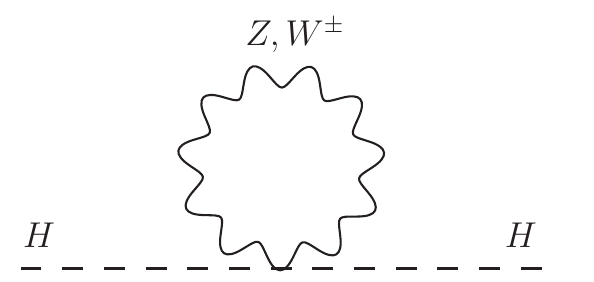}
\includegraphics[width=0.285\linewidth]{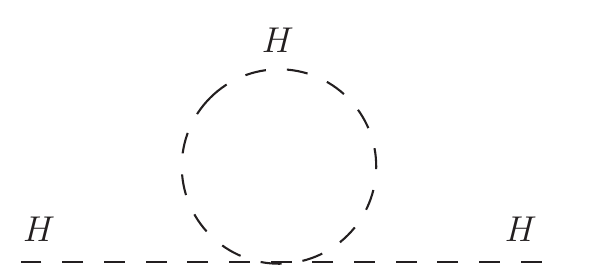}
\caption[Quadratically divergent corrections to the Higgs mass.]{Quadratically divergent corrections to the Higgs mass which are the perturbative root of the HP. Here $f$ represents any fermion.}
\label{fig:diagsHP}
\end{figure}

This argumentation is a bit misleading in that the cutoff, as presented, has no physical meaning and is nothing but an artifact used in order to make sense of intermediate calculations. The precise way to introduce the HP is through an effective field theory formalism, in which case $\Lambda$ becomes a physically meaningful Wilsonian cutoff~\cite{Aoki:2012xs}. Another way to perceive the inaccuracy of the previous rationale comes from trying to replicate it using, instead of cutoff, dimensional regularization. In that case, the HP does not manifest and one is tempted to say, as is often done, that it {\it becomes hidden}. This makes no strict sense, and the correct claim is that the HP can only be addressed in effective constructions. With this in mind, notice that, unlike in cutoff regularization which presents an explicit avatar of new physics ($\Lambda$), in dimensional regularization the new physics must be put in by hand, \textit{e.g.}, as new interactions with heavy particles. Indeed, to produce an example, we calculate the $\overline{\mathit{MS}}$ correction to the Higgs mass that results from adding a test heavy fermion, with $m_H/m_f \to 0$, to interact with the $H$ with Yukawa $\lambda_f$. We find, for physical characteristic energies of the order of the Higgs mass, that the squared mass of the scalar particle picks up a contribution of

\begin{equation}
\delta m_H^2(\mu) =  \Sigma^{\overline{\mathit{MS}}} = \frac{\lambda_f^2}{8 \pi^2} m_f^2\left[3 \log \left( \frac{m_f^2}{\mu^2} \right) - 1 \right].
\end{equation}
We see that $\delta m_H^2 \sim m_f^2$, which is how the `quadratic scaling of the renormalized mass with the new physics parameter' appears (which one could perhaps argue from dimensional reasoning alone).

Summarizing, we emphasize that the HP is a real issue -- the problem, however, is not `the top loop', but the sensitivity to higher scaled physics. Furthermore, it is an intrinsic issue of \textit{scalar} masses, as fermion masses are protected by the chiral symmetry which is restored if they vanish; and vector boson masses, on renormalizable theories, are protected by the gauge symmetry in the same way. 

The regular strategy to solve the HP amounts to extending this mechanism to the scalar masses. In fact, the HP is one of the greatest motivations for supersymmetry~\cite{Martin:1997ns}, as it is responsible, in the Minimal Supersymmetric Standard Model (MSSM), for guarding the Higgs mass against quadratic corrections. Diagrammatically, this occurs through the loop with inverse statistics (and hence sign) induced by the stops, which exactly cancels the top loop independently of the masses of the supersymmetric partners, as long as the relation between couplings is maintained. Unfortunately, the MSSM, even if eventually proven the right theory, have already become a non-ideal solution of the HP because the lightest stop mass has been pushed to the TeV scale by experiment~\cite{ATLAS:2019lng,CMS:2019zmd}, which entails a leftover 1\% fine tuning on the Higgs mass -- this is the so-called Little Hierarchy Problem. 

Variant attempts at solving the HP comprise models in which the physical Higgs arises as a pseudo Nambu-Goldstone boson (pNGB), such as Twin~\cite{Chacko:2005pe,Chacko:2005vw} or Composite Higgs~\cite{Agashe:2004rs} models. The intuition behind it is that putting a pNGB mass to zero suffices to recover the explicitly broken symmetry, so that it itself is a technically natural parameter. One more possibility is the Folded Minimal Supersymmetric Standard Model (FMSSM)~\cite{Burdman:2006tz}, which has a low energy representation content similar to that of the MSSM, but in which the colored superpartners are charged not under QCD, but under a distinct, `dark', QCD$'$. In the FMSSM, the ultraviolet supersymmetry still protects the Higgs mass, but the phenomenological constraints on the stops scale are less severe.

\section{Strong $\mathit{CP}$ problem}

If one takes it seriously, the naturalness principle originates a second source of stress onto the legitimacy of the SM. The requirement that every dimensionless ratio is of order one implicate the expectation that every term allowed by the theoretical framework is included in the Lagrangian. Now, consider the kinetic operator of a gauge boson multiplet in a Yang-Mills theory, defined to contain every invariant quadratic in the field strength tensor (since bilinears on the gauge fields can only appear through this specific combination). In a theory with a semisimple local symmetry group, this can be shown to be equivalent to~\cite{Weinberg:1996kr}

\begin{equation}
\mathcal{L}_\text{kin} = -\frac{1}{4}F^a_{\mu\nu}F^{a\mu\nu} + \frac{\theta}{32 \pi^2} \epsilon_{\mu\nu\rho\sigma}F^{\mu\nu}F^{\rho\sigma}.
\end{equation} 
The first term is the usual gauge boson kinetic term, but the second is generally omitted. To understand why, we verify

\begin{equation}
\begin{split}
\frac{1}{8} \epsilon_{\mu\nu\rho\sigma}F^{\mu\nu}F^{\rho\sigma} &= \tr{F \wedge F} =\tr{(\mathrm{d}A + A\wedge A) \wedge (\mathrm{d}A + A\wedge A)}  \\
&= \tr{\mathrm{d}(A \wedge \mathrm{d}A) + \frac{2}{3}\mathrm{d}(A \wedge A \wedge A)}  \\
&= \mathrm{d} \, \tr{A \wedge \mathrm{d}A + \frac{2}{3}A \wedge A \wedge A},
\end{split}
\end{equation}   
where $\mathrm{d}$ is the exterior derivative and we have used the ciclicity of the trace, the anti-commutativity of the wedge product, and $\text{d}^2=0$. This shows that the so-called $\theta$-term is a total derivative, and, thus, it is perturbatively inconsequential. 

Notwithstanding, the topological defect formalism shows that if euclidean spacetime is identified with the 3-sphere (which amounts to the compactification of spacetime by identifying infinity with a point), then for pure gauge configurations~\cite{Rajaraman:1982is,Manton:2004tk,Nastase:2019pzh}

\begin{equation}
 g_{\mu}\big\rvert_{S^3} = iU(x)\partial_\mu U^\dagger(x) \Rightarrow \int_{S^3} d^4x \frac{\theta}{32 \pi^2} \epsilon_{\mu\nu\rho\sigma}F^{\mu\nu}F^{\rho\sigma} = \theta n,
\end{equation}
where $g_\mu$ is the gluon field, $U(x)$ is a local $SU(3)$ gauge transformation and $n$ is an integer. This implies that a $\theta$-term generates (or is generated by, if one inverts the reasoning) non-perturbative solutions of the field equations which interpolates between inequivalent vacua. Moreover, consider the intuitive abelian electromagnetic situation, in which case we define the electric and magnetic fields as $E_{i} \equiv F_{0i}$ and $B_{i} \equiv - \frac{1}{2}\epsilon_{ijk}F^{jk}$. In this context, one has

\begin{equation}
\epsilon_{\mu\nu\rho\sigma}F^{\mu\nu}F^{\rho\sigma} \propto \vectorB{E} \cdot \vectorB{B}.
\end{equation}
This is the easiest way to arrive at the conclusion that the $\theta$-term violates $\mathit{P}$ and $\mathit{CP}$ if one recalls from electrodynamics that $\vectorB{E}$ is a $T$-odd vector while $\vectorB{B}$ is a $T$-even pseudovector, and that the gluon field is real. 

Being non-perturbative, one has no hope to evaluate exact phenomenological repercussions of the $\theta$-term with the fundamental degrees of freedom, but the fact that it is the only possible source of $\mathit{CP}$ violation in the strong sector prompts us to seek evidence of this non-conservation in pure QCD processes in order to estimate the value of $\theta$. Indeed, this is done via measurement of the neutron electric dipole moment (EDM), as the existence of such a permanent dipole of a spin-1/2 particle implicates in $C$ and $T$ violation, and, thus, that also $\mathit{CP}$ is violated. The connection is made through chiral perturbation theory~\cite{Crewther:1979pi} (for a more modern approach in lattice QCD, see~\cite{Shintani:2008nt}), which allows a bound on the neutron EDM to be translated into a bound on $\theta$.

The current result for the EDM bound is~\cite{Abel:2020pzs} 

\begin{equation}
d_n<1.12\times 10^{-26}\, e \hspace*{0.08em} \si{cm},
\end{equation}
implying a severe fine tuning on the physical parameter of the order~\cite{DiLuzio:2020wdo} 

\begin{equation}
\theta \lesssim 10^{-10},
\end{equation}
which states the strong $\mathit{CP}$ problem. 

Notice that, in principle, similar effects could be present within the electroweak sector -- specifically, a $\theta$-term for the non-abelian $SU(2)$ is also possible. To see why it is unphysical in that context, recall that by virtue of the anomalous nature of the chiral symmetry, a rotation of the form $\psi_f \to e^{i\alpha \gamma_5} \psi_f$ on a Dirac fermion $\psi_f$ induces an increment on the effective action of the form 

\begin{equation}
\delta \mathcal{L}_\text{YM} =  -\frac{\alpha}{16 \pi^2} \epsilon_{\mu\nu\rho\sigma}F^{\mu\nu}F^{\rho\sigma},
\end{equation}
which translates into a redefinition $\overline{\theta} = \theta -2\alpha$. Such a rotation, however, is of consequence within the scalar sector, introducing a complex phase on mass terms of the form $m_i \to e^{2 i \alpha_i} m_i$ for a diagonal matrix. In order to keep masses real, by convenience, one generally chooses to move all $\mathit{CP}$ violating effects of this type to the $\theta$-term. Now, in the electroweak theory, the chiral anomaly may be used to remove the $\theta$-term without repercussion on the mass terms, since any phase matrix may be absorbed into the right-handed fields, which are weak isospin singlets and cannot give rise to anomaly corrections (in fact, the most formal way to prove that an electroweak $\theta$-term is unobservable is to show that the instantonic configurations posses fermionic zero-modes~\cite{Anselm:1993uj}). This is not the case of QCD, since it is vector-like.

The traditional solution of the strong $\mathit{CP}$ problem is that of the Peccei-Quinn (PQ) mechanism~\cite{Peccei:1977hh}, which, usually in the context of extended scalar sectors, propose a spontaneously broken PQ global symmetry, which gives rise to a (pseudo) Nambu-Goldstone boson, the axion~\cite{DiLuzio:2020wdo}, that, through its effective potential corrections, dynamically sets $\overline{\theta} = 0$.

\section{Dark matter}

The next call for new physics is interdisciplinary. Several observations indicate the necessity of an exotic species of matter with specific characteristics, conventionally called dark matter (DM)~\cite{Profumo:2017hqp}, for reasons to become apparent. Among the evidences which originated this understanding, stand out the rotation curves of galaxies, whose circular velocity radial pattern generally indicate the existence of unseen spherically symmetric massive halos, much larger than the primary gaseous disk. Another is the Cosmic Microwave Background (CMB) anisotropies, which are too small to account for the currently observed large scale structure of the universe. The CMB spherical oscillations can be made remarkably compatible with the observations if we include into the description an additional component of matter which decoupled from the thermal bath before recombination of electrical subatomic particles into atoms.

The defining qualities of this specific type of matter make it

\begin{itemize}
\item \textbf{Neutral:} In other words, it must be \textit{dark}. Technically, it is required that DM was effectively decoupled from the baryon-photon plasma at recombination.

\item \textbf{Stable:} Clearly, DM must be incapable to decay in cosmological time scales if it is to account for its observable gravitational effects since early epochs.

\item \textbf{Cold:} In fact, there may be a DM component which is hot, but the totality must include a large cold component, because, if DM was predominantly hot, the small scale universal structure could not be explained, as ultra-relativistic almost inert matter would have free steamed out of dense regions.

\item \textbf{Massive:} Another obvious feature, as DM must interact gravitationally. The constraints on its mass are {\it highly} model dependent and vary immensely, but we quote conservative bounds which state that the DM mass is ${} >\SI{70}{eV}$~\cite{Randall:2016bqw} if it is fermionic and ${} \gtrsim \SI{10^{-22}}{eV}$~\cite{Nadler_2019} if it is bosonic, with upper limits going up to several masses of the sun.
\end{itemize}

The standard cosmological description assumes that at some point in the early history the reactions between DM and ordinary matter became too rare, which caused DM to fall out of chemical equilibrium and remain an independent component of the universal energy density, changing on large scales only due to the spacetime expansion. This process is called freeze-out. Employing the Einstein equations together with an ansatz for the spacetime geometry (the Robertson-Walker metric), another ansatz for the state equation of the gas components along the evolution of the universe and the assumption that the variations in the local comoving particle densities, on average, are due only to collisions (through the Boltzmann equation), it is possible to calculate the current relic density of dark matter (which makes up $\sim \hspace*{-0.18em}26$\% of the total energy budget), given a particle physics model, in terms of the freeze-out temperature and equilibrium density of DM.

One of the most motivated and important classes of DM candidate is composed by WIMPs, or Weakly Interacting Massive Particles. This is because it has been show that the DM relic density is naturally accounted for by a DM species which coannihilate with cross sections of electroweak order and masses within the approximate range $10 - \SI{100}{GeV}$. SM neutrinos are the immediate option for WIMP contender, but are known to have a mass which is too small and could account for only a small fraction of the relic density~\cite{de__Salas_2016}. This settles the SM as unable to propose a DM description. 

Popular BSM solutions include electroweak sector extensions that present more cosmologically effective WIMPs, such as, for instance, sterile heavy neutrinos (although this possibility is troubled -- see~\cite{Cerdeno:2009zz} and references therein) or the lightest neutralino in supersymmetric theories~\cite{doi:10.1142/S0217751X04018154}. Another possibility is given by axionic models~\cite{Adams:2022pbo}, already mentioned as a possible solution to the strong $\mathit{CP}$ problem, with the axion being a viable DM candidate.

\section{Matter-Antimatter assymetry}

The last important theoretical shortcoming of the SM that we shall mention is, again, of cosmological nature. Both experimental (such as the study of the diffuse $\gamma$-ray background~\cite{Cohen_1998}) and trivial everyday interactions with the universe of today show that the amount of existing antimatter is negligible. This is in contrast with the natural expectation that the primordial, hot, dense and in equilibrium universe would produce similar fractions of charge conjugate types of matter. 

The departure from this matter-antimatter `democracy' may be measured by the ratio of the matter-antimatter asymmetry to the number density of photons

\begin{equation}
\eta_\gamma \equiv \frac{n_B - n_{\bar{B}}}{n_\gamma},
\end{equation}
which is expected to be constant since early epochs because both numerator and denominator only scale with the expanding length parameter $a^3$. Another useful, sometimes preferred, quantity, is the ratio to total entropy density\footnote{Unfortunately, the same letter $\eta$ is usually used to denote both quantities, even though they differ in value.}~\cite{Bodeker:2020ghk}

\begin{equation}
\eta_s \equiv \frac{n_B - n_{\bar{B}}}{s},
\end{equation}
also stationary. To see this, recall, from basic thermodynamics, $dU = TdS - PdV + \mu dN$. Now use $U =  \rho V$ and the Friedmann equation $\dot{\rho} + 3H(\rho + P) = 0$ to obtain $dS = - \frac{\mu}{T} dN$. This implies that, for non-degenerate matter $\left(\mu/T \ll 1\right)$ or when reactions are impossible $\left(dN=0\right)$, the entropy is conserved. These two quantities are related by $\eta_s = \eta_\gamma \frac{n_\gamma}{s}$, where the proportionality factor may be calculated to be $\frac{n_\gamma}{s} \approx \frac{1}{7.04 k_B}$~\cite{Gannouji:2022ovy}.

This input parameter has been consistently calculated both by cosmic microwave background analysis~\cite{refId0} and by Big Bang nucleosynthesis to be 

\begin{equation}
\eta_\gamma = \frac{n_B - n_{\bar{B}}}{n_\gamma} \simeq \frac{n_B}{n_\gamma} \approx 6.12 \times 10^{-10},
\end{equation}
which cannot be accounted for by the standard cosmological particle model. 

The conditions upon which a matter-antimatter symmetric primordial universe could give rise to the highly asymmetrical current one had been laid out in a classical paper by Sakharov~\cite{Sakharov:1967dj}. The requirements are 

\begin{enumerate}
\item \textit{$C$ and $\mathit{CP}$ violation:} This is an obvious requirement, as a $C$-conserving interaction would compensate any conversion from matter to anti-matter with the opposite reaction. In the presence of chiral matter, $C$ violation is not enough, as a $C$ violating but $\mathit{CP}$ conserving interaction could still keep the matter-antimatter balance while only turning right-handed particles into left-handed ones or vice-versa. 

\item \textit{Baryon number violation:} Another trivial necessity, as, if baryon number was exactly conserved, no individual process that can produce an asymmetry can exist. 

\item \textit{Departure from thermal equilibrium:} If thermal equilibrium is maintained, $\mathit{CPT}$ would assure that no asymmetry can be generated even if the two first conditions are met.
\end{enumerate}

The SM can, in principle, accommodate all the Sakharov criteria for baryogenesis. $C$ and $\mathit{CP}$ violation are well understood to exist within the electroweak sector. A departure from thermal equilibrium is possible through the expansion of the electroweak phase transition bubbles supplied by the symmetry breaking -- we shall not discuss this further and direct the reader to the nice summary (on this and most topics in this subsection) in \cite{Cline:2006ts}.

Finally, let us discuss condition 2 at greater detail. Baryon number is a classical symmetry of the SM and generates a vector rotation of the quark fields of the form $q_{L(R)} \to e^{i \theta/3} q_{L(R)}$, which leaves the action invariant even in the presence of mass terms. It turns out that this symmetry is anomalous, as is lepton number. In fact, we have 

\begin{equation}\label{eq:BLCurrents}
\partial_\mu J_B^\mu = \partial_\mu J_L^\mu = n_f \frac{ g^2}{32 \pi^2} F_{\mu \nu}^a \tilde{F}^{a\mu \nu},
\end{equation}
where $n_f$ is the number of families, and $g$ and $F^a_{\mu \nu}$ are the $SU(2)_L$ coupling and field strength tensor, respectively\footnote{Note that the $B$ generator is defined with a factor $1/3$ with relation to the $L$ one, which is compensated by the three quark colors.}.

However, as has already been pointed out, a pure right-chiral rotation can be performed to eradicate the theoretical signs and hence, if the theory is accurate as it stands, the observability of any effect resulting of the anomaly. Nonetheless, the WKB formalism shows that if such effects were existent (as can be made to happen by the inclusion of explicit baryon violating interactions~\cite{MCLERRAN2012301}) the rate of instantonic tunneling between vacua that would mediate a change in $B$ is proportional to~\cite{tHooft:1976rip} 

\begin{equation}
\Gamma_\text{inst}(\Delta B) \propto e^{-S_\text{E}^\text{inst}} \sim e^{-16 \pi^2/g_{2L}^2} \approx  e^{-164},
\end{equation} 
which is to say it is impossible in practice and cannot account for baryogenesis. If this zero-temperature type process cannot help us, however, there is another type of topological solution which, at high temperatures, could possibly provide the desired effect. 

Note that the anterior pure Yang-Mills theory has no intrinsic scale, and as such contains a collective coordinate related to its extension. In fact, this is what saves the theory from the Derrick's theorem and allows for topological solutions -- in particular, this means that, although a barrier to be tunnelled through exists, it can be scaled to be arbitrarily small. Consider now a spontaneously broken gauge theory comprised of a gauge field and a scalar multiplet with a potential. Now there is a dimensionful parameter in the theory (the Higgs VEV), and a fixed energy barrier between classical vacua -- in particular, there is no exact instanton solution. If one can calculate such potential height, they can understand under which conditions a thermally induced transition between vacua may occur. 

The solutions corresponding to such transitions may be found as follows~\cite{Manton:2004tk,Klinkhamer:1984di}. Consider every path $\mathcal{C}$ in configuration space which connects consecutive topologically inequivalent vacua of the theory. Find, in each path, the configuration ({\it i.e.}, point) of maximal energy $x^*_\mathcal{C}$. The lowest such point $(x^*_\mathcal{C})_\mathrm{min}$, {\it i.e.}, the point obeying $\mathcal{E}((x^*_\mathcal{C})_\mathrm{min}) = \mathrm{min}\{\mathcal{E}(x^*_\mathcal{C})\}$, is, by construction, a strict saddle point of the euclidean action and thus a solution of the field equations. The saddle point solutions are static, unstable, finite action configurations which live in the top of the barrier between classical $n$-vacua, and explicit calculation shows they possess winding number $N_{\mathrm{CS}}=1/2$. Such a construction may be possible or not, and has been shown to exist in the standard electroweak model~\cite{Klinkhamer:1984di}.

Note that although $J_B^\mu$ and $J_L^\mu$ are individually non-conserved, it coincidentally occurs that $\partial_\mu\left(J_B^\mu-J_L^\mu\right)=0$ and $B-L$ remains conserved in the quantum theory. Furthermore, from (\ref{eq:BLCurrents}), using that that the integral of $(g^2/32\pi^2)F_{\mu \nu}^a \tilde{F}^{a\mu \nu}$ is the Chern-Simons number $N_\text{CS}=1/2$, we obtain, for the simplest sphaleron decay (between consecutive vacua), $\Delta (B+L) =N n_f$, for integer $N$. An example reaction induced by the sphaleron explosion, with $\Delta (B+L)= 6$ is

\begin{equation}
\bar{u} + \bar{d} + \bar{c} \, \to \, d + 2s + 2b + t + \nu_e + \nu_\mu + \nu_\tau.
\end{equation}
Specifically, the sphaleron height has been calculated to be about $\SI{10}{TeV}$.

As in the case of dark SM neutrinos, though, the asymmetry generated within pure SM is not enough. To start, the $\mathit{CP}$ violation provided by the quark mixing is too weak. Furthermore, to produce the observed asymmetry a much more violent first phase transition than that provided by the SM (which might be merely a crossover) is necessary~\cite{Kuzmin:1992up,Cohen:1993nk,Rubakov:1996vz}.

This topic is the most complex and is less direct to catalogue solutions, which come in the form of many specific models and mechanisms. These, however, usually involve Lagrangians with stronger sources of $\mathit{CP}$ violation. Additionally, the thermal transition may be strengthened by some exotic particle decay, such as a right-handed neutrino (which could also trigger an assymetry through {\it leptogenesis}). This electroweak first order phase transition is also boosted in supersymmetric or grand unified extensions.

\cleardoublepage


\pagebreak

\chapter{The Minimal 3-3-1 Model} \label{Chapter4}

One set of new physics we shall present regards the phenomenology of an exotic sector composed by a neutral scalar $s$, a doubly-charged scalar $Y^{\pm\pm}$ and a doubly-charged vector boson $U^{\pm \pm}$. A theory which accommodates at least one representative of each of this classes is the Minimal 3-3-1 Model (m331). Later, we shall directly tackle a problem of this theory: a thorough analysis of its exact Renormalization Group predictions. Although our results will prove to be useful in a larger scope -- and the phenomenological subject is, in fact, approached in a model independent manner --  it is paramount to give a detailed exposition of the basics of the model. That is the objective of this chapter, which begins with a discussion of our motivations and of why it is interesting to focus on the $U^{\pm \pm}$.

\section{Particles, particles, particles}

As expressed in the last Chapter, understanding the problems of the SM is detrimental in the quest of theorizing the physics out of its grasp.  This is, of course, not a deep philosophical statement (as if the SM was not specifically understood to be flawed, there would be no need for the theorizing of supposedly better theories), but it calls attention to the importance of considering every subtle hint that the failures of the SM can provide while building or studying new models. 

A not so subtle hint that can be gained from the discussions carried in that chapter is that there {\it probably} exists exotic particles in the universe which are not contemplated by the SM. Almost all usual solutions of its problems imply their existence: Neutrino masses imply either new right-handed degrees of freedom or exotic scalars (or both); Solutions to the hierarchy problem imply particles capable to protect the Higgs mass against its quadratic sensitivity to high scales, such as sfermions in supersymmetric theories or scalars in twin Higgs models; The Pecci-Quinn mechanism, which deals with the Strong $\mathit{CP}$ problem, is usually implemented on enlarged scalar sectors; Within the particle content of the SM there is no viable and sufficient candidate for Dark Matter, so that a solution itself amounts to the introduction of some exotic species; And, finally, models which strengthen thermal phase transition, thus allowing sufficient baryogenesis, also contain an enlarged particle spectrum. In fact, the situation is more drastic since {\it realistic} models which feature such mechanisms in order to solve one or, ideally, many of the issues, generally imply an even larger particle content, with, many times, extra spin-0, spin-1/2 {\it and} spin-1 particles simultaneously.

\begin{figure}[t!]
\centering
\makebox[0pt][c]{%
\includegraphics[trim={0 0 20mm 0}, clip, width=0.45\linewidth]{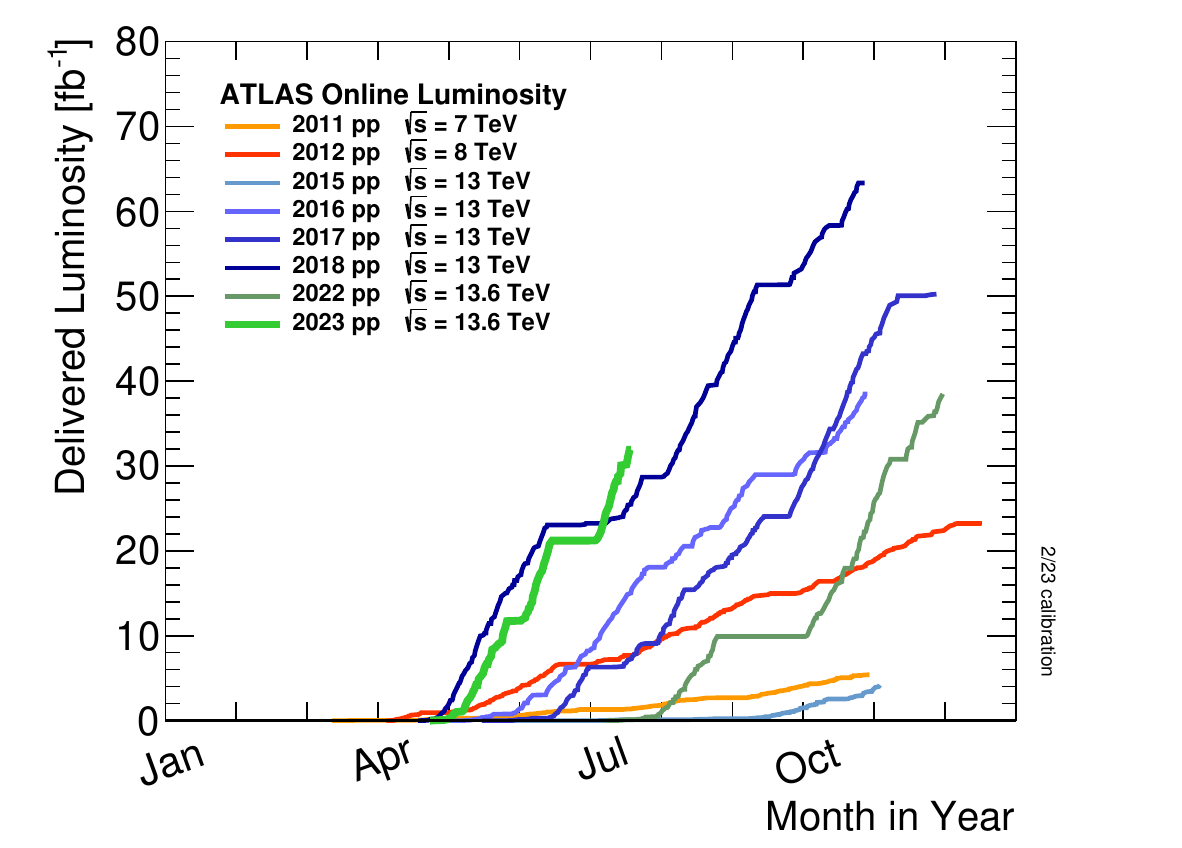}
\includegraphics[width=0.54\linewidth]{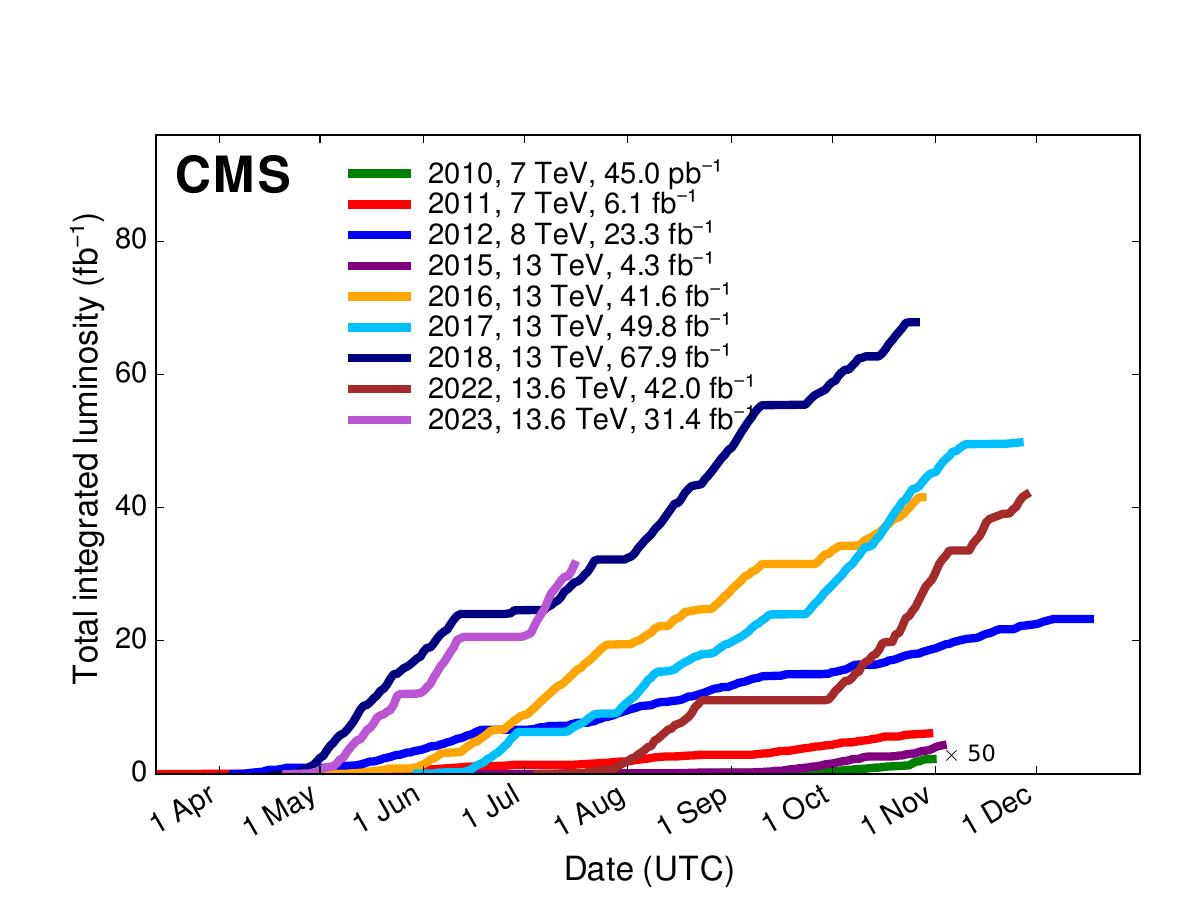}}
\caption[LHC cumulative integrated luminosity.]{Cumulative integrated luminosity of the Atlas and CMS experiments along each working year. Taken from~\cite{atlaslumi} and~\cite{CMSlumi}, respectively.}
\label{fig:lumi}
\end{figure}

The problem is that there is no direct evidence -- such as a previously unobserved excess of events indicating a resonance -- of these new elementary particles. Such experimentation is performed at industrial capacity at the LHC, currently at Run 3 colliding hadrons with a center-of-mass energy of $\sqrt{s}=\SI{13.6}{TeV}$. The cumulative integrated luminosity along the three runs within the Atlas and CMS experiments (the largest groups operating at the LHC) can be seen in Figure~\ref{fig:lumi}. The total luminosity delivered by the LHC along all runs sums around $\SI{266}{fb^{-1}}$. A better notion of the amount of collision data produced by this experiment, by far the most energetic available, can be obtained through the expected average luminosity for run 3: it implies a data production, at the end of works, of $600$ petabytes. Within all the information accumulated thus far there is no signs of an exotic species.

In this difficult scenario, it remains to perform exploratory phenomenology, taking advantage of the ever increasing bounds on physical quantities (particularly, the pure and direct cross sections on processes) to impose restrictions on the parameters of any hypothesized particle, theory or generic concept. Doing this, we may constrain the viable theory space to a smaller sector, hopefully guiding both theoretical and experimental efforts towards (or at least closer to) the right direction. This is exactly the spirit of the first part of this thesis, in which we do not deal directly with the SM issues, but, rather, perform skeptical, model-independent particle phenomenology.

Among the many members of the set of motivated hypothetical particles, one stands out because of its rarity in BSM models: the doubly-charged vector bilepton. While most of the other species -- specially bosons -- are present in several ultraviolet complete or effective models, the doubly-charged spin-1 boson is known to be contained within the m331 and in a $SU(15)$ grand unification theory -- and that is all. The fact that this species is so unique in terms of its existence within interesting models means that constraining its parameter space represents crucial information gain. This entity is the central object of our phenomenological inquiries. 

At this point, because it is the only `low energy' model which contains a $U^{\pm \pm}$ and because the second part of this text will regard it directly, an introductory review of the m331 is in order.

\section{3-3-1 Model: original motivations}

Although the 3-3-1 group had already been considered as a symmetry for the electroweak interactions in the 70s~\cite{Lee:1977qs} (similarly, a $SU(3)$ unified electroweak interaction had been proposed~\cite{Fritzsch:1976dq}), it was not until many years later that the model was fully constructed with the modern representation content. To understand its original motivation, consider the diagram of Figure~\ref{fig:nunu_tchann}, of the first order contribution to the process $\nu_e \bar{\nu}_e \to W^+ W^-$ in a phenomenologic 1-family theory of leptons. The diagram in the left, portraying the $t$-channel electron exchange, is known to violate unitarity in the production of a longitudinally polarized $W$ pair, with a cross section behaving at high energies as

\begin{equation} \label{eq:UnitViol}
\sigma(\nu \bar{\nu} \to W_0^+ W_0^-) \xrightarrow[s \gg M_W]{} \frac{G_F^2 s}{3 \pi}, 
\end{equation} 
where $G_F=g^2/4\sqrt{2}M_W^2=v_W^2/\sqrt{2}$ is the Fermi constant and $s$ is the center-of-mass energy squared. In the SM, gauge invariance comes to the rescue, and the diagram in Fig.~\ref{fig:nunu_schann}, of the $s$-channel exchange of a $Z$-boson, exactly cancel the high energy divergent $s$-dependence of the cross section above. 

\begin{figure}[t!]
	\centering
	\begin{subfigure}{0.33\linewidth}
	\centering
		\includegraphics[width=\linewidth]{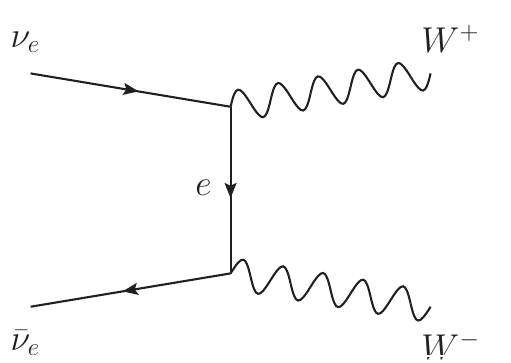}
	\caption{ } \label{fig:nunu_tchann}
	\end{subfigure}
	\begin{subfigure}{0.45\linewidth}
	\centering
		\includegraphics[width=\linewidth]{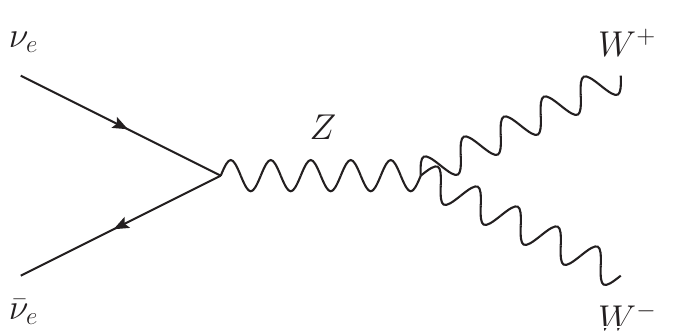}
	\caption{}	\label{fig:nunu_schann}
	\end{subfigure}
	\caption[Representation of how unitarity is saved in the production of a $W_0 W_0$ pair by the existence of the $Z$.]{(a): $t$-channel $e$ exchange contribution to the production of a pair of longitudinally polarized $W$ bosons. This graph is non-unitary if considered alone. (b): $s$-channel $Z$ boson exchange which saves the unitarity of the model at high energies.}
\end{figure}

Now, suppose that an exotic singly charged vector boson $W^{\prime \pm}$, henceforth called $V^\pm$, exists, and that it couples to right-handed SM currents or, equivalently, violates lepton number. Then, our BSM model predicts the first order contribution to the process $e^- e^- \to W^- V^-$  that is shown in Fig.~\ref{fig:ee_tchann}. The vertex and phase space structure of this graph is of the same form of that of~\ref{fig:nunu_tchann}, hence the theory inherits the same violation of unitarity depicted by Eq.~(\ref{eq:UnitViol}). It also follows that the high energy perturbativity of the theory may be saved by the same mechanism of the $s$-channel exchange of a vector boson belonging to the gauge group of $V^{\pm}$ (since a gauge invariant theory is unitary). The corresponding graph appears in Fig.~\ref{fig:ee_schann}. In this case, however, the spin-1 intermediate particle must be doubly-charged.

This mechanism to cure the non-unitarity in right-handed currents of exotic charged vector bosons motivated Pisano and Pleitez~\cite{Pisano:1992bxx} to propose the m331 as an extension of the electroweak theory, independently realized by Frampton~\cite{Frampton:1992wt}, whose interest was in producing a theory with a doubly-charged vector bilepton, interesting in itself.

\begin{figure}[t!]
	\centering
	\begin{subfigure}{0.33\linewidth}
	\centering
		\includegraphics[width=\linewidth]{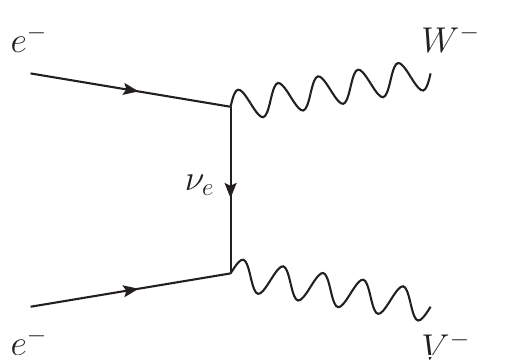}
	\caption{ } 	\label{fig:ee_tchann}
	\end{subfigure}
	\begin{subfigure}{0.45\linewidth}
	\centering
		\includegraphics[width=\linewidth]{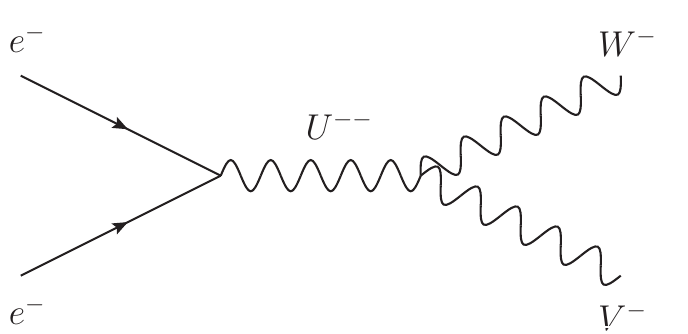}
	\caption{} 	\label{fig:ee_schann}
	\end{subfigure}
	\caption[The badly divergent $t$-channel $WV$ production amplitude and the $U$ $s$-channel exchange that cures it.]{The same mechanism that cured non-unitarity in the $W_0 W_0$ production in the SM now working to cure the bad high energy behaviour in the production of an exotic $V^\pm$. The intermediate $s$-channel particle, however, now must be doubly-charged.}
\end{figure}

\section{The representation content of the minimal version}

We are interested mainly in the minimal version of the 3-3-1 and begin describing its representation content. The first novelty relative to the SM is that all the leptonic degrees of freedom (of a same generation) are grouped within a same multiplet

\begin{equation}
L_{\ell}\equiv \begin{pmatrix}
\nu_\ell \\
\ell \\
\ell^c
\end{pmatrix}_{L} \thicksim (\mathbf{1},\mathbf{3},0), \;\;\;\; \ell=e,\mu,\tau.
\end{equation}
Note the explicit charge conjugation on the third components: this is the source of some subtleties in deriving the Feynman rules for the model.

In the quark sector, occurs the introduction of three exotic Dirac fermions: two $j_i$, of electric charge $-4/3$, and the $J$, of electric charge $5/3$. The complete representation content is given by

\begin{alignat}{2}
Q_{iL}&\equiv \begin{pmatrix}
d_i \\
-u_i \\
j_i
\end{pmatrix}_{L} \hspace{-1.4mm} \thicksim (\mathbf{3},\mathbf{\overline{3}},-1/3),&& \;\;\;\;\; Q_{3L}\equiv\begin{pmatrix}
u_3 \\
d_3 \\
J
\end{pmatrix}_{L} \hspace{-1.4mm} \thicksim (\mathbf{3},\mathbf{3},2/3)   \nonumber \\ 
&\;\;\;\;\; u_{\alpha R} \thicksim \left(\mathbf{3},\mathbf{1},2/3\right), &&d_{\alpha R} \thicksim \left(\mathbf{3},\mathbf{1},-1/3\right) \\
&\;\;\;\;\;J_{R} \thicksim \left(\mathbf{3},\mathbf{1},5/3\right), &&j_{i R} \thicksim \left(\mathbf{3},\mathbf{1},-4/3\right)  \nonumber \\
& \;\;\;\;\;\;\;\;\;\;\;\;\;\;\;\;\; i=1,2;  \;\;\;\, \alpha=&&1,2,3 \nonumber. 
\end{alignat}
Note that the fermion content is not democratic as the left-handed triplets of the first two quark generations form an anti-fundamental representation, contrary to the other triplets which are fundamental.

The minimal scalar sector~\cite{Foot:1992rh,Ng:1992st} is now understood to be composed of three triplets

\begin{equation}
\begin{split}
\eta\equiv \begin{pmatrix}
\eta^0 \\
\eta_1^- \\
\eta_2^+
\end{pmatrix}_{L} \hspace{-1.4mm} \thicksim (\mathbf{1},\mathbf{3},0), \;\;\; \rho\equiv\begin{pmatrix}
\rho^+ \\
\rho^0 \\
\rho^{++}
\end{pmatrix}_{L} \hspace{-1.4mm} \thicksim (\mathbf{1},\mathbf{3},1), \;\;\; 
 \chi&\equiv\begin{pmatrix}
\chi^- \\
\chi^{--} \\
\chi^0
\end{pmatrix}_{L} \hspace{-1.4mm} \thicksim (\mathbf{1},\mathbf{3},-1),
\end{split}
\end{equation}
and a sextet

\begin{equation}
S \equiv \begin{pmatrix}
\sigma_1^0 & \frac{h_2^+}{\sqrt{2}} & \frac{h_1^-}{\sqrt{2}} \\
\frac{h_2^+}{\sqrt{2}} & H_1^{++} & \frac{\sigma_2^0}{\sqrt{2}} \\
\frac{h_1^-}{\sqrt{2}} & \frac{\sigma_2^0}{\sqrt{2}} & H_1^{--}
\end{pmatrix} \thicksim (\mathbf{1},\mathbf{\overline{6}},0).
\end{equation}

The electric charge operator, from which the assigned values may be confirmed, is defined (in units of $e$) as the following combination of the diagonal generators

\begin{table}[t!]
\centering
\caption{3-3-1 gauge groups and their relevant symbols. The SM symbols appear as well since the theory is an intermediate stage of the 3-3-1 spontaneous symmetry breaking pattern. Note that gauge bosons of different groups are referred to by the same letter (this is a strict abuse of notation) because there is little opportunity for confusion. Notice, also, that the left-handed couplings are now called $g_{3L}$ and $g_{2L}$ (instead of simply $g$), making explicit reference to the group they relate to -- and the same goes for $g_X$ and $g_Y$ (in place of $g^\prime$).}\label{Tab:3-3-1symbols}
\begin{tabular}{cccc}
\toprule
			 & Gauge boson & Coupling constant & Generator
 \\ \midrule
   $SU(3)_c$ &  $g^a_\mu$    &  $g_s$        &  $\frac{1}{2}\lambda_c^a$ \\
   $SU(3)_L$ &  $W^a_\mu$    &  $g_{3L}$          &  $\frac{1}{2}\lambda^a$ \\
   $U(1)_X$  &  $b_\mu$    &    $g_X$   &  $X$ \\
   $SU(2)_L$ &  $W^a_\mu$    &  $g_{2L}$          &  $\frac{1}{2}\sigma^a$ \\
   $U(1)_Y$  &  $b_\mu$    &  $g_Y$   &  $Y$ \\
   $U(1)_\text{EM}$ &  $A_\mu$    &  $e$   &  $Q$ \\ \toprule
\end{tabular}
\end{table}

\begin{equation}\label{eq:Charge3-3-1}
Q=T_3-\beta T_8+X.
\end{equation} 
The $\beta$ parameter is defined by the version under consideration and, in particular, the minimal model is obtained by $\beta=\sqrt{3}$. The generators, as every other symbol parametrizing the gauge group of the 3-3-1, appear in Table~\ref{Tab:3-3-1symbols}.

It is useful to write down the general 3-3-1 covariant derivative

\begin{equation}\label{eq:3-3-1cov}
D_\mu=\partial_\mu - ig_s\frac{\lambda_c^a}{2} g^a_\mu- ig_{3L}\frac{\lambda^a}{2} W^a_\mu - ig_X X B_\mu,
\end{equation}
in which appear the vector adjoint multiplets that (besides the unphysical ghost degrees of freedom) complete the theory:

\begin{equation}
\begin{split}
g_\mu^a  & \thicksim (\mathbf{8},\mathbf{1},0) \\
W_\mu^a  & \thicksim (\mathbf{1},\mathbf{8},0) \\
b_\mu  & \thicksim (\mathbf{1},\mathbf{1},0).
\end{split}
\end{equation}

To finalize this section, we note how the full representation content is projected onto the SM symmetry in Table~\ref{tab:SMProjection}.

\begin{table}[t!]
\caption{Representation content of the m331 projected onto the SM symmetry.}
\begin{center}
\begin{tabular}{cc}
\toprule
Source multiplet	 & Projected multiplets
 \\ \midrule
$L_\ell$ & $L^{(2)}_{\ell}\equiv
\begin{pmatrix}
\nu_\ell \\
\ell 
\end{pmatrix}_{L} \hspace{-1.4mm} \thicksim (\mathbf{1},\mathbf{2},-1/2) \;\;\;\; \ell_R \thicksim \left(\mathbf{1},\mathbf{1},-1\right)$ \\ \midrule
$Q_{iL}$ &  \makecell{$Q^{(2)}_{iL}\equiv \begin{pmatrix}
d_i \\
u_i 
\end{pmatrix}_{L} \hspace{-1.4mm} \thicksim (\mathbf{3},\mathbf{\overline{2}},1/6) \;\;\;\; u_{i R} \thicksim \left(\mathbf{3},\mathbf{1},2/3\right) \;\;\;\; d_{i R} \thicksim \left(\mathbf{3},\mathbf{1},-1/3\right)$ \\
$j_{i L} \thicksim \left(\mathbf{3},\mathbf{1},-4/3\right) \;\;\;\; j_{i R} \thicksim \left(\mathbf{3},\mathbf{1},-4/3\right)$} \\ \midrule
$Q_{3L}$ &  \makecell{$Q^{(2)}_{3L}\equiv \begin{pmatrix}
u_3 \\
d_3 
\end{pmatrix}_{L} \hspace{-1.4mm} \thicksim (\mathbf{3},\mathbf{2},1/6) \;\;\;\; u_{3 R} \thicksim \left(\mathbf{3},\mathbf{1},2/3\right) \;\;\;\; d_{3 R} \thicksim \left(\mathbf{3},\mathbf{1},-1/3\right)$ \\ 
$J_{ L} \thicksim \left(\mathbf{3},\mathbf{1},5/3\right) \;\;\;\; J_{R} \thicksim \left(\mathbf{3},\mathbf{1},5/3\right)$} \\ \midrule
$\eta$   &  $\phi_\eta \equiv \begin{pmatrix}
\eta^0 \\
\eta_1^- 
\end{pmatrix} \hspace{-1.4mm}\thicksim (\mathbf{1},\mathbf{2},-1/2) \;\;\;\; \eta^{+}_2 \thicksim (\mathbf{1},\mathbf{1},1)$ \\ \midrule
$\rho$   &   $\phi_\rho\equiv \begin{pmatrix}
\rho^+ \\
\rho^0
\end{pmatrix} \hspace{-1.4mm} \thicksim (\mathbf{1},\mathbf{2},1/2) \;\;\;\; \rho^{++} \thicksim (\mathbf{1},\mathbf{1},2)$ \\ \midrule
$\chi$    &    $\phi_\chi\equiv\begin{pmatrix}
\chi^- \\
\chi^{--}
\end{pmatrix} \hspace{-1.4mm} \thicksim (\mathbf{1},\mathbf{2},-3/2) \;\;\;\; \chi^{0} \thicksim (\mathbf{1},\mathbf{1},0)$ \\ \midrule
$S$   &   \makecell{$\phi_S\equiv \begin{pmatrix}
h_1^- \\
\sigma_2^{0}
\end{pmatrix} \hspace{-1.4mm} \thicksim (\mathbf{1},\mathbf{\overline{2}},-1/2) \;\;\;\; H^{--} \thicksim (\mathbf{1},\mathbf{1},-2)$ \\
$\Phi_S \equiv \begin{pmatrix}
\sigma_1^0 & \frac{h_2^+}{\sqrt{2}} \\
\frac{h_2^+}{\sqrt{2}} & H_1^{++}
\end{pmatrix} \thicksim (\mathbf{1},\mathbf{\overline{3}},1)$}
\\ \toprule
\end{tabular}\label{tab:SMProjection}
\end{center}
\end{table}

\section{Spontaneous symmetry breaking}

The scalar potential is the most general gauge invariant, renormalizable functional of the scalar degrees of freedom. The true maximal functional that fulfil these requisites, however, is not the most frequently postulated because it prevents the possibility of conferring a consistent lepton number to each scalar. The popular potential which does allow for generalized lepton number conservation is

\begin{equation}
\begin{split}
V_1(\eta,\rho,\chi,S)&=\mu_1^2 \eta^\dagger \eta + \mu_2^2 \rho^\dagger \rho + \mu_3^2 \chi^\dagger \chi + \lambda_1 (\eta^\dagger \eta)^2 +\lambda_2 (\rho^\dagger \rho)^2 \\
&+\lambda_3 (\chi^\dagger \chi)^2 +\lambda_4 (\eta^\dagger \eta)(\rho^\dagger \rho) + \lambda_5 (\eta^\dagger \eta)(\chi^\dagger \chi) + \lambda_6 (\rho^\dagger \rho)(\chi^\dagger \chi) \\
&+\lambda_7 (\eta^\dagger \rho)(\rho^\dagger \eta) +\lambda_8 (\eta^\dagger \chi)(\chi^\dagger \eta) +\lambda_9 (\chi^\dagger \rho)(\rho^\dagger \chi) \\
&+ f_1(\eta\rho\chi + \text{H.C.}) \\
&+\mu_4^2 \text{Tr}(S^\dagger S)+\lambda_{10}[\text{Tr}(S^\dagger S)]^2+\lambda_{11}\text{Tr}[(S^\dagger S)^2] \\
&+ [\lambda_{12}(\eta^\dagger \eta)+\lambda_{13}(\rho^\dagger \rho)+\lambda_{13}(\chi^\dagger \chi)]\text{Tr}(S^\dagger S) \\
&+ f_2(\rho\chi S+ \text{H.C.}) \\
&+\lambda_{15}\eta^\dagger S S^\dagger \eta +\lambda_{16}\rho^\dagger S S^\dagger \rho + \lambda_{17}\chi^\dagger S S^\dagger \chi  \\
&+(\lambda_{19}\rho^\dagger S \rho \eta + \lambda_{20}\chi^\dagger S \chi \eta + \lambda_{21}\eta^2 S S+\text{H.C.}),
\end{split}
\end{equation}
where each term schematically represent the invariant contraction -- for instance, the term  $\eta\rho\chi$ is expanded out as $\epsilon^{ijk}\eta_i \rho_j \chi_k$, which is invariant since $\epsilon^{ijk}\eta_i \rho_j \sim (\mathbf{3}\otimes \mathbf{3})_A = \bar{\mathbf{3}}$, where the $A$ subscript here indicates that the irreducible anti-symmetric part is taken. The last two lines of the potential above are usually omitted, although there is no good first principle reason to do so (additional discrete symmetries that could be enforced to forbid that portion of the potential would accidentally forbid other terms and harmfully enhance the symmetry of the potential, generating further unwanted Goldstone bosons). The elimination of some parameters through minimization of the potential above and the obtaining and diagonalization of the mass matrices are straightforward and, because they do not pertain to our study and, in particular, are not necessary for the understanding of the symmetry breaking pattern, are not performed here. For a complete treatment of these matters we refer the reader to~\cite{Diaz:2003dk}. 

To get a first sense of what to expect of the SSB, let us recall the projection of the m331 scalars capable of acquiring a VEV onto the SM:

\begin{equation}
\begin{gathered}
\phi_\eta \equiv \begin{pmatrix}
\eta^0 \\
\eta_1^- 
\end{pmatrix}_{L} \hspace{-1.4mm}\thicksim (\mathbf{1},\mathbf{2},-1/2), \;\;\; \phi_\rho  \equiv \begin{pmatrix}
\rho^+ \\
\rho^0
\end{pmatrix}_{L} \hspace{-1.4mm} \thicksim (\mathbf{1},\mathbf{2},1/2), \;\;\; 
 \phi_S\equiv \begin{pmatrix}
h_1^- \\
\sigma_2^{0}
\end{pmatrix}_{L} \hspace{-1.4mm} \thicksim (\mathbf{1},\mathbf{\overline{2}},-1/2) \\
\Phi_S \equiv \begin{pmatrix}
\sigma_1^0 & \frac{h_2^+}{\sqrt{2}} \\
\frac{h_2^+}{\sqrt{2}} & H_1^{++}
\end{pmatrix} \thicksim (\mathbf{1},\mathbf{\overline{3}},1), \;\;\;\; \chi^{0} \thicksim (\mathbf{1},\mathbf{1},0).
\end{gathered}
\end{equation}
Looking above, it may be realized that the condensation of $\phi_\eta,\phi_\rho,\phi_S$ and $\Phi_S$ all trigger the pattern $SU(2)_L \times U(1)_Y \to U(1)_\mathrm{EM}$, {\it i.e.}, the SM SSB. In the other hand, $\chi^0$ is absolutely neutral from the SM point of view and cannot contribute to the SSB. This hints that the $\chi$-triplet alone is responsible for the m331 descent into the SM. 

With this finding in mind, consider the theory at high energies, in which case the active symmetry is the full $SU(3)_c \times SU(3)_L \times U(1)_X$. The VEV to be examined is

\begin{equation}
\langle \chi \rangle = \frac{1}{\sqrt{2}}\begin{pmatrix}
0 \\
0 \\
v_\chi
\end{pmatrix}\hspace{-1.4mm} \thicksim (\mathbf{1},\mathbf{3},-1).
\end{equation}
and the complete set of generators of the electroweak symmetry read

\begin{alignat*}{3}
T_1 &= \frac{1}{2}\begin{pmatrix} 0 & 1 & 0 \\ 1 & 0 & 0 \\ 0 & 0 & 0 \end{pmatrix}, \qquad
T_2 &&= \frac{1}{2}\begin{pmatrix} 0 & -i & 0 \\ i & 0 & 0 \\ 0 & 0 & 0 \end{pmatrix},\qquad
T_3 &&= \frac{1}{2}\begin{pmatrix} 1 & 0 & 0 \\ 0 & -1 & 0 \\ 0 & 0 & 0 \end{pmatrix} \\
T_4 &= \frac{1}{2}\begin{pmatrix} 0 & 0 & 1 \\ 0 & 0 & 0 \\ 1 & 0 & 0 \end{pmatrix},\qquad
T_5 &&= \frac{1}{2}\begin{pmatrix} 0 & 0 & -i \\ 0 & 0 & 0 \\ i & 0 & 0 \end{pmatrix} \\
T_6 &= \frac{1}{2}\begin{pmatrix} 0 & 0 & 0 \\ 0 & 0 & 1 \\ 0 & 1 & 0 \end{pmatrix}, \qquad
T_7 &&= \frac{1}{2}\begin{pmatrix} 0 & 0 & 0 \\ 0 & 0 & -i \\ 0 & i & 0 \end{pmatrix},\qquad
T_8 &&= \frac{1}{2\sqrt{3}} \begin{pmatrix} 1 & 0 & 0 \\ 0 & 1 & 0 \\ 0 & 0 & -2 \end{pmatrix} \\
&X = \begin{pmatrix} 1 & 0 & 0 \\ 0 & 1 & 0 \\ 0 & 0 & 1 \end{pmatrix} .&&
\end{alignat*}
It is imediate to see that 

\begin{equation}
\begin{split}
T_1 \langle \chi \rangle &= 0 \\
T_2 \langle \chi \rangle &= 0 \\
T_3 \langle \chi \rangle &= 0 \\
\left(-\sqrt{3} T_8 +X\right)\langle \chi \rangle &= 0. \\
\end{split}
\end{equation}
The four above are the only generators which annihilate the vaccum, and the symmetry generated by the five remaining ones is spontaneously broken. 

Now, the conserved $\{T_1,T_2,T_3\}$ manifestly generate a $SU(2)$ restricted to the uppermost two components of the triplets. Furthermore, from the definitions of the charge operators within the m331 and the SM as seen in Eqs.~(\ref{eq:ChargeSM}) and~(\ref{eq:Charge3-3-1}), respectively, we find that the embedding of the weak hypercharge into the m331 is given by 

\begin{equation} \label{eq:charge}
Y = - \sqrt{3} T_8 + X,
\end{equation}
which corresponds precisely to the fourth conserved generator!

We have thus found that the SSB of the m331 proceeds through the acquisition of a VEV by $\chi$, triggering $SU(3)_c \times SU(3)_L \times U(1)_X \to SU(3)_c \times SU(2)_L \times U(1)_Y$. And, through the VEVs of every other neutral scalar, the usual SM breaking is guaranteed.

%

\section{Vector boson eigenstates and masses}\label{sec:3-3-1VectorMasses}

The representation of the adjoint charge operator may be readily obtained as

\begin{equation}
\mathfrak{ad}_Q = 
\begin{pmatrix}
0 & -i &   &    &   &     &    &   &  \\
i & 0  & 0 &    &   &     &    &   &  \\
  & 0  & 0 & 0  &   &     &    &   &  \\
  &    & 0 & 0  & i &     &    &   &  \\
  &    &   & -i & 0 & 0   &    &   &  \\
  &    &   &    & 0 & 0   & 2i &   &  \\
  &    &   &    &   & -2i & 0  & 0 &  \\
  &    &   &    &   &     & 0  & 0 & 0 \\
  &    &   &    &   &     &    & 0 & 0 
\end{pmatrix}
\end{equation}
where the basis have been ordered as $\{T_1,T_2,\cdots,T_8,X\}$. $\mathfrak{ad}_Q$ is easily diagonalized, with eigenvectors

\begin{equation}
\begin{split}
T_{\pm} &= \frac{ T_1 \pm i T_2}{\sqrt{2}} \\
T^\prime_{\pm} &= \frac{ T_4 \mp i T_5}{\sqrt{2}} \\
T_{\pm \pm} &= \frac{ T_6 \mp i T_7}{\sqrt{2}},
\end{split}
\end{equation}
plus a degenerate three-dimensional eigenspace of neutral generators. Now, the gauge boson eigenstates associated to the generators above can be found by decomposing the corresponding sector of the algebra valued gauge multiplet $W^i_\mu T_i$ in terms of the generators with definite charge:

\begin{equation}
\begin{split}
\smash[b]{\sum_{T_i \in B_{\mathfrak{g}_c}} 
T^{}_i W^i_{\mu} }
&\equiv  T^{}_{+}W^+_{\mu} + T^{}_{-}W^-_{\mu}  \\
& \quad + T'_{+}V^+_{\mu}   + T'_{-}V^-_{\mu}   \\
& \quad + T^{}_{++}U^{++}_{\mu} + T^{}_{--}U^{--}_{\mu}\,,
\end{split}
\end{equation}
where $B_{\mathfrak{g}_c} = \{T_1,T_2,T_4,T_5,T_6,T_7\}$ is a basis of the charged sector of the algebra. The result for the charged gauge boson mass eigenstates in terms of the symmetry eigenstates, finally, reads

\begin{equation}
\begin{split}
W_\mu^\pm&=\frac{W_\mu^1\mp iW_\mu^2}{\sqrt{2}} \\
V_\mu^\pm&=\frac{W_\mu^4\pm iW_\mu^5}{\sqrt{2}} \\
U_\mu^{\pm\pm}&=\frac{W_\mu^6\pm iW_\mu^7}{\sqrt{2}}. \\
\end{split}
\end{equation}
Denoting the VEVs by $\langle \eta^0 \rangle = \frac{v_\eta}{\sqrt{2}}$, $\langle \rho^0 \rangle = \frac{v_\rho}{\sqrt{2}}$, $\langle \chi^0 \rangle = \frac{v_\chi}{\sqrt{2}}$, $\langle \sigma_2^0 \rangle = v_{s2}$  and $\langle \sigma_1^0 \rangle = \frac{v_s}{\sqrt{2}}$, their masses may be directly calculated from the scalar kinetic terms, to give

\begin{equation} \label{eq:chargedbmass}
\begin{split}
M_W^2&=\frac{1}{4}g^2 v_W^2 \\
M_V^2&=\frac{1}{4}g^2( v_\eta^2 + 2v_s^2+v_\chi^2)\\
M_U^2&=\frac{1}{4}g^2( v_\rho^2 + 2v_s^2+v_\chi^2).
\end{split}
\end{equation}

The diagonalization of the neutral sector is more involved, and there is no way to escape the mass matrix, given by 

\begin{equation}
M^2 = \frac{g^2 v_\chi^2}{4}\begin{pmatrix}
\bar{v}_W^2 & \frac{1}{\sqrt{3}}(\bar{v}_W^2-2\bar{v}_\rho^2) & -2t \bar{v}_\rho^2 \\
\frac{1}{\sqrt{3}}(\bar{v}_W^2-2\bar{v}_\rho^2) & \frac{1}{3}(\bar{v}_W^2 + 4) & 
\frac{2}{\sqrt{3}}t(\bar{v}_\rho^2 + 2) \\
-2t \bar{v}_\rho^2 & \frac{2}{\sqrt{3}}t(\bar{v}_\rho^2 + 2) & 4t^2(\bar{v}_\rho^2 + 1)
\end{pmatrix},
\end{equation}
where the overbar indicates the ratio by $v_\chi$ as $\bar{v}_\alpha\equiv \frac{v_\alpha}{v_\chi}$. To enhance clarity, in this chapter we avoid abbreviations expressing the relation between mass and symmetry eigenstates, which we calculate to be

\begin{equation}
\begin{split}
Z_{1\mu} &= \frac{Z_{13} W_{3\mu} + Z_{18} W_{8\mu} + Z_{1B} B_{\mu}}{N_1} \\
Z_{2\mu} &= \frac{Z_{23} W_{3\mu} + Z_{28} W_{8\mu} + Z_{2B} B_{\mu}}{N_2} \\
A_{\mu} &= \frac{A_{3} W_{3\mu} + A_{8} W_{8\mu} + A_{B} B_{\mu}}{N_A}, \\
\end{split}
\end{equation}
where $Z_{1\mu}$ components are given by 

\begin{equation}
\begin{split}
Z_{13} &= \frac{-\bar{v}_W^4+\left[(3t^2+2)\bar{v}_\rho^2 +3t^2+1 + R \right]\bar{v}_W^2-2(3t^2 +1)\bar{v}_\rho^4}{t\left[(3t^2+1)\bar{v}_\rho^4 +\left(3t^2+1 - R\right)\bar{v}_\rho^2 - \bar{v}_W^2 \right]} \\
Z_{18} &= \frac{-\bar{v}_W^4+\left[(3t^2+2)\bar{v}_\rho^2 +3t^2-1 + R \right]\bar{v}_W^2-2\bar{v}_\rho^2\left( R - 3t^2-1\right)}{\sqrt{3}t\left[(3t^2+1)\bar{v}_\rho^4 +\left(3t^2+1 - R\right)\bar{v}_\rho^2 - \bar{v}_W^2 \right]} \\
Z_{1B} &= 2;
\end{split}
\end{equation}
and the $Z_{2\mu}$ ones read

\begin{equation}
\begin{split}
Z_{23} &= \frac{-\bar{v}_W^4+\left[(3t^2+2)\bar{v}_\rho^2 +3t^2+1 - R \right]\bar{v}_W^2-2(3t^2 +1)\bar{v}_\rho^4}{t\left[(3t^2+1)\bar{v}_\rho^4 +\left(3t^2+1 + R\right)\bar{v}_\rho^2 - \bar{v}_W^2 \right]} \\
Z_{28} &= \frac{-\bar{v}_W^4-\left[-(3t^2+2)\bar{v}_\rho^2 -3t^2+1 + R \right]\bar{v}_W^2+2\bar{v}_\rho^2\left( R + (3t^2+1)\right)}{\sqrt{3}t\left[(3t^2+1)\bar{v}_\rho^4 +\left(3t^2+1 + R\right)\bar{v}_\rho^2 - \bar{v}_W^2 \right]} \\
Z_{2B} &= 2;
\end{split}
\end{equation}
for $A_\mu$, we have

\begin{equation}
\begin{split}
A_3 &= t \\
A_8 &= -\sqrt{3}t \\
A_B &= 1.
\end{split}
\end{equation}
The normalization factors have been naturally defined as

\begin{equation}
N_X = \sqrt{c_{X3}^2+c_{X8}^2 +c_{XB}^2}, 
\end{equation} 
for $X_\mu = (c_{X3}W_{3\mu}+c_{X8}W_{8\mu}+c_{XB}B_{\mu})/N_X$. The $R$ factor, appearing in most expressions, is given by

\begin{multline}
R \equiv \left\{\left[\bar{v}_W^2 - (3+6t^2)\bar{v}_\rho^2-1-6t^2\right]\bar{v}_W^2\right. + \\
\left. {} +(1+3t^2)\left[ 3(1+t^2)\bar{v}_\rho^4+6t^2\bar{v}_\rho^2 +1 +3t^2\right] \right\}^{1/2},
\end{multline}
with which the masses may be written

\begin{equation} \label{eq:neutralMasses}
\begin{split}
M_A^2 &= 0 \\
M_{Z_1}^2 & = \frac{g_{3L}^2 }{6}v_\chi^2\left( \bar{v}_W^2 + 3t^2\bar{v}_\rho^2 + 1 + 3t^2 -R\right) \\
M_{Z_2}^2 & = \frac{g_{3L}^2 }{6}v_\chi^2\left( \bar{v}_W^2 + 3t^2\bar{v}_\rho^2 + 1 + 3t^2 +R\right).
\end{split}
\end{equation}

\section{Yukawa interactions and fermion masses}

The Yukawa Lagrangian is composed by the most general renormalizable, gauge invariant functional bilinear on fermion fields and linear on scalars. Its primordial objective is that of generating fermion masses through SSB in a renormalizable fashion. Let us begin treating the leptonic Yukawa sector, which is instructive because it sheds light over the scalar sector of the m331. 

Note first that a mass term could be projected out of the product $\mathbf{3} \otimes \mathbf{3} = \bar{\mathbf{3}}_A \oplus \mathbf{6}_S$, which already represents great motivation for the existence of triplets and sextets within the scalar sector of the model. Furthermore, notice that leptons are $X=0$ representations of $U(1)_X$, which means they can only form an invariant with $\eta$ and $S$. Since $\bar{\mathbf{3}}$ is the anti-symmetric part of $\mathbf{3} \otimes \mathbf{3}$, the triplet term is given by

\begin{equation}
\begin{split}
G_{ab}^\eta \overline{(L_{ai})^c}L_{bj} \epsilon^{ijk} \eta_k + \mathrm{H.C.}&=G^\eta_{ab}\begin{pmatrix}
\overline{\nu_{aR}^c} & \overline{\ell_{aR}^c} & \overline{\ell_{aR}}
\end{pmatrix} \begin{pmatrix}
0 & 0 & 0 \\
0 & 0 & 1 \\
0 & -1 & 0
\end{pmatrix} \begin{pmatrix}
\nu_{bL} \\ 
\ell_{bL} \\
\ell_{bL}^c
\end{pmatrix}
\frac{v_\eta}{\sqrt{2}} \\
&=G_{ab}^\eta \frac{v_\eta}{\sqrt{2}} \left[\overline{\ell^c_{aR}}\ell^c_{bL}-\overline{\ell_{aR}}\ell_{bL}\right] + \mathrm{H.C.}.
\end{split}
\end{equation}
where we have projected out the vacuum component of $\eta$. In the expression above, $G^\eta$ is one {\it a priori} arbitrary Yukawa matrix which, from the symmetry structure of the starting expression, may be taken symmetric. Notice also that

\begin{equation}
\begin{split}
\overline{\ell_{aR}^c}\ell^c_{bL} &=\overline{\ell_{aL}^{\hphantom{aL}c}}\ell^{\hphantom{bR}c}_{bR} = \left(\overline{\ell_{aL}^{\hphantom{aL}c}}\ell^{\hphantom{bR}c}_{bR}\right)^T = -\left(\ell^{\hphantom{bR}c}_{bR}\right)^T \gamma_0^T \left(\ell_{aL}^{\hphantom{aL}c}\right)^* \\
& = -\left(\ell^{\hphantom{bR}c}_{bR}\right)^T \gamma_0^T C^* \left(\overline{\ell_{aL}}^T\right)^* = -\left(\ell^{\hphantom{bR}c}_{bR}\right)^T \gamma_0^T C^* \gamma_0^\dagger \ell_{aL}  \\
& = -\left(\ell^{\hphantom{bR}c}_{bR}\right)^T C C^{-1} \gamma_0^T C \gamma_0^\dagger \ell_{aL} = 
\left(\ell^{\hphantom{bR}c}_{bR}\right)^T C \ell_{aL}  \\
& =\overline{\ell_{bR}} \ell_{aL},
\end{split}
\end{equation}
where we have used $(\ell_L)^c=(\ell^c)_R$, and $\ell^c = C \bar{\ell}^T$. The charge conjugation matrix obeys $C^{-1} \gamma_\mu C = -\gamma_\mu^T$ and $C = C^*=-C^T=-C^\dagger=-C^{-1}$. Using the equation above, together with the anti-symmetry of $G^\eta$, we find

\begin{equation}
G_{ab}^\eta \overline{(L_{ai})^c}L_{bj} \epsilon^{ijk} \eta_k+ \mathrm{H.C.}=-\frac{2v_\eta}{\sqrt{2}} G_{ab}^\eta \left(\overline{\ell_{aR}}\ell_{bL}+ \mathrm{H.C.}\right),
\end{equation}
stating clearly that it gives a mass term.

Moving on to the term generated by the sextet, we have

\begin{equation} \label{eq:masssim}
\begin{split}
G_{ab}^S \overline{(L_{ai})^c}L_{bj} S^{ij} &=G^S_{ab}\begin{pmatrix}
\overline{\nu_{aR}^c} & \overline{\ell_{aR}^c} & \overline{\ell_{aR}}
\end{pmatrix} \begin{pmatrix}
\frac{v_{s}}{\sqrt{2}} & 0 & 0 \\
0 & 0 & \frac{v_{s_2}}{\sqrt{2}} \\
0 & \frac{v_{s_2}}{\sqrt{2}} & 0
\end{pmatrix} \begin{pmatrix}
\overline{\nu_{bL}} \\ 
\overline{\ell_{bL}} \\
 \overline{\ell_{bL}^c}
\end{pmatrix} \\
&=G_{ab}^S \frac{v_{s_2}}{\sqrt{2}} [\overline{(\ell^c_{a})_R}(\ell^c_{b})_{L}+\overline{\ell_{aR}}\ell_{bL}] + 
G_{ab}^S \frac{v_{s}}{\sqrt{2}} \overline{(\nu^c_{a})_R}\nu_{bL},
\end{split}
\end{equation}
which, with the same manipulations of the last case, morphs into

\begin{equation}
\begin{split}
G_{ab}^S \overline{(L_{ai})^c}L_{bj} S^{ij}+ \mathrm{H.C.}&=\frac{2v_{s_2}}{\sqrt{2}} G_{ab}^S \left(\overline{\ell_{aR}}\ell_{bL}+ \mathrm{H.C.}\right) \\
&\quad {} +G_{ab}^S \frac{v_{s}}{\sqrt{2}} \overline{(\nu^c_{a})_R}\nu_{bL}+ \mathrm{H.C.},
\end{split}
\end{equation}
so that the sextet generates a contribution to the leptonic Dirac masses and, if $v_s \neq 0$, a Majorana neutrino mass. The leptonic Yukawa Lagrangian as may then be written as

\begin{equation}\label{eq:YukawaLeptons}
\mathcal{L}_\ell^\text{Y}= \frac{1}{2}G_{ab}^\eta \overline{(L_{ai})^c}L_{bj} \epsilon^{ijk} \eta_k + \frac{1}{2}G_{ab}^S \overline{(L_{ai})^c}L_{bj} S^{ij}+ \text{H.C.}
\end{equation}
from which, and from the equations above, arises the lepton mass matrix

\begin{equation}
M_\ell= G_{ab}^\eta \frac{v_\eta}{\sqrt{2}} + G_{ab}^S \frac{v_{s2}}{2}.
\end{equation}
This matrix is, in principle, general, and may be diagonalized by a biunitary transformation like

\begin{equation}
\mathrm{diag}(m_e,m_\mu,m_\tau)=V_R^{\ell \dagger}M_\ell V_L^{\ell},
\end{equation}
where $V_{L(R)}^\ell$ relates symmetry and mass eigenstates as 

\begin{equation}
\ell'_{L(R)}=V^\ell_{L(R)}\ell_{L(R)}.
\end{equation}

This is a good point to seek insight into the scalar build of the model. Consider a theory in which the sextet has been omitted. In that case, $M_\ell$ is anti-symmetric and hence has a spectrum of the form $\{0,-m,m\}$, where the minus sign of the second eigenvalue may be removed by a chiral rotation. It is thus impossible to fit the three nonzero and non-equal lepton masses in a theory without this multiplet. 

Eq.~(\ref{eq:YukawaLeptons}) also gives rise to the following Majorana mass matrix for the neutrinos

\begin{equation}
M_\nu=G_{ab}^S \frac{v_{s}}{\sqrt{2}},
\end{equation}
by which another model building fact may be perceived. Suppose that the $X=0$ triplet had not been introduced, or that $v_\eta=0$ had been set. Then the lepton and neutrino mass matrices would be proportional $M_\ell \propto M_\nu$, and would thus be diagonalized by the same biunitary transformation. This, in turn, implies that the PMNS matrix obeys $V_\text{PMNS}=V_L^{\ell \dagger}V_L^\nu=\mathbb{1}$. Since this is phenomenologically unacceptable, the $\eta$-triplet is indispensable. 

To finalize the lepton sector discussion, let us consider the addition of a right-handed neutrino to the model. One may then add the extra piece to the Lagrangian 

\begin{equation}
\mathcal{L}_{\nu_R}^\text{Y}= \frac{1}{2}G_{ab}^{\nu}\overline{L_{ai}}\nu_{bR}\eta_i + \frac{1}{2}m^{\nu_R}_{ab}\overline{(\nu_{aR})^c}\nu_{bR} + \text{H.c.},
\end{equation}
giving, in total, the following neutrino mass matrix 

\begin{equation}
(M_\nu)=\begin{pmatrix}
  G^S_{3\times 3} \frac{v_{s}}{\sqrt{2}}&  G_{3\times 3}^{\nu} \frac{v_{\eta}}{\sqrt{2}}\\
 G_{3\times 3}^{\nu} \frac{v_{\eta}}{\sqrt{2}}& m_{3\times 3}^{\nu_R}
\end{pmatrix},
\end{equation} 
in the $\{\nu_L,\nu_R^{\hphantom{R}c}\}$ basis.

Once the lepton masses are fitted, it becomes clear that an extra triplet is necessary to arrange the quark masses. With this understood, we simply mention the results for the colored particles. The Yukawa Lagrangian is given by 

\begin{equation}
\mathcal{L}_q^\text{Y}=\overline{Q_{mL}'} [G_{m\alpha}^u U'_{\alpha R} \rho^* + G_{m\alpha}^d D'_{\alpha R} \eta^*] + \overline{Q_{3L}'} [F_{3\alpha}^{u} U'_{\alpha R} \rho + F_{3\alpha}^d D'_{\alpha R} \eta],
\end{equation}
where $G^{u(d)}$ are $2 \times 3$ and $F^{u(d)}$ $1 \times 3$ matrices. The resulting mass matrix for the known up-type quarks in the basis $\{-u,-c,t\}$ is

\begin{equation}
M_u=\begin{pmatrix}
r G^u_{11} & r G^u_{12} & r G^u_{13} \\
r G^u_{21} & r G^u_{22} & r G^u_{23} \\
F^u_{31} & F^u_{32} & F^u_{33} \\
\end{pmatrix} v_\eta;
\end{equation}
and the one for the down-type quarks in the $\{d,s,b\}$ basis reads

\begin{equation}
M_d=\begin{pmatrix}
r^{-1} G^d_{11} & r^{-1} G^d_{12} & r^{-1} G^d_{13} \\
r^{-1} G^d_{21} & r^{-1} G^d_{22} & r^{-1} G^d_{23} \\
F^d_{31} & F^d_{32} & F^d_{33} \\
\end{pmatrix} |v_\rho|;
\end{equation}
where we have defined $r\equiv v_\rho/v_\eta$.

The Yukawa lagrangian for the exotic quarks is easiest to construct 

\begin{equation}
\mathcal{L}_{j_m,J}^\text{Y} = K^j_{ab} \overline{Q_{aL}'} \chi^* j_{bR} + y^J \overline{Q_{3L}'} \chi J_R,
\end{equation}
giving the following $2\times 2$ mass matrix for the $j_i$

\begin{equation}
M^j_{ab} = K^j_{ab} \frac{v_\chi}{\sqrt{2}} 
\end{equation}
and, for the $J$ mass, 

\begin{equation}
M^J = y^J \frac{v_\chi}{\sqrt{2}}.
\end{equation}


\section{Anomaly cancellation}

Finally, it is instructive to verify that the gauge invariance of the m331 is unharmed by anomalies. Repeating the process of the end of Section~\ref{sec:SMReview}, we consider every triangle at a time

\begin{enumerate}[label={\bfseries\arabic*.}]
\item $\left[SU(3)_c\right]^3$:

This combination trivially vanishes by arguments already exhaustively explored.

\vspace*{0.7cm}

\item $\left[SU(3)_c\right]^2 \left[U(1)_X\right]$:

The corresponding anomaly factor reads

\begin{equation}
\begin{split}
\mathfrak{a}_{Xbc} &= \sum_f \frac{1}{4}\gamma_5 X\tr{\{\lambda_b,\lambda_c\}} =\sum_q \frac{1}{2}\gamma_5 X \delta_{bc}  \\
&=\frac{1}{2}\delta_{bc} \left[{} -2\times 3 \times \left(-\frac{1}{3}\right)-1\times 3\times \frac{2}{3}+3\times 1 \times \frac{2}{3} \right.\\
&\hspace*{1.4cm} \left. {} +3\times 1 \times \left(-\frac{1}{3}\right)+1\times 1 \times \frac{5}{3}+2\times 1 \times \left(-\frac{4}{3}\right)\right]=0,
\end{split}
\end{equation}
where handedness is this time included in the presign and the other factors have been ordered as

\begin{equation}
(\mathrm{\# \hspace*{0.25em} of \hspace*{0.25em} multiplets})\times (\mathrm{multiplicity}) \times (X);
\end{equation} 
The sum is only over quark fields. In particular, it may be noticed that the anomaly vanishes for any generation in isolation, even though they are treated differently.

\vspace*{0.7cm}

\item $\begin{cases} &\left[SU(3)_c\right]^2 \left[SU(3)_L \right] \\
 &\left[SU(3)_c\right] \left[SU(3)_L \right]^2 \\
 &\left[SU(3)_c\right] \left[SU(3)_L \right]\left[U(1)_X\right] \\
 &\left[SU(3)_c\right] \left[U(1)_X\right]^2 \\
 &\left[SU(3)_L\right] \left[U(1)_X\right]^2:
\end{cases}$

\vspace*{0.6em}

The combinatorics above all vanish because they are of the form 

\begin{equation}
\mathfrak{a} \propto \tr{\sigma} \hspace*{0.4em} \mathrm{or} \hspace*{0.42em} \tr{\lambda}.
\end{equation}

\vspace*{0.7cm}

\item $\left[SU(3)_L\right]^2 \left[U(1)_X\right]$:

This is similar to case 2, but now the sum runs over left-handed triplets

\begin{equation}
\begin{split}
\mathfrak{a}_{Xbc} &= \sum_f \frac{1}{4}\gamma_5 X\tr{\{\lambda_b,\lambda_c\}} =\sum_L \frac{1}{2}\gamma_5 X \delta_{bc}  \\
&=\frac{1}{2}\delta_{bc} \left[{}- 2\times \left(-\frac{1}{3}\right)-1 \times \frac{2}{3} {} \right]=0,
\end{split}
\end{equation}

\vspace*{0.7cm}
\item $\left[U(1)_X\right]^3$:

Straightforwardly:

\begin{equation}
\begin{split}
\mathfrak{a} &= \sum_f \gamma_5 \, X^3  \\
&= {} - 2\times 3 \times \left(-\frac{1}{3}\right)^3 - 1\times 3 \times \left(\frac{2}{3} \right)^3 + 3\times 1 \times \left(\frac{2}{3}\right)^3   \\
&\quad + 3\times 1 \times \left(-\frac{1}{3} \right)^3 + 2\times 1 \times \left(-\frac{4}{3} \right)^3 + 1 \times 1 \times \left(\frac{5}{3}\right)^3=0.
\end{split}
\end{equation}

\vspace*{0.7cm}
\item $\left[SU(3)_L\right]^3$:

Lastly, to analyse this case, consider

\begin{equation}
\begin{split}
\tr{\gamma^5 \bar{t}^a \left\{\bar{t}^b,\bar{t}^c\right\}} &= -\tr{\gamma^5 \left(t^a\right)^T \left\{\left(t^b\right)^T,\left(t^c\right)^T\right\}      }  \\
&= - \tr{\gamma^5 \left( \left\{t^c,t^b\right\}t^a\right)^T}= - \tr{\gamma^5 \left\{t^c,t^b\right\}t^a}\\
&= -\tr{\gamma^5 t^a \left\{t^b,t^c\right\}}.
\end{split}
\end{equation}
Above, $\bar{t}$ are the generators of the anti-fundamental, given by $\bar{t}^a=-\left(t^a\right)^T$, and we have used the invariance of the trace by transposition. What this shows is that components of an anti-fundamental contribute the opposite of those of a fundamental representation. The m331 is built in such a manner that the number of fermionic triplets equals that of anti-triplets: in the lepton sector, three triplets exist; within the colored spectrum, we have three triplets and six anti-triplets. This guarantees the vanishing of this 3-point function amplitude.

\end{enumerate}

The matter of anomaly cancellation in the m331 differs from it in the SM in a fundamental aspect: it only works when the three generations are taken into account. The last triangle we evaluated, corresponding to three $SU(3)_L$ currents, is enough to arrive at this conclusion. Note that, to achieve a version of the theory with equal numbers of fundamental and anti-fundamentals, the balance of two $\bar{\vectorB{3}}$ to one $\vectorB{3}$ must be maintained within the quark sector. This implies that the theory is only renormalizable if the number of generations is a multiple of three. Furthermore, to secure asymptotic freedom at high energies the number of quark flavours must be smaller than 16. This fixes the number of generations of the m331 to three, as six families, the next multiple of three available, corresponds to eighteen quarks and already surpasses the upper limit set by asymptotic freedom. In this way, the m331 offers a (at least partial) solution to the arbitrariness of the number of families.

\section{Closing the m331 numerically at the electroweak scale}\label{sec:3-3-1Solution}

The Equations~(\ref{eq:chargedbmass}) through~(\ref{eq:neutralMasses}), of masses and rotations between vector boson eigenstates, are highly dependent on the exotic scalar VEVs. Not only them, but also vertices and effective couplings between the various particles are reliant on these quantities. These objects, $v_\chi, v_\eta, v_\rho, v_{s_1}, v_{s_2}$, are all free parameters of the theory, with yet unknown values to be fitted through experimental data. They are not completely free, however, as they parametrize, in the m331, quantities that are dominated by pure SM phenomena. 

In this spirit, Dias {\it et al} found that imposing a relation between the VEVs identically causes known neutral gauge boson masses and neutral current parameters, at the electroweak scale, to descend into their known values~\cite{Dias:2006ns}. The solution to the {\it closing of the 3-3-1 symmetry at the electroweak scale}, as the authors called it, is given by 

\begin{equation} \label{eq:soldef}
v_\rho^2=\frac{1-4s_W^2}{2c_W^2}v_W^2.
\end{equation}
If the equation above is plugged into the exact expressions of the neutral gauge boson masses and neutral current (NC) parameters within the m331, their extremely involved forms descend into the experimentally verified SM predictions. Because some values differ from those presented in~\cite{Dias:2006ns}, we write down all the NC parameters, both general and constrained by the solution, in Appendix~\ref{app:couplings}. In particular, an interesting, albeit useless, consequence of the solution is that it sets a {\it theoretical} lower bound on the absolutely free parameter $v_\chi$, being $v_\chi > \SI{54}{GeV}$~\cite{Barela:2021phu}.

The RHS of Eq.\refEq{eq:soldef} involves only known numerical quantities and gives

\begin{equation}
v_\rho \approx \SI{54}{GeV}
\end{equation}
Moreover, $v_W$ is actually defined by fitting the Higgs mass and is given by $v_W^2 = v_\eta^2 + v_\rho^2 + v_{s_1}^2 + v_{s_2}^2$. Now, $v_{s_1}$ is expected to be small similarly to the VEV of any triplet extension of the SM~\cite{Montero:1999su}, and $v_{s_2}$ is assumed small simply because it does not contribute to the quark masses and this is the most immediate way to fit the lepton ones. Within this most common and reasonable benchmark, one obtains 

\begin{equation}
v_\eta \approx \SI{240}{GeV}.
\end{equation}

In this thesis, this solution will be the preferred choice whenever a numerical ansatz for the scalar VEVs is needed.

\cleardoublepage

\chapter{LHC phenomenology of flavour violating $U^{\pm\pm}$ processes}
\label{chap:LHCTrimuons}

This section is devoted to the first novel results from the project underlying this thesis. We perform a preliminary study of the LHC phenomenology of the vector bilepton, with the mixing between lepton flavours properly implemented. We start with a short catalogue of previous works and then move on to define a framework and derive results.

\section{Review of existing phenomenological literature}

In order to conduct a useful novel analysis, it is necessary to understand which
aspects of the $U^{\pm\pm}$ phenomenology have already been explored and which points have
not been focused on yet. This is to say that a brief review of the existing literature is necessary. Wanting to keep this recollection to a reasonable length, we focus only on LHC results, and for some great early pure leptonic analysis point the reader to~\cite{Tully:1999yg,Willmann:1998gd,Pleitez:1999ix}. Furthermore, because it is impossible to quote the full set of results that could be extracted from each work, we quote the points of exclusion or discovery of highest $M_U$, with the requirement that the experimental parameters are at most the current LHC characteristic ones of $\sqrt{s} \sim \SI{13}{TeV}$ and $\mathcal{L} \sim \SI{140}{fb^{-1}}$. 

In~\cite{Dutta:1994pd}, the authors use the process $ug \to U^{++}j_1$ to qualitatively examine the prospects of discovery of the $U^{++}$ and $j_1$. The remaining exotic quarks are deemed heavy $J, j_2 \gg U^{++}, j_1$. The object reaction occurs through $s$-channel exchange of an up quark or through $t$-channel exchange of a $j_1$. The detectable signal originate from the subsequent decay chains which, if $m_{j_1} > M_U$, are given by $U^{++} \to \ell^+ \ell^+$ and $j_1 \to U^{--} u \to \ell^- \ell^-u$. If $M_U > m_{j_1}$ , $U^{++}$
acquires the decay mode to $j_1 u$, in which case the $j_1$ proceeds to decay into charged leptons
via off-shell bileptons. The signal is again composed by four charged leptons, but
with a peak on only one pair. The authors calculate the cross section times branching ratio of subsequent decay for the $U^{++}$ and $U^{--}$ as a function of mass. They then perform a qualitative analysis of the LHC
background and behaviour of the signal for four benchmark $m_{j_1}$, remarking that the difference in the position of the intermediate resonance in the invariant mass of pairs of leptons (like-sign {\it vs}. different charge lepton pairs) plus the lower transverse momentum of the $Z$ mediated background are enough to guarantee spectacular signal. Finally, they state that,
for $\sqrt{s} = \SI{13}{TeV}$ and $\mathcal{L} = \SI{100}{fb^{-1}}$, masses of up to $M_U,m_{j_1} < \SI{1.5}{TeV}$ could be
explored at the LHC through around 100 events.

In~\cite{Dion:1998pw}, a variety of processes are considered in a model independent analysis which
evokes the m331 when necessary. The subset of their investigations which concern us look at the hard processes $q\bar{q} \to U^{++}U^{--}$, $q\bar{q} \to U^{++}e^-e^-$ and $u\bar{d} \to U^{++}V^-$, where
the first two reactions differ in that, although the signal is made up of four leptons in
both cases, in the first both $U$ are real, whereas in the second one of them is virtual. In
order to preserve unitarity, the authors add an exotic $Z^\prime$ to the analysis. They evaluate the full hadronic cross sections and show the results for four benchmark $M_{Z^\prime}$.  The authors define as discovery criteria a minimum cross section of $\SI{2.5}{fb}$ corresponding to around 25 events. The most useful process turns out to be $q\bar{q} \to U^{++}U^{--}$, and the most powerful findings indicate that bileptons with masses as high as $\SI{1}{TeV}$ may be explored at the LHC for $M_{Z^\prime}=1.8\sim \SI{3}{TeV}$.

\cite{Meirose:2011cs} carries a study of the specific four leptons signal $pp \to e^{\mp}e^{\mp}\mu^{\pm}\mu^{\pm}X$ at the LHC.
The authors fix the $Z^\prime$ mass, again included by unitarity, to $M_{Z^\prime} = \SI{1}{TeV}$, and chose
three benchmark values for all the exotic quark masses, considered the same. In view
of the reconstruction efficiency they set, the authors consider a $5\sigma$ discovery criteria to
match the requirement of five events, and use a bayesian technique to find a maximum
95\% exclusion cross section equivalent to the requirement of zero events of data. The
results show that, for an LHC run at $\sqrt{s} = \SI{14}{TeV}$ and of $\mathcal{L} = \SI{110}{fb^{-1}}$, with the
highest considered exotic quarks mass $M_Q = \SI{800}{GeV}$, bileptons of $\SI{1}{TeV}$ may be
excluded.

The object of Ref.~\cite{RamirezBarreto:2013edk} is basically the same. The authors consider pair production of doubly-charged vector bileptons through $s$-channel exchange of $\gamma, Z,Z^\prime$, {\it i.e.}, the full hard process given by $q\bar{q} \to \gamma,Z,Z^\prime \to U^{++}U^{--} \to e^{\mp}e^{\mp}\mu^{\pm}\mu^{\pm}$. The study focus specifically on the m331 Model, scanning over $M_U$ which, in this theory, is enough to fix the masses of the exotic $Z^\prime$ as well. As remarked in the earlier works, the observables of the invariant mass of like-sign leptons and lepton transverse momentum are verified to allow for distinction between signal and background. The authors plot the luminosity that is needed to discover the bilepton at a given mass for $\sqrt{s} = 7,8,\SI{14}{TeV}$, with the result that masses of $\sim \hspace*{-0.18em}\SI{700}{GeV}$ could be discovered at $\sqrt{s} = \SI{14}{TeV}$ with $\mathcal{L} = \SI{1}{fb^{-1}}$ of data.

Reference~\cite{Nepomuceno:2016jyr} considers for the first time results of the Run II, using the limits on the cross section for the process $pp \to \mu^{\mp}\mu^{\mp}\mu^{\pm}\mu^{\pm}X$ to explore bilepton phenomenology. The authors first analyse what was learned from Run I, exploiting the $\SI{4.7}{fb^{-1}}$ of data provided by the ATLAS experiment to derive exclusion contours on the $M_U \times m_Q$ plane, and find that for the highest $m_Q$ considered, of $\SI{600}{GeV}$, bileptons of up to $\SI{520}{GeV}$ are excluded by the LHC at $\sqrt{s} = \SI{7}{TeV}$. The authors then predicted that, for $m_Q = \SI{800}{GeV}$ and $M_{Z^\prime} = \SI{3}{TeV}$, the Run II at $\SI{13}{TeV}$ would improve this bounds up to way above $\SI{1}{TeV}$ for $\SI{100}{fb^{-1}}$ of integrated luminosity.

\begin{table}[t!]
\centering
\caption{Summary of the results of the literature we have reviewed. Most of the works considered several benchmarks, in which cases we have chosen the strongest $M_U$ bounds or, if the bounds are similar, the highest benchmark masses. Some benchmarks involve extra less important choices that we omit. Moreover, we select the exclusion bounds if presented, otherwise the discovery limits are shown (these are highlighted by an asterisk). }\label{Tab:Uliterature}
\begin{tabular}{>{\centering\arraybackslash}m{0.7in} >{\centering\arraybackslash}m{1.3in} >{\centering\arraybackslash}m{1.2in} >{\centering\arraybackslash}m{0.8in} >{\centering\arraybackslash}m{1.5in} @{}m{0pt}@{}
}
\toprule
			     & Process & Benchmarks & Exp. param. & Excluded
 \\ \midrule
\multicolumn{1}{c}{\raisebox{-2.35mm}[0pt][0pt]{\cite{Dutta:1994pd}}} & 
$\!\!\!\!\!\!\begin{aligned}[t]
ug &\to U^{++}j_1 \\
&\to (4\ell)u
\end{aligned}$ &  
\multicolumn{1}{c}{\raisebox{-2.35mm}[0pt][0pt]{$m_{j_1},m_J \gg m_{j_1}$}}       &  
$\!\!\!\!\!\!\begin{aligned}[t]
\sqrt{s} &= \SI{13}{TeV} \\
\mathcal{L} &= \SI{100}{fb^{-1}}
\end{aligned}$ & 
\multicolumn{1}{c}{\raisebox{-2.35mm}[0pt][0pt]{$M_U,m_{j_1}<\SI{1.5}{TeV} \hspace*{0.8mm} *$}}  \\[12mm]
\multicolumn{1}{c}{\raisebox{-2.35mm}[0pt][0pt]{\cite{Dion:1998pw}}} & 
$\!\!\!\!\!\!\begin{aligned}[t]
q\bar{q} &\to U^{++}U^{--} \\
&\to 4\ell
\end{aligned}$ &  
$\!\!\!\!\!\!\begin{aligned}[t]
 m_Q &= \SI{600}{GeV} \\
 M_{Z^\prime} &= \SI{3.0}{TeV}
\end{aligned}$     &  
$\!\!\!\!\!\!\begin{aligned}[t]
\sqrt{s} &= \SI{14}{TeV} \\
\mathcal{L} &= \SI{10}{fb^{-1}}
\end{aligned}$ & 
\multicolumn{1}{c}{\raisebox{-2.35mm}[0pt][0pt]{$M_U<\SI{1}{TeV} \hspace*{0.8mm} *$}}  \\[12mm]
\multicolumn{1}{c}{\raisebox{-2.35mm}[0pt][0pt]{\cite{Meirose:2011cs}}} & 
$\!\!\!\!\!\!\begin{aligned}[t]
pp &\to U^{\mp \mp}U^{\pm\pm}X \\
&\to e^{\mp}e^{\mp}\mu^{\pm}\mu^{\pm}X
\end{aligned}$ &  
$\!\!\!\!\!\!\begin{aligned}[t]
 m_Q &= \SI{800}{GeV} \\
 M_{Z^\prime} &= \SI{1}{TeV}
\end{aligned}$     &  
$\!\!\!\!\!\!\begin{aligned}[t]
\sqrt{s} &= \SI{14}{TeV} \\
\mathcal{L} &= \SI{110}{fb^{-1}}
\end{aligned}$ & 
\multicolumn{1}{c}{\raisebox{-2.35mm}[0pt][0pt]{$M_U<\SI{1}{TeV}$}}  \\[12mm]
\multicolumn{1}{c}{\raisebox{-2.35mm}[0pt][0pt]{\cite{RamirezBarreto:2013edk}}} & 
$\!\!\!\!\!\!\begin{aligned}[t]
pp &\to U^{\mp \mp}U^{\pm\pm} \\
&\to e^{\mp}e^{\mp}\mu^{\pm}\mu^{\pm}
\end{aligned}$ &  
%
\multicolumn{1}{c}{\raisebox{-2.35mm}[0pt][0pt]{$\!\!\!\!\!\! M_{Z^\prime} = \SI{2.6}{TeV}$}} &  
$\!\!\!\!\!\!\begin{aligned}[t]
\sqrt{s} &= \SI{14}{TeV} \\
\mathcal{L} &= \SI{1}{fb^{-1}}
\end{aligned}$ & 
\multicolumn{1}{c}{\raisebox{-2.35mm}[0pt][0pt]{$M_U<\SI{700}{GeV} \hspace*{0.8mm} *$}}  \\[12mm]
\multicolumn{1}{c}{\raisebox{-2.35mm}[0pt][0pt]{\cite{Nepomuceno:2016jyr}}} & 
$\!\!\!\!\!\!\begin{aligned}[t]
pp &\to U^{++}U^{--}X \\
&\to \mu^{+}\mu^{+}\mu^{-}\mu^{-}X
\end{aligned}$ &  
$\!\!\!\!\!\!\begin{aligned}[t]
 m_Q &= \SI{800}{GeV} \\
 M_{Z^\prime} &= \SI{3}{TeV}
\end{aligned}$ &  
$\!\!\!\!\!\!\begin{aligned}[t]
\sqrt{s} &= \SI{13}{TeV} \\
\mathcal{L} &= \SI{100}{fb^{-1}}
\end{aligned}$ & 
\multicolumn{1}{c}{\raisebox{-2.35mm}[0pt][0pt]{$M_U<\SI{1.1}{TeV}$}}  \\[12mm]
\multicolumn{1}{c}{\raisebox{-2.35mm}[0pt][0pt]{\cite{Corcella:2017dns,Corcella:2021mdl}}} & 
$\!\!\!\!\!\!\begin{aligned}[t]
pp &\to U^{++}U^{--}jj \\
&\to (4\ell)jj
\end{aligned}$ &  
$\!\!\!\!\!\!\begin{aligned}[t]
 m_Q &= \SI{1700}{GeV} \\
 M_{Z^\prime} &= \SI{3.229}{TeV}
\end{aligned}$ &
\multicolumn{1}{c}{\raisebox{-2.35mm}[0pt][0pt]{$\sqrt{s} = \SI{13}{TeV}$}} &
\multicolumn{1}{c}{\raisebox{-2.35mm}[0pt][0pt]{$M_U\approx\SI{874}{GeV} \hspace*{0.8mm} *$}}
\\[0.8cm] \toprule
\end{tabular}
\end{table}

References~\cite{Corcella:2017dns,Corcella:2021mdl} qualitatively check the reach of the LHC at $\sqrt{s} = \SI{13}{TeV}$ to observe scalar and vector bileptons for a fixed single benchmark choice. In the first work, the authors consider double bilepton production together with two jets, $pp \to U^{++}U^{--}jj \to \ell^+\ell^+\ell^-\ell^-jj$, and choose a single benchmark point for the complete m331 Model, including every exotic heavy quark and scalar parameter, which is coherent with every known quantity and the current phenomenological standing. This benchmark tests the doubly-charged vector bilepton mass of $M_U = \SI{873.3}{GeV}$. They carry a profound analysis of the possible discriminant observables to enhance signal strength, and predict that it is able to overcome the $ZZjj$ SM background and that a $\SI{873.3}{GeV}$ $U^{\pm \pm}$, in the single point chosen in parameter space, could not only be excluded, but discovered. The second work extends this investigation, considering also the contributions of the doubly-charged scalars of the model and vetoing final state jets.

We compile the strongest bounds of each reviewed paper in Table~\ref{Tab:Uliterature} for easy consult. One characteristic that is shared by all these works is that they ignore the mixing among leptons induced by the $V_U$ matrix in their interaction. Even if the processes that are described involve diagonal interactions, $U$ could still couple to different flavours with different strength. 

\section{Process and background}

In this work, the mass of every possibly contributing exotic particle other than the $U$ will be deemed heavy. In the m331, for instance, this first simplification amounts to the reasonable assumption $m_J,m_{j_i} \gg M_U$. In a process with two initial state particles, with the heavy quarks forbidden to participate, it is  clear that the most economical way to probe the $U$ come from signal final states with four charged leptons. The literature we have just reviewed considers lepton flavour conserving reactions or lepton flavour violating ones but with only diagonal $U$ interactions. The former are summarized, at parton level, by the simplest

\begin{equation}
q \bar{q} \to \ell^+ \ell^{+}\ell^{-} \ell^{-}.
\end{equation} 
This group has background, and SM contributions for this type of process appear on the diagrams of Figure~\ref{fig:bgLHC}.

\begin{figure}
\begin{center}
\includegraphics[width = 0.36\linewidth]{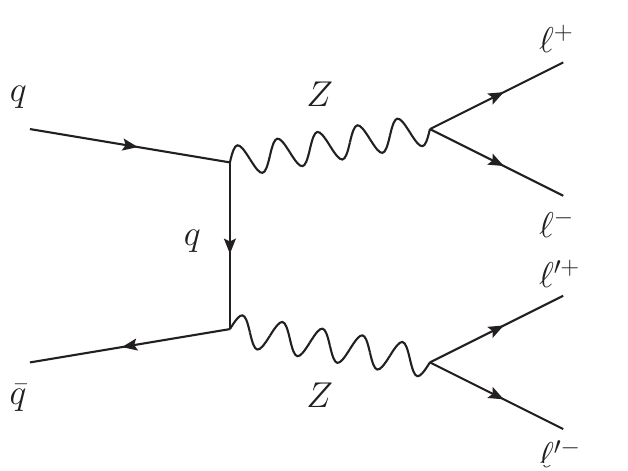}
\includegraphics[width = 0.42\linewidth]{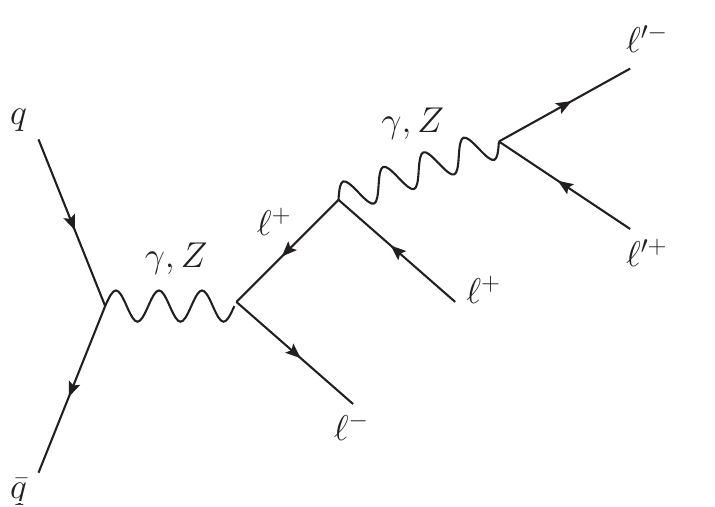}
\end{center}
\caption[SM contributions to $pp \to \ell^+ \ell^- \ell^{\prime +} \ell^{\prime -}$]{Diagrams contributing to the production of a (lepton flavour conserving) final state of 4 charged leptons in the SM.}\label{fig:bgLHC}
\end{figure}
To test the importance of this irreducible background, we evaluate the cross section at $\sqrt{s} = \SI{13}{TeV}$, with the usual kinematic cuts, from the individual contributions, for the specialized $4e$ final state $pp \to e^+e^-e^+e^-$. The results are

\begin{equation}
\sigma_{t: q} = \SI{5.3}{fb},
\end{equation}
for the cross section containing $t$-channel exchange of a quark and 

\begin{equation}
\sigma_{s: \ell} = \SI{0.83}{fb},
\end{equation}
for the cross section that features one $s$-channel exchange of a lepton. The total SM prediction for the rate of this process, featuring interference effects between the two types of contributions above, is given by

\begin{equation}
\sigma_\mathrm{total} = \SI{6.2}{fb}.
\end{equation}
With the luminosity usually associated with the LHC on phenomenological analysis in the present, $\SI{140}{fb^{-1}}$, this corresponds to about 900 events. Now, it has been shown that the distributions of observables such as invariant mass of pairs of leptons, transverse momentum and others hint that such SM contributions may be clearly distinguished from exotic physics. Nonetheless, a background production rate of this magnitude is an enourmous hindrance on the discovery potential of any weak scale physics.

Now, in general, the interaction of the $U^{\pm \pm}$ with leptons allows for a CKM-like matrix that mix different flavours. In particular, in the m331, such a matrix is almost unavoidable as a consequence of the fact that the lepton masses come from two distinct sources: $G^\eta$ and $G^s$. This prompts us to investigate CLFV processes, and if they could provide a possible smoking gun. Specifically, since these particles are more easily observable, our signal is comprised mostly by muons in the form of the trimuon process 

\begin{equation}
pp \to \mu^{\pm}\mu^{\pm}\mu^{\mp}e^{\mp}.
\end{equation}
As an aside, in the late seventies some events resulting from neutrino and anti-neutrino exposure  ending with two or three muons have been found~\cite{Holder:1977gp,Benvenuti:1977zp,Benvenuti:1977fb,Barish:1977bg} and the latter were called trimuon as well. No further events of this type appear to have been detected. 

The reason this process, in principle possible through $U^{\pm\pm}$ mediation, represents a strong road to information gain is because such reactions, as any with charged lepton flavour violation, is forbidden to all orders in perturbation theory in the SM (more on this in the next section). The $U$ contributions to this exotic process that interest us are given by the diagrams in Figure~\ref{fig:trimuonexotic}.  This signal is free of irreducible backgroud. By completeness, Figure~\ref{fig:exQuarksSignal} shows the contributions that would be given by the charge $-4/3$ and $5/3$ exotic quarks of the m331. These are assumed negligible by our condition on the masses of any additional exotic particles. 

\begin{figure}
\begin{center}
\includegraphics[width = 0.39\linewidth]{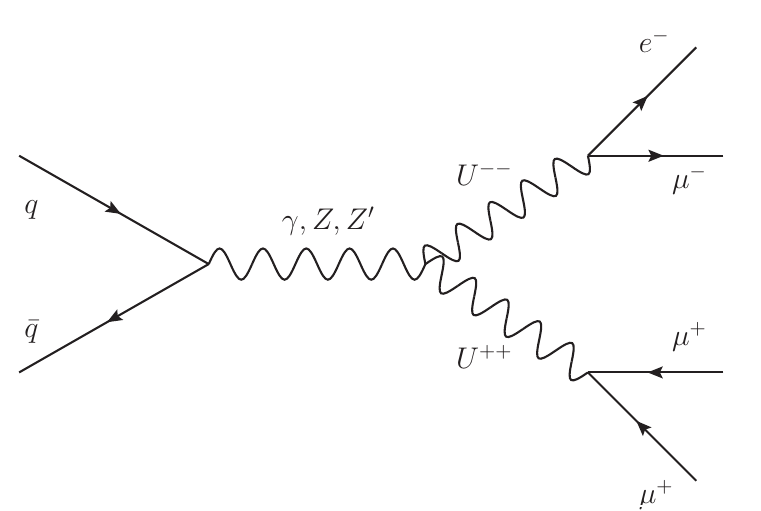}
\includegraphics[width = 0.39\linewidth]{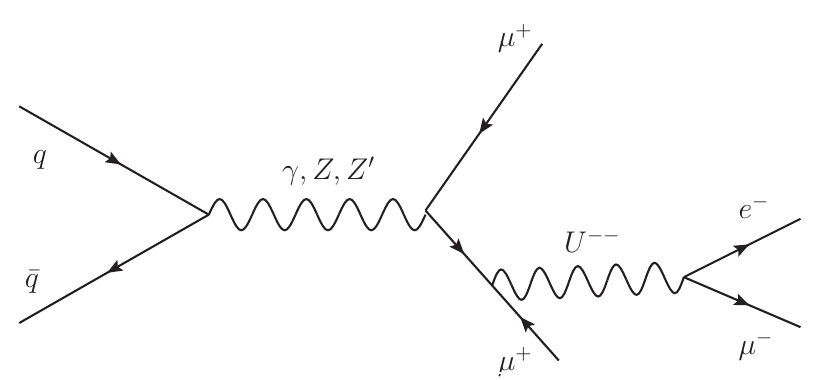}
\end{center}
\caption[m331 contributions to the trimuon process.]{m331 contributions to the trimuon process. Allowed permutations in the outgoing legs must also be considered.}\label{fig:trimuonexotic}
\end{figure}

\begin{figure}
\begin{center}
\includegraphics[width = 0.37\linewidth]{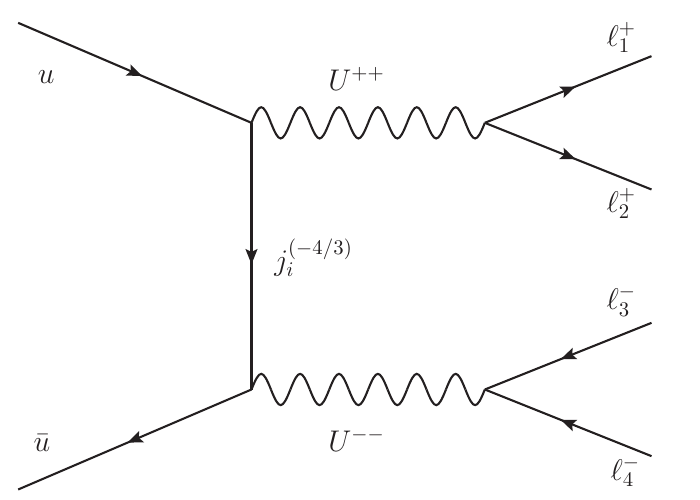}
\includegraphics[width = 0.37\linewidth]{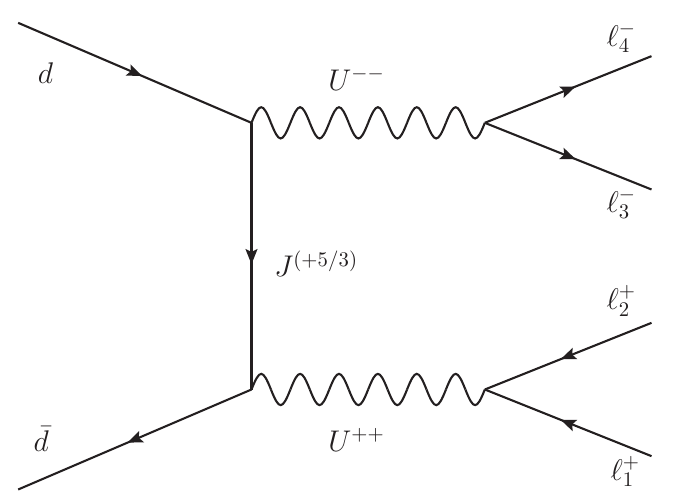}
\end{center}
\caption[Contributions of the m331 exotic quarks to $pp \to \ell^+_1 \ell^-_2 \ell^+_3 \ell^-_4$.]{$t$-channel exchange of exotic quarks that contribute to any four lepton final state at the LHC in the m331.}\label{fig:exQuarksSignal}
\end{figure}

\section{Parametrizing and numerically defining the necessary interactions}

We shall start with perfectly general interactions and then justify every numerical choice. Although our results are not fully model independent, they should cover an interesting sector of theory space. The most general form for the $U\ell \ell$ interaction that may be inferred from Lorentz and electromagnetic invariance is

\begin{equation}
\mathcal{L}_{U\ell\ell}=\sum_a g_U \overline{\ell_{a\phantom{c}L}^{\prime c}}\gamma^\mu \ell'_{aL}\thinspace U_\mu^{++}
+ g_U\overline{\ell'_{aL}}\gamma^\mu \ell^{\prime c}_{a\phantom{c}L}\thinspace U_\mu^{--}.
\end{equation}
The Lagrangian above is diagonal on symmetry eigenstates, the primed fields, because it supposedly comes from a minimally coupled kinetic term. In fact, if the underlying, ultraviolet complete model is not lepton universal, then there could be relative factors between flavours, but in this work we shall ignore this possibility. Note that a hand mirrored term $g^\prime_U \overline{\ell_{a\phantom{c}R}^{\prime c}}\gamma^\mu \ell'_{aR} U_\mu^{++}$ encompass all the same degrees of freedom. In fact, $g^\prime_U \overline{\ell_{a\phantom{c}R}^{\prime c}}\gamma^\mu \ell'_{aR} U_\mu^{++} = -g^\prime_U \overline{\ell_{a\phantom{c}L}^{\prime c}}\gamma^\mu \ell'_{aL} U_\mu^{++}$, so that adding the right-handed term would amount to a mere redefinition of the coupling (we shall come back to this and the following facts in the coming chapters).

The biunitary transformation whith rotates symmetry to mass eigenstates is defined by

\begin{equation}
\begin{split}
\ell'_{L} &\equiv V^\ell_{L} \ell_{L} \\
\ell'_{R} &\equiv V^\ell_{R} \ell_{R},
\end{split}
\end{equation}
from which one obtains
\\
\begin{equation}
\begin{split}
\overline{\ell'_{L}}&=\overline{\ell_{L}}V_L^{\ell\dagger} \\
(\ell^{\prime c})_L&=V_R^* (\ell^{c})_L \\
\overline{(\ell^{\prime c})_L}&=\overline{(\ell^{c})_L} V_R^T,
\end{split}
\end{equation}
and similarly for right-handed fields. With this, we write the $U\ell \ell$ interaction for mass eigenstates as

\begin{equation}\label{eq:UllMass1}
\mathcal{L}_{U\ell\ell}=\sum_{a,b} g_U\bar{\ell_a^c}\gamma^\mu P_L (V_U)_{ab} \ell_b\thinspace U_\mu^{++} + g_U\bar{\ell_a}\gamma^\mu P_L (V_U^\dagger)_{ab} \ell_b^c\thinspace U_\mu^{--},
\end{equation}
where we have defined the general mixing matrix 

\begin{equation}
V_U \equiv V_R^T V_L.
\end{equation}.

In Eq.\refEq{eq:UllMass1}, $g_U$ is an absolutely arbitrary coupling, except for possible unitarity bounds. Our first benchmark corresponds to using the m331 value for this parameter. This choice is natural, as this model is the only motivated TeV scale theory to feature such particle. Consider, in this theory, the algebra valued electroweak gauge field with the non-diagonal, charged sector in the physical basis

\begin{equation} \label{eq:Mkin}
\begin{split}
-i& g_{3L}\lambda^i W_\mu^i-ig_X X B_\mu \equiv -ig_{3L} \mathcal{M} \\
&=-i\frac{g_{3L}}{2}
\begin{pmatrix}
 W_\mu^3+\frac{1}{\sqrt{3}}W_\mu^8+2t_X XB_\mu & \sqrt{2}W_\mu^+ & \sqrt{2}V_\mu^- \\
 \sqrt{2}W_\mu^- & -W_\mu^3+\frac{1}{\sqrt{3}}W_\mu^8+2t_X XB_\mu & \sqrt{2}U_\mu^{--} \\
 \sqrt{2}V_\mu^+ & \sqrt{2}U_\mu^{++} & -\frac{2}{\sqrt{3}}W_\mu^8+2t_X XB_\mu
\end{pmatrix}.
\end{split}
\end{equation}
This matrix appears in the kinetic term of the left-handed leptons $-ig_{3L}\bar{L^\prime} \mathcal{M} L^\prime$, from which the form of the $U\ell \ell$ interaction may be extracted to be 

\begin{equation}\label{eq:LUllUll}
\mathcal{L}_{U\ell\ell} = -i\frac{g_{3L}}{\sqrt{2}}\overline{\ell_{L}^c}\gamma^\mu V_U \ell_{L}U^{++}_\mu + \mathrm{H.C.},
\end{equation}
in the mass basis. From this, we set

\begin{equation}
g_U = \frac{g_{3L}}{\sqrt{2}}.
\end{equation}

Another required interaction is the one which connects a pair of $U$ to the known $Z$ boson, which should generically be of the form  

\begin{equation}
\begin{split}
\mathcal{L}_{UUZ}&= f(g,v)\left\{ U^{++}_\mu \left[U^{--}_\alpha (\partial_\mu Z_\alpha )-Z_\alpha (\partial_\mu U^{--}_\alpha )\right] 
 \right.  \\
&\hspace*{18mm} {}+  U^{--}_\nu \left[Z_\alpha (\partial_\nu U^{++}_\alpha )-
U^{++}_\alpha (\partial_\nu Z_\alpha )\right]  
 \\
&\left. \hspace*{19.4mm} {}+ Z_\alpha \left[ U^{++}_\mu (\partial_\alpha U^{--}_\mu )-U^{--}_\mu (\partial_\alpha U^{++}_\mu )\right]\right\}.
\label{eq:UUZ}
\end{split}
\end{equation}
Above, $f(g,v)$ is a dimensionless effective coupling which, when the interaction above descends from an ultraviolet complete model, is a function of the various VEVs and gauge couplings of the theory. There are two natural options for the numerical definition of this quantity: ({\it i}) Setting $f(g,v) = g_{2L}c_W$ reproduces the $WWZ$ vertex of the SM. This represents an interesting choice as it would correspond to a preferred practice in the parametrizations of LHC searches;  ({\it ii}) Another possibility is to make use of the m331 prediction once more. In the m331 constrained by the {\it Solution to the closure} of Eq.~(\ref{eq:soldef}), the $f$ factor is given by $f(g,v) = (4s_W^2 - 1)/c_W$. Notice, in particular, that in this configuration the $U$ is $Z$-phobic since $s_W^2 \approx 0.23$. Only this latter alternative will be treated. 

It must be noted that to achieve the selected expression as the $UUZ$ vertex strength in the m331 the following implicit simplification must be made

\begin{equation}\label{eq:txFirstMention}
t_X^2 = \frac{s_W^2}{1-4s_W^2}.
\end{equation} 
This relation is ubiquitously utilized in m331 phenomenological studies, not always sensibly. This matter will be discussed in the last part of this thesis. 

A second triple gauge vertex that we must fixate corresponds to the $UU\gamma$ interaction. This time we are able to give it in a truly model independent way as

\begin{equation}
\begin{split}
\mathcal{L}_{UU\gamma}&= 2 Q_e(g) \left\{ U^{++}_\mu \left[U^{--}_\alpha (\partial_\mu A_\alpha )-A_\alpha (\partial_\mu U^{--}_\alpha )\right] 
 \right.  \\
&\hspace*{18.5mm} {}+  U^{--}_\nu \left[A_\alpha (\partial_\nu U^{++}_\alpha )-
U^{++}_\alpha (\partial_\nu A_\alpha )\right]  
 \\
&\left. \hspace*{19.9mm} {}+ A_\alpha \left[ U^{++}_\mu (\partial_\alpha U^{--}_\mu )-U^{--}_\mu (\partial_\alpha U^{++}_\mu )\right]\right\}.
\label{eq:UUA}
\end{split}
\end{equation}
It is easy to be convinced that the corresponding interaction must be of this form in any given model. This is because, from the point of view of the photon (which carries the force corresponding to a conserved symmetry), the $U$ is identical to the $W$, but with double the charge. As consequence, the Lagrangian must be analogous but with $Q_e \to 2Q_e$. The $Q_e(g)$ factor represents the fundamental electric charge of the theory as given in terms of the gauge couplings -- in the SM, $e = g_{2L} g_Y/\sqrt{g_Y^2 + g_{2L}^2}$.

Now, the $U$ could also couple to the SM Higgs, which would give an extra contribution to the process at hand. It is expected, however, that the vector mediated contributions dominate over the scalar ones. Usual arguments that lead to this conclusion are based upon the fact that there are less possibilities in spin space for the scalar channel, and upon the Goldstone Equivalence Theorem~\cite{Valencia:1990jp}. Nonetheless, since there is no reason to assume that the $UUH$ generic coupling strength is suppressed, let us take a closer look at how these contributions could be taken into account and at their importance within the m331.  

Even within this model, a strong benchmark is necessary, which amounts to the fixing of the unitary rotation $O$ between symmetry ($x_i^0$) and mass ($h_i^0$ -- unless otherwise stated, we denote the SM Higgs by $h^0_0$) eigenstates of $\mathit{CP}$-even neutral scalars, $\Ree x_i^0=O_{ij}h_j^0$. As the simplest construction to be used as avatar, consider the Higgs to correspond exactly to the $\mathit{CP}$-even part of one of the neutral symmetry eigenstates of the theory that originate from the triplets: $\chi^0$, $\eta^0$ or $\rho^0$. In order to find the best choice we mind each in turn, considering first the $\chi^0$. It cannot be an important $h^0$ component because it is absolutely neutral, sterile, from the SM perspective, and its only role is to break the 3-3-1 symmetry at a high scale. As for the $\eta^0$, consider the portion of the Lagrangian as parametrized in the ultraviolet completion that could originate the $UU\eta^0$ interaction. It comes from the scalar kinetic terms and reads 

\begin{equation}
(-ig\mathcal{M}^\mu \eta)^\dagger(-ig\mathcal{M}_\mu \eta) \propto 
\begin{pmatrix}
 \hphantom{\sqrt{2}U_\mu^{--}}& \hphantom{\sqrt{2}U_\mu^{--}} &  \hphantom{\sqrt{2}U_\mu^{--}}\\
 \hphantom{\sqrt{2}U_\mu^{--}} & \hphantom{\sqrt{2}U_\mu^{--}} & \sqrt{2}U_\mu^{--} \\
 \hphantom{\sqrt{2}U_\mu^{--}} & \sqrt{2}U_\mu^{++} & \hphantom{\sqrt{2}U_\mu^{--}}
\end{pmatrix}
\begin{pmatrix}
\frac{v_\eta+\text{Re}\, \eta^0+\text{Im} \, \eta^0}{\sqrt{2}} \\
\hphantom{\frac{v_\eta+\text{Re} \eta^0+\text{Im} \eta^0}{\sqrt{2}}} \\
 \hphantom{\frac{v_\eta+\text{Re} \eta^0+\text{Im} \eta^0}{\sqrt{2}}}
\end{pmatrix}=0.
\end{equation} 
In brief, a particle which is mostly $\Ree \eta^0$ does not interact with the $U$. The remaining alternative must then be the chosen one, and we identify $h_0^0 = \rho^0$. The interaction may be read from $g^2 (\mathcal{M}^\mu \eta)^\dagger (\mathcal{M}_\mu \eta)$ to be 

\begin{equation}\label{eq:UUHUUH}
\mathcal{L}_{UUH} = \frac{g^2}{2}v_{\rho} HU_\mu^{++} U^{--\mu}. 
\end{equation}
Now, as a ({\it very}) crude estimate of the relation of importance of this and the other contributions, we evaluate the widths of the channels $H \to U^{++} U^{--}$ and $\gamma \to U^{++} U^{--}$. Directly to the point, the $H$ involving process is more than 8 orders of magnitude smaller. With such a disparity, we may safely discard any Higgs involving contribution to such bilepton processes, at least to the current precision. 

Finally, the effective model appropriate to investigate the prospect of bilepton discovery at the LHC, through the trimuon processes, is given by 

\begin{equation}
\mathcal{L}_{e3\mu} \equiv \mathcal{L}_\mathrm{SM} + \mathcal{L}_{U^{\pm\pm}}+ \mathcal{L}_{U \ell \ell} + \mathcal{L}_{UUA} + \mathcal{L}_{UUZ} + \mathcal{L}_{UUH}.
\end{equation}
where

\begin{equation}
\mathcal{L}_{U \ell \ell} = -\frac{1}{2}(\partial_\mu U^{++}_\nu-\partial_\nu U^{++}_\mu)(\partial^\mu U^{--\nu}-\partial^\nu U^{--\mu})+M_U^2 U^{++}U^{--},
\end{equation}
and the other pieces may be verified in Eqs.\refEq{eq:LUllUll} through\refEq{eq:UUHUUH}.

As a last, possibly obvious, remark, we stress that the Lagrangian above is not unitary or gauge invariant. In particular, we do not include the $Z^\prime$ in the analysis, usually added to the effective model to preserve unitarity in some bilepton processes. Although this is good practice, we make the choice to focus on the sole $U$ contributions. The dangerous high energy behaviour should not cause any problems at the fixed partonic energies, and considering the $U$ alone simply turns our predictions into conservative bounds.

\section{Parameter space and simulation}

There are essentially ten free parameters that must be dealt with in $\mathcal{L}_\mathrm{e3\mu}$: $M_U$ plus the nine degrees of freedom of $V_U$.  It would be tempting to assume that this number is, in fact, much smaller because only the elements $(V_U)_{\mu \mu},(V_U)_{e \mu},(V_U)_{\mu e}$ enter the hard process. All the other ones, however, enter the calculation of the total width $\Gamma_U$, which does influence the results. In any case, the number must be reduced, and we start by considering $V_U$ real, ignoring the five phases of the entire unitary $V_U$. This turns $V_U$ into an orthogonal matrix and we are left with exactly four real parameters. Now an ansatz is needed to further cut this number down. 

By its apparent reasonableness, we chose that of considering a symmetric $V_U$. To understand the interplay between this and the unitarity requirement over the degrees of freedom of a general $3\times 3$ matrix, let us put forward a simple argument. First notice that orthogonality guarantees that there are four possible eigenvalue signatures, the `sum' of which generates our entire space. Let us analyse each signature in turn: ({\it i}) $\{ 1,1,1\}$: in this case there are no degrees of freedom as there is a single corresponding matrix -- the identity.  ({\it ii}) $\{-1,-1,-1\}$: Analogous to ({\it i}) but with $V_U = -\mathbb{1}$. ({\it iii}) $\{1,1,-1\}$ or $\{-1,-1,1\}$: Here there exists a two- and a one-dimensional eigenspaces. Now, the symmetry condition assures orthogonal diagonalization, {\it i.e.}, that the two eigenspaces are orthogonal. Hence, defining the one-dimensional eigenspace also fixes the two-dimensional one. This shows that a symmetric, orthogonal $3\times 3$ matrix features two degrees of freedom: the coordinates of a point in the unit sphere.

Our target parameter space may now be fully defined. $M_U$ is the most important quantity to be scanned over, and $V_U$ is reducible to two free elements, which are arbitrarily chosen to be $(V_U)_{11},(V_U)_{12}$, {\it i.e.}, $(V_U)_{ee}$ and $(V_U)_{e\mu}$. All the other matrix elements become numerically defined by a choice of these two through the orthogonality conditions (constrained by the symmetry requirement) given by

\begin{equation} \label{eq:orth}
\begin{split}
&(V_U)_{ee}^2+(V_U)_{e\mu}^2+(V_U)_{e\tau}^2=1 \\
&(V_U)_{ee}(V_U)_{e\mu}+(V_U)_{e\mu}(V_U)_{\mu\mu}+(V_U)_{e\tau}(V_U)_{\mu \tau}=0 \\
&(V_U)_{ee}(V_U)_{e\tau}+(V_U)_{e\mu}(V_U)_{\mu\tau}+(V_U)_{e\tau}(V_U)_{\tau \tau}=0 \\
&(V_U)_{e\mu}^2+(V_U)_{\mu\mu}^2+(V_U)_{\mu\tau}^2=1 \\
&(V_U)_{e\mu}(V_U)_{e\tau}+(V_U)_{\mu\mu}(V_U)_{\mu\tau}+(V_U)_{\mu\tau}(V_U)_{\tau \tau}=0 \\
&(V_U)_{e\tau}^2+(V_U)_{\mu\tau}^2+(V_U)_{\tau\tau}^2=1.
\end{split}
\end{equation} 

We explore this parameter space focusing on the bi-dimensional subspace composed by $M_U \times (V_U)_{e\mu}$. Although the existing literature already constrains masses of up to $\SI{1}{TeV}$, this work is innovative in that it takes into account the usually ignored mixing. Because of this, the scan in $M_U$ is effected starting from low masses. Masses are generated in the range $M_U \in \left(\SI{100}{GeV}, \SI{1200}{GeV}\right)$, in steps of $\SI{50}{GeV}$, and $(V_U)_{e\mu}$ is varied from 0.001 to 0.9 -- with the exception of when $(V_U)_{ee} = 0.9$, in which case it stops at $0.4$ -- through 12 strategically chosen points.. This 2D region is scanned four times, one for each $(V_U)_{ee}$ among $(V_U)_{ee} = 0.001,0.01,0.1,0.9$, which is the last free parameter to be dealt with.

To avoid the inconvenience and computational strain of manually doing the convolution of the hard partonic cross sections against the proton Parton Distribution Functions (PDF), we employ a Monte-Carlo generator. The $\mathcal{L}_{e3\mu}$ model is implemented through \texttt{FeynRules}~\cite{Alloul_2014,Degrande_2012}, which generates a {\it Universal FeynRules Output} that is then interfaced with the Monte-Carlo package \texttt{MadGraph}~\cite{Alwall_2014}. We generate $10^4$ events in each of the 1035 points in parameter space, subjected to the kinematic cuts 

\begin{equation} \label{eq:cuts}
\SI{1500}{GeV}>p_{T_\ell}>\SI{30}{GeV}, \;\;\; |\eta_\ell|<2.5, \;\;\; \Delta R_{\ell \ell}>0.4.
\end{equation}
$p_{T_\ell}$ is the transverse momentum (relative to the beam axis) of any charged lepton, and is limited from below to guarantee that the particles reach the detector and from above mainly to avoid (possibly uncertain) outliers. $\eta_\ell$ is the pseudorapidity and, in particle physics is defined as

\begin{equation}
\eta \equiv -\ln \tan\left(\frac{\theta}{2}\right) = \frac{1}{2} \ln \frac{|\vectorB{p}|+p_L}{|\vectorB{p}|-p_L},
\end{equation}
where $\theta$ is the polar angle with respect to the beam axis and $|\vectorB{p}|$ and $p_L$ are the magnitude of the momentum of the given particle and its longitudinal component, respectively. In other words, pseudorapidity is a pure measure of the polar angle and is limited from above by the familiar reason of avoiding that the species is lost along the track of the beam. Finally, $\Delta R_{\ell \ell}$ is the angular distance between any pair of leptons, defined as 

\begin{equation}
\Delta R \equiv \sqrt{\Delta \eta^2 + \Delta \phi^2},
\end{equation}
where $\Delta$ indicates the difference in the corresponding property between the two particles and $\phi$ is the azimutal angle. This definition uses $\Delta \eta$ over $\Delta \theta$ because it is invariant under longitudinal boosts. The same is clearly true for $\Delta \phi$ too, causing $\Delta R$ to be invariant under longitudinal Lorentz transformations as well. Finally, the reason for its lower limit is simply that exaggeratedly collinear particles are hard to distinguish.

\section{Statistics}

Now, our process has no background. This frees us from having to find a discriminant observable; arguing over resolution; and doing much formal statistics. The works we have reviewed in the beginning of this Chapter are separed by as much as 26 years, and also span a large range of choices for discovery criteria. There is one study which defines 25 events as necessary to announce a discovery, and another which seeks a single signal event as sufficient for an exclusion. Although we escaped from much of the statistics requirements, in order to at least motivate an exclusion criteria, let us perform a short, simplistic analysis (for amazing introductions to statistics in particle physics, see~\cite{Barlow:2019svl,Lista:2016chp}). 

In the border of the exclusion contours of a process without background, the number of events is, by construction, low. This forces us to employ a Poisson ansatz instead of a Gaussian one, which could work well for a large amount of events. The probability of an independent counting experiment (a single bin, if we were effecting a more complex analysis over a distribution in our case), with average count number $\mu$, to see $n$ events is given by 

\begin{equation}
P(n;\mu)=\frac{e^{-\mu}\mu^n}{n!}.
\end{equation} 
The most simple hypothesis testing method -- and recall that such methods are somewhat arbitrary, {\it i.e.}, although their motivations are easy to understand, they are not `proven' and their efficiency is not a mathematical fact -- would be to exclude models that fit the observation only outside their $95\%$ compatibility region. Specifically, assuming the separate expectations of signal ($\mu_s$) and background ($\mu_b$), we could then find what is the number of events $N_c$ corresponding to the threshold below which only $5\%$ ($\beta = 0.05$) of measurements would fall. In a binned analysis, this comes from

\begin{equation}\label{eq:CLsb}
\beta = \sum\limits_{N=0}^{N_c} P(N;\mu_B+\mu_S)= \sum\limits_{N=0}^{N_c} \frac{e^{-(\mu_B+\mu_S)}(\mu_B+\mu_S)^N}{N!}.
\end{equation}
This so-called $\mathit{CL}$ method, however, is oversensitive to downward flutuations of the background or, more simply, tends to exclude alternate hypothesis too strictly in cases with, for instance, small background. To remedy this obstacle, the $\mathit{CL}_s$ method is popularly used. It modifies the $\mathit{CL}$ by dividing it by the null hypothesis confidence level in the same distribution point, thus lowering the value of the threshold $N_c$. The formula becomes

\begin{equation}\label{eq:CLs}
\beta = \frac{\sum\limits_{N=0}^{N_c} P(N;\mu_B+\mu_S)}{\sum\limits_{N=0}^{N_c} P(N;\mu_B)}= \frac{\sum\limits_{N=0}^{N_c} \frac{e^{-(\mu_B+\mu_S)}(\mu_B+\mu_S)^N}{N!}}{\sum\limits_{N=0}^{N_c} \frac{e^{-\mu_B}\mu_B^N}{N!}}=\frac{\sum\limits_{N=0}^{N_c} \frac{e^{-\mu_S}(\mu_B+\mu_S)^N}{N!}}{\sum\limits_{N=0}^{N_c} \frac{\mu_B^N}{N!}}.
\end{equation}
To make it clear, $\beta$ is a predefined measure of certainty or, more specifically, relates to the confidence level as $\beta = 1 - \mathit{CL}$. 

We are interested in deriving exclusion, rather than discovery, contours on the $M_U \times (V_U)_{e\mu}$, and the recipe for that in particle physics suggests a 95\% confidence level (or $2\sigma$ in a gaussian analysis) band. Instead of inputting an expected $\mu_s$ and finding the threshold $N_c$, we invert the logic, to obtain what would be the bounds on $\mu_s$ if nothing unusual was observed, {\it i.e.}, with $N_c = \mu_b$. Finally, it is enough to numerically solve

\begin{equation}
0.05 = \frac{\sum\limits_{N=0}^{\epsilon} \frac{e^{-s}s^N}{N!}}{\sum\limits_{N=0}^{\epsilon} \frac{\epsilon^N}{N!}},
\end{equation} 
for $s$ to obtain an estimate of the excluded sector on parameter space. Above, we have estimated $\mu_s$ by the result of the Monte-Carlo pseudoexperiments $s$, and an arbitrarily small number $\epsilon$ is replaced for $\mu_b$. In the end, we have

\begin{equation}
s_c = -\ln 0.05 \approx 3,
\end{equation}
so that the points that are excluded by negative results at the LHC are those that predict more than three events. Technically, we have basically defined a single event as necessary for exclusion in a no-background situation, analogously to what is done in~\cite{Meirose:2011cs}, with the difference that they perform a bayesian analysis while our techniques are frequentist.

\section{Results}

The results are presented in Figure~\ref{fig:trimuonexclusion}. The number of events in each point is calculated as 

\begin{equation}
N=\sigma_\text{T} \times \mathcal{L}_\mathrm{int},
\end{equation}
and the points to the left of the curves are excluded at 95\% confidence level through $\mathcal{L}_\mathrm{int} = \SI{140}{fb^{-1}}$ of integrated luminosity (a more detailed distribution of the number of events appears in Figure~\ref{fig:trimuondetailed}).

\begin{figure}
\begin{center}
\includegraphics[width = 0.7\linewidth]{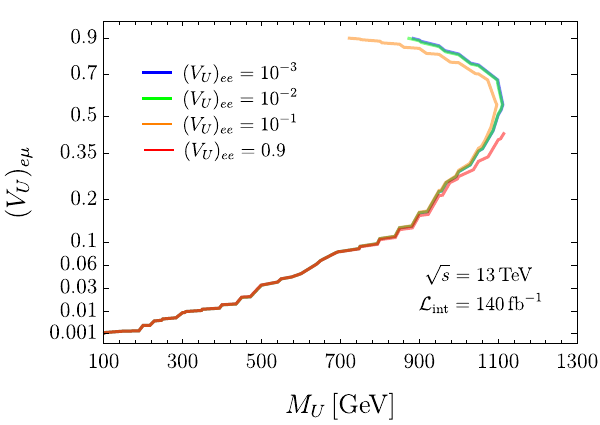}
\end{center}
\caption[95\% exclusion contour on the $M_U \times (V_U)_{e\mu}$ plane given by the trimuon process.]{95\% exclusion contour on the $M_U \times (V_U)_{e\mu}$ plane that would be implied by the non-observation of trimuon events in an LHC run of $\mathcal{L}_\mathrm{int} = \SI{140}{fb^{-1}}$. Four benchmark $(V_U)_{ee}$ are chosen, which arguably span the entire natural range for this parameter. In every $(V_U)_{e\mu}$ point, every other $V_U$ element becomes fixed by the orthogonality and ordinary matrix symmetry constraints.}\label{fig:trimuonexclusion}
\end{figure}

The first claim to be inferred from the plot is that the countour, in this specific arrangement, is close to independent from the choice of $(V_U)_{ee}$. To understand this, we note that, near the resonance, the amplitudes giving rise to the signal are proportional to 

\begin{equation}\label{eq:trimuonamp}
\mathcal{M}\left(q\bar{q}\to U^{++}U^{--} \to \mu^{\pm}\mu^{\pm}\mu^{\mp}e^{\mp}\right) \propto \frac{(V_U)_{\mu \mu}(V_U)_{e\mu}}{\Gamma_{U^{\pm \pm}}}.
\end{equation}
The leptonic part of the $U$ decay width, at least above the GeV scale, is proportional to 

\begin{equation}
\Gamma_{U^{\pm \pm}} \propto (V_U)_{ee}^2 + (V_U)_{\mu\mu}^2 + (V_U)_{\tau\tau}^2 + (V_U)_{e\mu}^2 + (V_U)_{e\tau}^2 + (V_U)_{\mu\tau}^2.
\end{equation} 
Together with the constraints of Equation\refEq{eq:orth}, this causes Eq.\refEq{eq:trimuonamp} to be resistant against a change in $(V_U)_{ee}$.

Finally, our results indicate that the $(V_U)_{e\mu}$ which delivers optimal resolution is given by $(V_U)_{e\mu} \sim 0.52$, and that the largest mass that can be excluded is around $\sim \hspace*{-0.18em} \SI{1100}{GeV}$.

For hierarchic matrices which are diagonal in the $\mu$ sector (or, possibly, favour the $\mu-\tau$ mixture to some extent), this process is essentially useless in constraining the $U^{\pm \pm}$ parameters, as can be seen in the lower end of the contours. This means that one should consider flavour diagonal $U$ processes simultaneously in order to be able to `corner' the excluded region into a high mass range.  

\begin{figure}
\begin{center}
\includegraphics[width = 0.78\linewidth]{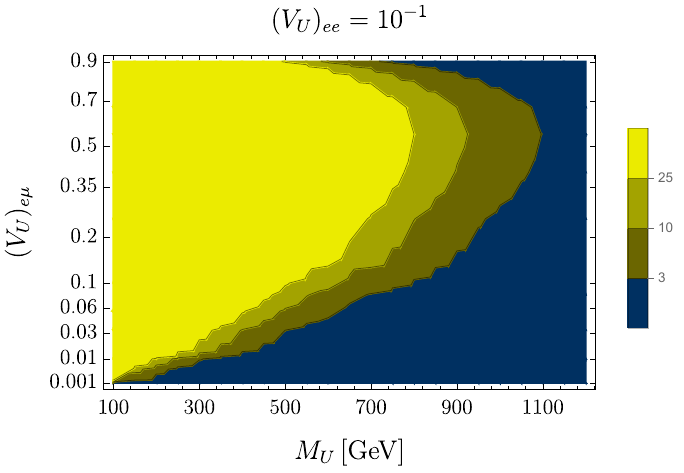}
\end{center}
\caption[Event thresholds on the $M_U \times (V_U)_{e\mu}$ plane.]{Event thresholds in the exclusion panel, for a more detailed analysis.}\label{fig:trimuondetailed}
\end{figure}

Our process is free from irreducible background, but we have ignored the unavoidable source of noise constituted by reducible one.  The most obvious example would be the failure to identify the missing energy in $p p \to  \mu^{+}\mu^{+}\mu^{-}e^{-} + \nu_\mu\nu_\mu\bar{\nu}_\mu\bar{\nu}_e$ (or the charge mirrored reaction). A more thorough analysis of the LHC physics involved in actually detecting our signal is thus the first way this study could be improved. Another style of analysis would be the hard adoption of a model, allowing us to fully employ its predictions, such as effective couplings and additional exotic particles. In the m331, this means the three heavy quarks $j_{i},J$ and the $Z^\prime$. A few benchmarks could be chosen for the parameters of these particles and a more reliable -- although, then, completely model dependent -- result for the bounds could be achieved. Finally, although the $V_U = V_U^T$ condition seems a reasonable simplification and a great place to start, other ansatz for the $V_U$ matrix could be considered as well. Or, in the same ansatz, our result showed how the results are nearly independent of the choice of $(V_U)_{ee}$, which gives room to fix this parameter to an arbitrary value and deepen the analysis in some other aspect.

\cleardoublepage

\chapter{Model independent constraints on exotic particles from flavour violating lepton decays: {\it Preparations}} \label{Chapter6}

Last section we began our investigations of the $U$ phenomenology by focusing on the regularly neglected $V_U$. In this section, we continue with the same goal, but with a much larger scope and greater complexity. The source of the constraints we seek to derive, now, are the purely leptonic decays $\tau^+ \to \ell^+_i\ell^+_j\ell^-_k$, where $\ell_{i,j,k} = e,\mu$, and $\mu^+ \to e^+ e^- e^+$. Without ceasing to pay attention to the mixing matrix, we now consider, in a model independent way, two exotic particles at a time, taken among the three species that can, at tree level, contribute to our channels. Our priority remains the $U$, and our primary results comprise the most conservative model independent lower bounds over its mass, given by the CLFV decays, in models where it is accompanied by an exotic neutral or doubly-charged scalar. In this chapter, the calculations and operational method are detailed, whereas the results are presented in the next one. 

\section{Phenomenological status of the 3-body lepton decays}

The LHC is, without contestation, the experimental particle apparatus with the farthest energy reach. Moreover, the bulk of data that it is capable to produce within each subsequent run increases at a fast pace. This sets the expectation of it eventually becoming a precision machine of the low TeV scale. Nonetheless, this chapter will test an alternative source of data to check if it can complement (or even overcome the usefulness of) the phenomenology arising from current LHC processes for the doubly-charged vector bilepton.

To initiate the qualitative discussion it should be noted that, in comparison with simple leptonic decays, the LHC introduces several complications: hadronic physics, which might relevantly change by the existence of new quarks or other alternative concepts; Detector properties and a large set of reducible background which causes the need for a profound analysis of the observables and definition of triggers and cuts; And the overall process `noise' that an LHC run inevitably produces. 

In the minimal SM, the processes of the form $\ell_0^+ \to \ell^+_i\ell^+_j\ell^-_k$, with $\ell_0$ heavy enough, are forbidden to all orders of perturbation theory. They are possible, however, in the SM with neutrino masses through PMNS mixing. The diagrams appear in Figure~\ref{fig:leptonDecaysSM}. Although there have been claims that the branching ratios (BR) associated with these processes could be calculated to give exceedingly high values of $10^{-14}\sim 10^{-16}$~\cite{Pham_1999,Esteban_2017}, the classical (and robust) result is that these BR stand well below $10^{-50}$~\cite{Petcov:1976ff,de_Gouv_a_2013,Heeck_2017,Hernandez-Tome:2018fbq}. One could thus say that the products of lepton decays are `clean' both in the experimental and theoretical sense, {\it i.e.}, there is still no relevant irreducible background and the reducible one is exaggeratedly smaller than that present in the LHC (although there is still beam induced background as we shall mention briefly). 

\begin{figure}
\begin{center}
\includegraphics[width = 0.42\linewidth]{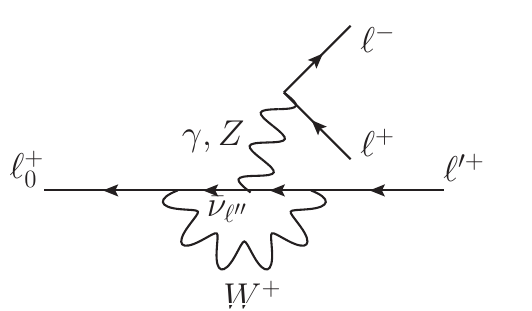}
\includegraphics[width = 0.42\linewidth]{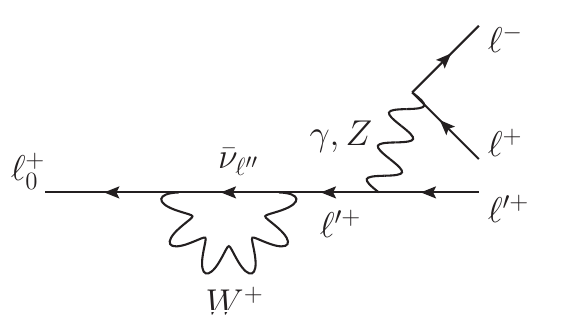}
\includegraphics[width = 0.42\linewidth]{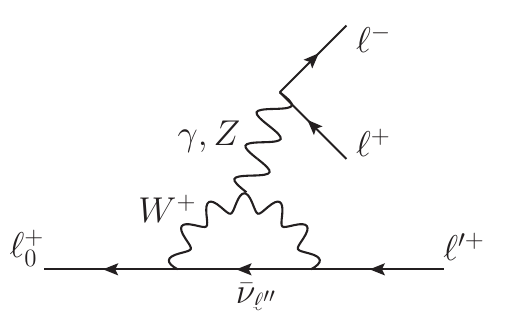}
\includegraphics[width = 0.42\linewidth]{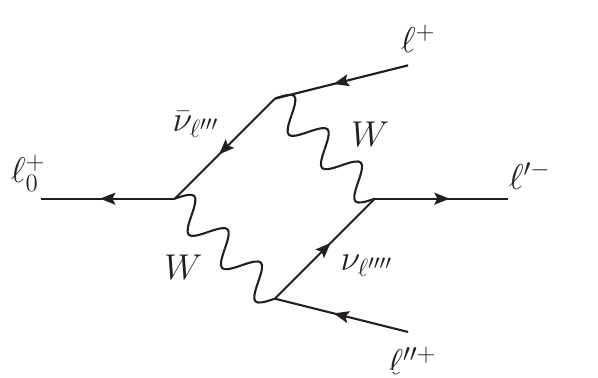}
\end{center}
\caption[SM contributions to the 3-body charged lepton decays]{Lowest order contributions to the 3-body lepton decays in the SM with neutrino masses. In this theory, this process is made possible thanks to the $\ell\bar{\nu}_{\ell^\prime} W$ vertex, which relies on one factor of $(V_\mathrm{PMNS})_{\ell \ell^\prime}$.}\label{fig:leptonDecaysSM}
\end{figure}

Besides highly efficient, processes with charged lepton flavour violation (CLFV) represent a great prospect for the inspection of BSM physics because it is, in one form or another, predicted by most kinds of exotic models, such as Supersymmetry~\cite{ROMAO1985295,PhysRevLett.57.961,Altmannshofer_2010}, two-Higgs-doublet (2HDM)~\cite{PhysRevD.81.095016,Dery_2013,Kopp_2014} and 3-3-1 models~\cite{PhysRevD.50.548,Machado_2019,cabarcas2013lepton,PhysRevD.101.015024}. 
\begin{table}[]
\caption{Current experimental limits on every 3-body lepton decay channel.}\label{Tab:explim}
\begin{center}
\begin{tabular}{lc}
\toprule
\multicolumn{1}{c}{Process}     &  BR \\ \midrule
$\mu^+  \to  e^+ e^- e^+$       &  $<1.0\times 10^{-12}$  \\
$\tau^+ \to  e^+ e^- e^+$       &  $<2.7\times 10^{-8}$   \\
$\tau^+ \to  e^+ \mu^- \mu^+$   &  $<2.7\times 10^{-8}$   \\
$\tau^+ \to \mu^+ e^- e^+$      &  $<1.8\times 10^{-8}$   \\
$\tau^+ \to \mu^+ \mu^- \mu^+$  &  $<2.1\times 10^{-8}$   \\
$\tau^+ \to \mu^+ e^- \mu^+$    &  $<1.7\times 10^{-8}$   \\
$\tau^+ \to  e^+ \mu^- e^+$     &  $<1.5\times 10^{-8}$   \\ \bottomrule
\end{tabular}
\end{center}
\end{table}
The most sensitive experiments -- which are equipped to generate the most stringent bounds -- are the ones relative to the simplest decay: $\mu^+ \to e^+e^+e^-$. Antimuons are used to avoid negative muon capture by nuclei, and because of the low, MeV scale energies involved, the detector must have excellent trackers to record the trajectories. The first source of reducible background comes from the unmeasured missing energy in $\mu^+ \to e^+ e^- e^+ \nu_e \bar{\nu}_\mu$. The second results from the coincidence of two or three muon decays, whose occurrence can be reduced by the use of a continuous beam. The standing bound is already more than 35 years old, and was given by the SINDRUM collaboration to be $\mathrm{BR}(\mu^+ \to e^+ e^- e^+)<10^{-12}$~\cite{SINDRUM:1987nra}. The Mu3e experiment at the Paul Scherrer Institute is the centerpiece of the global efforts to improve this limit, and intends to strengthen it by four orders of magnitude to $10^{-16}$~\cite{Wauters:2021ghc}. 

All 3-body purely leptonic decays of the $\tau$ are much less sensible. Notwithstanding, the $\tau$ is unstable and has a rich hadronic decay spectrum, which complicates its experimental dealing. The current bound on every one of the six possible channels was obtained by~\cite{HAYASAKA2010139} and appear in Table~\ref{Tab:explim}. Again, these limits are naturally much less sensitive and provide a much smaller prospect for the testing of exotic hypotheses than the $\mu \to 3e$ one. For additional discussions on the state of the art and improving sensitivity of the CLFV decay experiments see~\cite{BERNSTEIN201327,MORI201457,introduction_2018,Baldini_2018,
kraetzschmar2021results,Guzzi:2020fmg}.

\section{Exotic contributions and model independent interactions}\label{sec:LDecayInteractions}

Our main goal is to draw relevant exclusion contours on the masses of three species of exotic particles, constraining theory space by taking advantage of the simultaneous data of 3-body lepton decays. Despite the large disparity between the bound on the muon decay and the other channels, considering all the bounds together will prove useful. The
examined particles are the ones which can contribute to the relevant processes: Doubly-charged vector bileptons $U^{\pm \pm}$;
Doubly-charged scalars $Y^{\pm \pm}$; And flavour violation mediating neutral scalars $s$. The diagrams that these species generically induce to a given channel are displayed in Figure~\ref{fig:diagramsLDecays}.

\begin{figure}[t!]
	{\centering
	\includegraphics[width=0.4\linewidth]{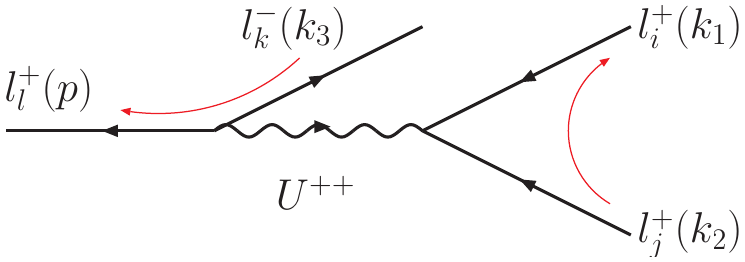}
	\hspace*{5mm}
	\includegraphics[width=0.4\linewidth]{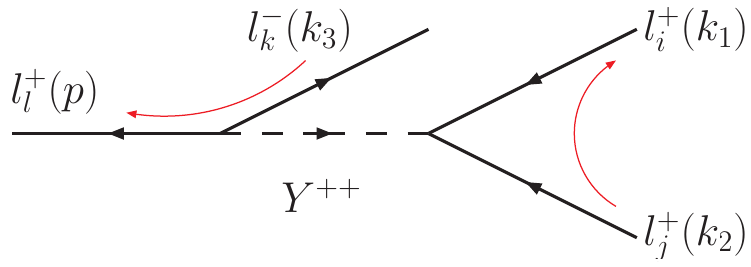} \\
	\vspace*{5mm}
	\includegraphics[width=0.4\linewidth]{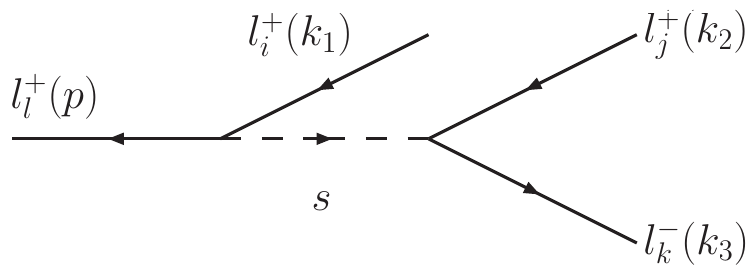}
	\hspace*{5mm}
	\includegraphics[width=0.4\linewidth]{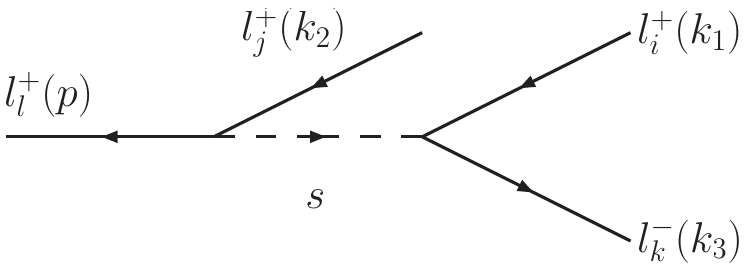}
	\caption[Exotic contributions to the 3-body charged lepton decays.]{Explicit diagrams contributing to the 3-body lepton decays through exchange of an exotic particle. The red arrow defines the direction of the Arbitrary Fermion Flow, and its function will become apparent soon.}
	\label{fig:diagramsLDecays}
	}
\end{figure}

Since we claim to, now, take every contributing species into account, it is an appropriate moment to recall an important remark made last chapter and expand on it. Besides the aforementioned particles, a neutral vector boson $Z^\prime$ of the type contained in the m331 could also contribute to our processes. This, however, can only happen in non-democratic underlying models, where distinct lepton families constitute different representations of the gauge group, otherwise the {\it a priori} diagonal kinetic terms result in a mixing matrix of the form $\mathcal{O}_{Z^\prime} = V_L^\dagger V_L = \mathbb{1}$. Because of this, since we avoid focusing on specific models and, furthermore, non-democratic leptonic sectors being rare, we overlook the possible role of an exotic neutral vector boson.

Besides simply checking the constraints that the 3-body lepton decays may conspire to impose, a crucial aspect that we wish to investigate is the dynamics of interference. In particular, we shall undertake the question of to what extent destructive interference between different species may weaken the bounds implied by the data over their exotic parameters. It is for this reason that we strive to take two species into account at once.

Again, one of the most substantial challenges of a study that seeks to impose limits, as model independent as possible, on the new particle parameters is an appropriate parametrization of every relevant interaction. As in the last study, we shall tackle each Lagrangian in turn, starting with the $U\ell \ell$ interaction. Commencing again from the intuitive form of the most general Lorentz and electromagnetic invariant Lagrangian that descends from an unknown, lepton universal, gauge invariant kinetic term:

\begin{equation}
\mathcal{L}_{U\ell\ell}=\sum_a g_U \overline{\ell_{a\phantom{c}L}^{\prime c}}\gamma^\mu \ell'_{aL}\thinspace U_\mu^{++}
+ g_U\overline{\ell'_{aL}}\gamma^\mu \ell^{\prime c}_{a\phantom{c}L}\thinspace U_\mu^{--},
\end{equation}
or, in the mass basis,

\begin{equation}\label{eq:UllMass12}
\mathcal{L}_{U\ell\ell}=\sum_{a,b} g_U\bar{\ell_a^c}\gamma^\mu P_L (V_U)_{ab} \ell_b\thinspace U_\mu^{++} + g_U\bar{\ell_a}\gamma^\mu P_L (V_U^\dagger)_{ab} \ell_b^c\thinspace U_\mu^{--}.
\end{equation}
Now, as an example, inspect the $U^{++}$ part of the interaction above. Additionally, isolate the terms which pertain to the $Ue\mu$ interaction, which read\footnote{For now we will be specially clear regarding the order in which operations should be performed, {\it e.g.}, in $\overline{e^c}$ one first takes the charge conjugate and then the Lorentz conjugate.}

\begin{equation}\label{eq:Uemu}
\mathcal{L}_{Ue\mu} = g_U \left[\overline{e^c} \gamma^\mu P_L (V_U)_{e\mu} \mu + \overline{\mu^c} \gamma^\mu P_L (V_U)_{\mu e} e\right] U_\mu^{++}.
\end{equation}
Unlike in any interaction which conserves lepton number\footnote{Depending on the scalar potential of the theory, it is still possible to assign a lepton number to the entire representation content of the theory in such a manner to keep the theory invariant under this {\it generalized} $L$.}, both terms in Eq.\refEq{eq:Uemu} feature the same degrees of freedom, {\it i.e.}, contain operators which annihilate (and create) the same states. This means that both contribute to the vertex, and in order for the rules to be actually derived, it is convenient to put the two spinor chains into the same form. To perform these transformations, let us start decomposing Lagrangian~\refEq{eq:UllMass12} into diagonal and non-diagonal parts

\begin{equation}
\mathcal{L}_{U\ell\ell} = \sum_{a \neq b} g_U\bar{\ell_a^c}\gamma^\mu P_L (V_U)_{ab} \ell_b\thinspace U_\mu^{++} + \sum_{a} g_U\bar{\ell_a^c}\gamma^\mu P_L (V_U)_{aa} \ell_a\thinspace U_\mu^{++} + \mathrm{H.C.},
\end{equation}
and proceed to treat the non-diagonal portion. For each $(a,b)$ pair, an ansatz is needed regarding the order in which the particles appear, and we chose to set the heaviest particle in the end -- that is, right -- of the chain. For this, it is enough to transpose each term of the pairs analogous to Eq.\refEq{eq:Uemu} which does not obey this requirement. The transposition may be performed without consequence as these terms are Lorentz scalars, and follows

\begin{equation}
\begin{split}
\bar{\ell_b^c}\gamma_\mu P_L (V_U)_{ba} \ell_a &= \left[\bar{\ell_b^c}\gamma_\mu P_L (V_U)_{ba} \ell_a\right]^T \\
&= -(V_U)_{ba} \, \ell_a^T P_L^T \gamma_\mu^T \left( \ell_b^T C\right)^T \\
&= -(V_U)_{ba} \, \ell_a^T P_L^T \gamma_\mu^T C^T \ell_b \\
&= -(V_U)_{ba} \, \ell_a^T C^{-1}CP_L^T C^{-1}C\gamma_\mu^T C^{-1} \ell_b \\
&=(V_U)_{ba} \, \ell_a^T C^{-1} P_L \gamma_\mu \ell_b \\
&=-(V_U)_{ba}\bar{\ell_a^c} P_L \gamma_\mu \ell_b \\
&= -(V_U)_{ba}\bar{\ell_a^c}  \gamma_\mu P_R\ell_b,
\end{split}
\end{equation} 
where we have used $\overline{\ell_b^c} = \ell_b^T C$, $C^{-1} \gamma_\mu C = -\gamma_\mu^T$, $C^{-1} \gamma_5 C = \gamma_5^T$ and $C = C^*=-C^T=-C^\dagger=-C^{-1}$ (most of these facts are basis independent, but some of the last equalities are specific to the Dirac representation).Using this we may rewrite the most general $U\ell\ell$ interaction as 

\begin{equation}\label{eq:UllFinal}
\begin{split}
\mathcal{L}_{U\ell\ell}&=\sum_{b> a} g_U \left\{\bar{\ell_a^c}\gamma^\mu [P_L (V_U)_{ab}-P_R (V_U)_{ba}] \ell_b\thinspace U_\mu^{++}
+\bar{\ell_a}\gamma^\mu [P_L (V_U^\dagger)_{ab}-P_R (V_U^\dagger)_{ba}]\ell_b^c \thinspace U_\mu^{--}\right\} \\
&+\sum_{a} g_U \left\{\bar{\ell_a^c}\gamma^\mu [P_L (V_U)_{aa}]\ell_a\thinspace U_\mu^{++}
+\bar{\ell_a}\gamma^\mu [P_L (V_U^\dagger)_{aa}] \ell_a^c\thinspace U_\mu^{--}\right\} ,
\end{split}
\end{equation}
where $a,b$ are generation labels. Not rarely, these manipulations cause some confusion and the Lagrangian above is rarely achieved. Notice that a property of the vector bilepton interaction is that the diagonal vertices are purely axial: $\bar{\ell_a^c}\gamma^\mu P_L  \ell_a=-\bar{\ell_a^c}\gamma^\mu \frac{\gamma^5}{2}  \ell_a$.

Now we turn to the doubly-charged scalar. While the previous interaction was originated from the covariant derivative of an unknown higher symmetry, this one is assumed to stem from Yukawa Lagrangians. We have as the most general effective interaction

\begin{equation}\label{eq:Yint1}
\mathcal{L}_{Y\ell\ell}=-\sum_{a,b}g_{Y}\left\{ \bar{\ell^c_a}(\mathcal{O}_Y)_{ab} P_L \ell_b\thinspace Y^{++} + \bar{\ell_a}(\mathcal{O}_Y^\dagger)_{ab} P_R \ell_b^c\thinspace Y^{--} \right\}.
\end{equation}
In the Lagrangian above, the fermions are already mass eigenstates and the interaction mixing matrix is arbitrary: it is related to one {\it a priori} (arbitrary) Yukawa matrix $G_Y$ as $\mathcal{O}_Y=V_R^{\ell T} G_{Y} V^{\ell}_L$. Again, it is not necessary to add a second handedness term. We perform the same transformations as in the $U\ell \ell$ Lagrangian to arrange any term involving a like-sign pair $\ell_a -  \ell_b$ of leptons into a reference spinor chain. The Lagrangian in its most useful form is then given by
\begin{equation}\label{eq:Yint2}
\begin{split}
\mathcal{L}_{Y\ell\ell}=&-\sum_{b>a}g_{Y}\left\{ \bar{\ell^c_a}\left[(\mathcal{O}_Y)_{ab}+(\mathcal{O}_Y)_{ba}\right] P_L \ell_b\thinspace Y^{++} + \bar{\ell_a}\left[(\mathcal{O}_Y^\dagger)_{ab}+(\mathcal{O}_Y^\dagger)_{ba}\right] P_R \ell_b^c\thinspace Y^{--} \right\} \\
&-\sum_{a=b}g_{Y}\left\{ \bar{\ell^c_a}(\mathcal{O}_Y)_{aa} P_L \ell_a\thinspace Y^{++} + \bar{\ell_a}(\mathcal{O}_Y^\dagger)_{aa} P_R \ell_a^c\thinspace Y^{--} \right\}.
\end{split}
\end{equation}

The interaction that is missing is that of the neutral scalar with leptons. Lorentz and electric charge invariance dictates it must be simply
\begin{equation}
\begin{split}
\mathcal{L}_{s\ell\ell}&= - g_{s}\bar{\ell}\mathcal{O}_s P_L \ell\thinspace s
-g_{sL}\bar{\ell}\mathcal{O}_s^\dagger P_R \ell\thinspace s\\
&=-\sum_{a,b}g_{s}\bar{\ell_a}\left[(\mathcal{O}_s)_{ab} P_L+(\mathcal{O}^\dagger_s)_{ab} P_R\right] \ell_b\thinspace s,
\end{split}
\end{equation}
where $\mathcal{O}_s$ is arbitrary and related to a Yukawa matrix as $\mathcal{O}_s=V_R^{\ell\dagger} G_{s} V^{\ell}_L$.

\section{On the free parameters}\label{sec:LDFreeParameters}

In order for the free particles to be defined, our effective model must also feature the kinetic terms 
\begin{equation}
\begin{split}
\mathcal{L}_{\text{kin}}=&{}-\frac{1}{2}U_{\mu\nu}^\dagger U^{\mu\nu}+M_U^2 U^{++} U^{--} \\
&{}+\partial_\mu Y^{++} \partial^\mu Y^{--} - M_Y^2 Y^{++} Y^{--} \\
&{}+ \frac{1}{2}\partial_\mu s\thinspace \partial^\mu s - \frac{1}{2}M_s^2\thinspace s^2,
\end{split}
\end{equation}
where $U_{\mu\nu}=\partial_\mu U^{++}_\nu-\partial_\nu U^{++}_\mu$. Three 2-particle {\it Scenarios} will be considered, each with a pair of exotic species that interfere. The corresponding Lagrangians are

\begin{equation}\label{eq:lagint}
\begin{split}
\mathcal{L}_{U-s}&=\mathcal{L}_{\text{kin}}+\mathcal{L}_{U\ell\ell}+\mathcal{L}_{s\ell\ell}\\
\mathcal{L}_{U-Y}&=\mathcal{L}_{\text{kin}}+\mathcal{L}_{U\ell\ell}+\mathcal{L}_{Y\ell\ell} \\
\mathcal{L}_{Y-s}&=\mathcal{L}_{\text{kin}}+\mathcal{L}_{Y\ell\ell}+\mathcal{L}_{s\ell\ell}.
\end{split}
\end{equation}

Now, notice that $g_s$ and $g_Y$ could be absorbed into their corresponding mixing matrices, and although we write them explicitly on analytical expressions (mostly for book keeping purposes), they will be effectively set to $1$ in all numerical evaluations. Notice also, checking Eq. (\ref{eq:Yint2}), that any element of $\mathcal{O}_Y$ only appears together with its symmetric partner, so that this effective mixing matrix may be taken symmetric. 

Since to effect numerical optimization with the number of free parameters that exists when considering the general case is impractical, we considerably reduce this number by restricting the analysis to real matrices. $V_U$ then becomes an orthogonal matrix, whose determinant may be chosen to be $1$ without loss of generality, and which we parametrize with Euler Angles:

\begin{equation}
V_U = \begin{pmatrix}
\cos\psi\cos\phi-\cos\theta\sin\phi\sin\psi & \cos\psi\sin\phi+\cos\theta\cos\phi\sin\psi & \sin\theta\sin\psi \\
-\sin\psi\cos\phi-\cos\theta\sin\phi\cos\psi & -\sin\psi\sin\phi+\cos\theta\cos\phi\cos\psi & \sin\theta\cos\psi \\
\sin\theta\sin\phi & -\sin\theta\cos\phi & \cos\theta
\end{pmatrix}.
\end{equation}

The number of free parameters in each {\it Scenario}, which includes masses and degrees of freedom of the applicable matrices, may be checked to be as appears in Table~\ref{Tab:nmbdof}.

\begin{table}[t!]
\caption{Number of parametric degrees of freedom contained within each {\it Scenario} if the mixing matrices are regarded as complex and real.}\label{Tab:nmbdof}
\begin{center}
\begin{tabular}{ccc}
\toprule
{\it Scenario}    &    Complex    &    Real         \\ \midrule
$U-Y$       &       23      &     11          \\
$U-s$       &       29      &     14          \\
$s-Y$       &       32      &     17          \\ \bottomrule
\end{tabular}
\end{center}
\end{table}

A few comments are now in order regarding the parametric structure of these interactions and how it relates to our objectives. It is true that if the elements of the 3 mixing matrices parametrizing the interaction Lagrangians could be arbitrarily small, any experimental constraint could be easily met; however, if these particles do exist (\textit{i}) elements too small are not desirable because of matters such as naturalness and; (\textit{ii}) more importantly, orthogonality of the $V_U$ matrix is powerful in inducing exclusion contours. 

Lastly, we should clarify the role of this matrix. Considering now the complete unitary case and referring to its definition $V_U \equiv V_R^T V_L$, there are two situations in which this mixing can be ignored in a natural way: (\textit{i}) If $V_R$ and $V_L$ could be set to $\mathbb{1}$. This can occur whenever the lepton mass matrix and every leptonic interaction can be simultaneously diagonalized, which is not general and relates to a small part of theory space. In fact, it is not the case of the m331, where the mass matrix receives a couple of different contributions. (\textit{ii}) A more general possibility is $V_R = V_L^*$. This implies that the mass matrix is diagonalized by an orthogonal transformation instead of by a biunitary one. In our special case of $V_U$ orthogonal, the condition becomes $V_R = V_L$. A squared mass matrix diagonalizable by a transformation of this type is symmetric and, therefore, not general. We consider a non-diagonal orthogonal $V_U$, which, apart from the missing phases, should be consistent with the general case.

\section{Feynman rules and diagrams from Lagrangians with explicit charge conjugation}

To correctly derive amplitudes from Lagrangians with explicit charge conjugation can be troublesome. The issue arises because when these fields make up the interaction there is generally more than a way to contract a spinor chain with initial and final states. In this case, simply writing the vertices with an explicit charge conjugation matrix is not by itself a well defined and unambiguous process. Therefore, we briefly discuss how to arrive at the amplitudes corresponding to the doubly-charged vector and scalar boson mediation, which suffer from this complication. We follow the algorithm and refer to the description of Refs.~\cite{Denner:1992vza,Denner:1992me}, but focus on the matter of dealing with Lagrangians with explicit charge conjugation, in the form as would naturally emerge from a renormalizable fundamental gauge theory, and strive to be didactic regarding the procedure instead of focusing on why this algorithm works.

We begin defining how to write down the spinor structure. Each spinor line in a diagram will come together with an Arbitrary Fermion Flow Arrow (AFFA) -- recall that the true fermion flow is not continuous in this type of graph. With reference to this arbitrarily drawn line and the true fermion flow arrow, the rules for external fermion lines are

\begin{figure}[H]
\label{fig:spinorsrules}
\centering
\includegraphics[width=0.82\linewidth]{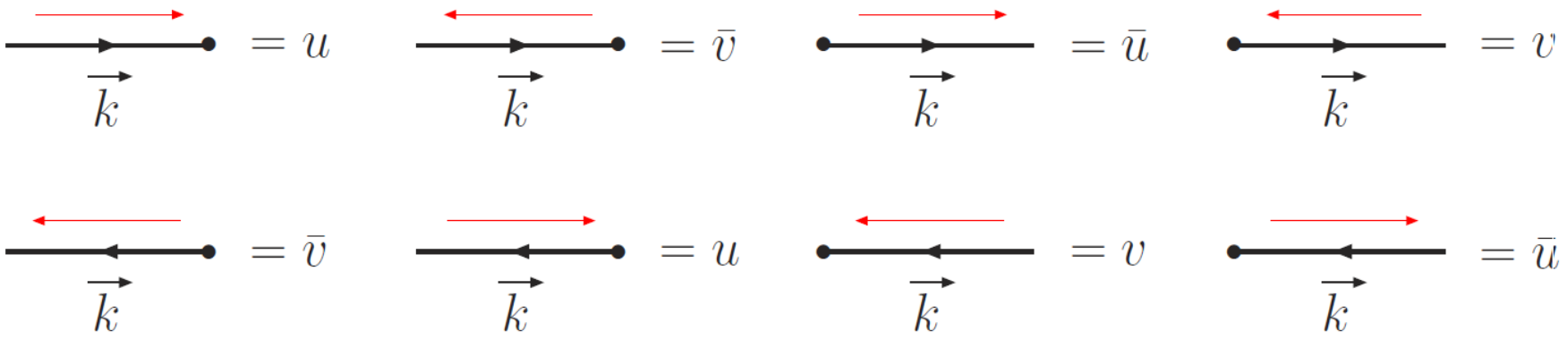}
\end{figure}

Now to read the vertices off of the Lagrangians (\ref{eq:UllFinal}) and (\ref{eq:Yint2}). Considering always incoming bosons, we have a first set of vertices, corresponding to the case in which the AFFA ends on the heaviest fermion, for each charge of the boson:

\begin{equation}
\includegraphics[height=2.3cm,valign=c]{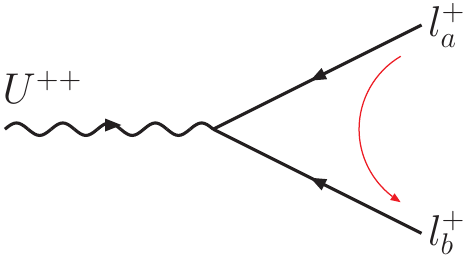} \equiv\Gamma^{U^{++}}_{ba}=g_U\gamma^\mu [P_L (V_U)_{ab}-P_R (V_U)_{ba}]
\end{equation}

\begin{equation}
\includegraphics[height=2.3cm,valign=c]{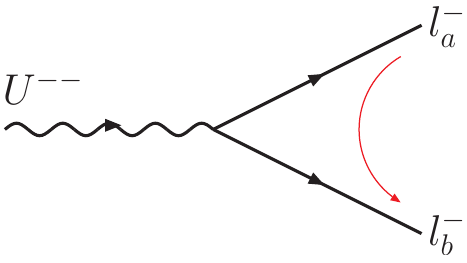} \equiv\Gamma^{U^{--}}_{ba}=g_U\gamma^\mu [P_L (V_U^\dagger)_{ab}-P_R (V_U^\dagger)_{ba}],
\end{equation}
The formulas above are valid even when $a=b$, which can be seen symmetrizing the diagonal part of Lagrangian~(\ref{eq:UllFinal}) as $\bar{\ell_a^c}\gamma^\mu P_L  \ell_a=\frac{1}{2}\left[\bar{\ell_a^c}\gamma^\mu P_L  \ell_a - \bar{\ell_a^c}\gamma^\mu P_R  \ell_a\right]$. \footnote{Notice again that the vector part of this interaction dies out.}The remaining relative factor of $1/2$ is compensated in the rule by a factor of $2$ due to the identical particles. These vertices are called \textit{regular}.

The seemingly innocuous choice of leaving the heaviest fermion on the right in the Lagrangians made in Section \ref{sec:LDecayInteractions} is what leads to the definition of the vertices above as the regular ones.

There is a second set of vertices for the $U^{\pm \pm}$, corresponding to graphs with the AFFA ending on the lightest lepton. The rule is obtained conjugating the original vertex by the charge conjugation matrix like $\Gamma^\prime=C \Gamma C^{-1}$ --  this recipe comes directly by transposition and manipulation of the reference spinor chain. In our case, this calculation gives

\begin{equation}
C\gamma^\mu [P_L (V_U)_{ab}-P_R (V_U)_{ba}]C^{-1}=\gamma^\mu [P_L (V_U)_{ba}-P_R (V_U)_{ab}],
\end{equation}
so that the new vertex rule is $\Gamma^\prime_{ab}=\Gamma_{ba}$. For completeness, we write the reversed vertices below (recall that we chose $\ell_b$ to symbolize the heaviest of the 2 leptons)

\begin{equation}
\includegraphics[height=2.3cm,valign=c]{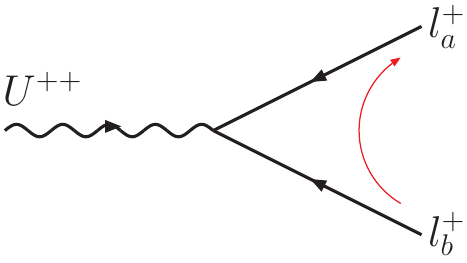}=\Gamma^{\prime U^{++}}_{ab}=ig_U\gamma^\mu [P_L (V_U)_{ba}-P_R (V_U)_{ab}]
\end{equation}

\begin{equation}
\includegraphics[height=2.3cm,valign=c]{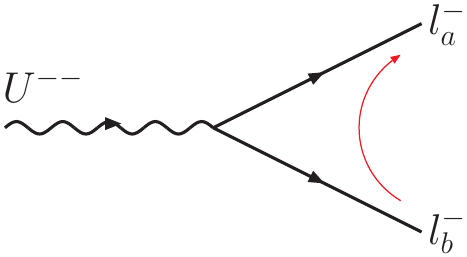}=\Gamma^{\prime U^{++}}_{ab}=ig_U\gamma^\mu [P_L (V_U^\dagger)_{ba}-P_R (V_U^\dagger)_{ab}].
\end{equation}

As an example up to this point, we write the rule corresponding to the two different choices of AFFA for the subdiagrams (and not vertex representations) below

\begin{equation}
\includegraphics[height=2.3cm,valign=c]{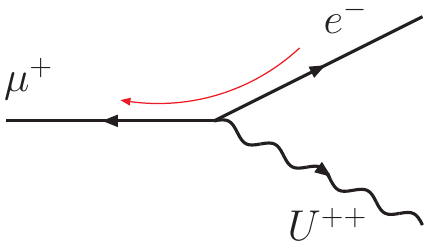}=\bar{v}_\mu \Gamma^{U^{--}}_{\mu e} u_e
\end{equation}

\begin{equation}
\includegraphics[height=2.3cm,valign=c]{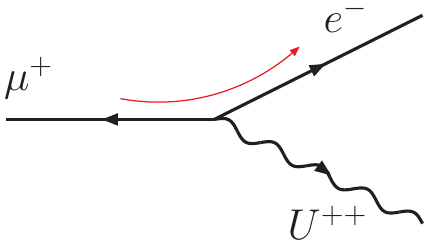}=\bar{v}_e \Gamma^{\prime U^{--}}_{e \mu} u_\mu.
\end{equation}

The complete set of vertices of the doubly-charged scalar read

\begin{equation}
\includegraphics[height=2.3cm,valign=c]{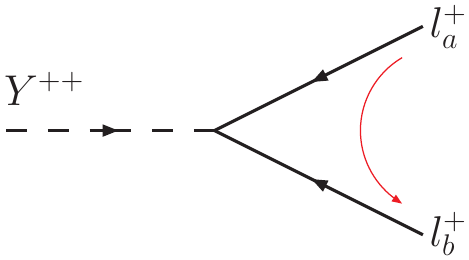}=\Gamma^{Y^{++}}_{ba}=-i g_{Y} \left[(\mathcal{O}_Y)_{ab}+(\mathcal{O}_Y)_{ba}\right] P_L
\end{equation}

\begin{equation}
\includegraphics[height=2.3cm,valign=c]{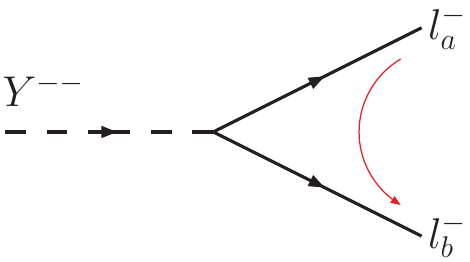}=\Gamma^{Y^{--}}_{ba}=-i g_{Y} \left[(\mathcal{O}_Y^\dagger)_{ab}+(\mathcal{O}_Y^\dagger)_{ba}\right] P_R
\end{equation}

\begin{equation}
\includegraphics[height=2.3cm,valign=c]{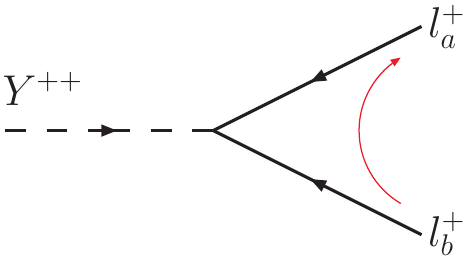}=\Gamma^{\prime Y^{++}}_{ab}=-i g_{Y} \left[(\mathcal{O}_Y)_{ba}+(\mathcal{O}_Y)_{ab}\right] P_L
\end{equation}

\begin{equation}
\includegraphics[height=2.3cm,valign=c]{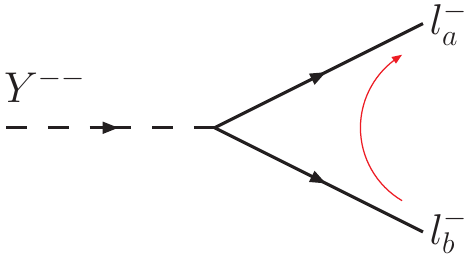}=\Gamma^{\prime Y^{--}}_{ab}=-i g_{Y} \left[(\mathcal{O}_Y^\dagger)_{ba}+(\mathcal{O}_Y^\dagger)_{ab}\right] P_R.
\end{equation}
We emphasize one last time that what defines if a vertex is regular or reversed is the direction of the AFFA with respect to fermion generation -- which, in turn, is a consequence of the conventional form of the Lagrangian.

With the vertices and how to write the exotic spinor chains now understood, the missing ingredient is the ability to find the relative sign between diagrams. This is the greatest reason for the necessity of an algorithm that substitutes the mere explicit use of the charge conjugation matrix. Within the algorithm, to find the relative signs amounts to simply comparing particle ``order'' -- more precisely, the order in which spinors appear in the chain -- with respect to the AFFA and identifying the order of the relating permutation. For a direct example, refer to our real diagrams of Figure~\ref{fig:diagramsLDecays}. The particle orders are (we label different particles by the momenta)

\begin{equation}
\begin{split}
R(\mathcal{M}_U)&=(p,k_3,k_1,k_2) \\
R(\mathcal{M}_Y)&=(p,k_3,k_1,k_2) \\
R(\mathcal{M}_{s1})&=(p,k_1,k_3,k_2) \\
R(\mathcal{M}_{s2})&=(p,k_2,k_3,k_1).
\end{split}
\end{equation}
One may identify that the only ordered set related to the referential $R(\mathcal{M}_U)$ by an odd permutation is $R(\mathcal{M}_{s1})$, so that this amplitude comes attached to an extra minus sign.

This concludes a sufficient description of how our amplitudes can be obtained from the given Lagrangians without having to appeal to an explicit analysis of the possible Wick contractions involved in the correlator.

\section{Amplitudes, phase space and method of evaluation}

Refer to Figure~\ref{fig:diagramsLDecays} for the diagrams that contribute to $\ell^+_l(p)\to \ell^+_i(k_1)\ell^+_j(k_2)\ell^-_k(k_3)$ at tree level. Following the rules and algorithm explained above, it is easy to write down the amplitudes:

\begin{equation}\label{eq:firstamp}
\begin{split}
i\mathcal{M}_U=\left(ig_U\right)^2&\bar{v}_{\ell_l}(p)\gamma^\mu (V_{Ukl}P_L-V_{Ulk}P_R)v_{\ell_k}(k_3)\frac{-i g_{\mu \nu}}{(k_1+k_2)^2-M_U^2} \\
\times&\bar{u}_{\ell_i}(k_1)\gamma^\nu (V_{Uij}P_L-V_{Uji}P_R)v_{\ell_j}(k_2)
\end{split}
\end{equation}

\begin{equation}
\begin{split}
i\mathcal{M}_Y=\left(-i g_Y\right)^2&\bar{v}_{\ell_l}(p)(\mathcal{O}_{Ylk}+\mathcal{O}_{Ykl})P_R v_{\ell_k}(k_3)\frac{i}{(k_1+k_2)^2-M_Y^2} \\
\times&\bar{u}_{\ell_i}(k_1) (\mathcal{O}_{Yij}+\mathcal{O}_{Yji})P_L v_{\ell_j}(k_2) \\
\end{split}
\end{equation}

\begin{equation}
\begin{split}
i\mathcal{M}_{s1}=(-1)\left(-i g_s\right)^2&\bar{v}_{\ell_l}(p)(\mathcal{O}_{sli}P_L+\mathcal{O}_{sil}P_R) v_{\ell_i}(k_1)\frac{i}{(k_2+k_3)^2-M_s^2} \\
\times&\bar{u}_{\ell_k}(k_3) (\mathcal{O}_{skj}P_L+\mathcal{O}_{sjk}P_R) v_{\ell_j}(k_2) \\
\end{split}
\end{equation}

\begin{equation}\label{eq:lastamp}
\begin{split}
i\mathcal{M}_{s2}=\left(-i g_s\right)^2&\bar{v}_{\ell_l}(p)(\mathcal{O}_{slj}P_L+\mathcal{O}_{sjl}P_R) v_{\ell_j}(k_2)\frac{i}{(k_1+k_3)^2-M_s^2} \\
\times&\bar{u}_{\ell_k}(k_3) (\mathcal{O}_{ski}P_L+\mathcal{O}_{sik}P_R) v_{\ell_i}(k_1).\\
\end{split}
\end{equation}
Nevertheless, we check the amplitudes above by generating them through \texttt{FeynRules}~\cite{Alloul:2013bka,Christensen_2011} in association with the \texttt{FeynArts}~\cite{Hahn:2000kx} and \texttt{FeynCalc} package~\cite{Shtabovenko_2016,Shtabovenko_2020}.

\begin{figure}
\begin{center}
\includegraphics[width = 0.7\linewidth]{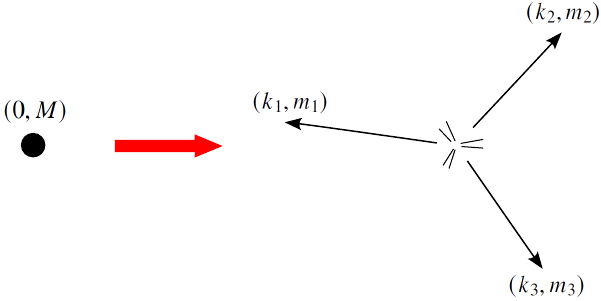}
\end{center}
\caption[Simple representation of the 3-body decay kinematics.]{Simple representation of the 3-body decay kinematics in the rest frame of the parent system. The labels represent the 3-momentum and mass of the particle, respectively.}\label{fig:kinDiag}
\end{figure}

The kinematics of the final state with three particles is immensely more complicated than the more usual $2\to 2$ phase space, and we make an aside to briefly discuss it. For a standard assessment of the usual parametrization, corresponding to the theory of Dalitz plots, see, for instance, the PDG~\cite{ParticleDataGroup:2022pth} -- but let us try to find a more intuitive route to describe these quantities.  Our goal is to understand the kinematics in order to write the invariant phase space element of the final state in a $1 \to 3$ process (refer to Figure~\ref{fig:kinDiag}). The first step is to favour one of the three end state particles and find the absolute allowed range for its energy. Focusing on particle 1, the minimum allowed energy is clearly obtained when this (massive) body is at rest, giving

\begin{equation}
E_1^{\mathrm{min}} = m_1.
\end{equation}
Now, energy conservation enforces $M = E_1 + E_2 + E_3$, and momentum conservation implies $\vectorB{k_1} + \vectorB{k_2} + \vectorB{k_3} = 0$. The analysis is facilitated by replacing particles 2 and 3 by a single system $Q$ with mass $m_Q$ and momentum $\vectorB{k_Q} = -\vectorB{k_1}$. The maximum $\vectorB{k_1}$ (which corresponds to maximum $E_1$) is attained at minimum $m_Q$. Since this mass is invariant, we may inspect its inequivalent configurations (which corresponds to inequivalent $k_2,k_3$ configurations) in the $Q$ rest frame. In this system, it becomes clear that the smallest energy (corresponding to smallest $m_Q$) is $m_Q = m_2 + m_3$ and occurs for particles 2 and 3 at rest. In turn, in the parent rest frame this corresponds to both particles moving at the same velocity and opposite direction to particle 1. With this information, energy conservation gives

\begin{equation}
\sqrt{m_1^2+\left\lvert \vectorB{k_1}^{\hspace*{-1.8mm}\mathrm{max}}\right \rvert^2}+\sqrt{(m_2+m_3)^2+\left\lvert\vectorB{k_1}^{\hspace*{-1.8mm}\mathrm{max}}\right \rvert^2} = M,
\end{equation} 
which, after some algebra, delivers 

\begin{equation}
E_1^\mathrm{max} = \frac{M^2 + m_1^2 - (m_2 + m_3)^2}{2M}.
\end{equation}

There is still one energy degree of freedom to be integrated over. This time we only need to find the limits on $E_2$ for a fixed $E_1$. Obviously enough, the minimum (maximum) $E_2$ corresponds to $k_3$ parallel (anti-parallel) to $k_2$. The conservation laws then give 

\begin{equation}
\label{eq:E2minmax}
\sqrt{E_2^{*2}-m_2^2} = \sqrt{E_1^{2}-m_1^2} \pm \sqrt{(M-E_1-E_2^*)^{2}-m_3^2},
\end{equation}
where $E_2^* = E_2^\mathrm{max(min)}$ for the plus (minus) sign. Now, by conformity, define the usual energy fractions $x \equiv \frac{2E_1}{M}$, $y \equiv \frac{2E_2}{M}$ and $z \equiv \frac{2E_3}{M}$,
%
%
and the squared mass ratios $\xi_i = m_i^2/M^2$. It is easy to show that the positive solution of each form of Eq.\refEq{eq:E2minmax} is also a solution to the quadratic equation

\begin{equation}
(1-x+\xi_1)y^2 + [x(2+\lambda - x)-2 \lambda]y + x(x+\xi_2 x - 2\lambda)+\lambda^2 - 4\xi_1 \xi_2 = 0.
\end{equation}
This can now be solved directly and we find

\begin{equation}
\begin{split}
y^{\mathrm{min}}(x) &= \frac{(2-x)(r_1+\xi_2 - \xi_3) - \sqrt{x^2 - 4\xi_1}\sqrt{\lambda(r_1,\xi_2,\xi_3)}}{2 r_1} \\
y^{\mathrm{max}}(x)& = \frac{(2-x)(r_1+\xi_2 - \xi_3) + \sqrt{x^2 - 4\xi_1}\sqrt{\lambda(r_1,\xi_2,\xi_3)}}{2 r_1},
\end{split}
\end{equation}
where we have defined $r_1 \equiv 1 + \xi_1 -x$ and $\lambda$ is the K{\"a}ll{\'e}n function, $\lambda(x,y,z) = (x-y-z)^2 - 4yz$.

Coming back now to the configuration $\ell^+_l(p)\to \ell^+_i(k_1)\ell^+_j(k_2)\ell^-_k(k_3)$, the angular inclusive differential partial width may be written in terms of the invariant phase space as

\begin{equation}
\begin{split}
d\Gamma(\ell^+_l\to \ell^+_i\ell^+_j\ell^-_k) &= \frac{1}{64 \pi^3 m_l}|\overline{\mathcal{M}}(\ell^+_l\to \ell^+_i\ell^+_j\ell^-_k)|^2 dE_1\,dE_1 \\
&=\frac{m_l}{256 \pi^3} |\overline{\mathcal{M}}(\ell^+_l\to \ell^+_i\ell^+_j\ell^-_k)|^2 dx dy.
\end{split}
\end{equation}
The final inclusive width is obtained as

\begin{equation}
\Gamma(\ell^+_l\to \ell^+_i\ell^+_j\ell^-_k) = \frac{m_l}{256 \pi^3}\int_{x^{\mathrm{min}}}^{x^{\mathrm{max}}}dx \int_{y^{\mathrm{min}}}^{y^{\mathrm{max}}} |\overline{\mathcal{M}}(\ell^+_l\to \ell^+_i\ell^+_j\ell^-_k)|^2 \, dy.
\end{equation}
For completeness, we show the expressions for all the invariants in terms of our parametrization\footnote{We associate the labels as $i \leftrightarrow 1$, $j \leftrightarrow 2$ and $k \leftrightarrow 3$.}

\begin{equation}
\begin{split}
k_1 \cdot k_2 &= \frac{m_l^2}{2}(-1 + x + y - \xi_1 - \xi_2 + \xi_3)\\
k_1 \cdot k_3 &= \frac{m_l^2}{2}(1 - y - \xi_1 + \xi_2 + \xi_3)\\
k_2 \cdot k_3 &= \frac{m_l^2}{2}(1 - x + \xi_1 - \xi_2 - \xi_3).
\end{split}
\end{equation}

Now, we are going to perform constrained optimization in order to find the best points in parameter space, which correspond to those which allow the smallest $M_U$, $M_s$ and $M_Y$ -- {\it i.e.}, those which do not exclude light particles. This procedure is quite demanding computationally, and to render it feasible we require an analytic expression for each BR. However, because of the complicated $1 \to 3$ phase space, $|\overline{\mathcal{M}}(\ell^+_l\to \ell^+_i\ell^+_j\ell^-_k)|^2$ is an enormous rational expression in terms of $x,y$. The conclusion is that some approximations are required, and we describe the operational method including them below. 

The hierarchy between mass scales in the processes we consider is 

\begin{equation}
M_U,M_s,M_Y \gg m_\tau \gg m_\mu \gg m_e.
\end{equation} 
The denominators of the squared matrix elements are of the form $M_1^2-m_\ell^2 f(x,y)$, where $M_1$ is any of the heavy boson masses and $m_\ell$ any of the lepton ones, and $f(x,y)$ is a small polynomial on the energy fractions. This hints that we could start ignoring the lepton masses {\it on natural denominators} of $|\overline{\mathcal{M}}(\ell^+_l\to \ell^+_i\ell^+_j\ell^-_k)|^2$. The integration in $y$ is performed with this simplified form of $\mathcal{M}$:

\begin{equation}
|\overline{\mathcal{M}}|_y^2 \equiv \int_{y^{\mathrm{min}}}^{y^{\mathrm{max}}} |\overline{\mathcal{M}}(\ell^+_l\to \ell^+_i\ell^+_j\ell^-_k)|^2 \, dy.
\end{equation}
Because of the complicated dependence of $y^{\mathrm{min}(\mathrm{max})}$ on $x$, this is a rational expression whose subsequent $x$ integration cannot be effected with ease. To remedy this situation we expand $|\overline{\mathcal{M}}|_y^2$ around $x = 0.5$ up to third order

\begin{equation}\label{eq:intyApprox}
\begin{split}
|\overline{\mathcal{M}}|_y^2(x) &= A_0(M_1,m_\ell) + A_1(M_1,m_\ell)(x-0.5) + A_2(M_1,m_\ell)(x-0.5)^2  \\
&\quad  + A_3(M_1,m_\ell)(x-0.5)^3 + O\left((x-0.5)^4\right).
\end{split}
\end{equation}
With this, the $x$ integration may be analytically performed. To validate the approximation above, we compare its results with those obtained by numerically performing the integration 

\begin{equation}
\int_{x^{\mathrm{min}}}^{x^{\mathrm{max}}} |\overline{\mathcal{M}}|_y^2(x) \, dx,
\end{equation}
with the exact $|\overline{\mathcal{M}}|_y^2(x)$. This comparison is made for 100 points strategically spread across parameters space, and lead us to conclude that the approximation may be safely maintained without any repercussions.

\cleardoublepage

\chapter{Model independent constraints on exotic particles from flavour violating lepton decays: {\it Results}} \label{Chapter7}

Finally, in this chapter we present the results for each 2-particle model, dubbed {\it Scenario}. The ranges for the non-massive parameters that are allowed in principle are

\begin{equation}
\begin{split}
0& \leq \phi,\,\psi < 2\pi \\
0& \leq \theta < \pi \\
-1& < \mathcal{O}_{Yij},\,\mathcal{O}_{sij} < 1.
\end{split}
\end{equation}
These are the limits that exhaust the $V_U$ space and keep the scalar interactions perturbative. As for the masses, we investigate

\begin{equation}
M_U,M_Y,M_s > 500.
\end{equation}
This is enough because smaller masses are extremely unlikely (to say the least) and their analysis would not change the qualitative aspects of the study.

Our objective is to find a solution, {\it i.e.}, a value for each of the free parameters (except masses) that allow the smallest possible masses to be phenomenologically possible. In the {\it Scenarios} where the $U$ is present, we search for solutions that prioritize its mass – {\it i.e.}, we seek sets of numbers which minimize $M_U$, with every other parameter, including $M_s,M_Y$, free.
We consider additional benchmark conditions to fix the lower bound on
the modulus of matrix elements {\it and/or} impositions on the diagonal couplings of the mixing matrices. 
Very
restrictive, these constraints are designed to check what are the lowest possible masses depending on of the level of naturalness of the model that correctly
describes nature.

\begin{figure}[t!]
\begin{center}
\adjustbox{center=10cm}{
	\makebox[1.2\linewidth]{
	\begin{subfigure}{0.5\linewidth}
	\centering
		\includegraphics[width=\linewidth,trim=0cm 0.4cm 0.8cm 1.2cm,clip]{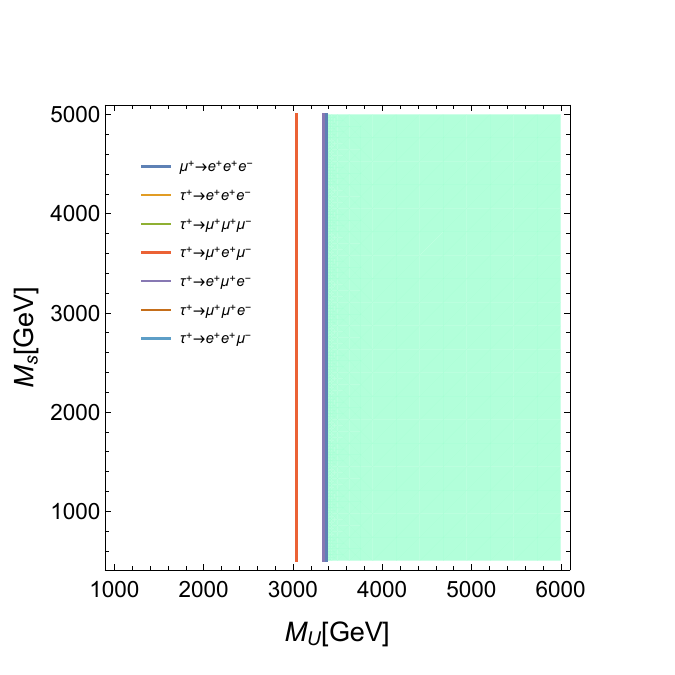}
		\caption{$|V_{Uij}|>10^{-3}$}\label{fig:contoursUa}
	\end{subfigure}
	\begin{subfigure}{0.5\linewidth}
	\centering
		\includegraphics[width=\linewidth,trim=0cm 0.4cm 0.8cm 1.2cm,clip]{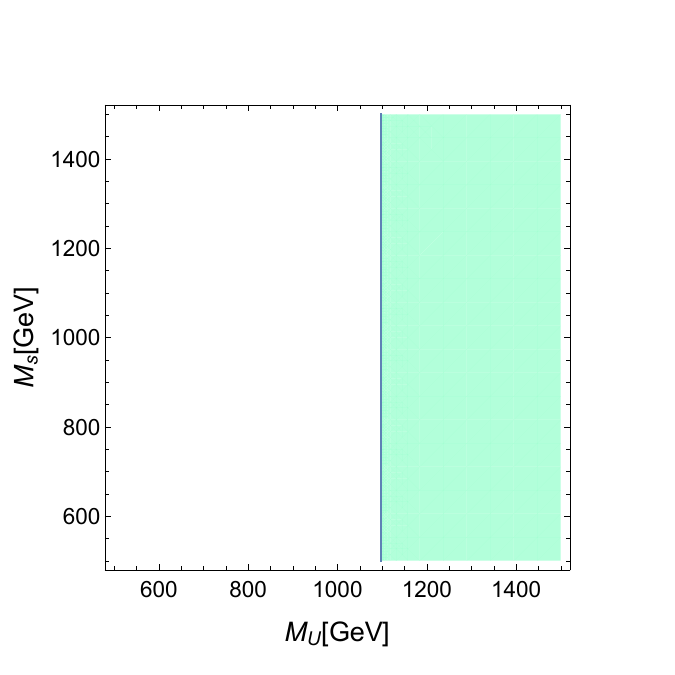}
		\caption{$|V_{Uij}|>10^{-4}$}\label{fig:contoursUb}
	\end{subfigure}}}  \\
	\vspace*{5mm}
\adjustbox{center=10cm}{
	\makebox[1.2\linewidth]{
	\begin{subfigure}{0.5\linewidth}
	\centering
		\includegraphics[width=\linewidth,trim=0cm 0.4cm 0.8cm 1.2cm,clip]{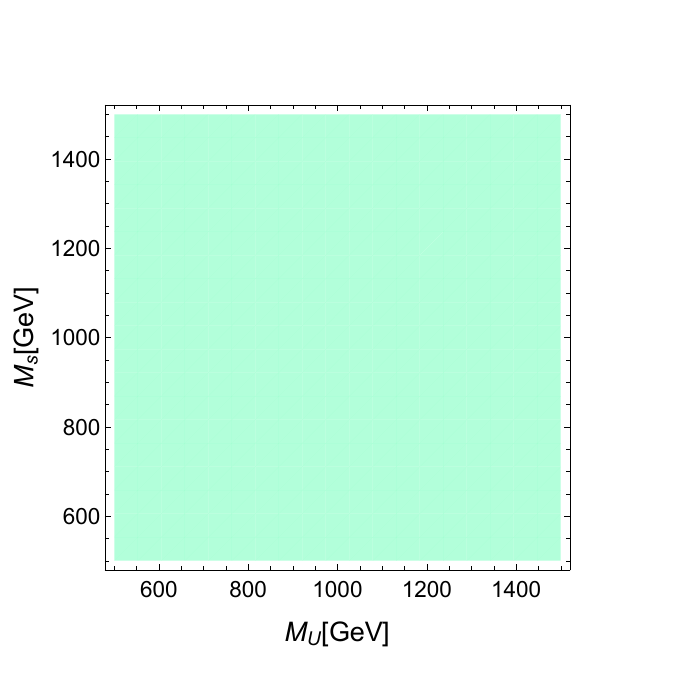}
		\caption{$|V_{Uij}|>10^{-5}$}
	\end{subfigure}
	\begin{subfigure}{0.5\linewidth}
	\centering
		\includegraphics[width=\linewidth,trim=0cm 0.4cm 0.8cm 1.2cm,,clip]{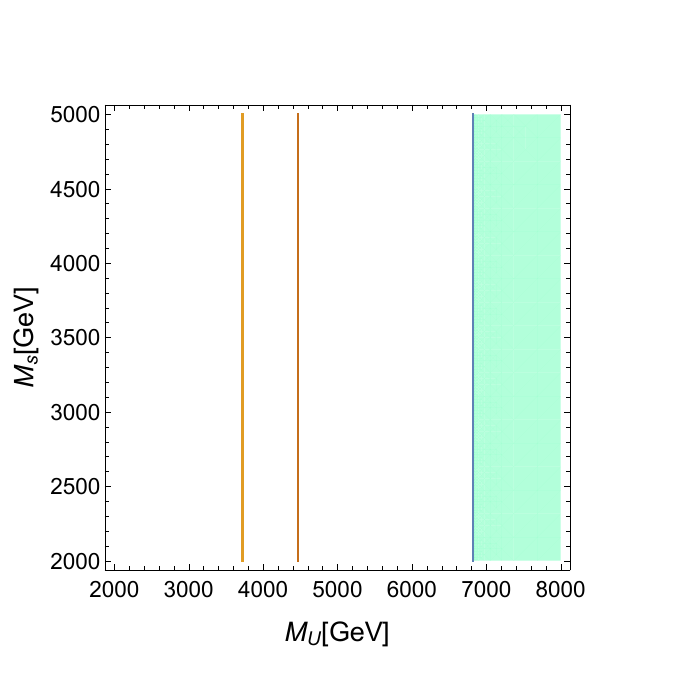}
		\caption{$|V_{Uij}|>10^{-5}$ and $V_{11}=0.85$}\label{fig:contoursUd}
	\end{subfigure}}}
	\caption[Exclusion contours in the pure $U$ {\it Scenario}.]{Exclusion ranges for $M_U$ generated by the experimental bounds on the various branching rations for leptonic decays, for each of our 4 benchmark cases.  The allowed region is painted green.}
	\label{fig:contoursU}
\end{center}
\end{figure}

The optimal points in the multi-dimensional parameters spaces are found through a numerical constrained global optimization routine, and the stability of the results are checked through its repetition with $10^2$ random seeds.

\section{Pure $U$ {\it Scenario}}

In order to set the stage for the 2-particle analysis, we begin examining the constraints resulting from a doubly-charged vector bilepton alone. The results are presented in Figure~\ref{fig:contoursU} and the solutions in Table~\ref{Tab:solU}. We show the contours in the $M_U \times M_s$ plane (even though there is no $M_s$ dependence) to facilitate comparison with the {\it Scenario} below.

As mentioned, we seek for solutions which minimize the allowed $M_U$ with {\it additional} requirements on the absolute values of every matrix element. These are meant to limit hierarchies and force interesting solutions to be kept over strongly diagonal ones. In the most liberal case, we allow for matrix elements as small as $10^{-5}$, corresponding to a hierarchy already as large as that of the SM quark sector.  Observing the results, we recognize that to allow for masses of the order of $\SI{1100}{GeV}$ (see Fig.~\ref{fig:contoursUb}) we need a hierarchy\footnote{Note that, in an orthogonal matrix, small elements imply the need for large ones.} within $V_U$ of four orders of magnitude, such that we approach a non-natural parameter regime. But the greatest feature to observe is that with small general hierarchy (Fig.~\ref{fig:contoursUa}) or with large \textit{but not maximal} diagonal coupling (Fig.~\ref{fig:contoursUd}) the constraints are strong, demanding $M_U> \SI{3200}{GeV}$ and $M_U> \SI{6900}{GeV}$, respectively.  It must be recognized that, in each instance, the contours result from a specific, not always evident, interplay between one or various BR bounds and the unitarity conditions of $V_U$.

\begin{table}[H]
\centering
\caption{Solutions of the pure $U$ {\it Scenario}, corresponding to the plots of Fig.~\ref{fig:contoursU}.}\label{Tab:solU}
\begin{center}
\begin{tabular}{c@{\hskip 1.5em}c@{\hskip 38pt}c@{\hskip 1.5em}c@{\hskip 38pt}r@{\hskip 1.5em}r}
\toprule
\multicolumn{4}{c}{Pure $U$: $\;|V_{Uij}|>10^{-3}$} \\ \midrule
$M_U$      &   $3380$       &  $\psi$     &   $1.48108$ \\
$\phi$     &   $3.03983$    &  $\theta$   &   $2.66989$ \\ \toprule
\multicolumn{4}{c}{Pure $U$: $\;|V_{Uij}|>10^{-4}$} \\ \midrule
$M_U$      &   $1100$       &  $\psi$     &   $0.78535$ \\
$\phi$     &   $5.49774$    &  $\theta$   &   $0.00014$ \\ \toprule
\multicolumn{4}{c}{Pure $U$: $\;|V_{Uij}|>10^{-5}$} \\ \midrule
$M_U$      &   $500$       &  $\psi$     &   $0.72066$ \\
$\phi$     &   $0.72067$   &  $\theta$   &   $3.13801$ \\ \toprule
\multicolumn{4}{c}{Pure $U$: $\;|V_{Uij}|>10^{-5}$, $V_{U11}=0.85$} \\ \midrule
$M_U$      &   $6830$       &  $\psi$     &   $4.74012$ \\
$\phi$     &   $4.68301$    &  $\theta$   &   $0.55504$ \\ \toprule
\end{tabular}
\end{center}
\end{table}

An important remark is that the $V_U = \mathbb{1}$ choice made in the phenomenological literature reviewed before may correspond, here, to the case of high hierarchy $|V_{Uij}|>10^{-5}$. This claim is not necessarily true, and to confirm that the solution to this benchmark imposition can be compared to the results of those studies, it must be checked that it is of the form $V_U \sim \mathbb{1}$ -- which is indeed the case. Now, the strongest bounds from the diagonal LHC literature are capable of excluding bilepton masses $M_U \lesssim \SI{1}{TeV}$. Our conclusion is that for the sector of theory space with a $V_U$ hierarchy of $10^{4}$ or lower, the CLFV lepton decays bounds should be considered, while for the flavour conserving sector, numerically and casually equivalent to tolerant hierarchies of $10^{5}$ or higher, the more energetic LHC phenomenology should be more appropriate.

Notice that the argument above is general: for not exceedingly low, improbable masses, the case of no mixing, $V_U = \mathbb{1}$ (which can be contained in a natural way within a theory as discussed in Sec.~\ref{sec:LDFreeParameters}) in which our processes do not occur at all, is represented by the most permissive case of high hierarchy $|V_{Uij}|>10^{-5}$.

\begin{figure}[t!]
	\centering
		\includegraphics[width=0.45\linewidth]{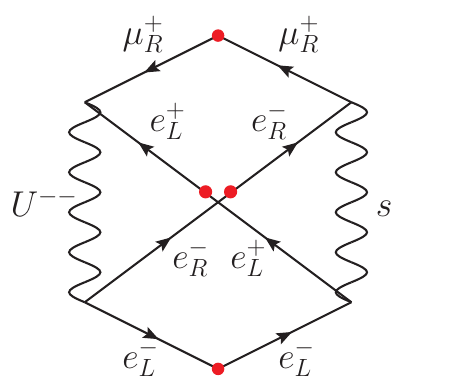}
		\caption[Heuristic demonstration of the existing interference between the $U$ and $s$ contributions to the 3-body CLFV decays.]{Mnemonic device which quickly shows that this {\it Scenario} features interference. Time flows from left to right and from top to bottom. The first step is choosing a handedness arrangement for the initial state, and we look at $\mu^-_L \to e^-_L e^-_R e^+_L$ (notice that, without any harm, we look at a negative decaying muon now). In the calculation of the squared matrix element, one of the amplitudes must be conjugated -- here, we choose the $s$ contribution. We then take advantage of $\mathit{CPT}$ invariance to transform the graphs in order to assure that one of them has only outgoing and the other only incoming lines. Finally, we attempt to join corresponding lines. This is possible for the $U$ and $s$ contributions, hence interference is present (the point in which lines are joined is highlighted by a red blob -- the lines do not touch at other crossings). }
	\label{fig:interferenceDiags}
\end{figure}

\section{$U-s$ {\it Scenario}}

The {\it Scenario} which postulates a model that contains a vector bilepton $U^{\pm \pm}$ and a neutral scalar $s$ is the most interesting one because it features interference and the influence of the unitary $V_U$. As a shortcut to realize that interference is present, we may employ the mnemonic device of checking if two diagrams can be `glued' together -- see Figure~\ref{fig:interferenceDiags}.

\begin{figure}[t!]
\begin{center}
\adjustbox{center=10cm}{
	\makebox[1.2\linewidth]{
	\begin{subfigure}{0.5\linewidth}
	\centering
		\includegraphics[width=\linewidth,trim=0cm 0.4cm 0.8cm 1.2cm,clip]{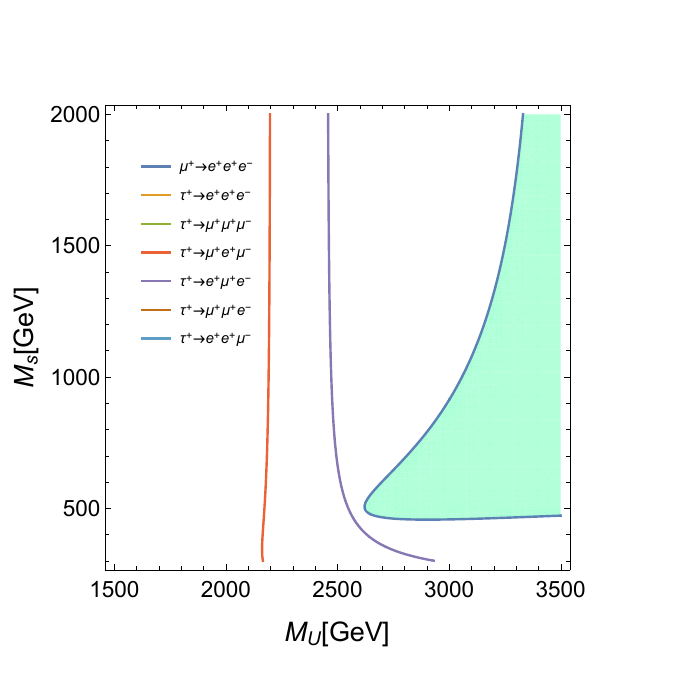}
		\caption{$|O_{sij},V_{Uij}|>10^{-3}$}
	\end{subfigure}
	\begin{subfigure}{0.5\linewidth}
	\centering
		\includegraphics[width=\linewidth,trim=0cm 0.4cm 0.8cm 1.2cm,,clip]{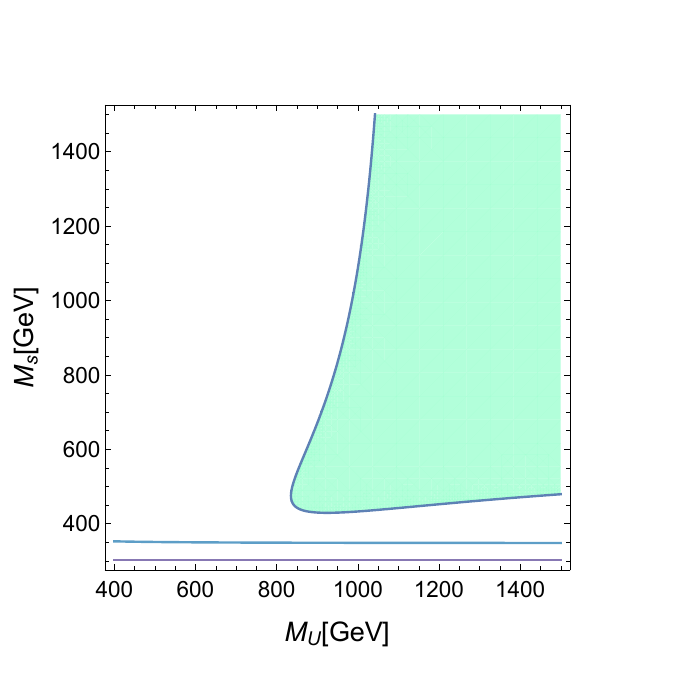}
		\caption{$|O_{sij},V_{Uij}|>10^{-4}$}
	\end{subfigure}}}  \\
	\vspace*{5mm}
\adjustbox{center=10cm}{
	\makebox[1.2\linewidth]{
	\begin{subfigure}{0.5\linewidth}
	\centering
		\includegraphics[width=\linewidth,trim=0cm 0.4cm 0.8cm 1.2cm,,clip]{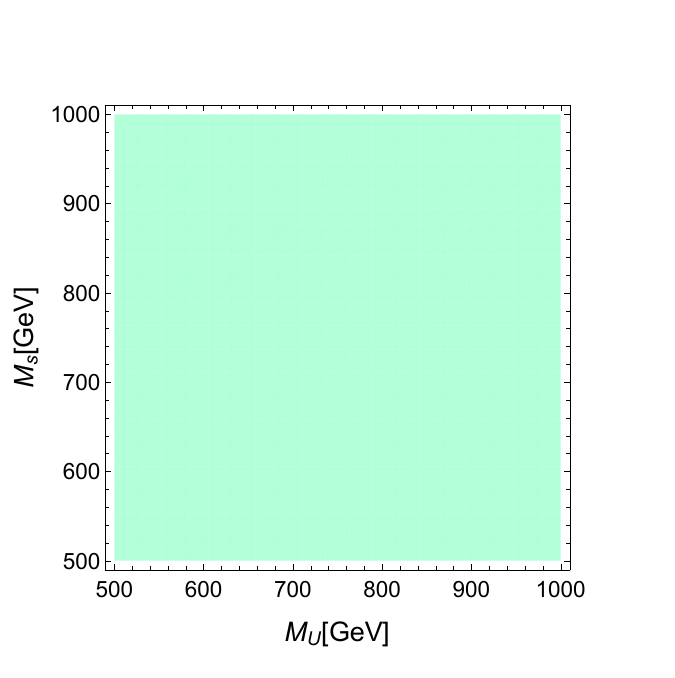}
		\caption{$|O_{sij},V_{Uij}|>10^{-5}$}
	\end{subfigure}
	\begin{subfigure}{0.5\linewidth}
	\centering
		\includegraphics[width=\linewidth,trim=0cm 0.4cm 0.8cm 1.2cm,,clip]{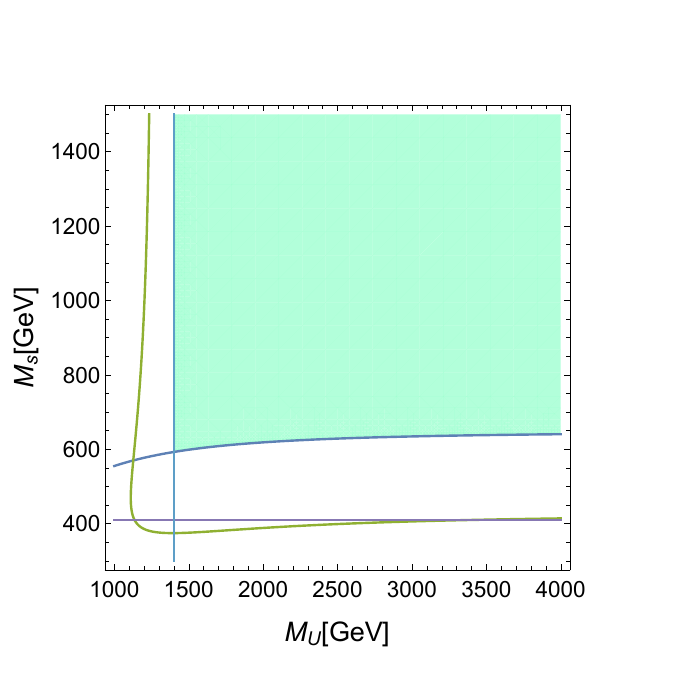}
		\caption{$|O_{sij},V_{Uij}|>10^{-5}$ and $O_{s11}=O_{s22}=1$
		}	\label{fig:contoursUsd}
	\end{subfigure}}}
	\caption[Exclusion contours on the $M_U \times M_s$ plane generated by the bounds on the 3-body CLFV decays.]{Exclusion contours on the $M_U \times M_s$ plane generated by the bounds on 3-body Lepton Flavour Violating decays. The allowed region is painted green.}
	\label{fig:contoursUs}
\end{center}
\end{figure}

\begin{table}[t!]
\caption{Solutions of the $U-s$ {\it Scenario}, respective of the plots shown in Fig.~\ref{fig:contoursUs}}\label{Tab:solUs}
\begin{center}
\begin{tabular}{c@{\hskip 1.5em}c@{\hskip 38pt}c@{\hskip 1.5em}c@{\hskip 38pt}r@{\hskip 1.5em}r}
\toprule
\multicolumn{6}{c}{$U-s$: $\;|V_{Uij},\mathcal{O}_{sij}|>10^{-3}$} \\ \midrule
$M_U$      &   $2650$       &  $\mathcal{O}_{s11}$  &   $3.11564 \times 10^{-3}$ & $\mathcal{O}_{s22}$   &  $1.43763 \times 10^{-3}$  \\
$M_s$      &   $500$        &  $\mathcal{O}_{s12}$  &   $3.01898 \times 10^{-3}$ & $\mathcal{O}_{s23}$   &  $4.06257 \times 10^{-2}$  \\
$\phi$     &   $6.26303$    &  $\mathcal{O}_{s13}$  &   $4.20596 \times 10^{-2}$ & $\mathcal{O}_{s31}$   &  $-1.32129 \times 10^{-1}$ \\
$\psi$     &   $1.55218$    &  $\mathcal{O}_{s21}$  &   $3.19852 \times 10^{-3}$ & $\mathcal{O}_{s32}$   &  $1.51094 \times 10^{-1}$  \\
$\theta$   &   $2.90919$    &                       &  \\ \toprule
\multicolumn{6}{c}{$U-s$: $\;|V_{Uij},\mathcal{O}_{sij}|>10^{-4}$} \\ \midrule
$M_U$      &   $840$       &  $\mathcal{O}_{s11}$  &   $\hphantom{-}2.57667 \times 10^{-3}$ & $\mathcal{O}_{s22}$   &  $-2.14209 \times 10^{-3}$  \\
$M_s$      &   $500$       &  $\mathcal{O}_{s12}$  &   $-3.29400 \times 10^{-3}$ & $\mathcal{O}_{s23}$  &  $-6.05650 \times 10^{-2}$  \\
$\phi$     &   $1.45901$   &  $\mathcal{O}_{s13}$  &   $\hphantom{-}4.05147 \times 10^{-1}$ & $\mathcal{O}_{s31}$   &  $-4.81308 \times 10^{-2}$ \\
$\psi$     &   $1.45911$   &  $\mathcal{O}_{s21}$  &   $-3.34363 \times 10^{-3}$ & $\mathcal{O}_{s32}$  &  $-1.44602 \times 10^{-1}$  \\
$\theta$   &   $3.13998$    &                       &  \\ \toprule
\multicolumn{6}{c}{$U-s$: $\;|V_{Uij},\mathcal{O}_{sij}|>10^{-5}$} \\ \midrule
$M_U$      &   $<500$       &  $\mathcal{O}_{s11}$  &   $1.00000 \times 10^{-5}$ & $\mathcal{O}_{s22}$   &  $1.00000 \times 10^{-5}$  \\
$M_s$      &   $<500$       &  $\mathcal{O}_{s12}$  &   $1.00000 \times 10^{-5}$ & $\mathcal{O}_{s23}$  &  $1.00000 \times 10^{-5}$  \\
$\phi$     &   $0.72067$   &  $\mathcal{O}_{s13}$  &   $1.00000 \times 10^{-5}$ & $\mathcal{O}_{s31}$   &  $1.00000 \times 10^{-5}$ \\
$\psi$     &   $0.72066$   &  $\mathcal{O}_{s21}$  &   $1.00000 \times 10^{-5}$ & $\mathcal{O}_{s32}$  &  $1.00000 \times 10^{-5}$  \\
$\theta$   &   $3.13801$    &                       &  \\ \toprule
\multicolumn{6}{c}{$U-s$: $\;|V_{Uij},\mathcal{O}_{sij}|>10^{-5}$, $\mathcal{O}_{s11}=\mathcal{O}_{s22}=1$} \\ \midrule
$M_U$      &   $1800$      &  $\mathcal{O}_{s11}$  &  $1.00000$                 & $\mathcal{O}_{s22}$  &  \multicolumn{1}{c}{$1.00000$}  \\
$M_s$      &   $580$       &  $\mathcal{O}_{s12}$  &  $-1.12685 \times 10^{-5}$ & $\mathcal{O}_{s23}$  &  $-8.60549 \times 10^{-4}$  \\
$\phi$     &   $0.00020$   &  $\mathcal{O}_{s13}$  &  $-1.19611 \times 10^{-3}$ & $\mathcal{O}_{s31}$  &  $1.83141 \times 10^{-4}$ \\
$\psi$     &   $0.00024$   &  $\mathcal{O}_{s21}$  &  $-1.13022 \times 10^{-5}$ & $\mathcal{O}_{s32}$  &  $2.55491 \times 10^{-3}$  \\
$\theta$   &   $3.09077$   &                       &  \\ \toprule
\end{tabular}
\end{center}
\end{table}

The exclusion contours are shown in Figure~\ref{fig:contoursUs} and the solutions appear on Table~\ref{Tab:solUs}. From the contours, we learn that to allow for bilepton masses of the order of $M_U<\SI{1}{TeV}$, at least some effective couplings $g_{\text{eff}}\sim g_U V_{Uij}$ must be set as low as $<10^{-4}$, meanwhile the entire parameter space is possible if the matrix elements are allowed to become as small as $\sim \hspace*{-0.18em} 10^{-5}$, showing, again, the complementarity between the phenomenology of CLFV decays and LHC processes, which cover the non-diagonal and diagonal $V_U$ models, respectively. Additionally, Figure~\ref{fig:contoursUsd} shows that if $\mathcal{O}_{s11}=\mathcal{O}_{s22}=1$, is enforced the bound on $M_U$ is strengthened from $M_U>\SI{500}{GeV}$ to $M_U>\SI{1500}{GeV}$ while virtually unchanging the bound on $M_s$. This just reasserts that the vector contribution is indeed dominant.

We would also like to answer the question of weather destructive interference can save interesting sectors of theory space (with light exotic vector particles) from exclusion. In practice, this question is translated into the need to investigate how the presence of a second particle may relieve naive constraints derived from single exotic particle Lagrangians. For this we show, in Figure~\ref{fig:interference}, a density plot of the ratio

\begin{equation}
\frac{\text{BR}_{U-s}+\text{BR}_{s}}{\text{BR}_{U}},
\end{equation}
where $\text{BR}_{X-Y}$ is the contribution of the interference between $X$ and $Y$ and $\text{BR}_{X}$ is the pure $X$ contribution to the BR.   We notice that, even if the scalar contribution is significantly smaller, it allows the solution to enhance destructive interference, which causes the distortion on the inferior left corner of the contour and contributes to, in the $|O_{sij},V_{Uij}|>10^{-3}$ case, rendering constraints softer by 19\% on $M_U$.

\begin{figure}[t!]
	\centering
		\includegraphics[width=0.6\linewidth]{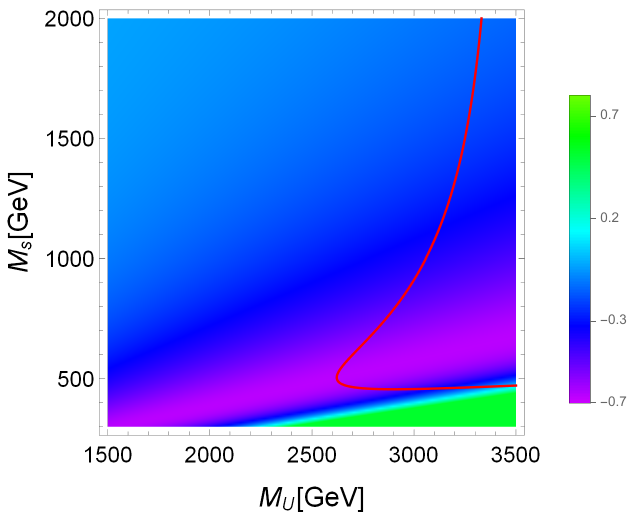}
		\caption[Density plot of the neutral scalar $s$ contributions to $\mu^+ \to e^+ e^- e^+$ in a model with a $U^{\pm \pm}$.]{Density plot of the extra contributions to $\mu^+ \to e^+ e^- e^+$ caused by the addition of the neutral scalar $s$ to a model with the doubly-charged vector bilepton $U^{\pm \pm}$. }
	\label{fig:interference}
\end{figure}

\section{$U-Y$ {\it Scenario}}

The {\it Scenario} with a doubly-charged vector and scalar bileptons is much simpler as a consequence of the absence of relevant interference. The contributions of interference between the $U^{\pm \pm}$ and the $Y^{\pm \pm}$ to this averaged fermionic process is proportional to the lepton masses (see Fig.~\ref{fig:interferenceDiagsUY}). In particular, this means that interference effects are absolutely negligible in the formation of the most powerful bound from $\mu \to 3e$. This indicates that possible solutions for this {\it Scenario} involve vector bilepton parameters identical to those of the pure $U^{\pm \pm}$ {\it Scenario} with $Y^{\pm \pm}$-related parameters very small in modulus -- the least allowed by the benchmark conditions. This guarantees that the  $Y^{\pm \pm}$ contribution is rendered insignificant and does not affect the exclusion contour, turned similar to those of Figure~\ref{fig:contoursU}.

\begin{figure}[t!]
	\centering
		\includegraphics[width=0.45\linewidth]{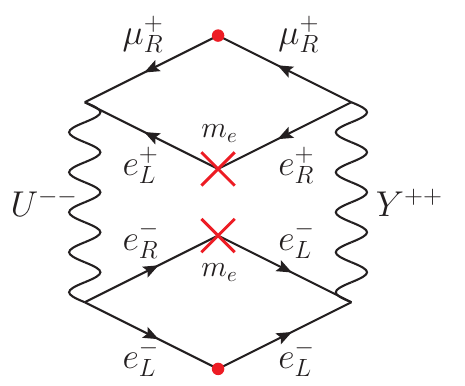}
		\caption[Heuristic demonstration that the interference between the $U$ and $Y$ contributions to the 3-body CLFV decays depend on $m_e$.]{Verification of the interference structure between the $U$ and $Y$ contributions. Now, the diagrams can only be joined together through $e$ mass terms.}
	\label{fig:interferenceDiagsUY}
\end{figure}

\begin{figure}[t!]
	\centering
		\includegraphics[width=0.6\linewidth]{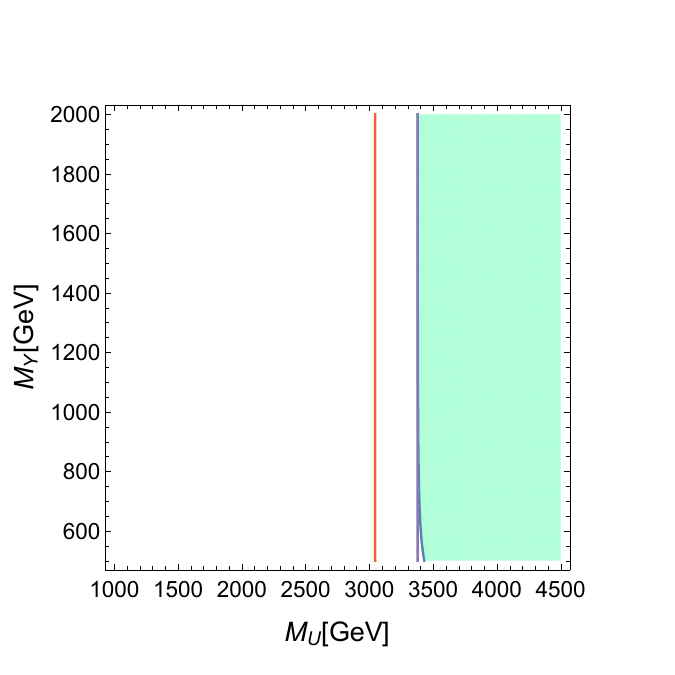}
		\caption[Exclusion contours on the $M_U \times M_Y$ plane generated by the bounds on the 3-body CLFV decays.]{Exclusion contour on the $M_U \times M_Y$ plane, showing that the solution subjected to the $|O_{Yij},V_{Uij}|>10^{-3}$ condition, prioritizing $M_U$, is analogous to the one of the pure $U^{\pm \pm}$ case, with negligible scalar contributions (except for the distortion at exceedingly low $M_s$). }
	\label{fig:contoursUY}
\end{figure}

To illustrate the point above, we show, in Figure~\ref{fig:contoursUY} the plot corresponding to the solution of Fig.~\ref{fig:contoursUa} together with $O_{Yij}=10^{-3}$. The results confirm the simple numerical thesis that, prioritizing the $U$ and forcing the scalar contributions to be as small as possible, the optimal bound is analogous to that of the pure $U$ {\it Scenario}.

\section{$Y-s$ {\it Scenario}}

The double scalar {\it Scenario} is even less involved. The interference is, again, proportional to $m_e$ for the strongest source of bounds and, in this case, there is no unitary mixing. Consequently, the structure of the solution is such that lower masses become allowed with diminishing couplings. We enforce that, in the optimal point, scalar masses are nearly degenerate. Then we see, from Figure~\ref{fig:contoursYs}, that couplings of the order $g_{\text{eff}} \sim O_{sij} \sim 10^{-2}$ allow for scalar masses of the order of \SI{2.5}{TeV}, while couplings as small as $10^{-3}$ are permissive of low masses. It is easy to notice that, in this case, since there is neither conditions tying different matrix elements together nor interference, the strongest bound, \textit{i.e.}, that of $\mu \to 3e$, is the only one that matters.

\begin{figure}[t!]
\begin{center}
\adjustbox{center=10cm}{
	\makebox[1.2\linewidth]{
	\begin{subfigure}{0.5\linewidth}
	\centering
		\includegraphics[width=\linewidth,trim=0cm 0.4cm 0.8cm 1.2cm,clip]{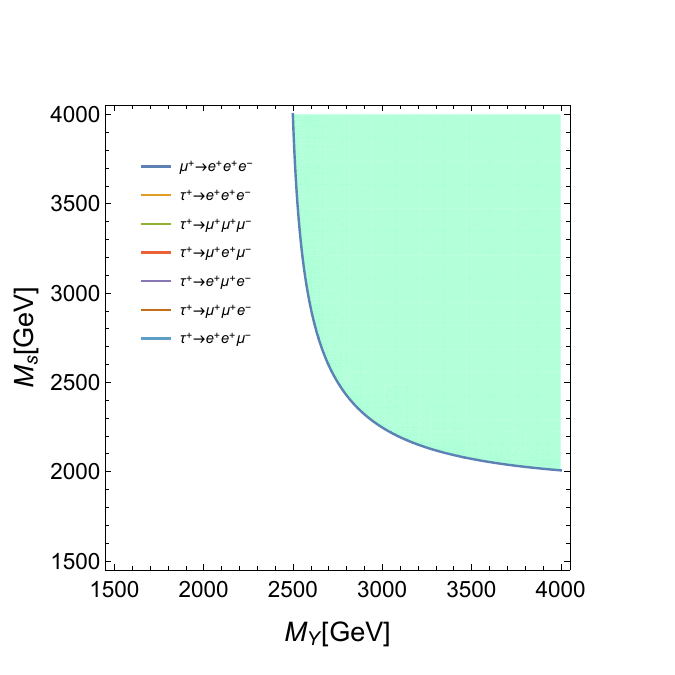}
		\caption{$|O_{Yij},O_{sij}|>10^{-2}$}
	\end{subfigure}
	\begin{subfigure}{0.5\linewidth}
	\centering
		\includegraphics[width=\linewidth,trim=0cm 0.4cm 0.8cm 1.2cm,,clip]{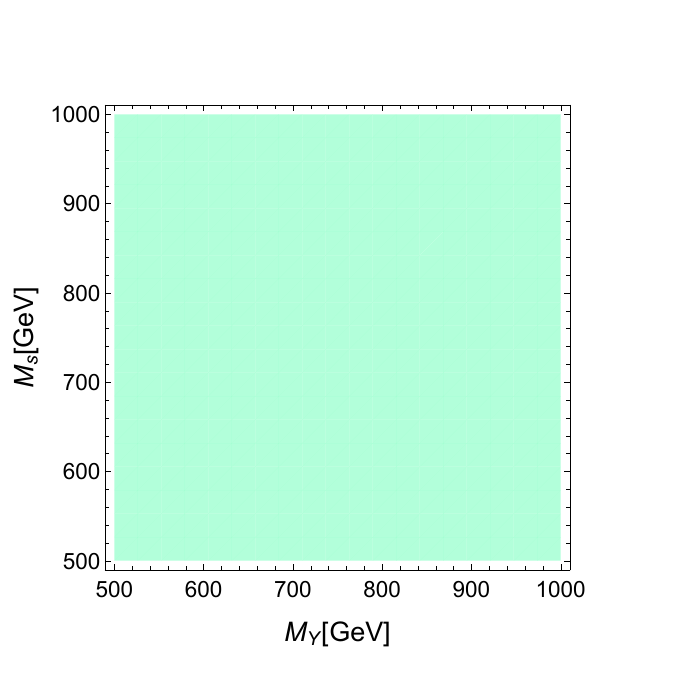}
		\caption{$|O_{Yij},O_{sij}|>10^{-3}$}
	\end{subfigure}}}
	\caption{Exclusion contours on the $M_Y \times M_s$ plane for bosons constrained to have similar masses on the optimal point.}
	\label{fig:contoursYs}
\end{center}
\end{figure}

\section{Analysis and conclusions}
\label{sec:LDconc}

A crucial aspect in the search for new physics is knowing where to look, which turns the constraining of theory space through its exotic parameters a central objective. The crown of this line of work is usually the obtaining of lower bounds on masses of hypothetical particles. The high explorable energies and the excited stage of high amount of data collection reached by the LHC make it one ideal tool in this quest.  As we have exhaustively discussed, it has been used to derive constraints on $M_U$, the mass of a doubly-charged vector bilepton, rare feature of BSM models,whose collective result is approximately well described by the bound $M_U \gtrsim \SI{1}{TeV}$.

Now, in any general phenomenological research, if purely leptonic processes (such as the 3-body decays) are able to provide for useful bounds, their operational advantages are manifest. In the specific case of the CLFV $U^{\pm\pm}$ physics, although the 3-body phase space is more computationally demanding than the LHC counterpart, the hadron physics and background analysis needed in LHC phenomenology are an immensely heavier complication.

We have showed that the simple bounds on the branching ratios of the 3-body lepton decays produce strong constraints on the bilepton mass and allow to explore general mixing matrices, regularly neglected. A fine representative of this is the pure $U^{\pm \pm}$ {\it Scenario}, in which we predict $M_U> \SI{3200}{GeV}$ if the hierarchy within $V_U$ is of the order or lesser than $10^{3}$. We conclude that the this data complement the LHC flavour-diagonal phenomenology, and is, furthermore, considerably more effective in the case of finite mixing if compared with our results of Chapter~\ref{chap:LHCTrimuons}, which considers a CLFV LHC process treating $V_U$ through a simplified construction.

We do not primarily intend to achieve new specific mass bounds for the scalars: the interactions these particles participate in are not governed by unitary mixing, and there are concrete (model-dependent) experimental bounds over the doubly-charged scalars \cite{ATLAS:2017xqs,ATLAS:2014kca}, and the neutral scalar is a well known and common particle, analogous to the Higgs boson, so that its phenomenology is well understood in most models where it is present \cite{ATLAS:2017eiz, CMS:2018rmh, ATLAS:2019nkf, ATLAS:2018bnv}. Nevertheless, we considered a pure scalar $Y-s$ {\it Scenario} and what we find is that for low masses to be possible after enforcement of the CLFV bounds, the effective coupling must be of order of $10^{-3}$. For comparison, the corresponding SM $H e\bar{e}$ and $H\tau \bar{\tau}$ couplings have strength $g_{He\bar{e}} \sim 2.07 \times 10^{-6}$ and $g_{H\tau\bar{\tau}} \sim 7.24 \times 10^{-3}$, indicating that it is certainly reasonable for an exotic flavour violating $C\!P$-even neutral scalar, generally associated with higher characteristic mass scales, to possess interactions parametrized by effective couplings of the order necessary for its mass to be low.

The addition of the scalar bosons to our analysis is mainly intended to aid us understand the part that secondary, non-dominant, particles can play altering the naive (single particle) exclusion contours of dominant degrees of freedom, which, in the present context, occurs when it is considered together with the $U^{\pm \pm}$. We observe that the balance between the $U^{\pm \pm}$ and $s$ contributions follows a trend in which, in the optimal interference region, the lower bound on the mass of the $U^{\pm \pm}$ is relieved by 20\%. Although it could be argued that such phenomenon can only happen in small, fine-tuned regions of parameter space, this behavior can happen in a general multi-particle scenario, specially in ones where a subset of parameters is constrained by exterior phenomenological or theoretical input, like the fitting of well measured distinct masses or mixing parameters.


\cleardoublepage

\chapter{The 331 perturbative regime}\label{Chapter8}

We have now deepened our understanding of the BSM particle $U^{\pm\pm}$ through two distinct phenomenological inquiries: first an analysis of its contribution to the LHC trimuon process~\cite{PhysRevD.101.015024}, focusing on the role of the unitary mixing. In turn, this motivated a study on how it is affected by the limits on the 3-body CLFV lepton branching ratios~\cite{Barela:2022sbb}, and whether this older data could supersede the utility of the LHC bounds under some conditions. Although all we have done is model independent to a high degree (specially the study over the 3-body decays), our investigations are highly relevant to the m331 and vice versa. This last part of the thesis takes a parallel path, explicitly focusing on this model. 

The m331 is understood to present arbitrary growth in the evolution of its $U(1)_X$ coupling (sometimes called a Landau pole) at characteristic energies of around $\SI{4}{TeV}$~\cite{Dias:2004dc,Martinez:2006gb,Santos:2017jbv,Doff:2023bgy}. If unchanged by additional theoretical mechanisms such as emerging states in the theory, this fact severely limits the usefulness of the model, as it implies that perturbativity is lost at nearly experimentable low TeV scales. This fact is usually derived through an approximation which considers the SM symmetry at intermediate scales. We escape this simplified method and explicitly perform the 1-loop renormalization of the heavy contributions to the running of constants. 
The results present a much safer model regarding its perturbativity regime.

\section{Symmetry Matching Conditions and Symmetric RGE}
\label{sec:matching}

%

Consider the breaking pattern $G_1 \times G_2 \times \cdots \times G_n \to G$, where all factors are simple, with respective coupling constants $g_1,g_2,\cdots g_n$ and $g$. In this case, at the breaking scale, the matching conditions for $g$ usually fall into one of the following two categories \cite{Georgi:1977wk}:

\begin{enumerate}
\item $G \subset G_1$

If the embedding is such that the lower group is entirely contained within a simple higher group factor, the condition becomes simply

\begin{equation}
g = g_1.
\end{equation}

\item $G \subset G_1 \times G_2 \times \cdots G_n$

Suppose, more generally, that $G$ is contained within the non-simple product and specialize to the case $G = U(1)$. The embedding may be parametrized as\footnote{Clearly, the product of simple groups in which $G$ is embedded could be a proper subset of the complete symmetry above the threshold, in which case $p_i = 0$ for some $i$.}

\begin{equation}
Z = \sum_{i=1}^n p_i T_i,
\end{equation}
where $Z$ is the $U(1)$ generator and $T_i$ collectively denote every generator of $G_1 \times G_2 \times \cdots G_n$. In this case, the matching condition reads\footnote{Notice that, in the sum above, every distinct $i$-term which corresponds to generators of a same group exhibits the same coupling factor, \textit{i.e.}, $g_i = g_j$ if $T_i,T_j$ belong to the algebra of the same simple group.}

\begin{equation} \label{eq:matchEq}
\frac{1}{g^2} = \sum_{i=1}^n \frac{p_i^2}{g_i^2}.
\end{equation}
\end{enumerate}

These conditions are not analogous to the finding of Wilson coefficients in the matching of effective theories at mass thresholds. They are, rather, simply a consequence of requiring that the theory can be described by Lagrangians with the expected symmetry and corresponding couplings at each energy range, and that at the breaking scale the theories coincide, as required by continuity. A simple example is the SM relation

\begin{equation} \label{eq:SMMatching}
\frac{1}{e^2} = \frac{1}{g_{2L}^2} + \frac{1}{g_{Y}^2},
\end{equation}
 which is a consequence of the definition of electric charge and that may, at the level of the lagrangian, be read from 

\begin{equation}
Q = T_3 + Y.
\end{equation}

Once the different couplings, defined to exist at distinct energy ranges, are correctly matched at the symmetry transition scales, they may be evolved through the RGE between thresholds or towards infinity according to

\begin{equation}
\beta(g) = \mu \frac{dg}{d\mu}.
\end{equation}
The $\beta$-function {\it in a fully gauge-symmetric theory} may be written as

\begin{equation} \label{eq:betaeq}
\beta(g) = -\frac{g^3}{(4 \pi)^2}b_1,
\end{equation}
where the $\beta$-function coefficient at one loop may be readily found from its related group theoretical quantities \cite{Roy:2019jqs}

\begin{equation} \label{eq:betacoeff}
b_1 = \frac{11}{3}C_2(\text{Gauge}) - \frac{4}{3}\kappa S_2 (\text{Fermion}) - \frac{1}{6} \eta S_2(\text{Scalar}),
\end{equation} 
where $C_2(R)$ and $S_2(R)$ are, respectively, the Casimir and Dynkin Index invariants of the representation $R$, and $\kappa=1/2(1)$ for Weyl(Dirac) components and $\eta = 1(2)$ for real(complex) scalars.

\section{Exotic mass scales and rotations in the m331}
\label{sec:331}

In order to more clearly define our framework, let us recall a few facts of the m331 model already laid down in Chapter~\ref{Chapter4}. To start, we have found that, after diagonalization, the electrically charged part of the spin-1 sector of the model is comprised by new singly and doubly charged bosons, written in terms of the symmetry eigenstates, along with the $W$-boson, simply as

\begin{equation}
\begin{split}
W_\mu^\pm&=(W_\mu^1\mp iW_\mu^2)/\sqrt{2} \\
V_\mu^\pm&=(W_\mu^4\pm iW_\mu^5)/\sqrt{2} \\
U_\mu^{\pm\pm}&=(W_\mu^6\pm iW_\mu^7)/\sqrt{2}, \\
\end{split}
\end{equation}
with masses

\begin{equation} \label{eq:chargedbmass2}
\begin{split}
M_W^2&=\frac{1}{4}g_{3L}^2 (v_\eta^2 + v_\rho^2 + 2v_s^2) \equiv \frac{1}{4}g_{3L}^2 v_W^2 \\
M_V^2&=\frac{1}{4}g_{3L}^2( v_\eta^2 + 2v_s^2+v_\chi^2)\\
M_U^2&=\frac{1}{4}g_{3L}^2( v_\rho^2 + 2v_s^2+v_\chi^2).
\end{split}
\end{equation}
Note that, to numerically fit the $W$ mass, $v_\eta^2 + v_\rho^2 + 2v_s^2$ must sum to the Higgs VEV $v_W^2$.

The last addition with respect to the SM is an extra neutral vector boson $Z^\prime$. The diagonalization of the neutral sector is immensely more complicated and given by

\begin{equation}
(W_{3\mu},W_{8\mu},B_{\mu})^T = \mathcal{O}\cdot (Z_{\mu},Z^\prime_{\mu},A_\mu)^T,
\end{equation}
where the $\mathcal{O}$ matrix (with basis ordered as above) is given exactly and in general by

\begin{equation}\label{eq:NPNGBRotation}
\mathcal{O} = \begin{pmatrix}
-N_1 a_1 & -N_2 a_2 & \frac{t_X}{\sqrt{4t_X^2+1}} \\
-\sqrt{3}N_1 b_1 &  -\sqrt{3}N_2 b_2 & -\frac{\sqrt{3}t_X}{\sqrt{4t_X^2+1}} \\
2t_X\left(1-\bar{v}_\rho^2\right)N_1 & 2t_X(1-\bar{v}_\rho^2)N_2 & \frac{1}{\sqrt{4t_X^2+1}}
\end{pmatrix},
\end{equation}
%
with

\begin{equation}
\begin{split}
a_{1(2)} &=  3 m_{2(1)}^2+\bar{v}_\rho^2-2\bar{v}_W^2 \\
b_{1(2)} &= m_{2(1)}^2+\bar{v}_\rho^2-\frac{2}{3}\bar{v}_W^2-\frac{2}{3}, 
\end{split}
\end{equation}
where the overbar indicates the ratio by $v_\chi$ as $\bar{v}_\alpha\equiv \frac{v_\alpha}{v_\chi}$, and $t \equiv \tan \theta_X \equiv \frac{g_X}{g_{3L}}$. The normalization factors are given by

\begin{equation}\label{eq:normfactors2}
\begin{split}
N_1^{-2}&=3 \left(2m_2^2+\bar{v}_\rho^2-\frac{4}{3}\bar{v}_W^2-\frac{1}{3}\right) + (\bar{v}_\rho^2-1)^2(4t_X^2+1) \\
N_2^{-2}&=3 \left(2m_1^2+\bar{v}_\rho^2-\frac{4}{3}\bar{v}_W^2-\frac{1}{3}\right) + (\bar{v}_\rho^2-1)^2(4t_X^2+1).
\end{split}
\end{equation} 
We have defined the factors

\begin{equation}\label{eq:massfactors2}
\begin{split}
A&=\frac{1}{3}\left[ 3t_X^2\left(\bar{v}_\rho^2+1 \right) + \bar{v}_W^2 + 1\right] \\
R&=\left\lbrace 1-\frac{1}{3A^2}\left(4t_X^2+1 \right)\left[\bar{v}_W^2\left(\bar{v}_\rho^2+1 \right) -\bar{v}_\rho^4\right]\right\rbrace^{1/2},
\end{split}
\end{equation}
and the dimensionless masses 

\begin{equation} \label{eq:neutralbmass2}
\begin{split}
m_{1}^2&=\frac{2 M^2_{Z_1}}{g_{3L}^2 v_\chi^2}=A(1-R) \\
m_{2}^2&=\frac{2 M^2_{Z_2}}{g_{3L}^2 v_\chi^2}=A(1+R).
\end{split}
\end{equation}
%

%
%
%

The mass matrices of the exotic quarks are of the form

\begin{equation}
M^J_{ab} = \frac{v_\chi}{\sqrt{2}}y^J, \;\;\;\; M^{j_i}_{ab} = \frac{v_\chi}{\sqrt{2}}K^{j}_{ab},
\end{equation}
which are proportional to the large $v_\chi$ and can be made arbitrarily massive by the free (besides possible phenomenological and unitarity constraints) Yukawa couplings $y^J,K^{j}_{ab}$. As for the scalars, the model contains, in total, four singly-charged, three doubly-charged, five $\mathit{CP}$-even neutral and three $\mathit{CP}$-odd physical scalars. The sheer amount of exotic parameters in the scalar sector together with the absence of TeV scale phenomenological input allow scalar masses to be large as well. Henceforth, the words {\it exotic} and {\it heavy} will be used interchangeably to qualify all the particles in the set

\begin{equation}
\left\{ Z^\prime, \; V^\pm, \; U^{\pm \pm}, \; j_1^{-4/3}, \; j_2^{-4/3}, \; J^{5/3}, \; \text{15 exotic scalars}\right\}.
\end{equation}

\section{Approximating the m331 by the SM symmetry}
\label{sec:ext}

\subsection{Matching and strategy}
\label{sec:extMath}

The usual manner to investigate the running structure of the gauge couplings takes advantage of the fact that, below the scale of importance of the exotic particles (all heavier), the m331 must be approximated by the SM. This is a consequence of the natural decoupling of the heavier particles and of the apparent absence of new physics up to the TeV scale. Furthermore, the SSB of the model automatically splits into two processes, a fact which can be used to describe the approximation mentioned above at a given energy range. From this perspective, the breaking pattern may be written as



\begin{equation} \label{eq:extbreaking2}
\begin{split}
\hspace*{-10cm} SU(&3)_c \times SU(3)_L \times U(1)_X \\
&\hspace*{8mm} \xrightarrow[E_\text{high}]{\langle \chi \rangle} SU(3)_c \times SU(2)_L \times U(1)_Y  \\
&\hspace*{11mm}\hspace{10mm}\xrightarrow[E_\text{low}]{\langle \eta \rangle,\langle \rho \rangle} SU(3)_c \times U(1)_{\text{EM}}.
\end{split}
\end{equation}
The schematics above represents that, at the unknown $E_\text{high}$ scale, the 3-3-1 symmetry is broken down to the SM one, after which, at $E_\text{low}$ (to be identified with the electroweak scale), the breakdown to the conserved sector occurs.

Referring to this, the full process induced by the RGE transformations is understood as follows: (\textit{i}) Nature is assumed to be well described by the $SU(3)_c \times SU(2)_L \times U(1)_Y$ symmetry immediately above an $E_\text{low}$ scale, below which the theory is supposed to be broken to the conserved $SU(3)_c \times U(1)_\text{EM}$; At this threshold, the couplings are fixed to their numerically known values; (\textit{ii}) $g_{2L},g_{Y}$ are evolved, according to the appropriate particle content (see below and in the next section), with increasing energy, up to $E_\text{high}$; (\textit{iii}) The second matching is performed to replace $g_{2L},g_{Y}$ by the emergent $g_{3L},g_{X}$; (\textit{iv}) The couplings are evolved again, now up to the pole, according to a second structure corresponding to the higher symmetry and larger set of particles suitable to the new energy range. Notice that $E_\text{high}$ is the scale in which the effects of the heavy particles become important, which causes the apparent symmetry to change.

At $E_\text{low}$, generically taken to be at the $Z$-pole from now on, $E_\text{low} = M_Z = \SI{91.1876}{GeV}$, the known SM couplings are simply fixed to their well measured numerical values \cite{ParticleDataGroup:2022pth}:

\begin{equation}
g_{2L}(E_\text{low}) = 0.62977, \;\;\;\;\; g_Y(E_\text{low}) =  0.34537.
\end{equation}

\begin{table}[t!]
\caption{$\beta$-function coefficients of the m331 couplings in the two energy ranges. Two configurations for the representation content are emphasized: `Full' refers to the entire m331 content, whereas only the SM degrees of freedom contribute to `SM'.}
\begin{center}
\begin{tabular}{ccccc}
\toprule
\hphantom{$E_\text{low}<\mu<E_\text{high}$} & $g_{Y(X)}$: Full & $g_{2L(3L)}$: Full & $g_Y$: SM & $g_{2L}$: SM
 \\ \midrule
$E_\text{low}<\mu<E_\text{high}$  &   $38$      &  $-2$    &  41/6   &  -19/6 \\
$\mu>E_\text{high}$              &   $22$      &  $-17/3$ &  &   \\ \toprule
\end{tabular}\label{Tab:bcoeffext}
\end{center}
\end{table}

At $E_\text{high}$, the two exotic couplings may be matched to the SM ones according to the formalism reviewed in Sec.~\ref{sec:matching}. Regarding $g_{3L}$, it is enough to realize that the relevant embedding structure obeys $SU(2)_L \subset SU(3)_L$, causing the necessary condition to be simply

\begin{equation}\label{eq:g3Lg2L}
g_{3L}(E_\text{high}) = g_{2L}(E_\text{high}).
\end{equation}

For $g_X$, we recall 

\begin{equation}
Y = -\sqrt{3} T_8 + X,
\end{equation}
which is a consequence of Eq.~(\ref{eq:charge}). Of inserting this information into formula (\ref{eq:matchEq}), results

\begin{equation}\label{eq:Matchingextbig}
\frac{1}{g_Y(E_\text{high})^2} = \frac{3}{g_{3L}(E_\text{high})^2} + \frac{1}{g_{X}(E_\text{high})^2},
\end{equation}
or, plugging Eq.~(\ref{eq:g3Lg2L}),

\begin{equation} \label{eq:gXFinalMatching}
g_X(E_\text{high}) = \frac{g_{2L}(E_\text{high}) \, g_Y(E_\text{high})}{\sqrt{g_{2L}(E_\text{high})^2 - 3 g_Y(E_\text{high})^2}}.
\end{equation}

Once more, it must be emphasized that these boundary conditions on the breaking scales are merely consistency requirements to guarantee the supposed symmetry structure and continuity. In fact, they obviously could be found -- and usually are -- by explicit brute force comparison of the two lagrangians with correct symmetry properties around the threshold. 

The $\beta$-function coefficients are straightforward to calculate through Eq.~(\ref{eq:betacoeff}) and are shown in Table~\ref{Tab:bcoeffext}. The values corresponding to the m331 with the exotic degrees of freedom removed are also presented. Note that removing the entire exotic sector within the 3-2-1 (SM symmetry) energy range amounts to eliminating closed multiplets, keeping the theory symmetric. This assures that the exact RGE may be solved directly, with the $\beta$-function coefficients given by the classic result of Eq.~(\ref{eq:betacoeff}).

\subsection{Results}

If the entire particle content is naively kept from very low energies, the explicit $g_X$ dependence on energy is given by

\begin{equation}
\begin{split}
g_X^{\text{full}}(\mu) = 2\pi (3&0.1 - 11.0 \log{E_\text{high}} + 22.0 \log{E_\text{low}} - 11.0 \log{\mu})^{-1/2}, \;\;\;\; \mu>E_\mathrm{high},
\end{split}
\end{equation}
with the immediate result that for $E_\text{high} \gtrsim \SI{370}{GeV}$
$g_X^{\text{full}}$ becomes imaginary in the entire higher range. This signifies that, for these choices of $E_\mathrm{high}$, the mandatory matching conditions are impossible to fulfil, rendering the theory senseless in the higher symmetry regime. The upper bound on $E_\text{high}$ then reads

\begin{equation}
E_\text{high} \lesssim \SI{370}{GeV}.
\end{equation}
However, such small values for the breaking scale are clearly not phenomenologically viable. 

This result does not come as a surprise, since in $\overline{\text{MS}}$ the decoupling of heavy particles is not automatic as happens in physical renormalization schemes, and must be introduced by hand \cite{Appelquist:1974tg}. If enforced, and only the known particles are kept at energies $\mu < E_\text{high}$, with the heavy degrees of freedom integrated out and only included above this threshold, the coupling depends on energy as

\begin{equation}
\begin{split}
g_X^{\text{SM}}(\mu) = 2\pi (3&0.1 + 2.83 \log{E_\text{high}} + 8.17 \log{E_\text{low}} - 11.0 \log{\mu})^{-1/2}, \;\;\;\; \mu>E_\mathrm{high}.
\end{split}\label{eq:gXknown}
\end{equation}
The corresponding behaviour for several benchmark $E_\text{high}$ is shown in Figure~\ref{fig:extSMdof}. It may be observed that, for breaking scales obeying $E_\text{high} \gtrsim \SI{3770}{GeV}$, the $g_X$ coupling is highly vertical, originating larger than $4 \pi$ and rapidly diverging, and there is no effectively perturbative window. The upper bound with the heavy particles decoupled becomes

\begin{equation}
E_\text{high} \lesssim \SI{3770}{GeV}.
\end{equation}
Unfortunately, even for the optimal $E_\text{high} = \SI{3770}{GeV}$ perturbativity is quickly harmed at around $\sim \hspace*{-0.4mm}\SI{4600}{GeV}$. This result is usually quoted without the corresponding choice of $E_\text{high}$, which, in turn, is usually thought to define the $v_\chi$ parameter. In summary, assuming the position for the pole in $g_X$ automatically induces phenomenological consequences.

\begin{figure*}[t!]
	\centering
		\includegraphics[width=0.7\linewidth]{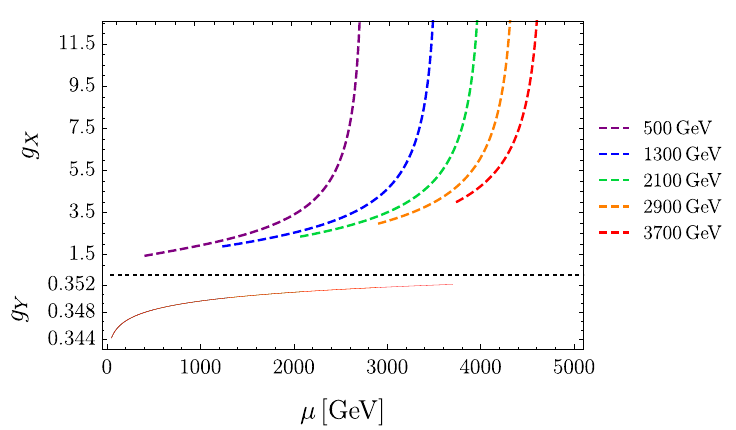}
		\caption[Running of $g_Y$ and $g_X$ in the SM approximation.]{Running of $g_Y$ (solid) and $g_X$ (dashed) in the SM approximation for five $E_\text{high}$ benchmarks.}
	\label{fig:extSMdof}
\end{figure*}

The conclusions of this section could be qualitatively understood by noting that, in the matching condition,

\begin{equation}
\frac{1}{g_X^2} =  \frac{1}{g_Y^2}-\frac{3}{g_{2L}^2},
\end{equation}
the RHS is smaller than one ($\sim \hspace*{-0.4mm}0.84$) already at the electroweak scale, and its reciprocal very sensitive to small changes in $g_Y$ and $g_{2L}$, which are increasing and decreasing, respectively.  

\section{Effective approach}
\label{sec:alt}

\subsection{Matching and strategy}
\label{sec:altMath}

The conclusion from the last section is that, to match an exact 3-2-1 -- the SM as embedded in the m331 -- to an exact 3-3-1 theory is only possible in a very low regime. Such attempts are justified by the reasoning that, if the exotic particles are ignored, interactions should be approximately parametrized as within the SM. This, in turn, should be possible within a model like the m331 because they are constructed with an SM embedding in mind, which guarantees that its good predictions are reproduced. To use the SM as a low energy description of an extension, however, is only an approximation (to be made around a given characteristic energy scale), and does not translate well to an RGE analysis. At zero temperature all that matters are the form and content of the non decoupled interactions, and the definition of an active intermediate symmetry, as in Eq.~(\ref{eq:extbreaking2}), is only a convenient manner to describe them in a theory with the embedded SM and every exotic effect decoupled. This, however, is not necessarily possible at an arbitrary mass regime. Note, in this sense, that in a general model it is not strictly necessary for the SSB pattern to decompose as in the m331: the VEV acquisition for distinct scalars (or subgroups of them) could each trigger the direct breaking to the electromagnetism or to arbitrary intermediate groups. Nonetheless, in principle, such a theory could still produce all the electroweak predictions of the SM.  

We shall now take on a more precise approach, in which the 3-3-1 parametrization is assumed from the start. What can be gained from this is the circumventing of the symmetry matching process (which is artificial and an approximation), and the downside is the need for a brute force, explicit approach, as Eq.~(\ref{eq:betacoeff}) is no longer valid when the heavy degrees of freedom are integrated out. 

%
%

To describe the coupling evolution process, notice that $g_{2L},g_Y$ are now undefined, and $g_{3L},g_X$ hold unrelated new values, to be numerically matched at $E_\text{low}$ with experimentally obtained quantities. For that,
match the strength of the $\gamma e \bar{e}$ and $W e \bar{\nu}$ interactions to their experimental counterpart. The requirements read

\begin{equation}\label{eq:EffMatching}
\begin{split}
\Gamma_\mu^{\gamma e \bar{e}} &= i\gamma_\mu \left(   -\frac{g_{3L}}{2} \mathcal{O}_{13}P_L + \frac{g_{3L}}{2\sqrt{3}} \mathcal{O}_{23}P_L + \frac{g_{3L}}{\sqrt{3}} \mathcal{O}_{23}P_R   \right) \overset{!}{=} -i|e|\gamma_\mu \\
\Gamma_\mu^{W e \bar{\nu}} &= i \frac{g_{3L}}{\sqrt{2}} \gamma_\mu \overset{!}{=} i \frac{g_{2L}}{\sqrt{2}} \gamma_\mu,
\end{split}
\end{equation}
where $\mathcal{O}$ rotates mass to symmetry eigenstates like $(W_{3\mu},W_{8\mu},B_{\mu})^T = \mathcal{O}\cdot (Z_{\mu},Z^\prime_{\mu},A_\mu)^T$, and is given explicitly in Eq.~(\ref{eq:NPNGBRotation}). In the end, the matching amounts to setting

\begin{equation}
g_{3L}(M_Z) = 0.63, \;\;\;\; g_{X}(M_Z) = 1.1.
\end{equation}

%

Let us now discuss how heavy particle decoupling may be implemented. In the last section, the theory that remained after elimination of the heavy particles continued symmetric in the SM approximation, and it was enough to modify the $\beta$-function coefficients to encompass the light multiplets alone. This was made possible because the SM projection of the m331 contains every exotic degree of freedom as a singlet, which may be removed without spoiling the symmetry. An analogous situation occurs if one inspects the RGE of the SM without the third generation of quarks. In the other hand, if the gauge symmetry is broken by disregarding a few particles, some brute force method becomes necessary. This is exemplified by the exclusion of the top quark from the SM, and corresponds to our problem at hand since the $SU(3)_L$ triplets and octet are broken once the heavy particles are removed. The theory is no longer symmetric below $E_\mathrm{high}$, and an explicit, specific calculation cannot be escaped. 

The procedure amounts to subtracting from the complete, symmetric $\beta$-functions the part resulting from the 1-loop contributions of the exotic particles (indicated by an $\mathscr{H}$ superscript), $\delta Z^\mathscr{H}_{g_{3L}},\delta Z^\mathscr{H}_{g_X}$, only activated above the $E_{\mathrm{high}}$ threshold. For that end, we construct a completely determined system of equations from which the counterterms of the couplings may be obtained. In order to build such a system, the counterterms for at least two vertex functions are required, and are conveniently chosen to be the $W u \bar{d}$ and $\gamma u \bar{u}$ ones. 

To summarize, the evolution process within the current construction follows as:  (\textit{i}) At $E_\text{low}$, we match $g_{3L},g_X$ to fit $Z$-pole interactions as in Eq.~(\ref{eq:EffMatching});  (\textit{ii}) We initially evolve the couplings with increasing energy according to an effective representation content where every exotic particle has been integrated out through some procedure; (\textit{iii}) At a parametrically free scale $E_\text{high}$, referring now to the one in which their effects become important, we include the heavy particles back into the theory, match to the model below, and evolve the couplings up to the pole. Notice that only $SU(3)_L \times U(1)_X$ quantities are ever mentioned. 

\subsection{Renormalization framework}

We renormalize the various quantities by making\footnote{Note the distinction between our convention and the common one which renormalizes the gauge vertex as $\Gamma_0^{V f_1 \bar{f_2}} = Z_g \Gamma^{V f_1 \bar{f_2}}$. In that alternative definition, one has $g_0 = \frac{Z_g}{Z_V^{1/2} Z^{1/2}_{f_1}Z^{1/2}_{f_2}}$~\cite{Srednicki:2007qs}.}

\begin{equation}\label{eq:RenDef1}
f_0 = Z_f^{1/2} f, \;\;\;\; V_0 = Z_V^{1/2} V, \;\;\;\; g_0 = Z_g g,
\end{equation}
where $f$ is any fermion, $V$ any vector boson field and $g$ any gauge coupling. The vertex functions to be renormalized read

\begin{equation}\label{eq:verticesRen}
\begin{split}
\Gamma_{m_1 m_2 \mu}^{W u \bar{d}} & = i \frac{g_{3L}}{\sqrt{2}} \delta_{m_1 m_2} \gamma_\mu P_L \\ 
\Gamma_{m_1 m_2 \mu}^{\gamma u \bar{u}} &= i\frac{g_{3L}}{2}\delta_{m_1 m_2} \gamma_\mu \left[ \left( \mathcal{O}_{13} - \frac{\mathcal{O}_{23}}{\sqrt{3}} - \frac{2\mathcal{O}_{33}}{\sqrt{3}} \right)P_L + \frac{4 t_X \mathcal{O}_{33}}{3}P_R \right],
\end{split}
\end{equation}
where $m_1,m_2$ are color indices. 

At this point, in order to solve for $\delta Z^\mathscr{H}_{g_{3L}},\delta Z^\mathscr{H}_{g_{X}}$, the set of functions 

\begin{equation}
\left\{\delta \Gamma^\mathscr{H}_{W u\overline{d}},\delta \Gamma^\mathscr{H}_{\gamma u\bar{u}},\delta Z^\mathscr{H}_{u},\delta Z^\mathscr{H}_{d},\delta Z^\mathscr{H}_{W},\delta Z^\mathscr{H}_{A}\right\},
\end{equation}
must be evaluated. Since the dependence of all these quantities on their individual 1-loop contributions is additive, they may be calculated referring only to the exotic diagrams, shown in Fig.~\ref{fig:diagrams}, and there are no crossed effects between those and the contributions of the pure SM. From Eq.~(\ref{eq:verticesRen}) the relations among the vertex function counterterms $\delta \Gamma^\mathscr{H}$ and the other ones may be found and read (omitting the Lorentz and color indices in the LHS for clarity)

\begin{equation}\label{eq:counterterms}
\begin{split}
\delta \Gamma^\mathscr{H}_{W u \bar{d}} &= i  \frac{g_{3L}}{2\sqrt{2}}\delta_{m_1 m_2} \gamma_\mu P_L\left( 2\delta Z^{\mathscr{H}}_{g_{3L}} + \delta Z^{\mathscr{H}}_{d} + \delta Z^{\mathscr{H}}_{u} + \delta Z^{\mathscr{H}}_{W} \right)\\
\delta \Gamma^\mathscr{H}_{\gamma u \bar{u}} &= i \frac{\gamma_\mu}{2}\delta_{m_1 m_2}  \left[g_{3L}P_L \left( \mathcal{O}_{13}-\frac{\mathcal{O}_{23}}{\sqrt{3}}\right)\delta_\mathrm{E}Z^{\mathscr{H}}_{g_{3L}}  \right. \\ 
& \left. {} \hspace*{22mm} + g_{X} \left( \frac{4\mathcal{O}_{33}}{3}P_R -\frac{2\mathcal{O}_{33}}{\sqrt{3}}P_L \right)\delta_\mathrm{E}Z^{\mathscr{H}}_{g_{3L}}  \right],
\end{split}
\end{equation}
where we have defined the `effective' coupling counterterms

\begin{equation}
\begin{split}
\delta_\text{E} Z^{\mathscr{H}}_{g_{3L}}&\equiv \delta Z^\mathscr{H}_{g_{3L}} + \delta Z^{\mathscr{H}}_u + \frac{\delta Z^{\mathscr{H}}_A}{2} \\
\delta_\text{E} Z^{\mathscr{H}}_{g_{X}}&\equiv\delta Z^\mathscr{H}_{g_{X}} + \delta Z^{\mathscr{H}}_u + \frac{\delta Z^{\mathscr{H}}_A}{2},
\end{split}
\end{equation}
where $Z_{x} \equiv 1 + \delta Z_x$.

The calculated wave function counterterms read

\begin{equation}
\begin{split}
\delta Z^\mathscr{H}_u &= -\frac{g_{3L}^2}{64 \pi^2 \epsilon} \\
\delta Z^\mathscr{H}_d &= -\frac{g_{3L}^2}{64 \pi^2 \epsilon} \\
\delta Z^\mathscr{H}_W &= \frac{5 g_{3L}^2 (2 \mathcal{O}_{12}^2 + 1)}{48 \pi^2 \epsilon} \\
\delta Z^\mathscr{H}_A &= \frac{3 g_{3L}^2 (5 \mathcal{O}_{13}^2 + 9 \mathcal{O}_{23}^2)+ 16\sqrt{3} g_{3L}g_{X} \mathcal{O}_{23}\mathcal{O}_{33}- 126 g_{X}^2 \mathcal{O}_{33}^2}{144 \pi^2 \epsilon}.
\end{split}
\end{equation}
With this, together with the vertex function counterterms, the system in Eq.\refEq{eq:counterterms} may be solved for $\delta Z^\mathscr{H}_{g_X},\delta Z^\mathscr{H}_{g_{3L}}$. Instead of giving $\delta \Gamma^\mathscr{H}_{W u \bar{d}},\delta \Gamma^\mathscr{H}_{\gamma u \bar{u}}$, we write down the solutions for the couplings directly:

\begin{equation}
\begin{split}
\delta Z^\mathscr{H}_{g_{3L}} &= -\frac{3g_{3L}^2(53 \mathcal{O}_{12}^2 + \mathcal{O}_{22}^2 + 25) + 4\sqrt{3}g_{3L}g_{X}\mathcal{O}_{22}\mathcal{O}_{32}+4g_X^2 \mathcal{O}_{32}^2}{576 \pi^2 \epsilon} \\
\delta Z^\mathscr{H}_{g_{X}} &= \frac{\frac{28}{3}g_X^3 F_1+ \frac{2}{3}g_{3L}g_X^2 F_2 - g_{3L}^2 g_X F_3 - g_{3L}^3 F_4}{192 \pi^2 g_X \mathcal{O}_{33} \epsilon},
\end{split}
\end{equation}
where, 

\begin{equation}
\begin{split}
F_1 &= \mathcal{O}_{33} (\mathcal{O}_{32}^2 + 9\mathcal{O}_{33}^2) \\
F_2 &= 2 \mathcal{O}_{32}\mathcal{O}_{33}(3\mathcal{O}_{12}-\sqrt{3}\mathcal{O}_{22})+3\mathcal{O}_{13}(2\mathcal{O}_{32}^2+ 63 \mathcal{O}_{33}^2)-\sqrt{3}\mathcal{O}_{23}(2\mathcal{O}_{32}^2+ 79 \mathcal{O}_{33}^2) \\
F_3 &= \mathcal{O}_{12}\left[ 6\mathcal{O}_{13}\mathcal{O}_{32}-2\sqrt{3}(\mathcal{O}_{22}\mathcal{O}_{33}+\mathcal{O}_{23}\mathcal{O}_{32})\right]  -4 \sqrt{3} \mathcal{O}_{13}(\mathcal{O}_{22}\mathcal{O}_{32}-4\mathcal{O}_{23}\mathcal{O}_{33}) \\
& \quad + \mathcal{O}_{33}\left[ \mathcal{O}_{22}^2+2\mathcal{O}_{23}^2+10\mathcal{O}_{13}^2 +3 (\mathcal{O}_{12}^2+1) \right] + 4 \mathcal{O}_{22} \mathcal{O}_{23} \mathcal{O}_{32} \\
F_4 &= 
-3 \mathcal{O}_{13}(-5 \mathcal{O}_{13} + 28\mathcal{O}_{12}^2 - \sqrt{3} \mathcal{O}_{12} \mathcal{O}_{22} + \mathcal{O}_{22}^2 - 9\mathcal{O}_{23}^2 + 23) + {}\\
&\quad + \mathcal{O}_{23}\left[-3\mathcal{O}_{12}\mathcal{O}_{22}+\sqrt{3}(28\mathcal{O}_{12}^2+\mathcal{O}_{22}^2 - 9\mathcal{O}_{23}^2-5\mathcal{O}_{13}^2+ 35)\right].
\end{split}
\end{equation}

To understand how these formulae can be used to find $\beta^\mathscr{H}$, start from the dimensionally regularized Eq.\refEq{eq:RenDef1}, or (with $g_X$ as representative)

\begin{equation}
\ln g_{X_0} = \ln Z_{g_X} g_X \tilde{\mu}^\epsilon.
\end{equation}
Where, above, $Z_{g_X} = Z_{g_X}(g_X,g_{3L})$. Exchanging the renormalization factor by the counterterm correction, taking the $\ln \tilde{\mu}$ derivative and multiplying by $g_X$, one obtains

\begin{equation}
0 = \left( 1 + g_X \delta Z_{g_X}^{(g_X)} \right) \frac{\partial g_{X}}{\partial \ln \tilde{\mu}} + g_X \delta Z_{g_X}^{(g_{3L})} \frac{\partial g_{3L}}{\partial \ln \tilde{\mu}} + g_X \epsilon,
\end{equation}
where, here, a $(y)$ superscript indicates $\partial/\partial y$. Assuming the counterterms to be perturbatively defined, the equation above gives, after linearization

\begin{equation}\label{eq:perturbativeConsistency}
\frac{\partial g_{X}}{\partial \ln \tilde{\mu}} = -g_X \epsilon \left( 1 - g_X \delta Z_{g_X}^{(g_X)}  - g_{3L} \delta Z_{g_X}^{(g_{3L})} \right).
\end{equation}
Denoting the Laurent expansion of the counterterms as 

\begin{equation}
\delta Z_{g_X} = \sum_{n=1}^\infty \frac{G_{X,n}}{\epsilon^n},
\end{equation}
the $\beta$-function is the part of the RHS of Eq.\refEq{eq:perturbativeConsistency} finite at $\epsilon \to 0$\footnote{The $\beta$-function results from requiring perturbative consistency at order $\epsilon^0$. Demanding the same consistency at every remaining (negative) power of the expansion give an infinite number of recursive relations.}. Explicitly, 

\begin{equation}
\beta_{g_X} = g_X^2 G_{X,1}^{(g_X)} + g_X g_{3L} G_{X,1}^{(g_{3L})}.
\end{equation}
Specializing to the $\beta$-function contribution of the heavy particles and adding the formula for $g_{3L}$, finally

\begin{equation}
\begin{split}
\beta^{\mathscr{H}}_{g_X} = g_X^2 G_{X,1}^{{\mathscr{H}}(g_X)} + g_X g_{3L} G_{X,1}^{{\mathscr{H}}(g_{3L})} \\
\beta^{\mathscr{H}}_{g_{3L}} = g_{3L}^2 G_{{3L},1}^{{\mathscr{H}}(g_{3L})} + g_X g_{3L} G_{{3L},1}^{{\mathscr{H}}(g_{X})}.
\end{split}
\end{equation}

With these formulae, the running is obtained, in the non-symmetric regime below the heavy particles threshold $E_\mathrm{high}$, by numerically solving the coupled system of differential equations 

\begin{align}
\frac{\partial g_{3L}(\mu)}{\partial \ln \mu} - \frac{g_{3L}(\mu)^3}{(4 \pi)^2}b_{g_{3L}} &- \left[- \beta^\mathscr{H}_{g_{3L}}\left(g_{3L}(\mu),g_{X}(\mu)\right) \right] = 0 \label{eq:NSolving1} \\
\frac{\partial g_{X}(\mu)}{\partial \ln \mu} - \frac{g_{X}(\mu)^3}{(4 \pi)^2}b_{g_{X}} &- \left[- \beta^\mathscr{H}_{g_{X}}\left(g_{3L}(\mu),g_{X}(\mu)\right) \right] = 0. \label{eq:NSolving2}
\end{align}

An important remark is that the exotic scalars are not included in the calculation of the counterterms (see Fig.~\ref{fig:diagrams}). This is because they come in fully exotic triplets, and can be correctly eliminated from the theory through the subtraction of their contributions to $b_{g_X},b_{g_{3L}}$. The exception is the SM scalar doublet, which must be kept. We consider it to be the one projected by the $\rho$-triplet, $\rho^{(2)}$  from Table~\ref{tab:SMProjection}, and conserve its contribution to the coefficients. The $b_{g_X},b_{g_{3L}}$ to be plugged into the symmetric term of the equations above, already accounting for the removal of the scalars other than $\rho^{(2)}$, are

\begin{equation}
b_{g_{X}} = \frac{62}{3}, \;\;\;\;\; b_{g_{3L}} = -\frac{41}{6}.
\end{equation}  


\subsection{Parametric structure}
\label{sec:parStr}

In order for Eqs.~(\ref{eq:NSolving1},\ref{eq:NSolving2}) to be solved for $g_X$ below $E_\mathrm{high}$, the quantities $v_\rho,v_\eta,v_\chi$ must be fixed. We use the {\it solution to the closure} mentioned near Eq.~(\ref{eq:soldef}) to set $v_\rho = \SI{54}{GeV}$ and $v_\eta = \SI{240}{GeV}$. Additionally, to fix the values of the exotic quarks masses, we set their three Yukawa eigenvalues to $0.5$, {\it i.e.},

\begin{equation}
y^J = 0.5, \;\;\;\; K^{j} = 0.5 \,\mathbb{1}.
\end{equation} 

With this, we examine four benchmarks for $v_\chi$, including the resulting masses for the exotic fermionic and vector boson particles 

\begin{equation}\label{eq:Benchmarks}
\begin{split}
&\mathrm{B1}\hspace{-0.1cm}: \quad v_\chi=\SI{3}{TeV}, \, M_U = \SI{945}{GeV}, \, M_V = \SI{948}{GeV}, \, M_{Z^\prime} = \SI{3476}{GeV},\\
&\hphantom{\mathrm{B1}\hspace{-0.1cm}: \quad v_\chi=\SI{3}{TeV},} \hspace*{2mm} M_j = \SI{1061}{GeV}, \,M_J = \SI{1061}{GeV};   \\[0.22em]
&\mathrm{B2}\hspace{-0.1cm}: \quad v_\chi=\SI{6.5}{TeV}, \, M_U = \SI{2048}{GeV}, \, M_V = \SI{2049}{GeV}, \, M_{Z^\prime} = \SI{7531}{GeV},\\
&\hphantom{\mathrm{B1}\hspace{-0.1cm}: \quad v_\chi=\SI{6.5}{TeV},} \hspace*{2mm} M_j = \SI{2298}{GeV}, \,M_J = \SI{2298}{GeV};   \\[0.22em]
&\mathrm{B3}\hspace{-0.1cm}: \quad v_\chi=\SI{9.5}{TeV}, \, M_U = \SI{2993}{GeV}, \, M_V = \SI{2993}{GeV}, \, M_{Z^\prime} = \SI{11}{TeV},\\
&\hphantom{\mathrm{B1}\hspace{-0.1cm}: \quad v_\chi=\SI{9.5}{TeV},} \hspace*{2mm} M_j = \SI{3359}{GeV}, \,M_J = \SI{3359}{GeV};   \\[0.22em]
&\mathrm{B4}\hspace{-0.1cm}: \quad v_\chi=\SI{13}{TeV}, \, M_U = \SI{4095}{GeV}, \, M_V = \SI{4096}{GeV}, \, M_{Z^\prime} = \SI{15.06}{TeV},\\
&\hphantom{\mathrm{B1}\hspace{-0.1cm}: \quad v_\chi=\SI{13}{TeV},} \hspace*{2mm} M_j = \SI{4596}{GeV}, \,M_J = \SI{4596}{GeV}.    
\end{split}
\end{equation}
B1, the most conservative benchmark point, is chosen with $M_U \approx \SI{1}{TeV}$ because this is a very conservative reasonable lower bound for the mass of the bilepton given by the joint phenomenology already produced for this particle~\cite{Meirose:2011cs,Nepomuceno:2016jyr,Corcella:2017dns,RamirezBarreto:2013edk,PhysRevD.101.015024,Barela:2022sbb} and, specially, by all that was discussed in this thesis.  

Besides $v_\chi$, $E_\mathrm{high}$ is yet to be fixed. This parameter is directly identifiable with the scale of importance of the heavy states. In each of the Benchmarks above, we set it to the mass of the vector bilepton $E_\mathrm{high}=M_U$ which, except for $M_{Z^\prime}$, corresponds to a good representative for the scale of all masses.

\subsection{Results}
\label{sec:resAlt}

The objective is to investigate whether the more precise effective approach can alleviate the stress that the $\SI{4}{TeV}$ pole, obtained through an approximation that makes use of an artificial intermediate exact symmetry, generates over the model. The operational method implies several steps, and, as in the other studies, many amplitudes of Figure~\ref{fig:diagrams} are derived from rules which originate from Lagrangians with explicit charge conjugation, which turns their obtaining into a subtle matter \cite{Denner:1992vza,Barela:2022sbb}. We check them using \texttt{FeynRules} \cite{Alloul:2013bka} paired with \texttt{FeynArts} \cite{Hahn:2000kx}. The loop calculations are performed with the help of \texttt{Package-X} \cite{Patel:2016fam} connected to \texttt{FeynCalc} through \texttt{FeynHelpers} \cite{Shtabovenko:2016whf}. The Feynman-'t Hooft gauge \cite{Fujikawa:1972fe} is employed when needed.

To validate our calculations, we derive the runnings through a second, approximate method of integrating the exotic particles out below $E_\mathrm{high}$. The procedure amounts to, as before, modifying the $b$ in an attempt to remove their effects. In principle, this could be done exactly for the $g_X$, as each contribution is computed from a single particle. However, because there is mixing between the $U(1)_X$ and the `diagonal' gauge bosons of $SU(3)_L$, the results could, in principle, differ, mainly because of the physical $Z^\prime$. This could also play a role in the form and strength of the interactions, possibly increasing the imprecision. The situation is more critical for the $g_{3L}$ running, since the quark triplets are broken by the removal of $j_i,J$. To find $b_{g_{3L}}$, we ignore this fact, and consider the contribution of the $S_2$ of each broken triplet to Eq.~(\ref{eq:betacoeff}) as if they were intact -- in presenting the results, we dub this method {\it doublet approximation}. Note that the $g_X$ $\beta$-function depends on $g_{3L}$, hence, again, although this procedure could be used expecting exact results for $g_X$, there could also be distortions caused by the $g_{3L}$ error. The $\beta$-function coefficients to be used in the $\mu< E_\mathrm{high}$ regime in this approximation are

\begin{figure*}[t!]
	\centering
	\hspace*{-0.11cm}\makebox[0pt][c]{%
		\includegraphics[width=0.5\linewidth]{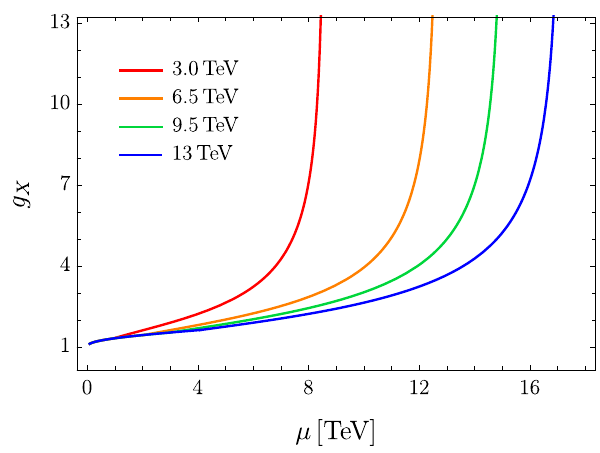}
		\includegraphics[trim=0 0.5mm 0 0, clip, width=0.52\linewidth]{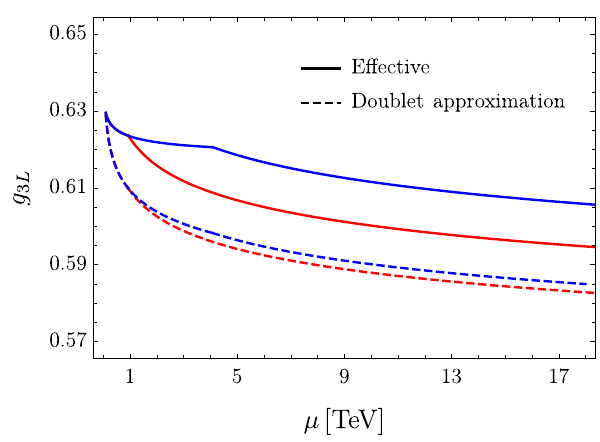}}
		\caption[Running of $g_X$ and $g_{3L}$ in the effective approach.]{\textit{Left}: Effective (exact) evolution of $g_X$ for the four benchmarks defined in Eq.~(\ref{eq:Benchmarks}), labeled by the free $v_\chi$. The plot attests that avoiding the SM approximation greatly enlarges the perturbativity regime of the model, depending on $v_\chi$. \textit{Right}:  Evolution of the left-handed 3-3-1 coupling (to avoid cluttering in the image, only the two extreme benchmarks appear). The dashed line corresponds to the running obtained through the {\it doublet approximation}.}
	\label{fig:altSMdof}
\end{figure*}

\begin{equation}
b_{g_{X}} = \frac{55}{9}, \;\;\;\;\; b_{g_{3L}} = -\frac{17}{3}.
\end{equation}

The results are shown in Figure~\ref{fig:altSMdof}. The first observation to be made is that, despite the discussion carried in the last paragraph, the {\it doublet approximation} is, for all purposes, perfect for the $g_X$ running, i.e., the curves coincide. One way to explain this is to realize that the the greatest source of `mixing' between $g_X$ and $g_{3L}$ comes from $Z^\prime$ effects, whose scale is exaggeratedly larger than $E_{\mathrm{high}}$, hence, in the scales of matching between the theory of light states and the complete model, such effects are negligible. The greatest result, however, is that considering the rightful parametrization of the model from very low energies extends its unitary range from $\sim \hspace*{-1.5mm} \SI{4.5}{TeV}$ (in the SM approximation) to $\sim \hspace*{-1.5mm} \SI{8.5}{TeV}$ in the most conservative benchmark. This range can be enlarged further with an increasing $v_\chi$ which, besides influencing the neutral spin-1 particles projection onto low energies, has as major consequence pushing the heavy particle threshold upwards. The left panel of Figure~\ref{fig:polePosition} shows the upper limit of the perturbative window of the model, depicted as the energy scale in which $g_X(\mu) = 4 \pi$, as a function of $v_\chi$. An interesting fact that we have verified is that this figure is not altered by a change in $v_\rho$ and $v_\eta$, for fixed $v_\chi$, at least as long as they obey $v_\rho^2 + v_\eta^2 = \SI{246^2}{GeV^2}$.    

\begin{figure*}[t!]
	\centering
	\hspace*{-0.11cm}\makebox[0pt][c]{%
		\includegraphics[width=0.5\linewidth]{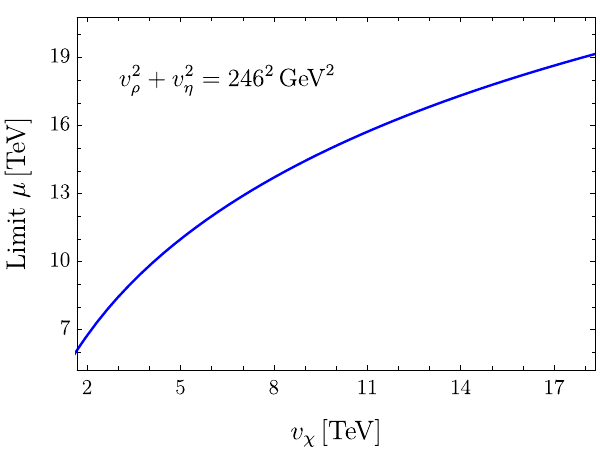}
		\includegraphics[trim=0 0.5mm 0 0, clip, width=0.5\linewidth]{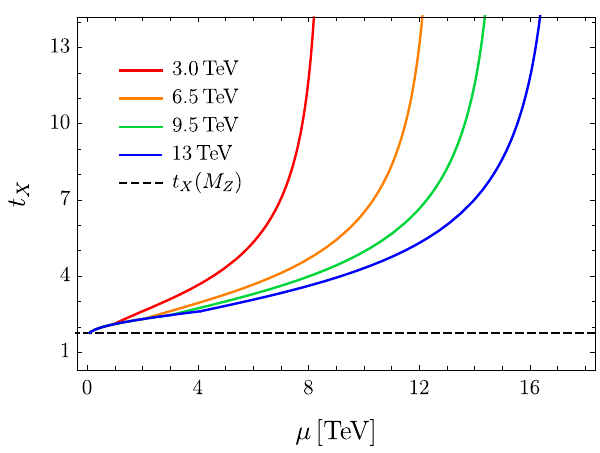}}
		\caption[Upper limit of the perturbative range of the m331 as a function of $v_\chi$.]{{\it Left}: Upper limit of the perturbative range of the m331 as a function of $v_\chi$. This curve is independent of $v_\rho,v_\eta$ as long as they belong to the circle $v_\rho^2+v_\eta^2 = \SI{246^2}{GeV^2}$, with $v_s \approx \SI{2}{GeV}$ (negligible in practice). {\it Right}: Running of the m331 symmetry parameter $t_X \equiv g_X/g_{3L}$ for our four benchmark points, to be compared with the usually employed value $t_X = t_X(M_Z) \approx 1.75$.}
	\label{fig:polePosition}
\end{figure*}

\section{Can we set $t_X^2 = \frac{s_W^2}{1-4 s_W^2}$?}

Finally, let us try to estimate a measure of the harm of generating TeV scale predictions, in the m331, without conducting RGE improved calculations. In general, the amplitudes for physical processes may be written in terms of $g_{3L}$ and the symmetry parameter $t_X = g_X/g_{3L}$. The right panel of Figure~\ref{fig:polePosition} shows the running of $t_X$ for our four benchmarks. In most 3-3-1 studies, this parameter is eliminated in favour of the known quantity $s_W = g_{2L}/\sqrt{g_{2L}^2+g_{Y}^2}$ through the relation

\begin{equation} \label{eq:tanX}
t_X^2 \equiv \frac{g_X}{g_{3L}} = \frac{s_W^2}{1-4 s_W^2},
\end{equation}
which is nothing but a matching condition between a 3-3-1 and the 3-2-1 SM. As such, it is required to hold at a single energy point (which, in fact, is $E_\mathrm{high}$, and not in the electroweak scale), not as an identity between functions of $\mu$. As this chapter has shown, even if one runs $s_W$ with energy, the SM approximation and the artificial matching turn out to be a great source of inaccuracy. In any case, to get a sense of the effects of this disparity, consider the exotic $Z^\prime$ mediated contribution to the hard partonic process $u\bar{u} \to Z^\prime \to e^+e^-$. The lowest order term in $m_u/\hat{s}$ of the averaged, angular inclusive, cross section is given by

\begin{equation}
\hat{\sigma}_{Z^\prime}\left(u\bar{u} \to e^+e^-\right) = \frac{5g_{3L}^4 \hat{s}}{2^8 3^4 \pi \left[\hat{s}^2 + M_{Z^\prime}^4 + M_{Z^\prime}^2(\Gamma_{Z^\prime}^2 - 2\hat{s})\right]} f(\theta_X),
\end{equation}
where $M_{Z^\prime},\Gamma_{Z^\prime}$ are the $Z^\prime$ mass and width, respectively, and $\hat{s}$ is the $u\bar{u}$-pair center-of-mass energy squared. The $f$ factor is given by

\begin{equation}
f(\theta_X) = \frac{60  t_X^6 + 20  t_X^4 + 4\sqrt{3t_X^2 + 1}  t_X^2 + 1}{(3t_X^2 + 1)^2}.
\end{equation}
Now, consider $g_{3L}$ slow varying and a partonic $\hat{s} = \SI{1.5}{TeV}$. In this scenario, one obtains that $\hat{\sigma}_{Z^\prime}$ picks up an extra, wrong factor of $\sim \hspace*{-1.5mm} 1/2$ if one uses $t_X = t_X(M_Z)$, i.e., the value obtained by matching to the SM at the $Z$-pole, the usual practice. This discrepancy increases fast with energy.

\section{Analysis and perspectives}
\label{sec:con}


The SSB of the m331 has two groups of contributions: the first, generated by the condensation of the $\chi$-triplet neutral component, triggers the descent of the 3-3-1 symmetry group to that of the SM. The second, originated from every other VEV, prompts the usual SM breaking. This is a mathematical construct of the model building, and its interpretation as a meaningful physical process is a conceptual simplification. Although it is a phenomenological necessity of any BSM theory to possess the SM as an effective approximation below the TeV scale, this does not imply that a 3-2-1 symmetry approximation is appropriate outside a narrow window within the electroweak regime. In fact, to force the SM symmetry as a physical feature of the m331 in intermediate scales is a strong simplification, which becomes strictly impossible for heavy particle thresholds above $\sim \hspace*{-1.5mm} \SI{3.5}{TeV}$, and badly imprecise way before it. We firstly review the prediction of the $\SI{4}{TeV}$ pole in this SM approximation, finding what is the greatest matching scale that leaves a small perturbative range available above it, which corresponds to around $3.7-\SI{3.8}{TeV}$, resulting in a pole at a little above $\SI{4.5}{TeV}$. 

A full account of the most precise, effective approach is then given. We define the heavy particle threshold along the mass of the vector bilepton, one of the most interesting features of the model and which gives a reasonable avatar for the general exotic mass scale. Other important free parameters are the triplet VEVs, $v_\chi, v_\rho, v_\eta$, which influence the projection of the neutral vector boson masses and interactions at low energies. The $v_\rho$ and $v_\eta$ are fixed through a numerical solution that fits the known neutral current parameters at the $Z$-pole, and four benchmarks are chosen for $v_\chi$, the lowest of which reproduce predictions of the current bilepton phenomenology. Thus, the full structure of the diagonalization of the neutral sector in the m331 is taken into consideration, and the exact physical states are removed below their scale of importance.  

Our main results show that, in the most conservative benchmark, coherent with a bilepton mass of $M_U = \SI{945}{GeV}$, the true perturbative range of the m331 extends up to $\SI{8.5}{TeV}$, already greatly reducing the stress generated onto the model by the usual assessments. For a heavy particle threshold around $M_U = \SI{2990}{GeV}$, a still viable phenomenological (in some sense, more natural as explored last Chapter) scenario, this window is increased further up to $\SI{15}{TeV}$. Our calculations are validated by reproducing the $\beta$-functions of general, non-abelian theories and, more importantly, by identically matching the {\it doublet approximation}, whose results for the coupling of the abelian $U(1)_X$ should be highly reliable. This is because the only sources of error in this approach for this interaction ultimately come from its mixing with the $SU(3)_L$ one.

Apart from assessing alternative versions of the model, left for posterior works, there are a few points in the analysis, ignored by simplicity, which could be addressed, starting by the possibility of splitting the threshold: we have considered a single heavy particle one. A more thorough analysis of the parametric structure of the model could be carried, considering more realistic, strategically chosen benchmark points which consider, for instance, different quark Yukawa couplings and the parameters of the scalar potential. With such a parametric map at hand, one would be able to perform fully realistic RGE analysis, integrating each particle out at their exact mass scale predicted in the given benchmark point of parameter space. In particular, a deep assessment of the scalar sector would define the projection of the SM physical Higgs onto the low energies, for which we used the simplest possible benchmark. 

Such new analysis are, however, intrinsically tied to experimental and phenomenological advances and should either not immensely vary the paradigm unveiled here or be tied to distinct and complementary premises. It must be understood that the Landau Poles, by themselves, do not condemn the model, but attest that, at the corresponding energy ranges, new degrees of freedom or theoretical mechanism must arise to protect the theory.

Also of importance is the review of a common practice which eliminates a free electroweak parameter of the model and should ideally be avoided from skeptical investigations. The electroweak angle $t_X \equiv g_X/g_{3L}$ \textit{should not} be equated to $s_W^2/(1-4 s_W^2)$, except if in a conscientious approximated ansatz around a fixed scale. Finally, the importance of our results is not limited to the perturbative qualities of the model, but calls attention to possible effects of modifying the theoretical status of the theory regarding the SSB, which urges RGE improved phenomenology comparing both scenarios to be performed.  

\newpage
\thispagestyle{empty}

\newgeometry{left=2.6cm,bottom=0.1cm,right=1.8cm,top=1.6cm}

\enlargethispage{50\baselineskip}
\begin{figure*}[h!]
	\begin{center}	
 	\begin{adjustbox}{minipage=0.93\linewidth}
 		\begin{center}
		\includegraphics[width=0.33\linewidth]{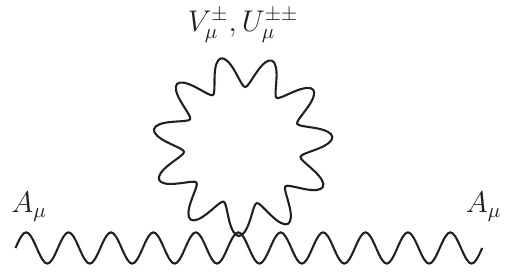}
		\includegraphics[width=0.33\linewidth]{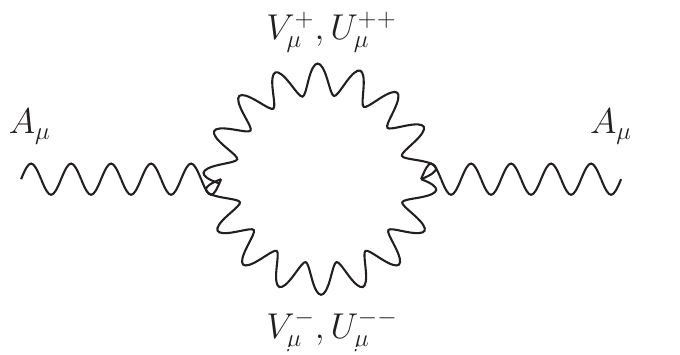} \\
		\vspace*{4mm}
		\includegraphics[width=0.33\linewidth]{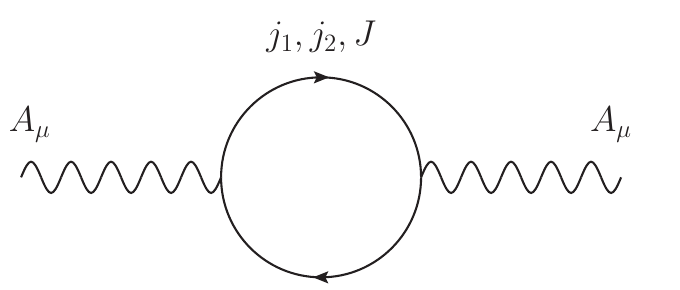}
		\includegraphics[width=0.33\linewidth]{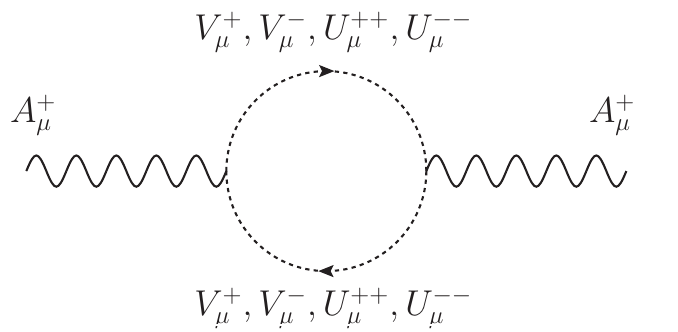} \\
		\vspace*{4mm}
		\includegraphics[width=0.32\linewidth]{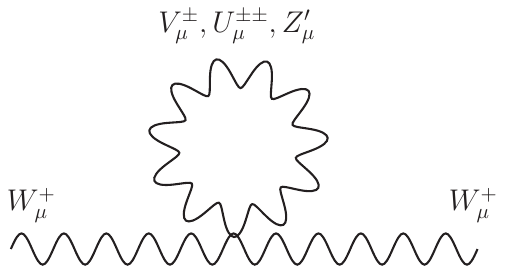}
		\includegraphics[width=0.32\linewidth]{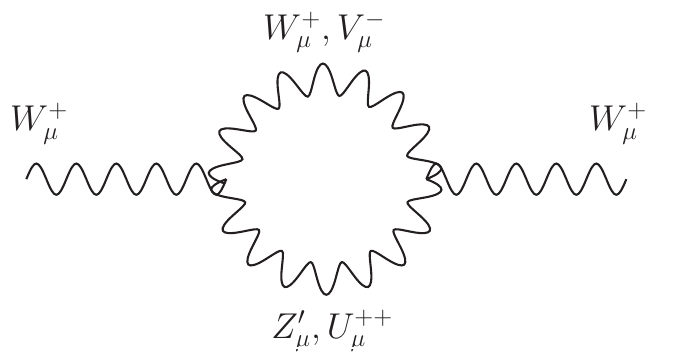}
		\includegraphics[width=0.32\linewidth]{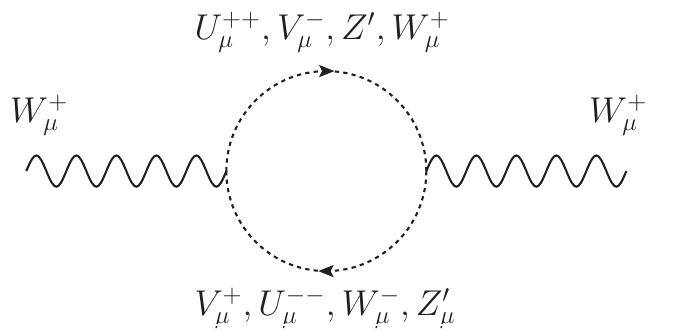} \\
		\vspace*{4mm}
		\includegraphics[width=0.33\linewidth]{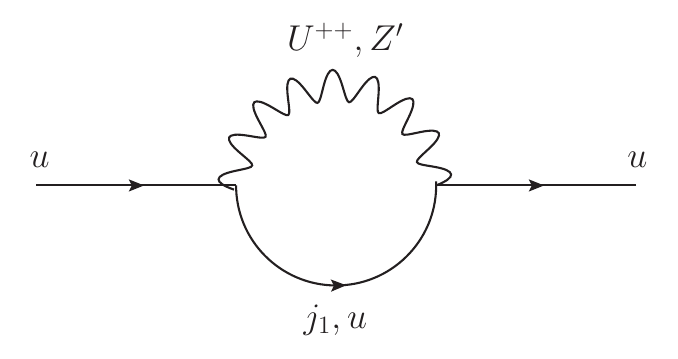}
		\includegraphics[width=0.33\linewidth]{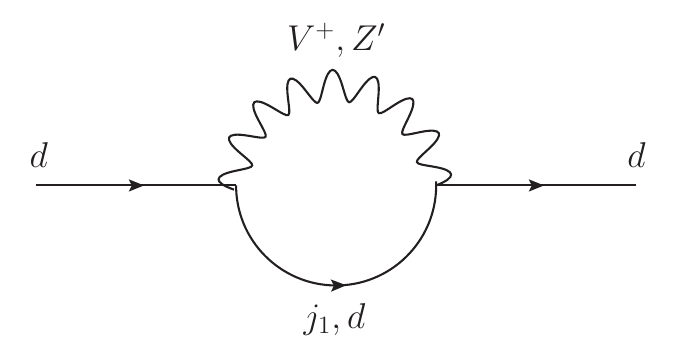} \\
		\vspace*{4mm}
		\includegraphics[width=0.33\linewidth]{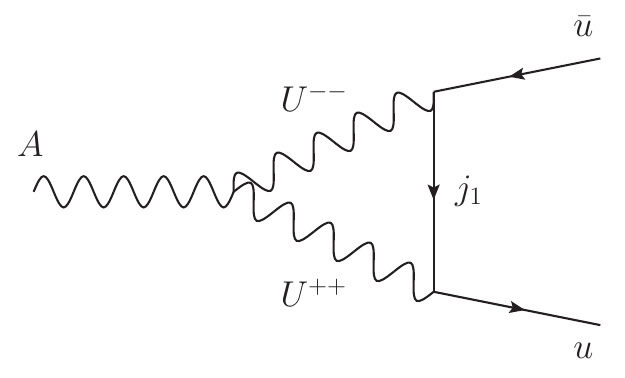}
		\includegraphics[width=0.33\linewidth]{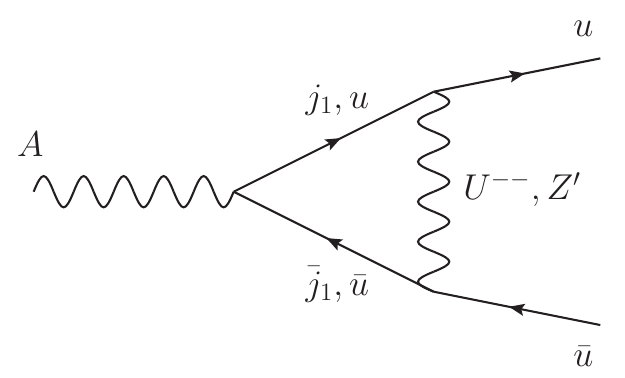}\\
		\vspace*{4mm}
		\includegraphics[width=0.33\linewidth]{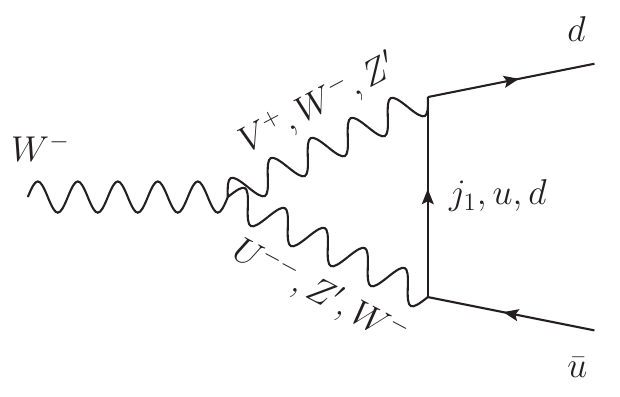}
		\includegraphics[width=0.33\linewidth]{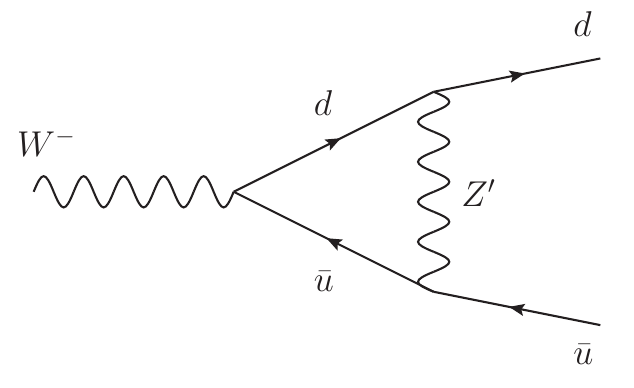}
		\end{center}
 	\end{adjustbox}
	\end{center}
	\caption[Complete set of exotic diagrams of the m331 relevant to the RGE analysis of the $\gamma u \bar{u}$ and $W d\bar{u}$ vertices.]{The complete set of exotic diagrams of the m331 relevant to the RGE analysis of the $\gamma u \bar{u}$ and $W d\bar{u}$ vertices. Time flows from left to right. To avoid introducing new symbols and since they only appear here, ghosts are denoted by their corresponding vector boson label, and are depicted by dotted lines. Diagrams with $N$ labels on internal lines should be read as $N$ diagrams, each with a combination of labels that should be paired as they are ordered. To allow for a sanity check, we note that the number of diagrams counts 31.}
	\label{fig:diagrams}
\end{figure*}

\pagebreak
\newpage
\newpage

}

{
\restoregeometry
{\fontsize{12.2pt}{15.1pt}\selectfont

\pagestyle{fancy}
\fancyfoot{}
\fancyhead[R]{\thepage}
\fancyhead[L]{{\it \leftmark}}
\fancypagestyle{plain}{ %
  \fancyhf{} 
  \renewcommand{\headrulewidth}{0pt} 
  \renewcommand{\footrulewidth}{0pt}
}

\cleardoublepage

\pagebreak

\chapter{Conclusions}\label{Chapter9}

As we have extensively emphasized, the Standard Model of elementary particles and interactions is an exceptional theory. The reality of its insufficiency, however, is unavoidable. The theoretical quest of the search for new physics, mainly through the building and testing of new and motivated models, has been made difficult by the fact that, despite the phenomenal experimental efforts, there is no abundance of post SM guiding data. Within this rough scenario, to seek insight from the SM problems and to deeply explore the established alternative models are obligatory paths in particle physics. 

This thesis is another small piece of such exploratory phenomenology and has focused on a specific species of exotic particle: the doubly-charged vector bilepton $U^{\pm \pm}$. There is no good theoretical reason to believe that such exotic charges are less likely to exist than the usual ones. Notwithstanding, the $U^{\pm \pm}$ is contained in only one low energy model: the m331. This theory is able to solve many of the SM issues at the cost of enlarging the gauge symmetry and introducing several new particles. 

The collective result of the previous works on the $U$ phenomenology attests that vector bileptons with masses larger than $1 \hspace{-0.8mm} \sim \hspace{-0.8mm} \SI{1.5}{TeV}$ are excluded.
This literature does not concern itself with non-diagonal interactions, and rarely treated CLFV at all. In particular, it neglected the effects of the $V_U$ matrix, whose influence on phenomenology is one of the focuses of the novel work presented in this thesis. 

We naturally started by exploring the LHC reach, through an analysis of the trimuon process $p p \to \mu^\pm \mu^\pm \mu^\mp e^\mp$, $V_U$-dependent and free of irreducible background. We narrowed the parameter space to consider only orthogonal, symmetric $3 \times 3$ matrices, which accommodate two degrees of freedom. Our results show that, in the optimal point (in this context meaning the point with the most stringent constraints), bileptons with masses of up to $\sim \hspace*{-0.18em} \SI{1.1}{TeV}$ are excluded, corroborating the established results. In the other hand, we have also found that there is a large parameter space available (with this single process as a constraint source) in which bileptons with very low masses are not excluded. Additionally, with $V_U$ having been parametrized in terms of the $(V_U)_{ee}$ and $(V_U)_{e\mu}$ elements, we verified that there is little dependence on $(V_U)_{ee}$, which opens up a clear research avenue. 

With these findings in mind, the following question have arisen: {\it could there be data emanating from a source other than the LHC that is useful to help constrain the $U$ parameter space?} In seeking for an answer, the set of 3-body CLFV decays was an obvious place to start. Our study considered every particle that could contribute to the relevant branching ratios in lepton universal models: neutral ($s$) and doubly-charged ($Y^{\pm\pm}$) scalars, besides the $U^{\pm\pm}$. For several benchmark limits on the absolute values of the $V_U$ elements, we computationally obtained the solution (for the plethora of free parameters) which allowed the smallest $U$ masses to remain not excluded by the current experimental bounds. Such solutions were obtained on multi-dimensional parameter spaces composed by those pertaining to each pair among the contributing particles. The summary of our discoveries is that, in an enormous sector of theory space, the purely leptonic decays can be much more powerful than LHC processes, showing an intricate complementarity between the two. For example, if nature is highly diagonal with regard to the $U\ell \ell$ interaction, then the LHC should be considered and restrains the $U$ to be heavier than $1 \sim \SI{1.5}{TeV}$. In the other hand, if the correct model has relevant non-diagonal features, than the CLFV decays should be chosen and, in natural regimes, constrain $M_U$ to be larger than  $\sim \hspace*{-0.18em} \SI{3}{TeV}$.

The analysis is relevantly complicated by the simultaneous inclusion of two exotic particles, which we chose to perform in order to go one step further from the usual analysis and check if the result could differ relevantly. Indeed, we found that the solution in the $U-s$ {\it Scenario} takes advantage of destructive interference to allow $U$ masses 20\% smaller.

One last project this thesis initiates seeks to produce RGE improved phenomenology of the m331 -- in particular, of the $U$. We discuss the Landau pole in the $g_X$ coupling, generally assumed to be around $\SI{4}{TeV}$ by a simplified derivation. We effect a more precise analysis, explicitly renormalizing the heavy contributions, to obtain that the lowest phenomenologically acceptable scale of the pole sits around $\SI{8.5}{TeV}$.

\appendix

\chapter{General exact neutral current couplings}
\label{app:couplings}

\section{Minimal Model}

\subsection{General exact neutral current couplings}

Consider the neutral current lagrangian parametrized as usual:

\begin{equation}
\mathcal{L}^\mathrm{NC}=-\frac{g}{2c_W}\sum\limits_{i}\bar{\Psi}_i \gamma^{\mu} [(g^i_V-g^i_A \gamma_5)Z_{1 \mu}+(f^i_V-f^i_A \gamma_5)Z_{2 \mu}]\Psi_i.
\end{equation}

We calculate, independently, the values of the neutral current couplings following from the representation content of the minimal version of the 3-3-1 Model. They are as follows (note that some values diverge from those presented in~\cite{Dias:2006ns})

\begin{equation} \label{eq:firstpar}
g_V^\nu=g_A^\nu=N_1 \left(2m_2^2+\bar{v}_{\rho}^2-\frac{4}{3}\bar{v}_{W}^2-\frac{1}{3}\right),
\end{equation}

\begin{equation}
g_V^\ell= -c_W N_1\left(1 -\bar{v}_{\rho}^2\right),
\end{equation}

\begin{equation}
g_A^\ell=-c_W N_1\left(2m_2^2 -\frac{4}{3}\bar{v}_{W}^2+\bar{v}_{\rho}^2-\frac{1}{3}\right),
\end{equation}

\begin{equation}
g_V^{u_m}= c_W N_1 \left[m_2^2-\frac{2}{3}\bar{v}_{W}^2+\frac{1}{3}-\frac{2}{3}t^2\left(1 -\bar{v}_{\rho}^2\right)\right],
\end{equation}

\begin{equation}
g_A^{u_m}=c_W N_1 \left[m_2^2-\frac{2}{3}\bar{v}_{W}^2+\frac{1}{3}+\frac{6}{3}t^2\left(1 -\bar{v}_{\rho}^2\right)\right],
\end{equation}

\begin{equation}
g_V^{d_m}= c_W N_1 \left[-2m_2^2+\frac{4}{3}\bar{v}_{W}^2-\bar{v}_{\rho}^2+\frac{1}{3}+\frac{4}{3}t^2\left(1 -\bar{v}_{\rho}^2\right)\right],
\end{equation}

\begin{equation}
g_A^{d_m}= c_W N_1 \left[-2m_2^2+\frac{4}{3}\bar{v}_{W}^2-\bar{v}_{\rho}^2+\frac{1}{3}\right],
\end{equation}

\begin{equation}
g_V^{j_m}= c_W N_1 \left[m_2^2+\bar{v}_{\rho}^2-\frac{4}{3}\bar{v}_{W}^2-\frac{2}{3}+\frac{10}{3}t^2\left(1 -\bar{v}_{\rho}^2\right)\right],
\end{equation}

\begin{equation}
g_A^{j_m}= c_W N_1 \left[m_2^2+\bar{v}_{\rho}^2-\frac{4}{3}\bar{v}_{W}^2-\frac{2}{3}-\frac{6}{3}t^2\left(1 -\bar{v}_{\rho}^2\right)\right],
\end{equation}

\begin{equation}
g_V^{u_3}= -c_W N_1 \left[-2m_2^2-\bar{v}_{\rho}^2+\frac{4}{3}\bar{v}_{W}^2+\frac{1}{3}+\frac{8}{3}t^2\left(1 -\bar{v}_{\rho}^2\right)\right],
\end{equation}

\begin{equation}
g_A^{u_3}=-c_W N_1 \left[-2m_2^2-\bar{v}_{\rho}^2+\frac{4}{3}\bar{v}_{W}^2+\frac{1}{3}\right],
\end{equation}

\begin{equation}
g_V^{d_3}=-c_W N_1 \left[m_2^2-\frac{2}{3}\bar{v}_{W}^2+\frac{1}{3}+\frac{2}{3}t^2\left(1 -\bar{v}_{\rho}^2\right)\right],
\end{equation}

\begin{equation}
g_A^{d_3}=-c_W N_1 \left[m_2^2-\frac{2}{3}\bar{v}_{W}^2+\frac{1}{3}+\frac{6}{3}t^2\left(1 -\bar{v}_{\rho}^2\right)\right],
\end{equation}

\begin{equation}
g_V^{J}= -c_W N_1 \left[m_2^2+\bar{v}_{\rho}^2-\frac{4}{3}\bar{v}_{W}^2-\frac{2}{3}+\frac{14}{3}t^2\left(1 -\bar{v}_{\rho}^2\right)\right],
\end{equation}

\begin{equation}
g_A^{J}=-c_W N_1 \left[m_2^2+\bar{v}_{\rho}^2-\frac{4}{3}\bar{v}_{W}^2-\frac{2}{3}-\frac{6}{3}t^2\left(1 -\bar{v}_{\rho}^2\right)\right],
\end{equation}

\begin{equation}
f_V^\nu=f_A^\nu=N_2 \left(2m_1^2+\bar{v}_{\rho}^2-\frac{4}{3}\bar{v}_{W}^2-\frac{1}{3}\right),
\end{equation}

\begin{equation}
f_V^\ell=-c_W N_2\left(1 -\bar{v}_{\rho}^2\right),
\end{equation}

\begin{equation}
f_A^\ell= -c_W N_2\left(2m_1^2 -\frac{4}{3}\bar{v}_{W}^2+\bar{v}_{\rho}^2-\frac{1}{3}\right),
\end{equation}

\begin{equation}
f_V^{u_m}= c_W N_2 \left[m_1^2-\frac{2}{3}\bar{v}_{W}^2+\frac{1}{3}-\frac{2}{3}t^2\left(1 -\bar{v}_{\rho}^2\right)\right],
\end{equation}

\begin{equation}
f_A^{u_m}=c_W N_2 \left[m_1^2-\frac{2}{3}\bar{v}_{W}^2+\frac{1}{3}+\frac{6}{3}t^2\left(1 -\bar{v}_{\rho}^2\right)\right],
\end{equation}

\begin{equation}
f_V^{d_m}= c_W N_2 \left[-2m_1^2+\frac{4}{3}\bar{v}_{W}^2-\bar{v}_{\rho}^2+\frac{1}{3}+\frac{4}{3}t^2\left(1 -\bar{v}_{\rho}^2\right)\right],
\end{equation}

\begin{equation}
f_A^{d_m}= c_W N_2 \left[-2m_1^2+\frac{4}{3}\bar{v}_{W}^2-\bar{v}_{\rho}^2+\frac{1}{3}\right],
\end{equation}

\begin{equation}
f_V^{j_m}= c_W N_2 \left[m_1^2+\bar{v}_{\rho}^2-\frac{4}{3}\bar{v}_{W}^2-\frac{2}{3}+\frac{10}{3}t^2\left(1 -\bar{v}_{\rho}^2\right)\right],
\end{equation}

\begin{equation}
f_A^{j_m}= c_W N_2 \left[m_1^2+\bar{v}_{\rho}^2-\frac{4}{3}\bar{v}_{W}^2-\frac{2}{3}-\frac{6}{3}t^2\left(1 -\bar{v}_{\rho}^2\right)\right],
\end{equation}

\begin{equation}
f_V^{u_3}= -c_W N_2 \left[-2m_1^2-\bar{v}_{\rho}^2+\frac{4}{3}\bar{v}_{W}^2+\frac{1}{3}+\frac{8}{3}t^2\left(1 -\bar{v}_{\rho}^2\right)\right],
\end{equation}

\begin{equation}
f_A^{u_3}=-c_W N_2\left[-2m_1^2-\bar{v}_{\rho}^2+\frac{4}{3}\bar{v}_{W}^2+\frac{1}{3}\right],
\end{equation}

\begin{equation}
f_V^{d_3}=-c_W N_2 \left[m_1^2-\frac{2}{3}\bar{v}_{W}^2+\frac{1}{3}+\frac{2}{3}t^2\left(1 -\bar{v}_{\rho}^2\right)\right],
\end{equation}

\begin{equation}
f_A^{d_3}=-c_W N_2 \left[m_1^2-\frac{2}{3}\bar{v}_{W}^2+\frac{1}{3}+\frac{6}{3}t^2\left(1 -\bar{v}_{\rho}^2\right)\right],
\end{equation}

\begin{equation}
f_V^{J}= -c_W N_2 \left[m_1^2+\bar{v}_{\rho}^2-\frac{4}{3}\bar{v}_{W}^2-\frac{2}{3}+\frac{14}{3}t^2\left(1 -\bar{v}_{\rho}^2\right)\right],
\end{equation}

\begin{equation} \label{eq:lastpar}
f_A^{J}=-c_W N_2 \left[m_1^2+\bar{v}_{\rho}^2-\frac{4}{3}\bar{v}_{W}^2-\frac{2}{3}-\frac{6}{3}t^2\left(1 -\bar{v}_{\rho}^2\right)\right],
\end{equation}
where we have introduced the abbreviations

 \begin{equation}\label{eq:massfactors}
\begin{split}
A&=\frac{1}{3}\left[ 3t^2\left(\bar{v}_\rho^2+1 \right) + \bar{v}_W^2 + 1\right] \\
R&=\left\lbrace 1-\frac{1}{3A^2}\left(4t^2+1 \right)\left[\bar{v}_W^2\left(\bar{v}_\rho^2+1 \right) -\bar{v}_\rho^4\right]\right\rbrace^{1/2},
\end{split}
\end{equation}
the dimensionless masses

\begin{equation} \label{eq:neutralbmass}
\begin{split}
m_{1}^2&=A(1-R) \\
m_{2}^2&=A(1+R),
\end{split}
\end{equation}
and the normalization factors

\begin{equation}\label{eq:normfactors}
\begin{split}
N_1^{-2}&=3 \left(2m_2^2+\bar{v}_\rho^2-\frac{4}{3}\bar{v}_W^2-\frac{1}{3}\right) + (\bar{v}_\rho^2-1)^2(4t^2+1) \\
N_2^{-2}&=3 \left(2m_1^2+\bar{v}_\rho^2-\frac{4}{3}\bar{v}_W^2-\frac{1}{3}\right) + (\bar{v}_\rho^2-1)^2(4t^2+1).
\end{split}
\end{equation}

\subsection{Neutral current couplings of the known fermions to the SM $Z$ within the closing solution}
\label{app:couplingssol}

When constrained by the solution given in Eq.\refEq{eq:soldef}, the general values from last section, respective of the known fermions, reduce to 

\begin{align}
g_V^\nu=g_A^\nu=
&\begin{cases}
\frac{1}{2}, & \text{if} \ v_\chi \geq \sqrt{\frac{1-4s_W^2}{2 c_W^2}}v_W \\
\frac{1}{2}\sqrt{\frac{1-4s_W^2}{3}}, & \text{otherwise}
\end{cases} \\
g_V^\ell=
&\begin{cases}
-\frac{1}{2}+2s_W^2, & \text{if} \ v_\chi \geq \sqrt{\frac{1-4s_W^2}{2 c_W^2}}v_W \\
\frac{1}{2}\sqrt{3(1-4s_W^2)}, & \text{otherwise}
\end{cases} \\
g_A^\ell=
&\begin{cases}
-\frac{1}{2}, & \text{if} \ v_\chi \geq \sqrt{\frac{1-4s_W^2}{2 c_W^2}}v_W \\
-\frac{1}{2}\sqrt{\frac{1-4s_W^2}{3}}, & \text{otherwise}
\end{cases}
\end{align}

\begin{align}
g_V^{u_m}=
&\begin{cases}
-\frac{1}{2}+2s_W^2, & \text{if} \ v_\chi \geq \sqrt{\frac{1-4s_W^2}{2 c_W^2}}v_W \\
\frac{1}{2}\frac{-1+6s_W^2}{\sqrt{3(1-4s_W^2)}}, & \text{otherwise}
\end{cases} \\
g_A^{u_m}=
&\begin{cases}
\frac{1}{2}, & \text{if} \ v_\chi \geq \sqrt{\frac{1-4s_W^2}{2 c_W^2}}v_W \\
\frac{1}{2}\frac{-1-2s_W^2}{\sqrt{3(1-4s_W^2)}}, & \text{otherwise}
\end{cases} \\
g_V^{d_m}=
&\begin{cases}
\frac{1}{6}(-3+4s_W^2), & \text{if} \ v_\chi \geq \sqrt{\frac{1-4s_W^2}{2 c_W^2}}v_W \\
-\frac{1}{2}\frac{1}{\sqrt{3(1-4s_W^2)}}, & \text{otherwise}
\end{cases} \\
g_A^{d_m}=
&\begin{cases}
-\frac{1}{2}, & \text{if} \ v_\chi \geq \sqrt{\frac{1-4s_W^2}{2 c_W^2}}v_W \\
-\frac{1}{2}\sqrt{\frac{1-4s_W^2}{3}}, & \text{otherwise}
\end{cases} \\
g_V^{u_3}=
&\begin{cases}
\frac{1}{6}(3-8s_W^2), & \text{if} \ v_\chi \geq \sqrt{\frac{1-4s_W^2}{2 c_W^2}}v_W \\
\frac{1}{2}\frac{1+4s_W^2}{\sqrt{3(1-4s_W^2)}}, & \text{otherwise}
\end{cases} \\
g_A^{u_3}=
&\begin{cases}
\frac{1}{2}, & \text{if} \ v_\chi \geq \sqrt{\frac{1-4s_W^2}{2 c_W^2}}v_W \\
\frac{1}{2}\sqrt{\frac{1-4s_W^2}{3}}, & \text{otherwise}
\end{cases} \\
g_V^{d_3}=
&\begin{cases}
\frac{1}{6}(-3+4s_W^2), & \text{if} \ v_\chi \geq \sqrt{\frac{1-4s_W^2}{2 c_W^2}}v_W \\
-\frac{1}{2}\frac{1-2s_W^2}{\sqrt{3(1-4s_W^2)}}, & \text{otherwise}
\end{cases} \\
g_A^{d_3}=
&\begin{cases}
-\frac{1}{2}, & \text{if} \ v_\chi \geq \sqrt{\frac{1-4s_W^2}{2 c_W^2}}v_W \\
\frac{1}{2}\frac{1+4s_W^2}{\sqrt{3(1-4s_W^2)}}, & \text{otherwise}
\end{cases}
\end{align}
from which we may observe that the SM value is reproduced in every case when the solution is satisfied.

\section{Model with right handed neutrinos}

In~\cite{Dias:2006ns}, the authors also find a solution for the closing of the version of the 3-3-1 with right-handed heavy neutrinos to the electroweak scale. The solution, in this case, reads

\begin{equation}
v_\rho^2 = \frac{1-2s_W^2}{2c_W^2}v_W^2.
\end{equation}  

By completeness, we also show that imposing this requirement, on that model, causes neutral current parameters to descend to the SM prediction -- taking the opportunity to catalogue these quantities in this other version of the theory.

\subsection{General exact neutral current couplings}
\label{app:couplingsrhn}

The general neutral current parameters within this model are found to be

\begin{equation} 
g_V^\nu=-\frac{1}{2}c_W N_1 \left(F_{31}+\sqrt{3}F_{81}\right),
\end{equation}

\begin{equation} 
g_A^\nu=-\frac{1}{6}c_W N_1 \left(3 F_{31}-\sqrt{3}F_{81}-4B_1 t\right),
\end{equation}

\begin{equation}
g_V^\ell=\frac{1}{6}c_W N_1 \left(3 F_{31}-\sqrt{3}F_{81}+8B_1 t\right),
\end{equation}

\begin{equation}
g_A^\ell=\frac{1}{6}c_W N_1 \left(3 F_{31}-\sqrt{3}F_{81}-4B_1 t\right),
\end{equation}

\begin{equation}
g_V^{u_m}= \frac{1}{6}c_W N_1 \left(-3 F_{31}+\sqrt{3}F_{81}-4B_1 t\right),
\end{equation}

\begin{equation}
g_A^{u_m}=\frac{1}{6}c_W N_1 \left(-3 F_{31}+\sqrt{3}F_{81}+4B_1 t\right),
\end{equation}

\begin{equation}
g_V^{d_m}= \frac{1}{6}c_W N_1 \left(3 F_{31}+\sqrt{3}F_{81}+2B_1 t\right),
\end{equation}

\begin{equation}
g_A^{d_m}= \frac{1}{6}c_W N_1 \left(3 F_{31}+\sqrt{3}F_{81}-2B_1 t\right),
\end{equation}

\begin{equation}
g_V^{D_m}= \frac{1}{3}c_W N_1 \left(-\sqrt{3}F_{81}+B_1 t\right),
\end{equation}

\begin{equation}
g_A^{D_m}= \frac{1}{3}c_W N_1 \left(-\sqrt{3}F_{81}-B_1 t\right),
\end{equation}

\begin{equation}
g_V^{u_3}= -\frac{1}{6}c_W N_1 \left(3F_{31}+\sqrt{3}F_{81}+6B_1 t\right),
\end{equation}

\begin{equation}
g_A^{u_3}=-\frac{1}{6}c_W N_1 \left(3F_{31}+\sqrt{3}F_{81}-2B_1 t\right),
\end{equation}

\begin{equation}
g_V^{d_3}=\frac{1}{6}c_W N_1 \left(3F_{31}-\sqrt{3}F_{81}\right),
\end{equation}

\begin{equation}
g_A^{d_3}=\frac{1}{6}c_W N_1 \left(3F_{31}-\sqrt{3}F_{81}-4B_1 t\right),
\end{equation}

\begin{equation}
g_V^{U}=\frac{1}{3}c_W N_1 \left(\sqrt{3}F_{81}-B_1 t\right),
\end{equation}

\begin{equation}
g_A^{U}=\frac{1}{3}c_W N_1 \left(\sqrt{3}F_{81}+B_1 t\right),
\end{equation}

For the values of the fermionic couplings with $Z_2$, make the replacements $N_1 \to N_2, \; F_{31} \to F_{32}, \; F_{81} \to F_{82}, \; B_{1} \to B_{2}$ in the formulae above. The components are explicitly given by

\begin{equation}
\begin{split}
F_{31}&=3(3+t^2)\bar{v}_W^4+6(3+t^2)\bar{v}_\rho^4-3\bar{v}_W^2(3+R+t^2+6\bar{v}_\rho^2-t^2\bar{v}_\rho^2) \\
\frac{F_{81}}{\sqrt{3}}&=(3+t^2)\bar{v}_W^4-2(3+t^2-R)\bar{v}_\rho^2-\bar{v}_W^2[-3+R+6\bar{v}_\rho^2+5t^2(1+\bar{v}_\rho^2)] \\
B_1&=-2t[(3+t^2)\bar{v}_W^4+\bar{v}_W^2(-6-R+t^2-6\bar{v}_\rho^2+4t^2\bar{v}_\rho^2)]+\bar{v}_\rho^2[-R+(3+t^2)(1+3\bar{v}_\rho^2)] \\
F_{32}&=-3(3+t^2)\bar{v}_W^4-6(3+t^2)\bar{v}_\rho^4-3\bar{v}_W^2[-3+R-6\bar{v}_\rho^2+t^2(-1+\bar{v}_\rho^2)] \\
\frac{F_{82}}{\sqrt{3}}&=(3+t^2)\bar{v}_W^4-2(3+t^2+R)\bar{v}_\rho^2-\bar{v}_W^2[-3-R+6\bar{v}_\rho^2+5t^2(1+\bar{v}_\rho^2)] \\
B_2&=-2t[(3+t^2)\bar{v}_W^4+\bar{v}_W^2(-6+R+t^2-6\bar{v}_\rho^2+4t^2\bar{v}_\rho^2)]+\bar{v}_\rho^2[R+(3+t^2)(1+3\bar{v}_\rho^2)],
\end{split}
\end{equation}
where

\begin{multline}
R=\{t^4(1+\bar{v}_W^2+3\bar{v}_\rho^2)^2+9[1+\bar{v}_W^4+3\bar{v}_\rho^4-\bar{v}_W^2(1+3\bar{v}_\rho^2)]+\\
+6t^2[1+\bar{v}_W^4+3\bar{v}_\rho^2+
6\bar{v}_\rho^4-\bar{v}_W^2(4+3\bar{v}_\rho^2)]\}^{1/2},
\end{multline}
with, still, $\bar{v}_\alpha\equiv \frac{v_\alpha}{v_\chi}$.

With this, the relation between symmetry and mass eigenstates is written

\begin{equation}
\begin{split}
W_3^\mu&=N_1 F_{31} Z_1^\mu + N_2 F_{32} Z_2^\mu + \frac{t}{\sqrt{1+\frac{4}{3}t^2}} A^\mu \\
W_8^\mu&=N_1 F_{81} Z_1^\mu + N_2 F_{82} Z_2^\mu - \frac{t}{\sqrt{3+4t^2}} A^\mu \\
B^\mu&=N_1 B_{1} Z_1^\mu + N_2 B_{2} Z_2^\mu + \frac{1}{\sqrt{1+\frac{4}{3}t^2}} A^\mu,
\end{split}
\end{equation}
where the normalization factors are as expected

\begin{equation}
\begin{split}
N_1=&\frac{1}{\sqrt{F_{31}^2+F_{81}^2+B_{1}^2}} \\
N_2=&\frac{1}{\sqrt{F_{32}^2+F_{82}^2+B_{2}^2}}.
\end{split}
\end{equation}

\subsection{Neutral current couplings of the known fermions to the SM $Z$ within the closing solution}
\label{app:couplingssolrhn}

The SM $Z$ parameters simplified by the solution now are

\begin{align}
g_V^\nu=
&\begin{cases}
\frac{1}{2}, & \text{if} \ v_\chi > \frac{v_W}{\sqrt{2}c_W} \\
0, & \text{otherwise}
\end{cases} \\
g_A^\nu=
&\begin{cases}
\frac{1}{2}, & \text{if} \ v_\chi > \frac{v_W}{\sqrt{2}c_W} \\
0, & \text{otherwise}
\end{cases}
\end{align}

\begin{align}
g_V^\ell=
&\begin{cases}
-\frac{1}{2}+2s_W^2, & \text{if} \ v_\chi > \frac{v_W}{\sqrt{2}c_W} \\
0, & \text{otherwise}
\end{cases} \\
g_A^\ell=
&\begin{cases}
-\frac{1}{2}+2s_W^2, & \text{if} \ v_\chi > \frac{v_W}{\sqrt{2}c_W} \\
0, & \text{otherwise}
\end{cases}\\
g_V^{u_m}=
&\begin{cases}
\frac{1}{6}(3-8 s_W^2), & \text{if} \ v_\chi > \frac{v_W}{\sqrt{2}c_W} \\
0, & \text{otherwise}
\end{cases}\\
g_A^{u_m}=
&\begin{cases}
\frac{1}{2}, & \text{if} \ v_\chi > \frac{v_W}{\sqrt{2}c_W} \\
0, & \text{otherwise}
\end{cases} \\
g_V^{d_m}=
&\begin{cases}
\frac{1}{6}(-3+4s_W^2), & \text{if} \ v_\chi > \frac{v_W}{\sqrt{2}c_W} \\
0, & \text{otherwise}
\end{cases} \\
g_A^{d_m}=
&\begin{cases}
-\frac{1}{2}, & \text{if} \ v_\chi > \frac{v_W}{\sqrt{2}c_W} \\
0, & \text{otherwise}
\end{cases} \\
g_V^{u_3}=
&\begin{cases}
\frac{1}{6}(3-8s_W^2), & \text{if} \ v_\chi > \frac{v_W}{\sqrt{2}c_W} \\
0, & \text{otherwise}
\end{cases} \\
g_A^{u_3}=
&\begin{cases}
\frac{1}{2}, & \text{if} \ v_\chi > \frac{v_W}{\sqrt{2}c_W} \\
0, & \text{otherwise}
\end{cases} \\
g_V^{d_3}=
&\begin{cases}
\frac{1}{6}(-3+4s_W^2), & \text{if} \ v_\chi > \frac{v_W}{\sqrt{2}c_W} \\
0, & \text{otherwise}
\end{cases} \\
g_A^{d_3}=
&\begin{cases}
-\frac{1}{2}, & \text{if} \ v_\chi > \frac{v_W}{\sqrt{2}c_W} \\
0, & \text{otherwise}
\end{cases}
\end{align} 
}
}

\bibliographystyle{JHEP}
\bibliography{biblio}
}
\end{document}